\documentclass{brownthesis}

	\usepackage{booktabs}	%

	\usepackage{graphicx}	%
	\usepackage{amsmath, amsthm, amssymb}
	\usepackage[font=small,labelfont=bf]{caption}
	\usepackage{wrapfig}
	\usepackage[position=b]{subcaption}
	\DeclareCaptionType{listing}[Listing][List of Listings]
	\DeclareCaptionSubType{listing}
	
	\usepackage{multirow}
	
	\usepackage{amsthm}
	\newtheorem{theorem}{Theorem}[section]
	\newtheorem{corollary}[theorem]{Corollary}
	\newtheorem{definition}[theorem]{Definition}
	\newtheorem{lemma}[theorem]{Lemma}
	
	\usepackage{listings}
	\lstdefinelanguage{pseudoscala}{
  	morekeywords={%
          abstract,case,catch,class,contract,def,do,else,extends,%
          false,final,finally,for,forSome,function,if,implicit,import,lazy,%
          match,new,null,object,override,package,private,protected,public,%
          return,sealed,super,this,throw,trait,true,try,type,%
          val,var,while,with,yield},
  otherkeywords={=>,<-,<\%,<:,>:,\#,@,:,=,|},
  sensitive=true,
  morecomment=[l]{//},
  morecomment=[n]{/*}{*/},
  morestring=[b]",
  morestring=[b]',
  morestring=[b]"""
}[keywords,comments,strings]
	\usepackage{program}
	
	\usepackage{proof}
	\usepackage{algorithm}
	\usepackage{algpseudocode}
	\usepackage{MnSymbol}
	\usepackage{harpoon}
	\usepackage{url}

	\algnewcommand{\parState}[1]{\State%
		\parbox[t]{\dimexpr\linewidth-\algorithmicindent-\algorithmicindent}{\hangindent\algorithmicindent\strut #1\strut}}
	
	\usepackage[inline]{enumitem}
	
	\usepackage{hyperref}
	\hypersetup{colorlinks=false,linktoc=all}

\begin{document}
	\title{Adapting Persistent Data Structures for Concurrency and Speculation}
	\author{Thomas D. Dickerson}
	\degrees{B.~S., Saint Michael's College, 2013\\
		Sc.~M., Brown University, 2016}
	\principaladvisor{Maurice Herlihy}
	\reader{Rodrigo Fonseca}
	\reader[(Department of Computer Science, Steven's Institute of Technology)]{Eric Koskinen}
	\dean{Andrew G. Campbell}
	\submitdate{May 2019}
	\copyrightyear{2017,2018,2019}
	\disclaimer{Parts of this thesis have previously been published elsewhere with Paul Gazzillo, Maurice Herlihy, and Eric Koskinen as co-authors.}
	\abstract{
		This work unifies insights from the systems and functional programming communities, in order to enable compositional reasoning about software which is nonetheless efficiently realizable in hardware.
		It exploits a correspondence between design goals for efficient concurrent data structures and efficient immutable persistent data structures, to produce novel implementations of mutable concurrent trees with low contention and an efficient snapshot operation to support speculative execution models.
		It also exploits commutativity to characterize a design space for integrating traditional high-performance concurrent data structures into Software Transactional Memory (STM) runtimes, and extends this technique to yield a novel algorithm for concurrent execution of so-called ``smart contracts'' (specialized programs which manipulate the state of blockchain ledgers).
	}
	\abstractpage
	\beforepreface
	\prefacesection{Acknowledgements}
		\vspace{-0.5cm}
	   I would first like to thank my wife, Courtney, for her love, support, and incredible patience, both in life generally and through my PhD process specifically. 
	   I don't have words to properly express my love and gratitude.
	   
	   I thank my friend and co-founder, Christopher Mitchell, for being the first machine-tribe victim of my $\lambda$-tribe propagandizing, for allowing his sleep schedule to bear the brunt of my final sprint towards completion, and for always keeping my graphics drivers updated.
	   I also thank my many friends and family for believing in me, and helping shape me into the person I am today, but especially my father, Matthew Dickerson, for inspiring and encouraging my love of computation.
	   
	   I would next like to personally thank a number of Brown faculty for their particular impact in shaping the way I think about various aspects of computation: Maurice Herlihy (concurrency and distributed systems), Paul Valiant (algorithms and machine learning), Shriram Krishnamurthi (programming languages and CS pedagogy), Michael Littman (game theory and reinforcement learning), and Daniel Ritchie (creative computing and procedural modeling).
	   I would also like to extend my thanks to the numerous other faculty, collaborators, and committee members who have shared their time, knowledge, and ideas with me during my time at Brown, as well as personal thanks to Lauren Clarke for taking excellent care of the graduate student population of our department.
	   Prior to my time at Brown, I would like to thank the faculty of Saint Michael's College (especially John Trono, Alain Brizard, Greta Pangborn, Jo Ellis-Monaghan, and Jim Hefferon) for helping prepare me to succeed as a scientist, mathematician, and researcher; and before that to the elementary, middle, and high-school teachers who gave me my first tastes of academic freedom (especially Gail Martin, Carol Kress, Tom Tailer, Paul Stetson, and Jim Brown) and the Middlebury College faculty (Amy Briggs, John Schmitt, and Tim Huang) and students who welcomed me into their classrooms.
	   	   
	   Additionally, I would like to thank the various communities that supported my physical, mental, emotional, and spiritual well-being throughout my PhD process: the Brown, RIPUL, and PCUT ultimate frisbee communities; Rock Spot Climbing; the Rhode Island Pok\'{e}mon Go community; Eli and QQ for two years of keeping me fed, letting me co-parent their cats, and making sure I turned the lights on occasionally; in Computer Science, my numerous office-mates over the years, my fellow PhD Recruiting Czars, the board-gamers, tea-drinkers, movie-goers, Taco Tuesday connoisseurs, and TGIFers; and finally Renaissance Church, Christ Church, GMCF, and the RUF and BCF undergrads who let a PhD student gate-crash their events.
	   
	  Finally, I am grateful for the financial support provided by the NSF and by Oracle.

	\afterpreface

	\chapter{\label{chap:intro}Introduction}
	  We know that there are multiple possible formalisms for describing computation, which are equivalent in expressive power.
Perhaps the two most influential are the $\lambda$-calculus and the Turing Machine (and its RAM-endowed successors). \cite{sep-church-turing}
In practice these formalisms yield very different ways of expressing, and reasoning about, computation.
Such differences make it easy to form isolated communities in which one representation or the other is the center of attention:
$\lambda$ folks tend to emphasize the benefits of reasoning about the composition of pure mathematical functions for designing software which is maintainable, whereas  machine folks tend to emphasize the benefits of clever manipulation of the concrete state of physically realized machines for designing software which is efficient.
\paragraph{Thesis Statement:}
Unifying insights from both communities will enable compositional reasoning about (concurrent) software which is nonetheless efficiently realizable in hardware, thereby also making more efficient use of the human developers of that software.

\subsection{Design Goals}
When designing a new concurrent data structure, it's helpful to have some high-level design goals in mind, so that when we're done, we can ask whether we were successful.
Three goals that this work treats as particularly important are:
\paragraph{1. Easy to Implement}
First, making sure that programmers who need to implement it down the road for a new language or runtime can actually understand what's going on --- a clever algorithm isn't particularly useful if nobody can tell whether an implementation is working correctly (or what's going wrong if it isn't).
\paragraph{2. Efficient Snapshots}
Second, making sure that it will be useful for more than a single niche application.
Many concurrent data structures have relatively limited operations compared to a sequential implementation of the same abstract data-type, particularly when it comes to reading or writing multiple elements, so supporting consistent iteration is important.
Similarly, it should make it as easy as possible for a thread to perform speculative operations and discard the results without corrupting the state of other threads, to support transactional APIs.
In fact, iteration and speculation really reduce to the same problem: being able to efficiently take a snapshot.
\paragraph{3. Scalable Performance}
In the face of concurrency, we ought to ensure that our data structure performs well under contention.

\section{Overview}
It is generally accepted that the design of concurrent software is a difficult problem, which can nevertheless be simplified somewhat by building libraries of reusable concurrent data structures.
Still, the task of designing even one (correct) concurrent data structure remains a difficult one, more so if it needs to support bulk operations while remaining linearizable.
We propose that modifying existing data structures from the purely functional literature will simplify the task further.

In Chapter \ref{chap:handoverhand} we consider the Braun Heap \cite{nipkow2016priority,okasaki1997three} as an exemplar of how existing persistent \cite{driscoll1989making} data structures can be rewritten, with minimally invasive changes, as mutable concurrent data structures with an efficient snapshot operation, using hand-over-hand locking.
Such snapshots enable support for both linearizable read-only iteration, and support speculative modes of execution with low memory overhead.

In Chapter \ref{chap:lockfree} we generalize the techniques developed in Chapter \ref{chap:handoverhand} to the lock-free setting, using the LLX/SCX synchronization primitives of Brown et al.~\cite{BrownLLXSCX}, continuing to use the Braun Heap as an example.
We describe an algorithm that supports efficient lock-free concurrent execution of a broad class of operations on trees, and develop a Scala library to automate implementations with only type-level rewriting of the pure-functional descriptions.
We also show how the use of ``helping'' common in lock-free settings makes it difficult for the original thread to retrieve results from a compound operation, and the restrictions that this imposes on the expressivity of our library. %
Finally, we discuss the challenges for implementing lock-free concurrent trees with structural sharing in a programming environment without automatic garbage collection, and suggest avenues for future work in that vein.

In Chapter \ref{chap:returns} we describe a modified commit protocol for the SCX primitive that allows the original thread to follow the execution of a compound operation in the face of ``helping'', and modify our library to support a broader class of operations.
We use this increased expressivity to implement a lock-free concurrent Hash Array Mapped Trie (HAMT).~\cite{bagwell2001ideal}
We show that the remaining class of tree operations which are unsupported by our library correspond to a class of concurrency bottlenecks under our execution strategy.
Finally, we discuss avenues for future work that could leverage commutativity specifications to improve concurrency in these cases.

In Chapter \ref{chap:proust} we show how commutativity specifications can be used to integrate traditional high-performance concurrent data structures with Software Transactional Memory (STM) runtimes.
We characterize a design space for such transactional wrappers in terms of \emph{lazy} versus \emph{eager} update strategies, and \emph{pessimistic} versus \emph{optimistic} synchronization primitives, and show that Transactional Boosting~\cite{HerlihyK2008} corresponds to the Eager/Pessimistic quadrant of our design space.
We further characterize the design space of wrappers in terms of which quadrants satisfy opacity against a taxonomy of STM runtime design decisions. 
We find that data structures supporting efficient snapshots, such as those discussed in Chapters \ref{chap:handoverhand}-\ref{chap:returns}, are well suited for implementing Lazy/Optimistic wrappers, which are needed to satisfy opacity under the strictest assumptions about STM runtime behavior.
Finally, we develop a Scala library implementation and benchmark several example wrappers.

In Chapter \ref{chap:smartcontracts} we show the same techniques used to implement commutativity-based transactional wrappers can be used to discover safe fork-join schedules from a job pool, and discuss the application to blockchains and smart contracts.
We emulate a blockchain network using Scala STM, and show that our algorithm performs well on common workloads for both miners and validators.

In Chapter \ref{chap:conclusion} we conclude with some brief remarks.

In the remainder of \emph{this} chapter, we discuss the conventions used throughout.

\section{Conventions}
Unless otherwise specified, we assume that all code is being executed on a modern Symmetric Multiprocessing (SMP) architecture, supporting at least a single-word Compare-and-Swap (CAS) instruction for synchronization.

\subsection{Pseudocode}
Pseudocode will be written in a simplified syntax inspired by Scala and Pyret.
This should be \textit{mostly} familiar to readers with experience in Java, C++, or similar languages, with a few key differences.

Parametric (generic) types will be written with square braces rather than angle-brackets: the type of a list containing integers will be typeset as \texttt{List[Int]}.
Types will be inferred where possible, and annotated (where necessary) in a post-fix notation.
Thus a variable declaration for an integer would look like \texttt{var a = 0} (or \texttt{var a:Int = 0} for clarity), rather than \texttt{int a = 0}.
Since most code will be in a functional, rather than imperative, style, most identifiers will actually be bound as constants, denoted as \texttt{val a:Int}.

Compound types will be written algebraically, rather than in an object-oriented style.
These are contrasted in Listing \ref{lst:typedefs}.
\begin{listing*}
\begin{sublisting}[lb]{0.45\textwidth}
\begin{lstlisting}[mathescape=true]
// Compact Data Definitions
type List[T] = Link(head:T, tail:List[T])
             | Nil

// Single definition of each method
def isEmpty[T](list:List[T])$\to$Boolean {
  list match {
   | case Link(head, tail) => false
   | case Nil => true
  }
}

// Will use same syntax for type-aliases
type List2D[T] = List[List[T]]
\end{lstlisting}
\caption{Example Algebraic Data Type definition.}
\label{sublst:adt}
\end{sublisting}
\hspace{0.05\textwidth}
\begin{sublisting}[rb]{0.45\textwidth}
\begin{lstlisting}[mathescape=true]
abstract class Node[T] {
  def isEmpty()$\to$Boolean
  def head()$\to$T
  def tail()$\to$Node[T]
}

class Link[T](private val head_:T,
               private val tail_:Node[T]) extends Node[T] {
  override def isEmpty()$\to$Boolean { false }
  override def head()$\to$T { head }
  override def tail()$\to$Node[T] { tail }
}

object Nil extends Node[Nothing] {
  override def isEmpty()$\to$Boolean { true }
  override def head()$\to$T { throw UnsupportedOperation }
  override def tail()$\to$Node[T] { throw UnsupportedOperation }
}
\end{lstlisting}
\caption{Example Object-Oriented Type definitions.}
\label{sublst:oop}
\end{sublisting}
\caption{Comparing algebraic compound type definitions (left) with a Java-style class-centric style (right).}
\label{lst:typedefs}
\end{listing*}
We also note here another key difference from Java or C++-style code: there are no explicit \texttt{return} statements, instead each compound statement implicitly returns the value of its last expression.

Finally, since we will often have functions as first-class values, we denote the type of the $n$-ary function from types $A_1 \cdots A_n$ to type $R$ as $\mathrm{(A_1, \ldots, A_n) \to R}$.

	\chapter{\label{chap:handoverhand}Concurrent Braun Heaps}
	   		\section{Introduction}
		Concurrent collection data structures are an important tool in modern software design, but they often provide a limited set of operations compared to their serial counterparts, due to the complexity of concurrent reasoning and implementation.
		The priority queue is a widely used data structure, that allows a user to efficiently query and remove its smallest element.
		These operations impose a weaker ordering requirement than other collections (e.g. an ordered map), and it is thus potentially more efficient to implement those operations.
		Serial priority queues are typically implemented as either a heap or a skiplist, and heaps have stronger guarantees of efficiency.
		Thus far, tree data structures have proven more difficult to efficiently parallelize, and so most concurrent priority queues are based on some form of skiplist.~\cite{shavit2000skiplist}
		In specialized circumstances, relaxing the priority queue constraint such that element removed is only approximately the smallest may also yield efficient constructions.~\cite{DBLP:conf/spaa/Alistarh0KLN18}
		
	\subsection{Motivation}
	The motivation for this chapter is two-fold.
	First, many concurrent algorithms are complex, and thus both difficult to implement from scratch, for new environments, and to reason about for purposes of ensuring correctness.
	In contrast, the data-structure presented in this chapter augments a well-studied serial data structure with a simple hand-over-hand locking protocol.
	At any point during its execution, its state can be verified as a function of purely local invariants.
	
	Second, optimistic or speculative execution of code is an important technique in concurrency, but aborted operations must be able to undo their effects.
	The authors of Transactional Boosting show that the \texttt{insert} and \texttt{removeMin} operations provided by a heap are not easily inverted.~\cite{HerlihyK2008}
	An alternate strategy for speculative executions involving a heap is to operate on local snapshot which can be modified and discarded along with the parent operation.
	Such functionality is a key component in wrapping concurrent data structures with a transactional API (Chapter \ref{chap:proust}) or implementing other kinds of reversible atomic objects.~\cite{antonopoulostheory}
	No extant heap or priority queue implementations appear to support this functionality.
	
	Even when speculative executions are unnecessary, snapshots can also be used to provide read-only iterator semantics.
		
	\subsection{Contributions}
	In this chapter, we define a new concurrent heap algorithm with a shape property\footnote{A \emph{shape property} is a constraint on the layout of nodes in a tree, usually to enforce a bound on the depth of the tree.} that is well-suited to concurrent updates,
	and augments it with a fast snapshot operation based on a copy-on-write methodology.
	
	The contributions can be summarized as follows:
	\begin{enumerate}
	\item Section \ref{sec:keyinsights} provides intuition as to why the Braun heap is well suited for both concurrency and efficient snapshots.~\cite{braun1983logarithmic}
	\item Section \ref{sec:implementation} describes the transformation from the purely functional Braun heap to a concurrent Braun heap with fast atomic snapshot.
	\item Section \ref{sec:correctness} discusses the correctness of this transformation.
	\item Section \ref{sec:perf} provides benchmarking results showing the scalability of both standard heap operations as well as atomic snapshot operations as compared to existing implementations.
	\end{enumerate}
	
	\subsection{Prior Work}
	\begin{figure*}
		\centering
		\begin{subfigure}[t]{0.3\textwidth}
			\includegraphics[width=\textwidth]{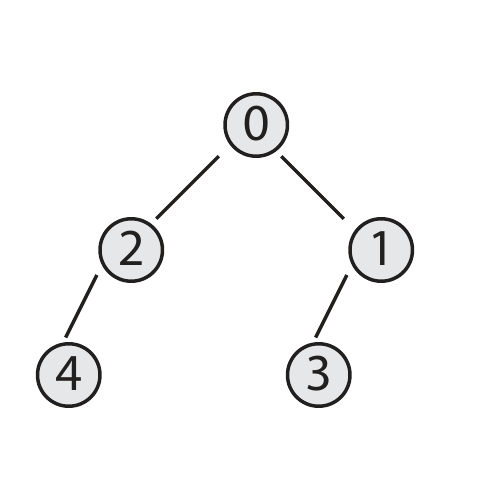}
			\caption{The initial state.}
			\label{fig:cbheapinit}
		\end{subfigure}
		\begin{subfigure}[t]{0.3\textwidth}
			\includegraphics[width=\textwidth]{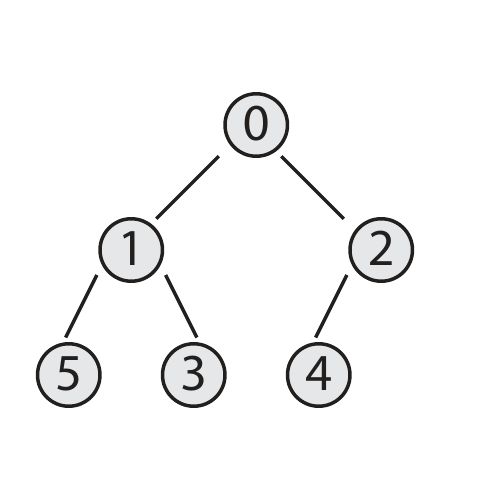}
			\caption{After \texttt{insert(5)}.}
			\label{fig:cbheapinsert}
		\end{subfigure}
		\begin{subfigure}[t]{0.3\textwidth}
			\includegraphics[width=\textwidth]{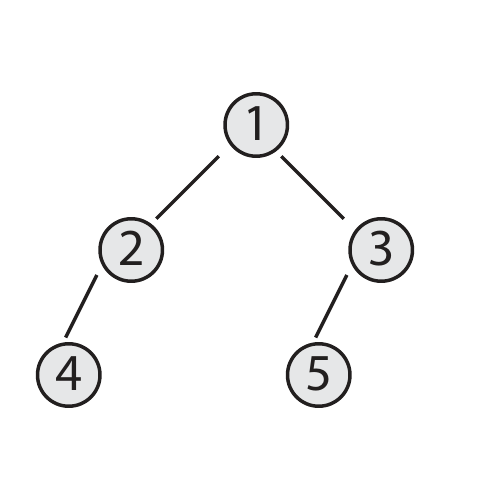}
			\caption{After \texttt{removeMin}.}
			\label{fig:cbheapdelete}
		\end{subfigure}
		\caption{A sequence of operations on a Braun heap, initialized by inserting the integers from $0$ to $4$, in order.}
	\end{figure*}
	This chapter builds on several prior lines of work.
	
	Efficient tree data structures rely on a shape (or balance) property to ensure that the depth of the tree grows only logarithmically in the number of elements.
	The Braun property states that for every node, $n$, in a tree, the size of its left and right subtrees, $l$ and $r$, are constrained such that $|r| \le |l| \le (|r|+1)$.
	It was first proposed for efficient implementations of immutable trees in a pure-functional setting \cite{braun1983logarithmic,hoogerwoord1993logarithmic}, and while relatively unknown in the broader community\footnote{Unlike many algorithms and data structures, there is as of yet no Wikipedia article on the subject, which seems a reasonable measure of obscurity}, Braun trees have long been known to be useful in implementing heaps.~\cite{nipkow2016priority,okasaki1997three}
	
	An example of operations on a Braun heap is provided in Figures \ref{fig:cbheapinit}, \ref{fig:cbheapinsert}, and \ref{fig:cbheapdelete}.
	When inserting a new element (Figure \ref{fig:cbheapinit} to Figure \ref{fig:cbheapinsert}), the recursive steps swap the left and right subtrees and descend into the new left subtree, thus preserving the Braun property.
	When remove an existing element (Figure \ref{fig:cbheapinsert} to Figure \ref{fig:cbheapdelete}), the recursive steps swap the right and left subtrees and descend into the new right subtree, mirroring insertions and preserving the Braun property.
	
	Collections data structures are of limited utility if their API only allows you to inspect or alter a single element at a time, yet bulk read operations are challenging in a concurrent context.
	Correctness of concurrent operations is often defined in terms of linearizability \cite{HerlihyW1990}, which ensures that all operations appear to happen instantaneously; however, a bulk read operation necessarily must capture the state of the entire data structure.
	Several recent works have leveraged a copy-on-write methodology to provide efficient linearizable snapshots for a number of data structures, including AVL trees \cite{BronsonPCBST}, queues \cite{ProkopecSnapQueue}, and hash tries \cite{ProkopecCTrie}. 
	Common techniques for snapshots involve either marking nodes as immutable \cite{ProkopecSnapQueue} or assigning nodes an explicit version (or generation) number \cite{BronsonPCBST, ProkopecCTrie}.
	Linearizable iterators are also an area of inquiry, recent work has implemented them for skip lists \cite{petrank2013lock} and other ordered key-value stores \cite{Basin:2017:KKM:3018743.3018761}, which are of interest for implementing priority queues.
	Recent work by Arbel-Raviv and Brown uses similar techniques, alongside Epoch-based Reclamation, to implement efficient range queries.~\cite{DBLP:conf/ppopp/Arbel-Raviv018}
	
	Additionally, there are a number of other concurrent skiplist variants, which might be used as the basis for a priority queue.~\cite{herlihy2007simple,pugh1990concurrent,dleaCSLM}
	
	\section{Overview}
	\subsection{\label{sec:keyinsights}Key Insights}
	It is unsurprising that importing insights from the purely-functional data structures community would be of benefit in designing new concurrent data structures, particularly those requiring a fast-snapshot operation.
	Without mutation, \textbf{every} reference to a data structure is intrinsically a snapshot (this property is known as persistence \cite{okasaki1996purely}), because every update operation returns a new data structure (potentially sharing references to sub-structures in the old one).
	Since allocation of new objects is a relatively expensive operation, it is beneficial to design data structures that do not require extensive modifications to maintain bookkeeping state or enforce algorithmic invariants.
	Moreover, every point where an algorithm on purely functional data structures would allocate a new object, a concurrent algorithm implementing the same update as a mutation will require synchronization, potentially resulting in a bottleneck.
	
	Furthermore, functional programming separates reasoning about the result of a computation from reasoning about its execution, allowing a program to be written correctly even if the result of a computation is delayed\footnote{In a single threaded setting, this is referred to as laziness. In a concurrent setting it is often referred to by the names ``future'' or ``promise'' \cite{friedman1976impact,baker1977incremental}.}.
	
	The Braun property in particular can be maintained by following two simple rules, without any bookkeeping.
	First, every insertion or deletion must swap the left and right subtrees.
	Second, insertions traverse towards the (new) left subtree, and deletions proceed towards the (new) right subtree.
	As an added benefit for concurrency, for every consecutive pair of updates, $u,v$, to a subtree rooted at a node $n$, the traversal paths of $u$ and $v$ will intersect only at $n$.
	Thus, the number of threads that can be working simultaneously doubles at each level of the tree.
	On the other hand, in a traditional binary heap, utilizing the fullness property, every insertion and deletion is in contention for the boundary position between filled and unfilled leaves, and, with high probability, every consecutive pair of updates will intersect for the majority of their path towards the root\footnote{Intuitively, they will intersect after 1 step with probability $\frac{1}{2}$, after 2 steps with probability $\frac{3}{4}$, etc.}.
	
	\subsection{\label{sec:implementation}Implementation}
	\begin{listing}
	\begin{lstlisting}[mathescape=true]
type HeapNode[E] = BNode(elem:E, left:HeapNode[E], right:HeapNode[E])
                   | Nil
                   
type BraunHeap[E] = Holder(root:HeapNode[E])
	\end{lstlisting}	
	\caption{Definition of the BraunHeap type (and its internal HeapNode).}
	\label{lst:bheaptypes}
	\end{listing}
	The structure for a sequential implementation of a Braun heap is provided in Listing \ref{lst:bheaptypes}, consisting of two classes: \texttt{HeapNode}s, containing a value and two child references, and the \texttt{BraunHeap} itself, containing a single reference to the root.
	\begin{listing}
	\begin{lstlisting}[mathescape=true]
def BNode[E]::update(elemN:E, leftN:HeapNode[E], rightN:HeapNode[E])$\to$BNode[E] {
  elem = elemN
  left = leftN
  right = rightN
  this
}
	
def HeapNode[E]::insert(elemN:E)$\to$HeapNode[E] {
  this match {
    | case Nil => HeapNode(elemN, Nil, Nil)
    | case BNode(elem, left, right) =>
      val (smaller, larger) = if (elemN < elem) { (elemN, elem) }
                               else { (elem, elemN) }
      
      update(smaller, right.insert(larger), left) // Swap and insert on new left
  }
}

def BNode[E]::pullUpLeft(isRoot:Boolean=false)$\to$(HeapNode[E],Maybe[E]) {
  left match {
    | case Nil => 
      (Nil, if(isRoot) { None } 
            else { Some(value) }))
    | case BNode(_, _, _) => 
       val (rightN, retVal) = left.pullUpLeft()
       
       (update(value, right, rightN), retVal) // Swap and pull up on new right.
  }
}

def BNode[E]::pushDown()$\to$BNode[E] {
  (left, right) match {
    | case (Nil, Nil) => this
    | case (BNode(elemL, leftL, rightL), Nil) =>
      if(elemL < elem) { update(elemL, left.update(elem, leftL, rightL), right) }
      else { this }
    | case (BNode(elemL, leftL, rightL), BNode(elemR, leftR, rightR)) =>
      if(elem <= elemL && elem <= elemR) { this }
      else {
        if(elemL <= elemR) {
          update(elemL, left.update(elem, leftL, rightL).pushDown(), right)
        } else {
          update(elemR, left, right.update(elem, leftR, rightR).pushDown())
        }
      } 
    | case _ => throw IllegalState("This tree is not Braun")
  } 
}			
	\end{lstlisting}
	\caption{Pseudo-code for the methods necessary to implement the HeapNode sequentially.}
	\label{lst:seqbheaphelpers}
	\end{listing}
	Listing \ref{lst:seqbheaphelpers} shows the \texttt{HeapNode} methods which implement the recursive logic to enforce both the Braun property and the heap property.
	We note that the \texttt{update} method defined on Lines 1 - 6 implements a transient version of the data structure, but that replacing it with a fresh allocation (instead of mutation) would be sufficient to convert all of the other helper methods into persistent versions.
	\begin{listing}
	\begin{lstlisting}[mathescape=true]
def Holder[E]::update(rootN:HeapNode) {
  root = rootN
  this
}

def Holder[E]::getMin()$\to$Maybe[E] {
  root match {
    | case Nil => None
    | case BNode(elem, _, _) => Some(elem)
  }
}

def Holder[E]::insert(v:E)$\to$Holder[E] {
  update(root.insert(v))
}
	
def Holder[E]::removeMin()$\to$(Holder[E],Maybe[E]) {
  root match {
    | case Nil => (this, None)
    | case BNode(elem, _, _) =>
      (root.pullUpLeft(true) match {
        | case (Nil, None) => this.update(Nil)
        | case (BNode(_, leftN, rightN), Some(elemN)) =>
          this.update(BNode(elemN, leftN, rightN).pushDown())
      }, Some(elem))
  }
}
	\end{lstlisting}
	\caption{Pseudo-code for the methods necessary to implement the BraunHeap sequentially.}
	\label{lst:seqbheap}
	\end{listing}
	Listing \ref{lst:seqbheap} shows the \texttt{BraunHeap} methods which serve as the public API for the data structure.
	As before, the data structure is transient, but we channel all of our mutations through the \texttt{update} method on Lines 1 - 4, and replacing its implementation would be sufficient to make all the other methods persistent.
	
	Our concurrent version preserves the essence of this implementation, while providing thread-safety and linearizable snapshots.
	With one exception, the effects of mutating operations only propagate leafward through the tree, so thread safety is achieved simply by following a hand-over-hand locking discipline (for readability, we do not provide any listings showing how to apply hand-over-hand locking, as this is both verbose and mechanical).
	In the case of \texttt{removeMin}, the mutations are separated into two phases, the first to maintain the Braun property (\texttt{pullUpLeft}, Lines 19 - 29 in Listing \ref{lst:seqbheaphelpers}), the second (\texttt{pushDown}, Lines 31 - 48 in Listing \ref{lst:seqbheaphelpers}) to maintain the heap property.
	In the first phase, nodes must be able to reach upwards to modify references to themselves held by their immediate parents.
	To ensure thread safety is not violated by this breach of hand-over-hand discipline, the first phase begins by acquiring reentrantly acquiring the mutex on the root twice: once for the first phase traversal, and once for the second phase traversal.
	This prevents new threads from arriving and observing nodes while the heap property is violated between the first and second phases.
	An example execution tracing concurrent calls to \texttt{insert} and \texttt{removeMin} is illustrated across Figures \ref{fig:proof-example-i-rm} and \ref{fig:proof-example-rm}.
	
	\begin{figure*}
		\centering
		\includegraphics[width=.5\textwidth]{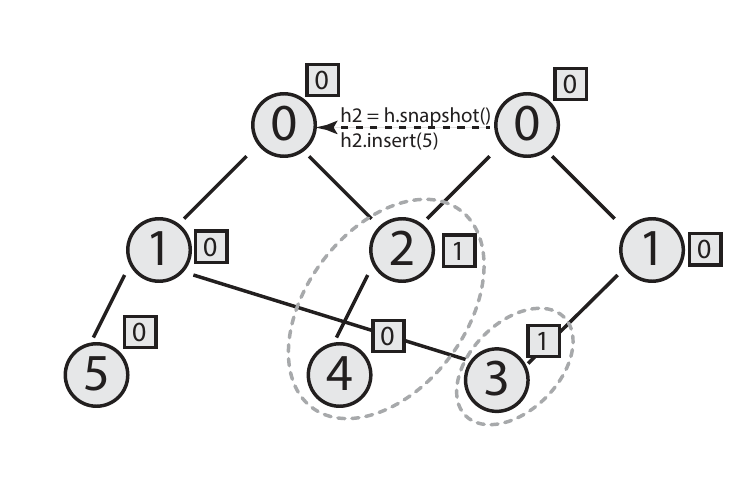}
		\caption{Structural sharing for the heaps from Figs \ref{fig:cbheapinit} and \ref{fig:cbheapinsert}, if the call to \texttt{insert(5)} is preceded by a call to \texttt{snapshot}. Each node is now annotated with its \texttt{snapCount} property, and the shared subtrees are surrounded by a dashed oval.}
		\label{fig:cbheapshare}
	\end{figure*}
	\begin{listing}
	\begin{lstlisting}[mathescape=true]
def HeapNode[E]::snapshot()$\to$BNode[E] {
  snapCount += 1
  this
}

def BNode[E]::unsnap()$\to$BNode[E] {
  if(snapCount > 0){
    val e = elem
    val l = left.snapshot()
    val r = right.snapshot()
    snapCount -= 1
    
    BNode(v, l, r)
  } else {
    this
  }
}

def BNode[E]::update(elemN:E, leftN:HeapNode[E], rightN:HeapNode[E])$\to$BNode[E] {
  // The .{ ... } notation here is to suggest an anonymous member method.
  // We could make an ``updateHelper'', but the correspondence to earlier
  // listings is clearer this way.
  this.unsnap().{
    elem = elemN; left = leftN; right = rightN; 
    this
  }
}
	\end{lstlisting}
	\caption{Pseudo-code for the methods necessary to implement the Copy-on-Write HeapNode.}
	\label{lst:snapbheaphelpers}
	\end{listing}
	The fast snapshot functionality is implemented as a reference-counting lazy copy-on-write (COW) scheme.
	Each node is augmented with a \texttt{snapCount} property, which tracks the number of snapshot references to that node (it is initialized to 0, since the copy in the ``original'' heap is not considered a snapshot).
	When a snapshot of a heap is requested, a new heap (recall that the heap class serves only to hold a reference to the root) is allocated with a reference to the same root node, and the \texttt{snapCount} of that node is incremented.
	Mutation of any node with \texttt{snapCount > 0} is forbidden, as this would lead to inconsistent state in the other heaps that reference it.
	Since we already used the \texttt{update(v,l,r)} method of Listing \ref{lst:seqbheaphelpers} to perform any mutation of node state, these changes are easily encapsulated.
	Listing \ref{lst:snapbheaphelpers} shows the new \texttt{update(v,l,r)} implementation, which cooperates with \texttt{snapshot()}.
	Instead, a new node is allocated with the updated values, the \texttt{snapCount}s of its left and right children are incremented (i.e. we store a snapshot reference to each child), and the \texttt{snapCount} of the original node, for which mutation was requested, is decremented.
	This has the effect of lazily ``peeling'' the snapshot away from the original (i.e. only as-needed).
	As a result of this lazy snapshot process, two different heaps (diverging from a common ancestor) may be subject to \emph{structural sharing} (just like the previously discussed purely-functional persistent data structures).
	This phenomenon is demonstrated in Figure \ref{fig:cbheapshare}.
	Note that the two heaps represent logically distinct trees, despite their shared state.
	
	In a language with deterministic destructors, it would be correct for a heap to decrement the \texttt{snapCount} of any snapshotted nodes when it exits scope; however, on the JVM \texttt{finalize} is often executed much later, and any unnecessary allocations resulting from an imprecise \texttt{snapCount} are likely to have already occurred.
	Thus, this implementation does not implement a \texttt{finalize} hook for that traversal.
	
	\section{\label{sec:correctness}Correctness}
	\begin{figure*}
		\centering
		\includegraphics[width=.22\textwidth]{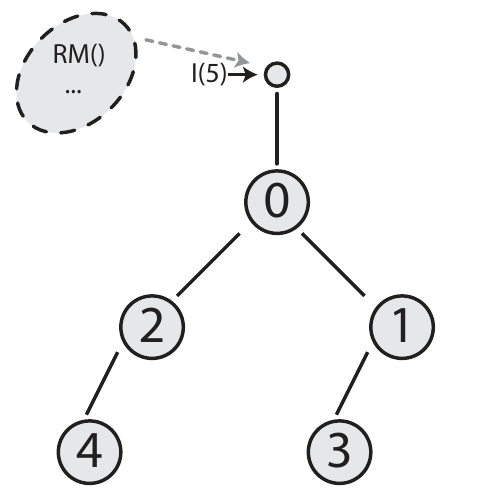}
		\includegraphics[width=.22\textwidth]{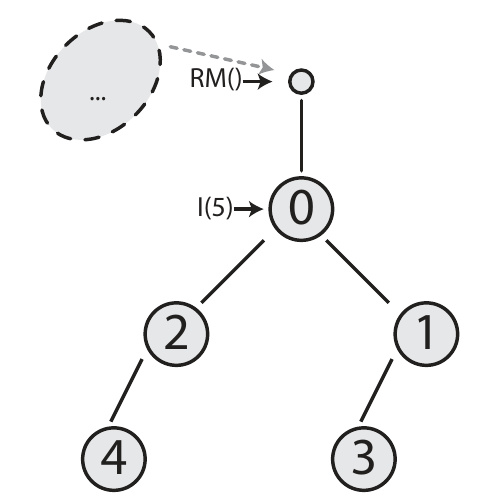}
		\includegraphics[width=.22\textwidth]{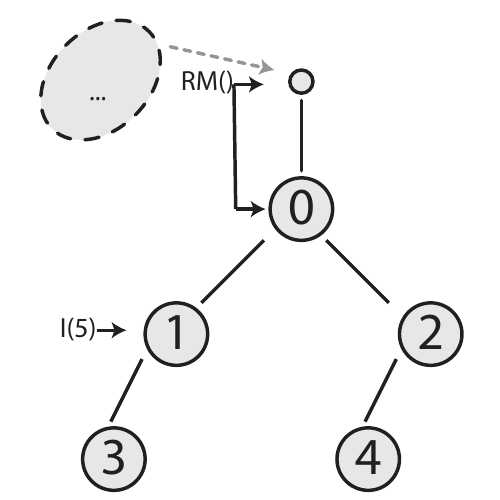}
		\includegraphics[width=.22\textwidth]{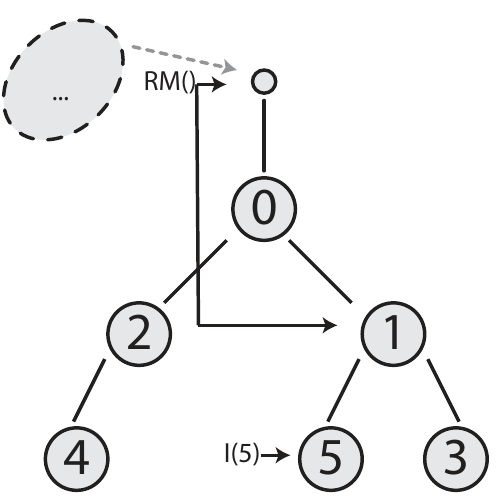}
		\caption{The heap from Fig. \ref{fig:cbheapinit} is updated with the \texttt{insert} and \texttt{removeMin} operations from Figs. \ref{fig:cbheapinsert} and \ref{fig:cbheapdelete}, now executing concurrently. The solid arrows represent locks held by a particular execution, and the dashed oval represents the set of pending operations which have not yet acquired any locks on the heap. \texttt{removeMin} must wait until the preceding \texttt{insert} has released each lock before proceeding. Note that during its \texttt{pullUpLeft} phase, \texttt{removeMin} retains a lock on the root.}
		\label{fig:proof-example-i-rm}
	\end{figure*}
	\begin{figure*}
		\centering
		\includegraphics[width=.22\textwidth]{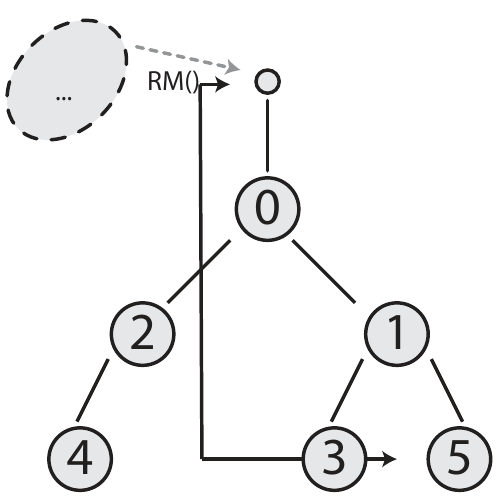}
		\includegraphics[width=.22\textwidth]{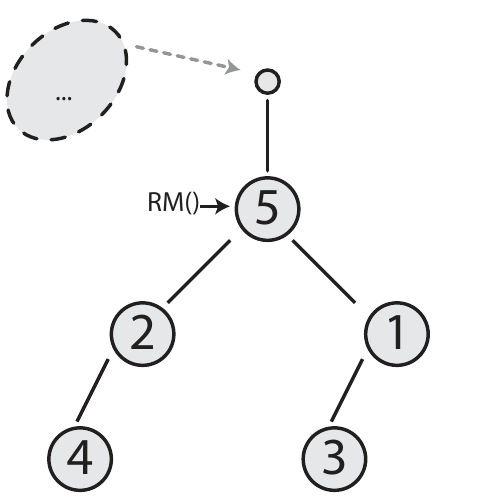}
		\includegraphics[width=.22\textwidth]{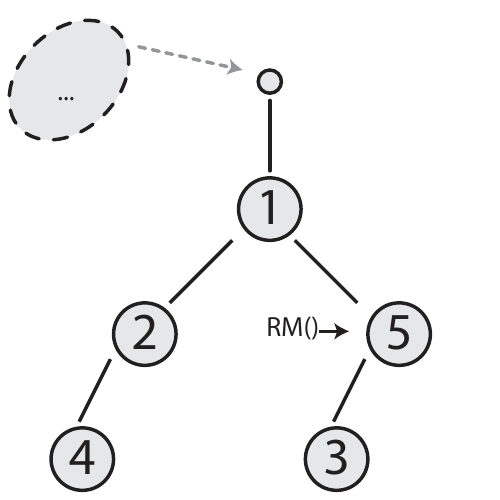}
		\includegraphics[width=.22\textwidth]{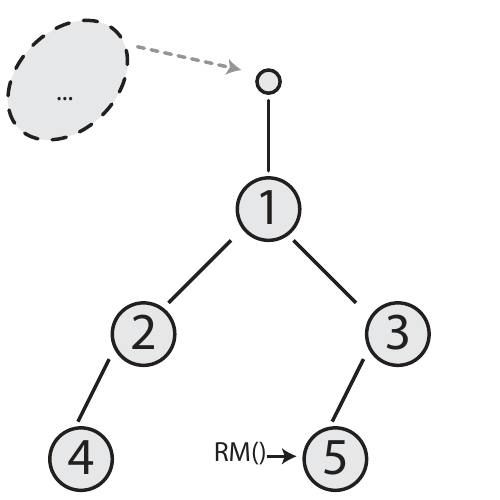}
		\caption{A continuation of the example execution in Fig. \ref{fig:proof-example-i-rm}, as \texttt{removeMin} executes its \texttt{pushDown} phase.}
		\label{fig:proof-example-rm}
	\end{figure*}
	
	The correct execution of operations on the concurrent Braun Heap is subject to three invariants.
	The Braun property has already been defined, but we restate it here for convenience, in Definition \ref{def:braun}.
	\begin{definition}[\label{def:braun}Braun Tree]
	A binary tree, $T$, is Braun iff for every node, $n \in T$,  the size of its left and right subtrees, $\mathsf{n.l}$ and $\mathsf{n.r}$, are constrained such that $\|\mathsf{n.r}\| \le \|\mathsf{n.l}\| \le (\|\mathsf{n.r}\|+1)$.
	\end{definition}
	The heap property is well known, but we use a non-standard formulation (Definition \ref{def:heap}) conducive to describing the behavior of threads traversing a tree.
	\begin{definition}[\label{def:heap}Heap]
	A tree, $T$, is a heap iff the values held by the nodes in every rootward path are nonincreasing.
	\end{definition}
	
	As previously discussed, taking a snapshot of a heap will result in two semantically independent heaps which nevertheless share their internal structure.
	As one, or both, heaps are updated over time their degree of structural sharing will decrease.
	Furthermore, as long as the two heaps share any structure, creating a snapshot of one will cause the fresh snapshot also sharing structure with the other.
	\begin{definition}[\label{def:structshare}Structural Sharing]
	Two heaps, $H_1$ consisting of nodes $N = \{n_1, \ldots, n_{\|H_1\|}\}$, and $H_2$ consisting of nodes $M = \{m_1, \ldots, m_{\|H_2\|}\}$, exhibit structural sharing iff $N \cap M \ne \emptyset$.
	\end{definition}
	
	Snapshot isolation is well studied in the database literature, and says that, intuitively, transactions operate on a consistent snapshot of the database and can not observe the effects of updates made after the snapshot was taken.~\cite{berenson1995critique,daudjee2006lazy} 
	Unlike for databases, our application is concerned with the evolution of data structures subject to linearizable operations, and not subjected to transactional semantics.
	Nevertheless, we appropriate the terminology and provide a closely related\footnote{An optimistic STM wrapper built using the techniques in Chapter \ref{chap:proust} would provide snapshot isolation to any transactions accessing the heap.} Definition \ref{def:snap}.
	\begin{definition}[\label{def:snap}Snapshot Isolation]
	Two heaps, $H_1$ consisting of nodes $N = \{n_1, \ldots, n_{\|H_1\|}\}$, and $H_2$ consisting of nodes $M = \{m_1, \ldots, m_{\|H_2\|}\}$, subject to structural sharing, are snapshot isolated, iff, without loss of generality, no modification of $H_1$ is reflected in the multiset of values held by $H_2$.
	
	In other words, a node $t$ must not be modified so long as $t \in N \cap M$.
	\end{definition}
	\begin{corollary}[\label{coro:snap}Snapshot Isolation]
	An operation, $f$, on a heap $H$ preserves snapshot isolation, iff, for any other heap, $H'$, sharing structure with $H$, executing $f(H)$ does not modify $H'$.
	\end{corollary}
	
	If every operation provided by the heap individually preserves the Braun, Heap, and Snapshot Isolation properties, as defined above, and appears to take place at an atomic linearization point, then the algorithm is correct.
	We will explicitly consider the \texttt{insert}, \texttt{removeMin}, \texttt{getMin}, and \texttt{snapshot} operations, as well as the \texttt{pullUpLeft} and \texttt{pushDown} helper methods that \texttt{removeMin} is built on.
	We assume that the sequential implementations shown in Listings \ref{lst:seqbheaphelpers} and \ref{lst:seqbheap} have been augmented to follow hand-over-hand locking, and use the \texttt{snapshot}-cooperating implementation of \texttt{update} given in Listing \ref{lst:snapbheaphelpers}.
	For each of these operations, will present a proof sketch.

	To develop a stronger intuition for the interplay between these operations, it may be convenient to refer to the example execution in Figures \ref{fig:proof-example-i-rm} and \ref{fig:proof-example-rm}.
	
	For purposes of reasoning about atomicity through a hand-over-hand traversal, we will consider the heap's lock mediating access to the root to be the base case, rather than the locks owned by the root itself.
	
	\begin{theorem}[Correctness of getMin] 
	\texttt{getMin} is linearizable, and preserves the Braun, Heap, and Snapshot Isolation properties.
	\end{theorem}
	\begin{proof}
	 \texttt{getMin} is a read-only operation, so Braun, Heap, and Snapshot Isolation are preserved. 
	 \texttt{getMin} acquires a read-lock on the heap, and subsequently on its root node, thus is linearized at the moment when the read-lock is acquired on the heap, as no conflicting operations can occur once that lock is acquired. \qed
	\end{proof}
	
	\begin{theorem}[Correctness of snapshot] 
	\texttt{snapshot} is linearizable, and preserves the Braun, Heap, and Snapshot Isolation properties.
	\end{theorem}
	\begin{proof}
	When a thread $t$ executes \texttt{snapshot} on a heap, $h$, rooted at node $n$, it increments the \texttt{snapCount} of $n$, and allocates a new heap, $h'$, with $h'.n = n$.
	
	 \texttt{snapshot} does not alter either a node's stored value, or its children, so Braun, Heap, and Snapshot Isolation are preserved.
	 \texttt{snapshot} acquires a write-lock on the heap, and subsequently on its root node, thus it linearizes at the moment when the write-lock is acquired on the heap, as no conflicting operations can occur once that lock is acquired. \qed
	\end{proof}
	
	\begin{theorem}[Correctness of insert] 
	\texttt{insert} is linearizable, and preserves the Braun, Heap, and Snapshot Isolation properties.
	\end{theorem}
	\begin{proof}
	 \texttt{insert} performs hand-over-hand locking along its leafward path.
	 This is initialized by acquiring a write-lock on the heap.
	 At step $i$, no previous conflicting operation can intrude, as the lock for node $n_i$ on the path is held, and no conflicting operation can be observed, as the lock for $n_{i+1}$ will be acquired before proceeding.
	 Note that this holds even when $n_i$ is structurally shared as the result of a snapshot operation.
	 If $n_i \in h$ and $n_i \in h'$, then competing threads performing inserts on $h$ and $h'$ must still acquire the same lock on $n_i$.
	 Since at each step, no previous conflicting operations can intrude and no conflicting observations can be observed, inductively, the entire traversal linearizes at the moment when the write-lock is acquired on the heap.
	 
	 When a thread executing \texttt{insert(v)} arrives at a subtree\footnote{
	 Recall that insertions on a Braun tree always swap left and right subtrees, then recurse into the \emph{new} left subtree.}, $t = \{t.v,t.l,t.r\}$, an \texttt{update} is performed such that, $$
	 t' = \{\mathtt{min}(t.v,v),t.r.\mathtt{insert}(\mathtt{max}(t.v,v)) ,t.l\}
	 $$and $t'$ is returned.
	 
	 If $t$ is a snapshot (i.e $t.\mathtt{snapCount} > 0$), then $t'$ is the result of allocating a new node (with snapshots of its children), otherwise $t$ is mutated directly, thus Snapshot Isolation is preserved.
	  
	 Assume that when the thread arrives at $t$, then $t$ is Braun.
	 Recall that Braun implies $|t.r| \le |t.l| \le |t.r|+1$.
	 Thus after \texttt{insert}, we have that $|t'.l| = |t.r|+1$ and $|t'.r| = |t.l|$.
	 Thus $|t'.r| \le |t'.l| \le |t'.r|+1$ and Braun holds for $t'$.
	 
	 Assume that when the thread arrives at $t$, then $t$ is a heap, implying $t.v \le t.r.v$.
	 Since $t'.v = \mathtt{min}(v, t.v)$ and $t'.l.v = \mathtt{min}(t.r.v,\mathtt{max}(v,t.v))$, then $t'.v \le t.v \le t'.l.v$.
	 Thus $t'$ is also a heap. \qed
	\end{proof}
	
	\begin{theorem}[Correctness of pullUpLeft]
	If invoked by \texttt{removeMin},  \texttt{pullUpLeft} is linearizable, and preserves the Braun, Heap, and Snapshot Isolation properties.
	\end{theorem}
	\begin{proof}
	\texttt{pullUpLeft} is called with a write-lock on the root and performs hand-over-hand locking along its leafward path.
	The linearizability argument is largely as for \texttt{insert}; however, upon arriving at the target leaf, the reference to that leaf in its immediate parent is deleted.
	Despite this violation of hand-over-hand locking, \texttt{pullUpLeft} maintains its linearizability if the invoking code retained a lock on the heap preventing any other threads from traversing down after it.
	
	When a thread executing \texttt{pullUpLeft} arrives at a non-leaf subtree, $t = \{t.v,t.l,t.r\}$, an \texttt{update} is performed such that,
	$t' = \{t.v,t.r,t.l\}$, and then \texttt{pullUpLeft} is invoked on $t'.r$.
	If $t'.r$ is a leaf, it is deleted, and $t'.r.v$ is returned.
	
	 If $t$ is a snapshot, then $t'$ is the result of allocating a new node (with snapshots of its children), otherwise $t$ is mutated directly, thus Snapshot Isolation is preserved.
	 
	 Assume that when the thread arrives at $t$, then $t$ is Braun.
	 Thus after \texttt{pullUpLeft}, we have that $|t'.r| = |t.l|-1$ and $|t'.l| = |t.r|$.
	 Thus $|t'.r| \le |t'.l| \le |t'.r|+1$ and Braun holds for $t'$.
	 
	 Assume that when the thread arrives at $t$, then $t$ is a heap.
	 No elements are inserted below $t$, thus every rootward path through $t$ remains nonincreasing, and \texttt{pullUpLeft} preserves heap. \qed
	\end{proof}
	
	\begin{theorem}[Correctness of pushDown]
	If invoked by \texttt{removeMin},  \texttt{pushDown} is linearizable, preserves the Braun and Snapshot Isolation properties, and restores the Heap property.
	\end{theorem}
	\begin{proof}
	\texttt{pushDown} is called with a write-lock on the root and performs hand-over-hand locking along its leafward path.
	The linearizability argument is as for \texttt{insert}.
	
	When a thread executing \texttt{pushDown} arrives at a non-leaf $t = \{t.v,t.l,t.r\}$, \texttt{update}s are performed such that,
	$t'.v = \mathtt{min}(t.v, t.l.v, t.r.v)$, $c'.v = \mathtt{max}(t.v,c.v)$, $t'.c = c'$, where $c$ selects whichever of $t.l$ and $t.r$ contains the smaller value.
	Then \texttt{pushDown} is invoked on $t'.c$.
	
	 If $t$ is a snapshot, then $t'$ is the result of allocating a new node (with snapshots of its children), otherwise $t$ is mutated directly.
	 If $c$ is a snapshot, then $c'$ is the result of allocating a new node (with snapshots of its children), otherwise $c$ is mutated directly.
	 Thus Snapshot Isolation is preserved.
	 
	 Assume that when the thread arrives at $t$, then $t$ is Braun.
	 Since no elements are added or removed, $t'$ is also Braun.
	 
	 Assume that when the thread arrives at $t$, the subtrees rooted at its children are heaps.
	 For children $c,d \in t$, assume without loss of generality that $c.v \le d.v$.
	 After \texttt{pushDown}, we have that $t'.v = \mathtt{min}(c.v,t.v)$, $c'.v = \mathtt{max}(c.v,t.v)$, $d'.v = d.v$.
	 Thus $t'.v \le c'.v$ and also $t'.v \le d'.v$, and inductively, \texttt{pushDown} restores heap. \qed
	\end{proof}
	
	\begin{theorem}[Correctness of removeMin]
	\texttt{removeMin} is linearizable, and preserves the Braun, Heap, and Snapshot Isolation properties.
	\end{theorem}
	\begin{proof}
	After acquiring the heap, \texttt{removeMin} reentrantly acquires the write-lock on the root, $r$, twice and updates the root, s.t. $r'.v = r.\mathtt{pullUpLeft}$.
	After this, \texttt{removeMin} executes $r'.\mathtt{pushDown}$.
	
	Since \texttt{pullUpLeft} ensures that $r'$ is not a snapshot, the direct assignment to $r'.v$ does not violate Snapshot Isolation.
	
	Since the lock on the root is held continuously between starting \texttt{pullUpLeft} and starting \texttt{pushDown}, \texttt{removeMin} is linearizable.
	
	Both subroutines are Braun preserving, so their composition is also Braun preserving.
	
	Since \texttt{pullUpLeft} is heap preserving,  $r'.l$ and $r'.r$ are heaps, \texttt{pushDown} is heap restoring, and \texttt{removeMin} is heap preserving. \qed
	\end{proof}
	
	\section{\label{sec:perf}Performance}
	We benchmarked the concurrent Braun heap implementation of this chapter against two other readily available concurrent priority queue implementations.
	\texttt{PriorityBlockingQueue} is part of the Java standard library, and its methods use a simple mutex around a high-performance single-threaded binary heap implemented with an array.~\cite{dleaPBQ}
	It provides no native snapshot operation; however, while the documentation asserts that its iterator is ``weakly consistent'', the source of the implementation reveals that it uses a mutex for a copy of the backing array, which provides strong consistency.
	The provided constructor for efficient construction from sorted bulk data is incompatible with the \texttt{Iterator} interface, so we implemented \texttt{snapshot} with successive inserts from the iterator.
	
	We built \texttt{SkipListPriorityQueue} from a variant of the Java standard library's \texttt{ConcurrentSkipListMap} that has been augmented with the lock-free linearizable iterator of Petrank and Timnat.~\cite{dleaCSLM,petrank2013lock}
	Simple CAS loops built on \texttt{putIfAbsent}, \texttt{replace}, and the key/value variant of \texttt{remove} were used to transform the ordered map into an ordered multiset supporting \texttt{insert} and \texttt{removeMin}.
	Their iterator was augmented with a wrapper to properly repeat each entry in the multiset.
	As with \texttt{PriorityBlockingQueue}, no native snapshot was supported, and no compatible bulk data constructor was available, so we implemented it with successive inserts from the iterator.
	
	\subsection{Methodology}
	The experiments were run on the Brown Department of Computer Science's grid, using the ``mblade12'' hostgroup, each providing 32 cores and 64GB of RAM, using an Oracle Java 8 runtime.
	For each sequence of tests, the priority queue implementation being benchmarked was initialized with a large quantity of random entries,
	then a series of warmup executions for the VM's JIT compiler were performed and discarded. Finally, each sequence of tests were performed and the times recorded and averaged over a number of iterations.
	The priority queues were initialized with $2^{20}$ entries, using 10 warmup runs and 40 experimental runs.
	
	In all, six tests were conducted, \texttt{sum}, \texttt{snapshot+insert}, \texttt{add/removeMin/sum}, \texttt{snapshot}, \texttt{insert}, and \texttt{removeMin}.
	For the \texttt{sum} task, all threads were tasked with individually calculating the sum of every entry in the priority queue.
	For the \texttt{snapshot+insert} task each of $t$ threads created its own snapshot and inserted $1344/t$ items\footnote{
	This number was selected because it is a common multiple of the thread counts used, to avoid rounding error}, whereas for the \texttt{snapshot}-only task, each of $t$ threads merely had to instantiate a snapshot.
	For the \texttt{add/removeMin/sum} task, each thread performed $1344/t$ random operations, where with probability $\frac{3}{8}$ the operation was an \texttt{insert}, or with probability $\frac{3}{8}$ the operation was an \texttt{removeMin}, or with the remaining $\frac{1}{4}$ probability, the operation was a \texttt{sum} of the first 1024 items in the priority queue.
	
	For the \texttt{insert} tests, 1344 randomly selected values were inserted into the priority queue, with each of $t$ threads responsible for $1344/t$ insertions.
	Similarly, for the \texttt{removeMin} tests, each of $t$ threads was tasked with executing RemoveMin $1344/t$ times.
	
	\subsection{Results}
	The experimental results are graphed jointly with the results in Chapter \ref{chap:lockfree}; however, we will discuss this chapter's results in detail here.
	Each graph shows thread count along the horizontal axis, and a normalized execution time in log-scale along the vertical axis.

	With one exception, the concurrent Braun heap (grey) implementation substantially outperformed both the \texttt{SkipListPriorityQueue} (orange) and the \texttt{PriorityBlockingQueue} (blue) on the tests involving iteration and snapshots (Figures \ref{fig:lfSumTest}, \ref{fig:lfSnapAndInsertTest}, \ref{fig:lfMixedTest}, and \ref{fig:lfSnapOnlyTest}).
	On the \texttt{sum} test, all threads acquire \texttt{PriorityBlockingQueue}'s single read/write lock without difficulty, and thus outperform all other priority queue implementations: this is expected behavior.
	Also of note is that Petrank and Timnat's lock-free iterators appear to pay a heavy performance penalty for linearizability in a lock-free setting without the benefit of fast snapshots (see Figure \ref{fig:lfSumTest} in particular).
	
	On the tasks where iteration or snapshots are not required (Figures \ref{fig:lfInsertTest} and \ref{fig:lfremoveMinTest}), the Braun Heap is outperformed.
	This is not unexpected as the cost for supporting a richer set of operations, and we believe that the mixed workloads of the \texttt{snapshot+insert} (Figure \ref{fig:lfSnapAndInsertTest}) and \texttt{add/removeMin/sum} (Figure \ref{fig:lfMixedTest}) tasks represent more realistic real-world conditions.
	
	\section{Conclusion}
	The algorithm for concurrent Braun heaps presented here is both easily implemented\footnote{The \texttt{ConcurrentSkipListMap} source code, on which the competing \texttt{SkipListPriorityQueue} was based, has $>2000$ lines of code, whereas the BraunHeap we implemented has $<200$.} and easily verified, while providing competitive performance scaling on traditional priority queue operations, and supporting fast and consistent snapshots and iteration.
	
	This chapter leads naturally to several lines of inquiry.
	Much of the algorithm lends itself to a lock-free approach based on CAS instructions; however, the two phases of the \texttt{removeMin} operation leave the heap in an inconsistent state if not executed atomically.
	An efficient approach to speculative executions of \texttt{removeMin} would be a useful next step in developing a lock-free concurrent Braun heap (there is also a fairness concern here, as long-running remove operations may consistently get delayed by fast inserts).
	The experimental results also suggest that without fast snapshots, iteration over concurrent data structures is inordinately expensive due to synchronization costs.
	Another direction would be to address this by leveraging copy-on-write semantics in the pursuit of efficient snapshot operations for other data structures, potentially including skip lists.
	These questions are addressed in Chapters \ref{chap:lockfree} and \ref{chap:returns}.
	 \chapter{\label{chap:lockfree}A Type-Level Transformation of Purely-Functional Trees into Lazy COWs}
	   
This chapter adapts the techniques used in the previous chapter to the lock-free setting.
In Sections \ref{sec:llxscx} and \ref{sec:shortops} we develop an intuition for the form that operations on lock-free concurrent trees ought to take.
Then, in Section \ref{sec:lockfreealg} we formalize this intuition, and describe an algorithm to execute operations that satisfy the constraints it imposes.
Next, in Sections \ref{sec:lockfreelib} and \ref{sec:braunheaplf} we implement and evaluate a library that partially automates the process of writing lock-free concurrent trees by a type-level transformation on purely-functional code. We conclude in Section \ref{sec:lockfreenogc} with a discussion of challenges to, and possible approaches for, implementing the lock-free trees with structural sharing in a language without automatic garbage collection.

\section{The Right Synchronization Primitives\label{sec:llxscx}}
The success of the hand-over-hand locking approach to concurrent Braun trees in Chapter \ref{chap:handoverhand} suggests that a similar ``sliding window'' approach will work in the lock-free setting: read a small amount of local context, decide how the tree must be modified, atomically apply some small local change (if our read-set is still valid) and proceed recursively down the tree.

At first appearances, Multiword Compare-and-Swap (MCAS), seems a good tool for the job: for an $n$-ary tree where each node stores a single mutable pointer to an immutable record containing both the node's stored value and its child pointers, then $(n+1)$-CAS would allow us to atomically change every field in a node and all its children; however, we will see that this is more expressive power than we actually need.
Moreover, this power comes at a price: $n$-CAS can be implemented in $O(n)$ time in terms of a native Double Compare-and-Swap (DCAS\footnote{Not to be confused with Doublewide Compare-and-Swap (DwCAS), which operates only on adjacent words.}) instruction, but only $O(n \cdot log(p))$ in terms of a native single-word CAS (where $p$ is the number of contending processes).~\cite{GreenwaldSynchrMCAS}
Whereas CAS is widely available in extant processors, very few support DCAS.
Similarly, MCAS can be emulated using a nested Load-Link/Store-Conditional (LL/SC); however, again, such primitives are not available in extant hardware.~\cite{fraser2004practical}

Instead, we propose to use the extended Load-Link/Store-Conditional (LLX/SCX) primitives described by Brown et al..~\cite{BrownLLXSCX}
LLX/SCX allows us to read from multiple words and then update a single word atomically if the read-set is still valid, and \emph{can} be efficiently implemented in terms of a hardware single-word CAS.
The SCX primitive further allows you to mark any elements of the read-set as immutable to future SCXs, which we will leverage to implement our lazy COW snapshots.
Finally, we note that LLX/SCX has already been used to implement a number of highly concurrent tree data structures.~\cite{BrownGeneralTrees,brown2017techniques}

\section{Some Design Constraints\label{sec:shortops}}
We observe that, assuming we can implement an efficient lazy COW snapshot operation on our data structure, we could perform a speculative update on the snapshot, and then atomically replace the publicly visible root with the root of the snapshot.
This is quite similar to Herlihy's universal construction, if perhaps slightly more efficient in the specifics.~\cite{herlihy1993methodology}
However, it clearly would share the same performance pathologies under contention.

Another option would be to mix a sliding-window strategy for operations which can execute in a single leaf-ward traversal from the root with the universal construction's replace-the-root strategy for the first $(n-1)$ passes of an operation that requires $n$ leaf-ward traversals from the root.
This is roughly analogous to the strategy used in the preceding chapter to handle the differing needs of \texttt{insert} and \texttt{removeMin} in the locking setting.

This mixed strategy would likely scale slightly better than the locking version for pure \texttt{insert} workloads, as each thread would waste less time in spurious sleep-wake cycles due to convoying.~\cite{Blasgen:1979:CP:850657.850659}
It would also likely scale slightly worse than the locking version for pure \texttt{removeMin} workloads, as each thread would waste $O(log(n))$ time on each failed CAS, and on a workload with more threads than available processors, that time could be spent sleeping long enough for another thread to achieve productive work.

However, on \emph{mixed} workloads, the mixed strategy would be disastrous: much like the \textit{starving elder} pathology \cite{Bobba:2007:PPH:1250662.1250674} in hardware transactional memory (HTM) systems, the operations (like \texttt{insert}) with an $O(1)$ read-set would cause the operations (like \texttt{removeMin}) with an $O(log (n))$ read-set to perpetually abort and retry.
We now model this with both an intuitive theoretical approximation and experimentally, and obtained qualitatively similar results from both models.

\subsection{Long Operations Will Starve\label{sec:nolongops}}
\paragraph{Theoretical Model}
Consider the execution of a program, $R$, which must execute $n$ instructions, each taking 1 clock-cycle.
If there is some uniform probability $p$ that the scheduler will choose to sleep $P$ for any given clock-cycle, then we can model the time to completion, $t$ as a negative binomial distribution, with PMF given by:
\begin{equation}
P(\mathrm{CO}(t))= \binom{t + m - 1}{t} \cdot (1 - p)^{m} \cdot p^{t}
\end{equation}
and CDF given by:
\begin{equation}
P(\mathrm{CB}(t))= 1 - I_{p} (t + 1, n)
\end{equation} (where $I$ is the regularized incomplete beta function).

Thus, given two processes $R$ (with $n_R$ instructions), and $Q$ (with $n_Q$ instructions), whose final instructions are conflicting CASes, and which are executing on two SMP processors, and, WLOG, taking $n_R \le n_Q$, then the probability that $Q$ is not aborted by $R$ is given by:
\begin{equation}
P(t_Q < (t_R + \epsilon)) = \sum_{t = 0}^{\infty} P(\mathrm{CO}_{Q}(t)) \cdot P(\mathrm{CB}_{R}(t - 1 + n_Q - n_R)) + \frac{1}{2} P(\mathrm{CO}_{Q}(t)) \cdot P(\mathrm{CO}_{R}(t + n_Q - n_R))
\end{equation} (or the probability that $Q$ took strictly less time, plus half the probability that they took exactly the same amount of time).

This model is idealized in several ways, particularly the assumption that the probability of sleeping is uniform at the required granularity\footnote{The assumption of equal-length instructions is made less unrealistic if we assume the workload consists primarily of atomic memory accesses, which will dominate the execution time of any small number of operations on registers.}.
Nonetheless, it should exhibit qualitatively similar behavior, particularly if the running times are long and the probability of sleeps is low.

We used numerical integration to evaluate this model, and selected results are presented below, combined with the experimental model.
Varying the parameters along different axes results in a family of multidimensional sigmoids of varying steepness.

\paragraph{Experimental Model}
To test the model experimentally, we wrote a simple test program with two threads: one performs $C$ LLXs, and one performs $C \cdot log(N)$ LLX's.
Here, $C$ corresponds to the branching factor, and $log(N)$ corresponds to the tree depth
Both end with a single SCX.
We then varied both $C$ and $log(N)$, and averaged the probability of the $log(N)$ thread finishing first over 16000 runs.

\paragraph{Combined Results}
\begin{figure*}
\centering
\includegraphics[width=\textwidth]{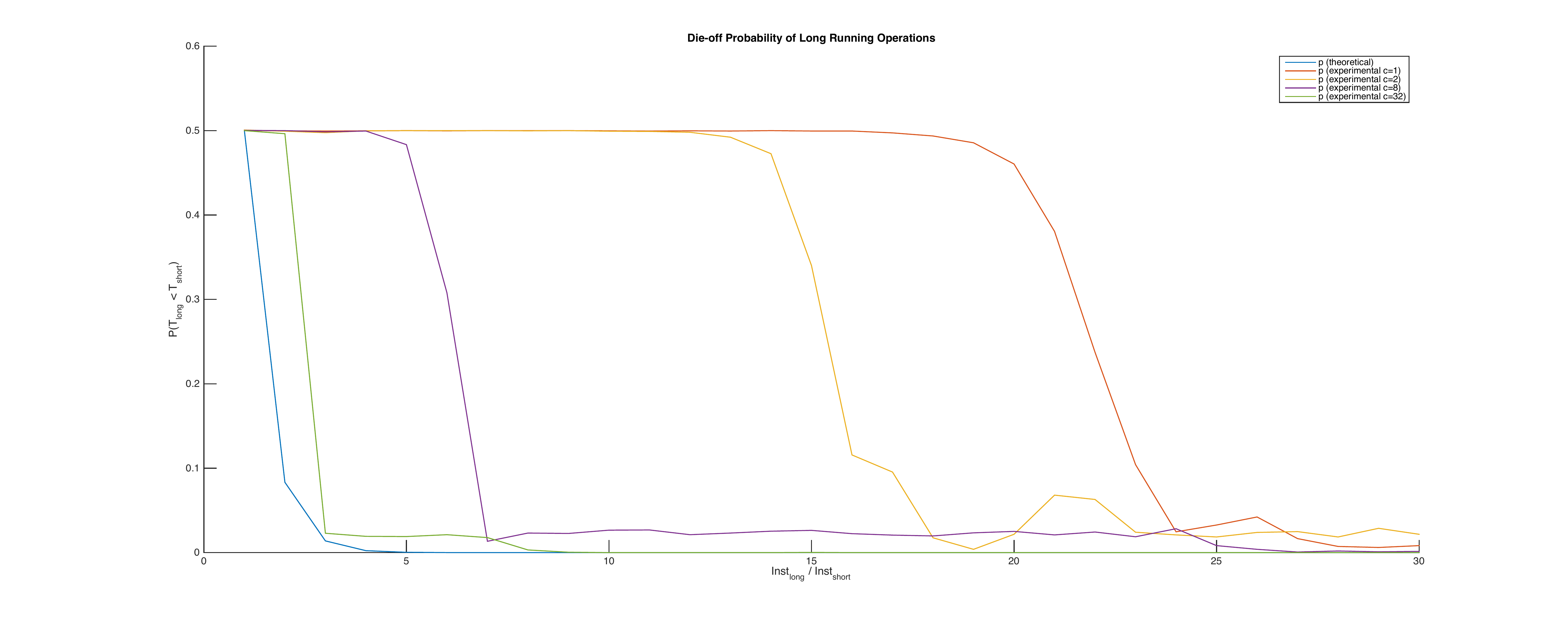}
\caption{Combined Experimental and Theoretical Model Results}
\label{fig:dieoff}
\end{figure*}
Figure \ref{fig:dieoff} shows the results of both models, with the theoretical results in blue, $c=1$ in red (a linked list), $c=2$ in orange (a binary tree), $c=8$ in purple, and $c=32$ in green (a typical branching factor for a HAMT~\cite{bagwell2001ideal} or RRB-Tree~\cite{bagwell2011rrb}).
Even the binary trees show near-extinction for a long-running operation competing against even a single short-running operation around a tree-depth of 15.

We also confirm that, as expected, as the read-set per node $C$ increases, the experimental results converge towards the theoretical model.

As a result, we conclude that any efficient lock-free concurrent tree operations must be separable into a number of small read-sets punctuated by writes.
This is compatible with both our intuition of applying LLX/SCX to a sliding-window strategy.

\section{One Weird Trick for Lock-Free Trees\label{sec:lockfreealg}}
To understand our approach to lock-free trees, we begin with a toy system, supporting two operations: \texttt{addOne} and \texttt{double}.
\paragraph{Simple Linearizability}
In the first version of our system, we simply have two counters (see Listing \ref{sublst:twocounters}), stored in memory cells that have been augmented to support the metadata used by LLX/SCX.
Our first version of \texttt{addOne} will increment the \texttt{right} counter if \texttt{left} is larger, and increment \texttt{left} otherwise.
Our first version of \texttt{double} does the same thing, except doubling instead of incrementing.
\begin{listing*}
\begin{sublisting}[c]{0.9\textwidth}
\centering
\begin{lstlisting}[mathescape=true]
val left = Cell(0); val right = Cell(0)
\end{lstlisting}
\caption{The two counters for our shared state.}
\label{sublst:twocounters}
\end{sublisting}
\begin{sublisting}[lb]{0.45\textwidth}
\begin{lstlisting}[mathescape=true]
def addOne()$\to$Unit {
  val leftNow = LLX(left)
  val rightNow = LLX(right)
  if(leftNow > rightNow){
    SCX(right, rightNow + 1)
  } else {
    SCX(left, leftNow + 1)
  }
}
\end{lstlisting}
\caption{\texttt{addOne}, executing in one thread.}
\label{sublst:counterAddOne}
\end{sublisting}
\hspace{0.05\textwidth}
\begin{sublisting}[rb]{0.45\textwidth}
\begin{lstlisting}[mathescape=true]
def double()$\to$Unit {
  val leftNow = LLX(left)
  val rightNow = LLX(right)
  if(leftNow > rightNow){
    SCX(right, rightNow * 2)
  } else {
    SCX(left, leftNow * 2)
  }
}
\end{lstlisting}
\caption{\texttt{double}, executing in a second thread.}
\label{sublst:counterDouble}
\end{sublisting}
\caption{Source code for the simple version of our toy system.}
\label{lst:counterToy}
\end{listing*}
If we assume that both operations are being executed in a loop by some parent thread, and that if any LLX or SCX fails the whole operation will simply retry, then it should be obvious that the operations are each linearizable: as long as LLX/SCX behaves correctly, the value of the counters can't change in between being read and being written by the matching SCX.

Going forward, we will refer to any linearizable operation consisting of a sequence of LLXs followed by a single SCX as satisfying \emph{simple linearizability}.

\paragraph{Compound Linearizability}
In the second version of our system, we instead will consider a \emph{list} of counters, and change \texttt{addOne} to increment the whole list, and \texttt{double} to double the whole list, using an SCX to update each node in order.
\begin{figure*}
\begin{subfigure}[c]{0.25\textwidth}
\includegraphics[width=\textwidth]{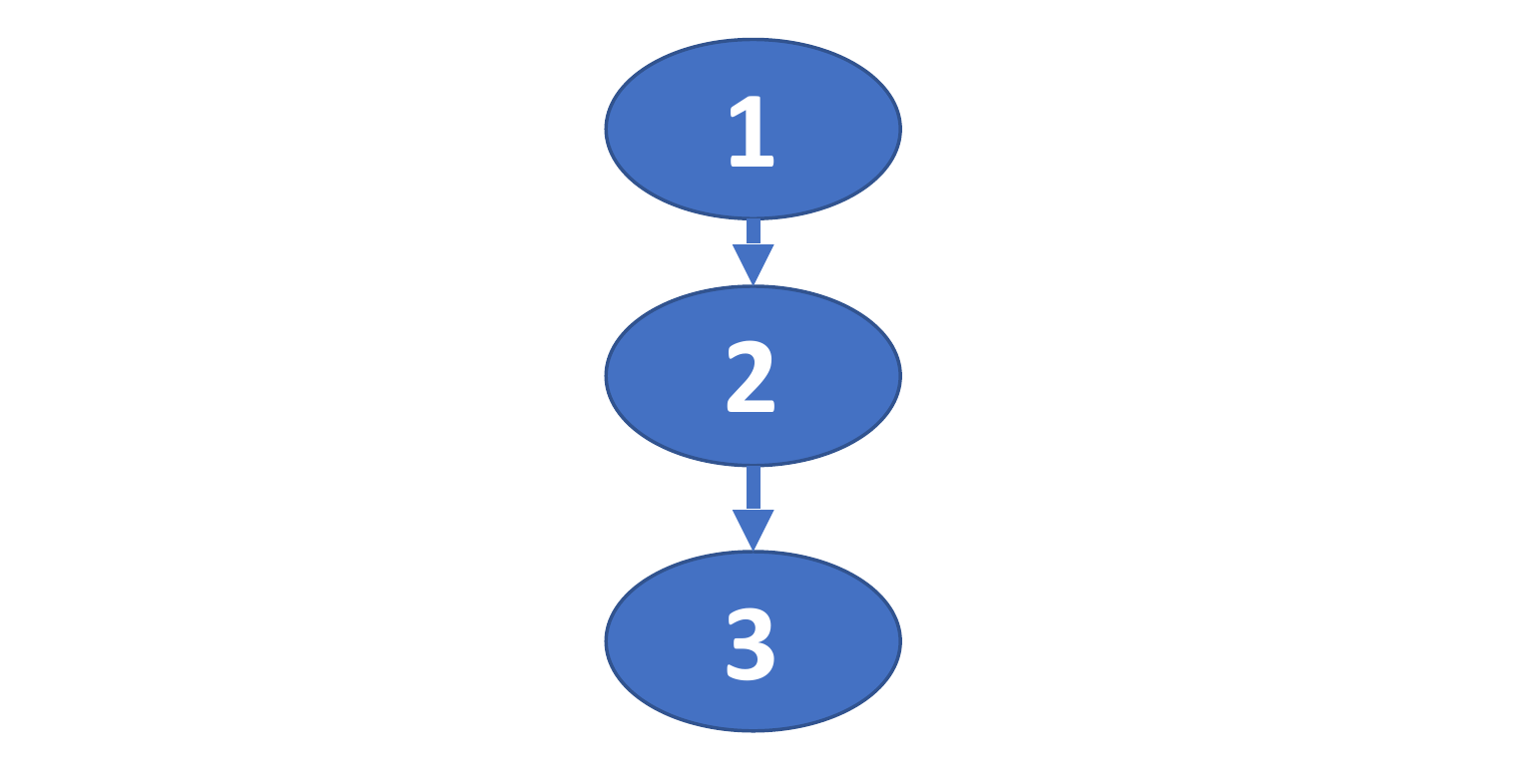}
\caption{} \label{subfig:extthunk1}
\end{subfigure}
\hspace{0.1\textwidth}
\begin{subfigure}[c]{0.25\textwidth}
\includegraphics[width=\textwidth]{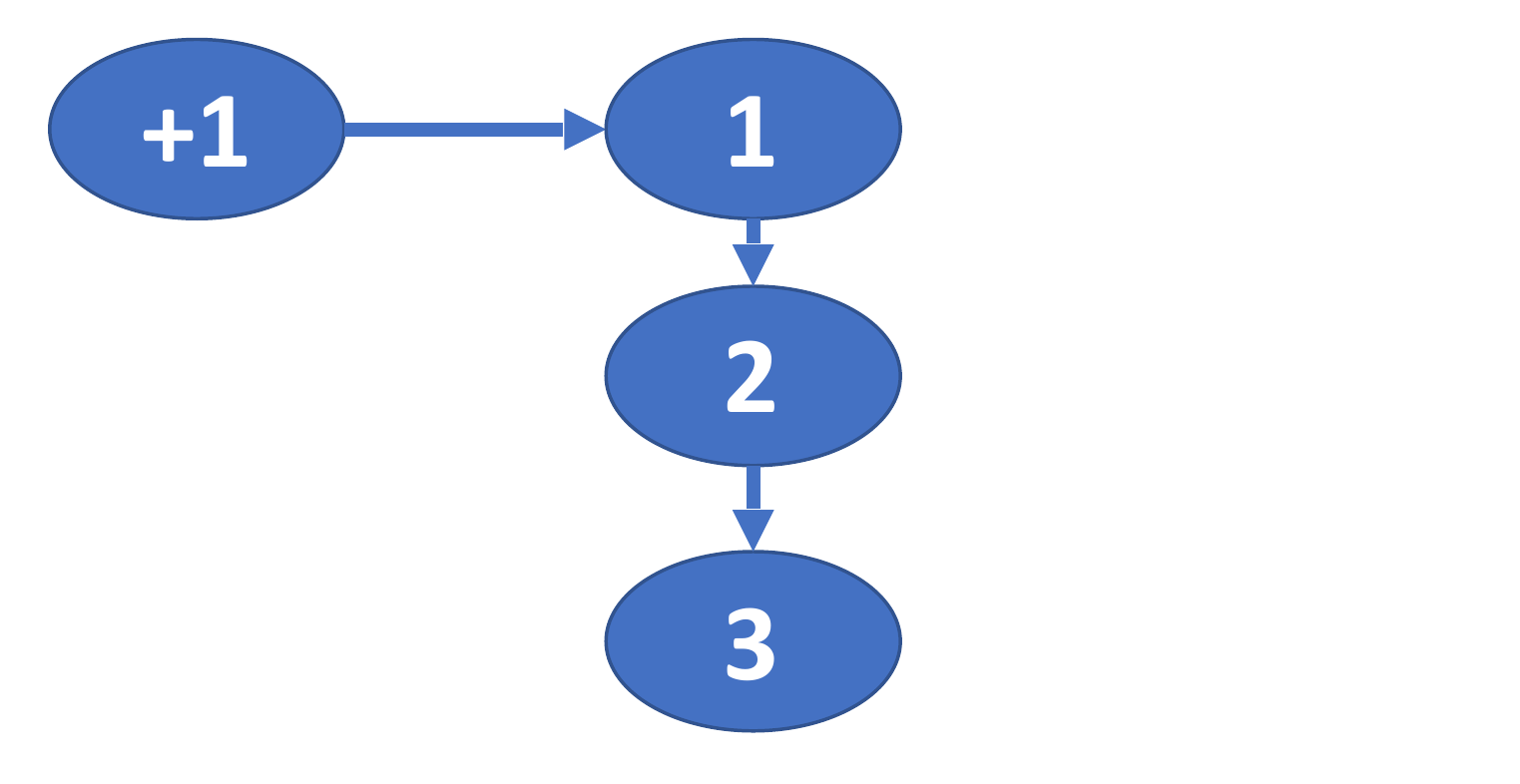}
\caption{} \label{subfig:extthunk2}
\end{subfigure}
\hspace{0.1\textwidth}
\begin{subfigure}[c]{0.25\textwidth}
\includegraphics[width=\textwidth]{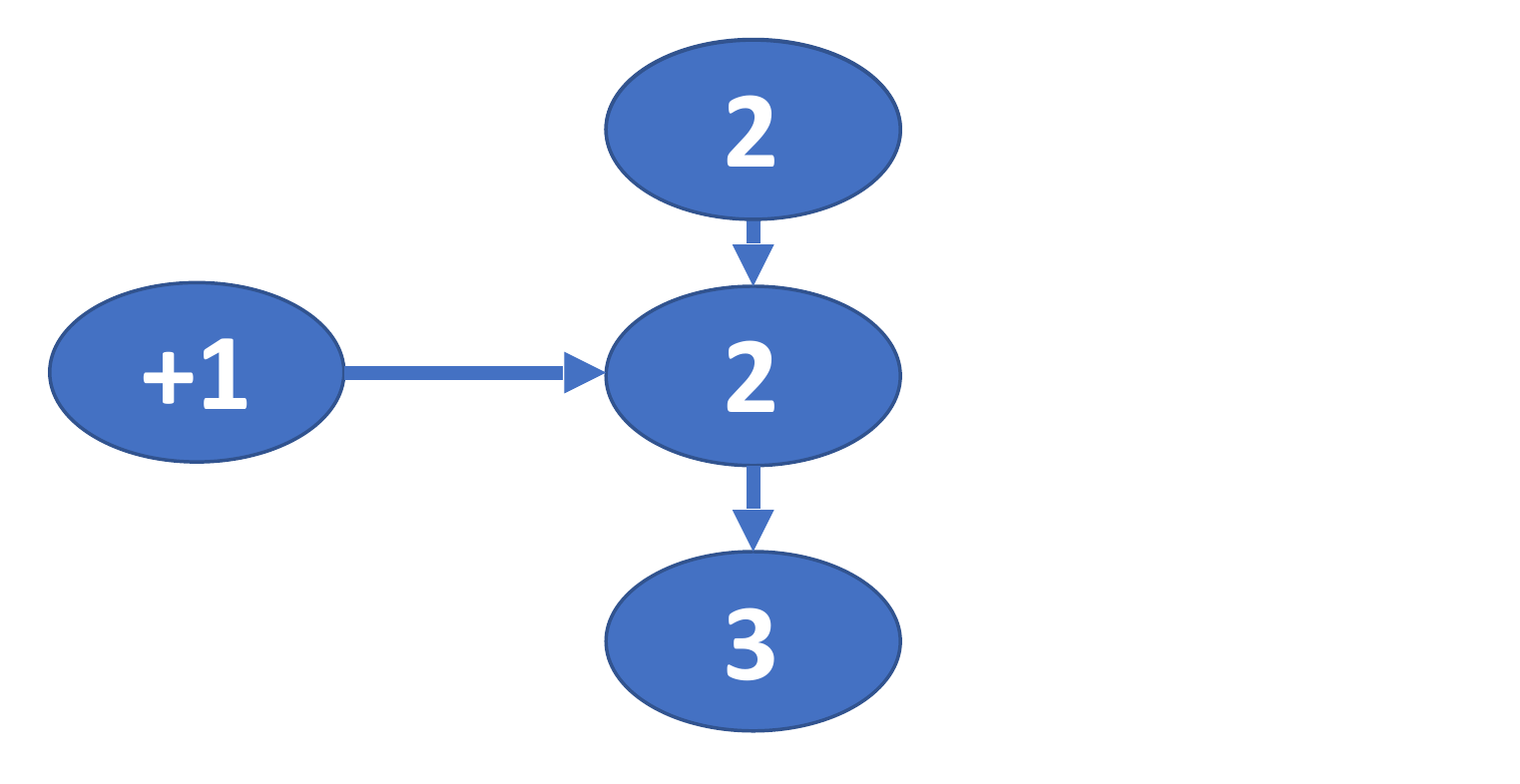}
\caption{} \label{subfig:extthunk3}
\end{subfigure}
\begin{subfigure}[c]{0.25\textwidth}
\includegraphics[width=\textwidth]{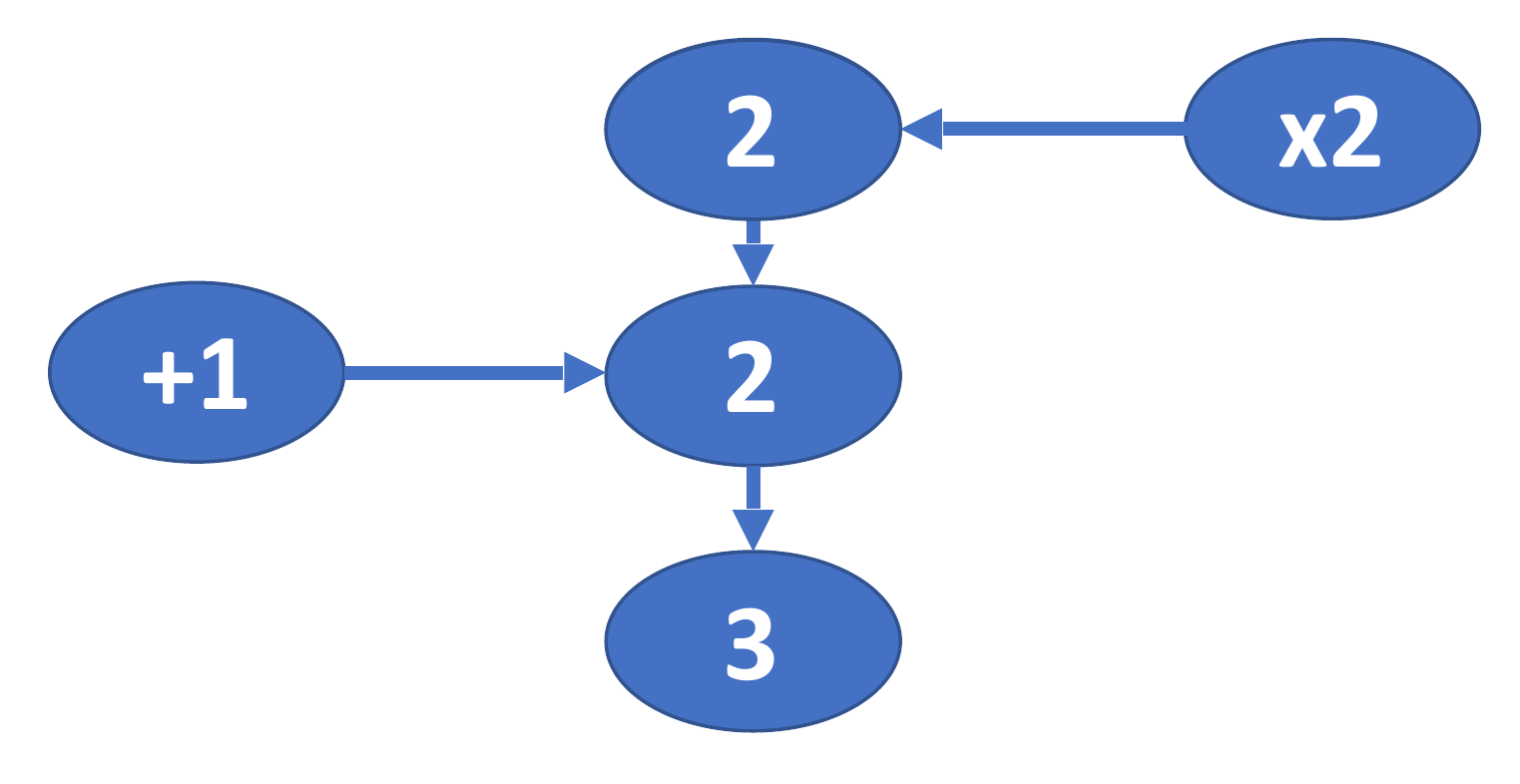}
\caption{} \label{subfig:extthunk4}
\end{subfigure}
\hspace{0.1\textwidth}
\begin{subfigure}[c]{0.25\textwidth}
\includegraphics[width=\textwidth]{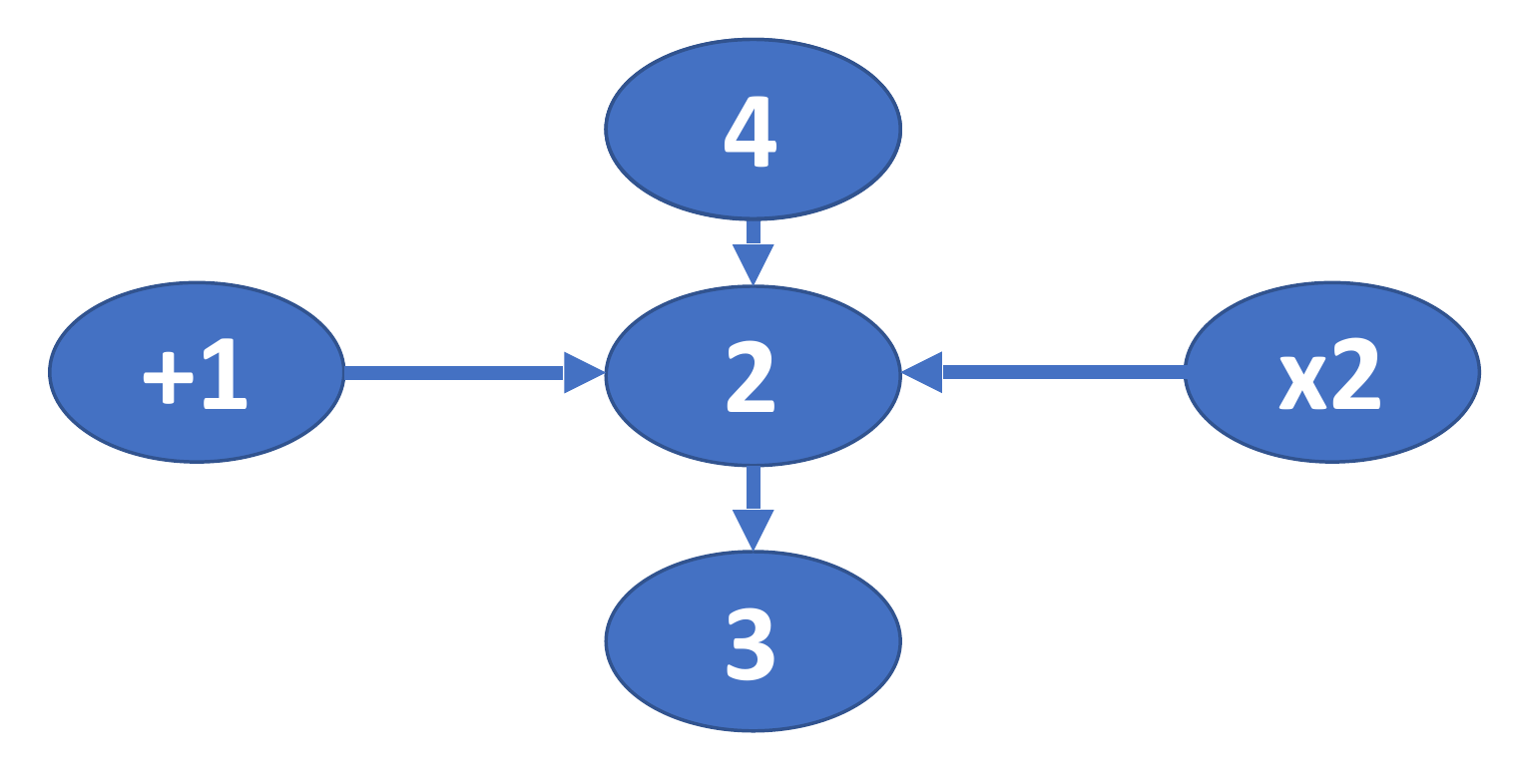}
\caption{} \label{subfig:extthunk5}
\end{subfigure}
\hspace{0.1\textwidth}
\begin{subfigure}[c]{0.25\textwidth}
\includegraphics[width=\textwidth]{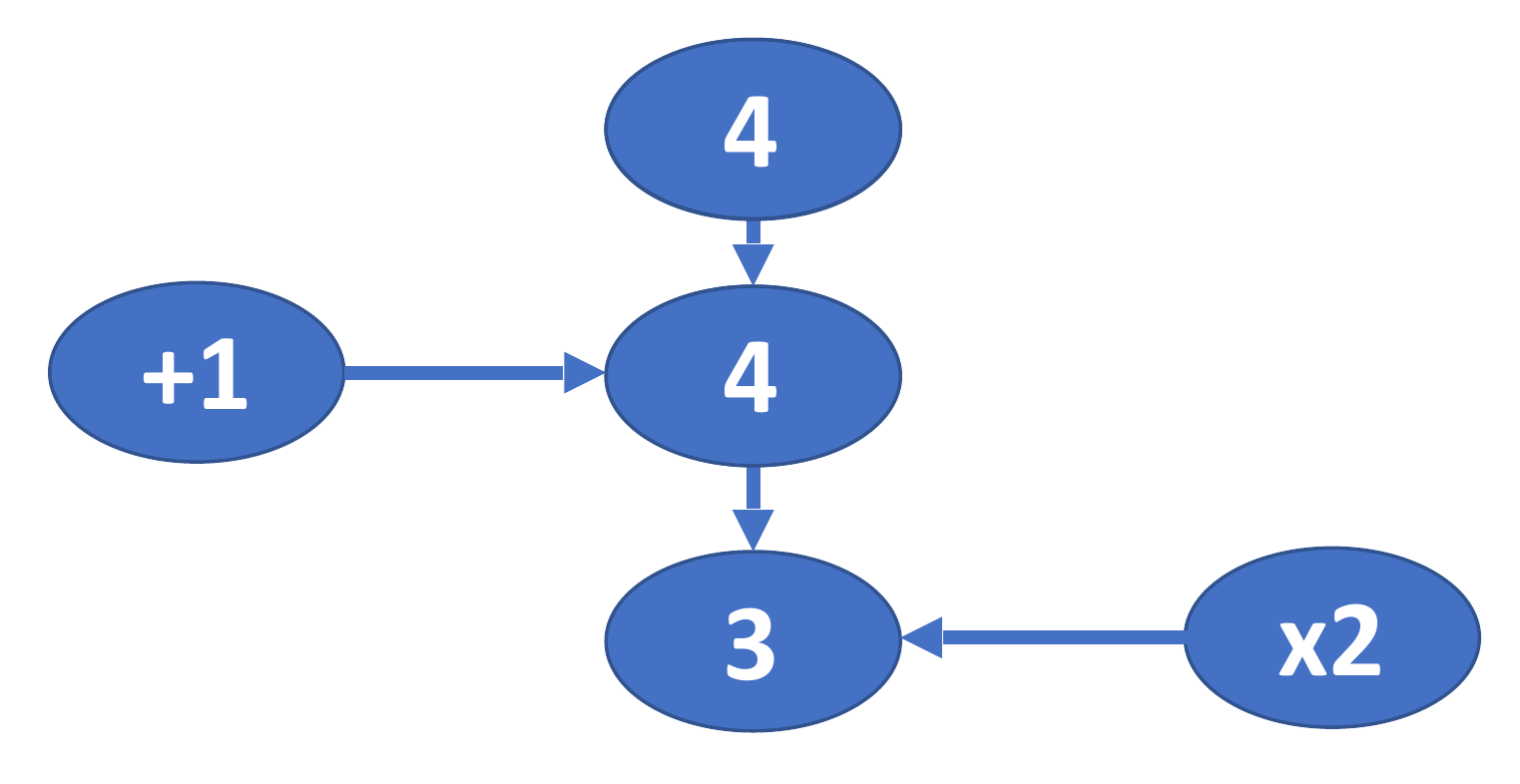}
\caption{} \label{subfig:extthunk6}
\end{subfigure}
\begin{subfigure}[c]{0.25\textwidth}
\includegraphics[width=\textwidth]{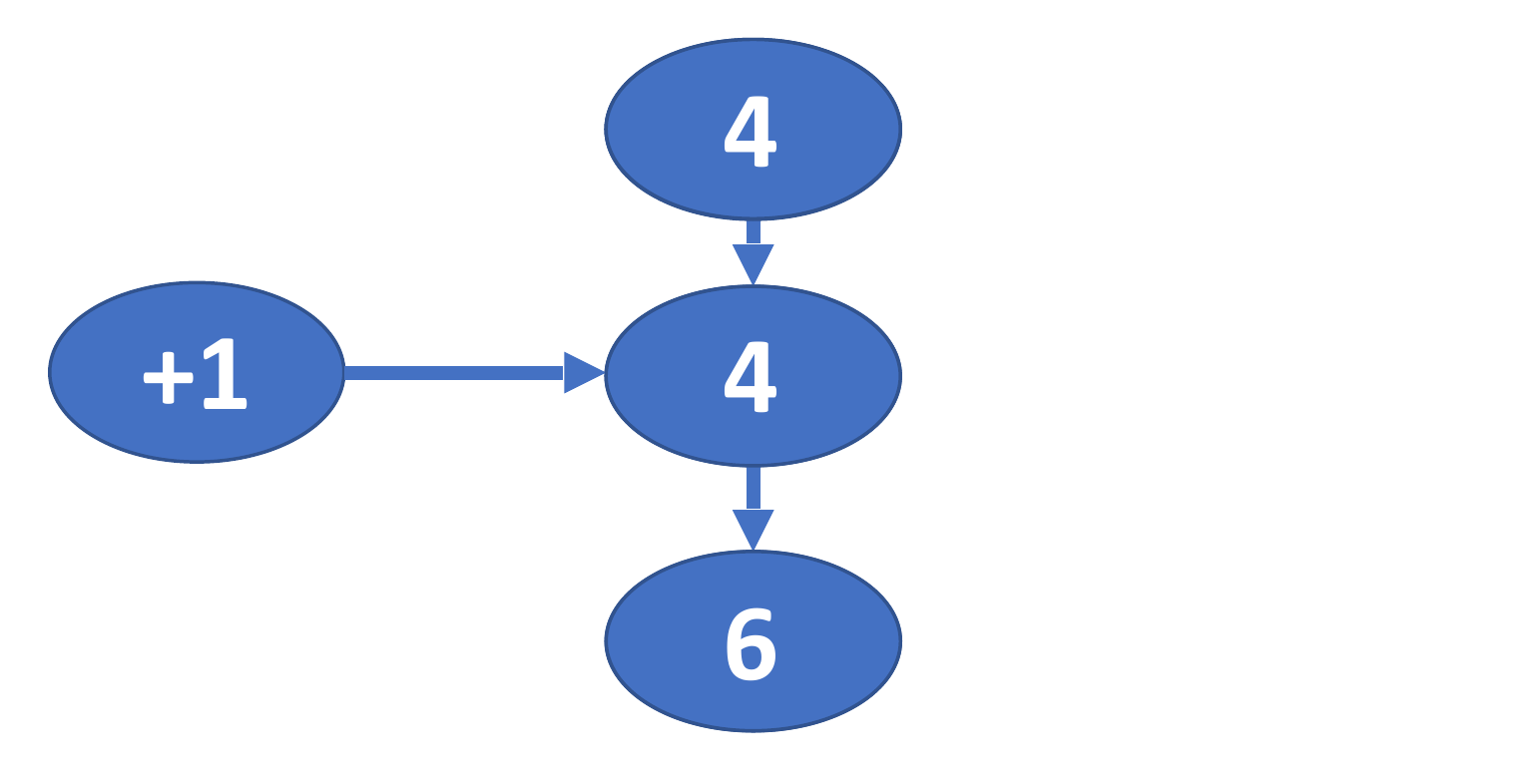}
\caption{} \label{subfig:extthunk7}
\end{subfigure}
\hspace{0.1\textwidth}
\begin{subfigure}[c]{0.25\textwidth}
\includegraphics[width=\textwidth]{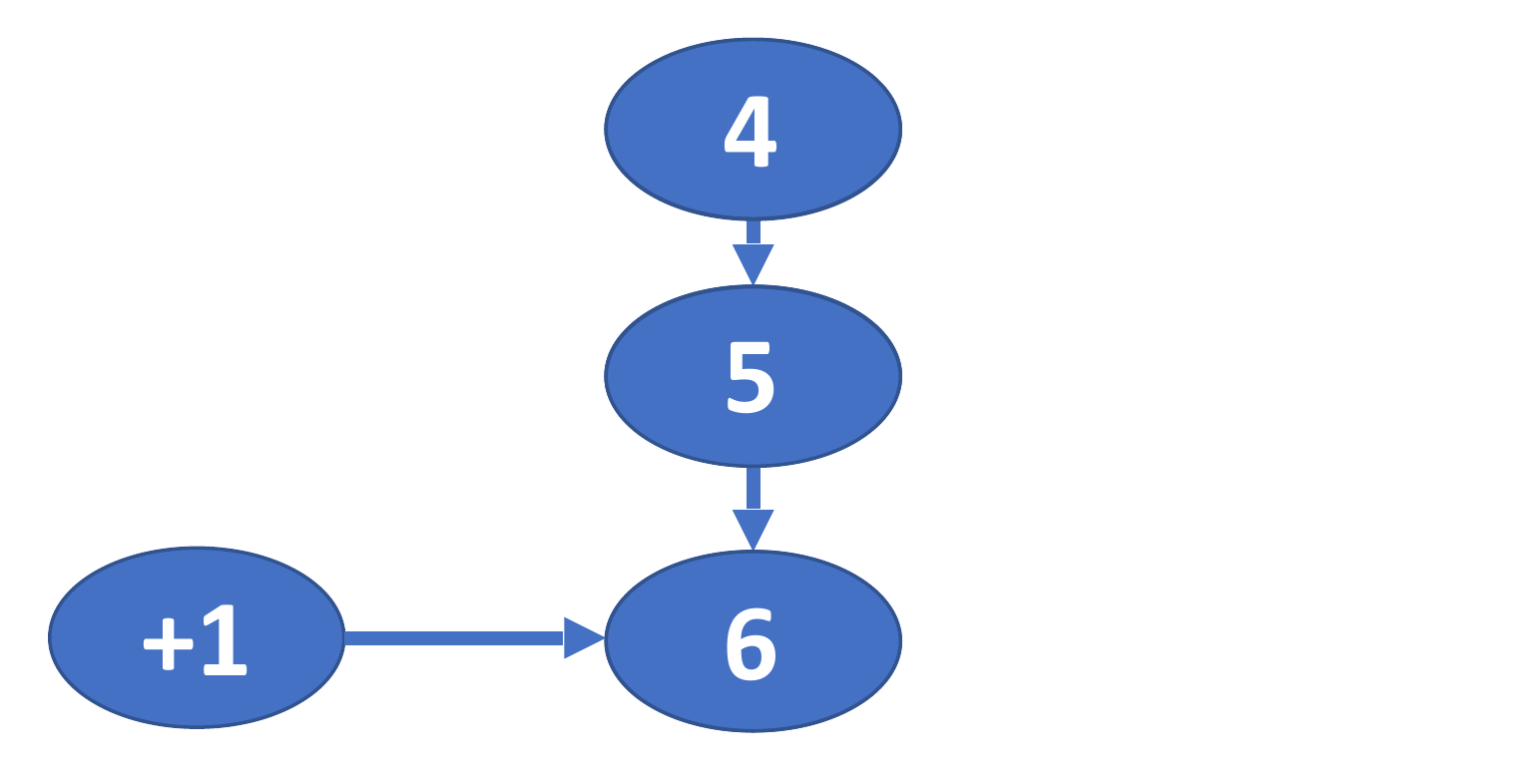}
\caption{} \label{subfig:extthunk8}
\end{subfigure}
\hspace{0.1\textwidth}
\begin{subfigure}[c]{0.25\textwidth}
\includegraphics[width=\textwidth]{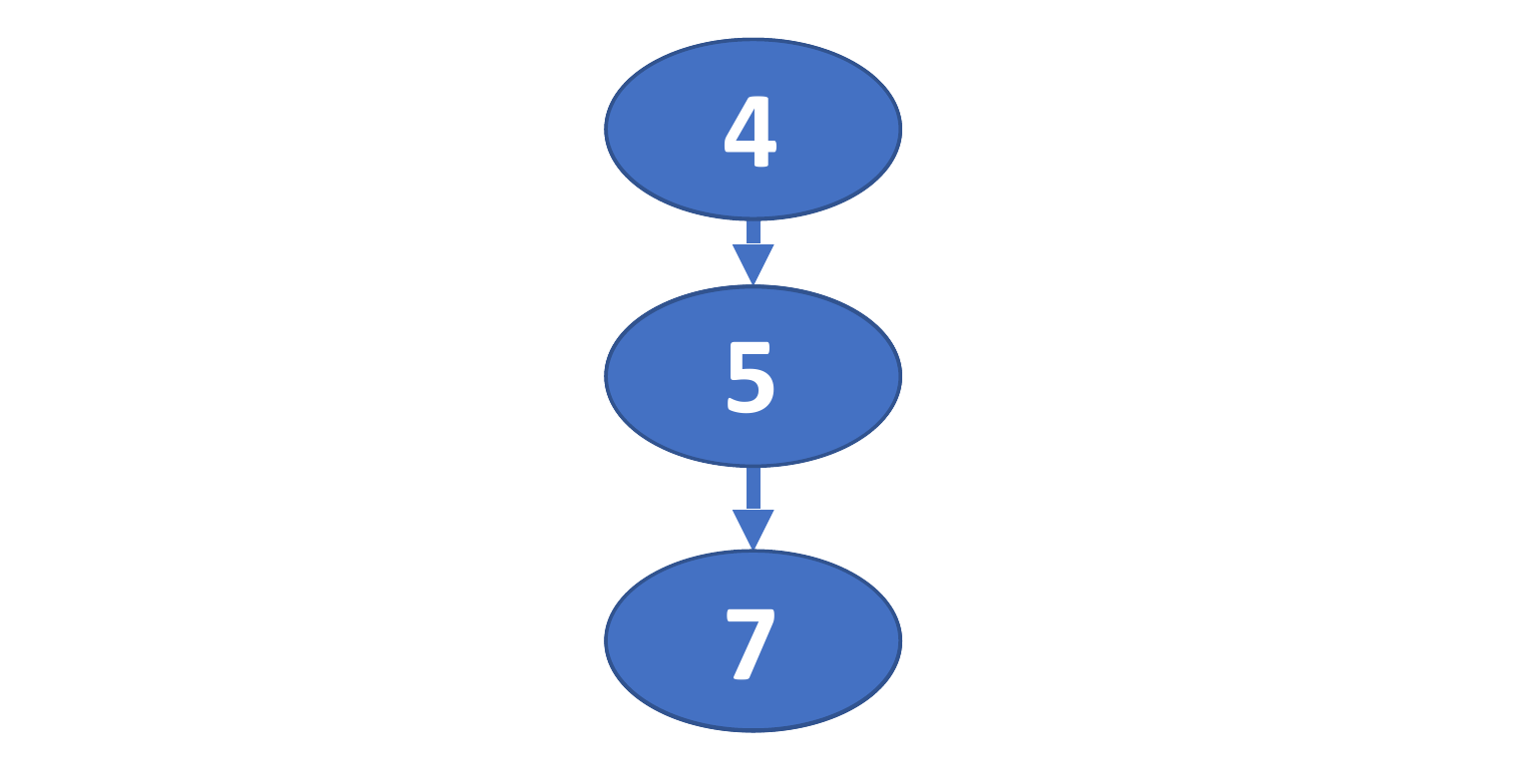}
\caption{} \label{subfig:extthunk9}
\end{subfigure}
\caption{An execution trace of an incorrect implementation for the compound version of our toy system.}
\label{fig:listToyWrong}
\end{figure*}
Before considering any source code, we will first consider the execution trace in Figures \ref{subfig:extthunk1} - \ref{subfig:extthunk9}.
We observe that even when each node is correctly updated in a linearizable way, problems can still arise.
Here we initialize the system as \texttt{List(1,2,3)} (Figure \ref{subfig:extthunk1}) and start both an \texttt{addOne} thread and a \texttt{double} thread.
The first thread appears (Figure \ref{subfig:extthunk2}), and successfully updates the first node (Figure \ref{subfig:extthunk3}),
but then the second thread appears (Figure \ref{subfig:extthunk4}), and races through ahead (Figures \ref{subfig:extthunk5} - \ref{subfig:extthunk7}), 
finally, the first thread continues (Figure \ref{subfig:extthunk8}), but the data structure is left in an inconsistent state (Figure \ref{subfig:extthunk9}).

We should have gotten either \texttt{List(4,6,8)} or \texttt{List(3,5,7)}.
The problem is that neither the \texttt{addOne} thread nor the \texttt{double} thread had any way of knowing that the other one was working, and so they were unable to coordinate their efforts.
Here we find it useful to separate reasoning about the result of a computation, from reasoning about its implementation (and in particular, when it executes).

This $\lambda$-tribe insight has already benefited the systems community in the form of futures and fork-join as primitives for building concurrent programs (\cite{friedman1976impact}, \cite{baker1977incremental}), but it's a useful lens, even when writing synchronization code by hand: we can decompose a tree-like data structure into a ready-now root, and ready-later subtrees.
Under hand-over-hand locking, we view any thread waiting to access a data-structure behind a lock as forcing the evaluation of a (possibly lazy) computation, in whichever thread holds the lock.
In a lock-free setting, threads try to help in forcing the computation, instead of waiting for someone else to do it for them.

\begin{figure*}
\centering
\begin{subfigure}[c]{0.4\textwidth}
\includegraphics[width=\textwidth]{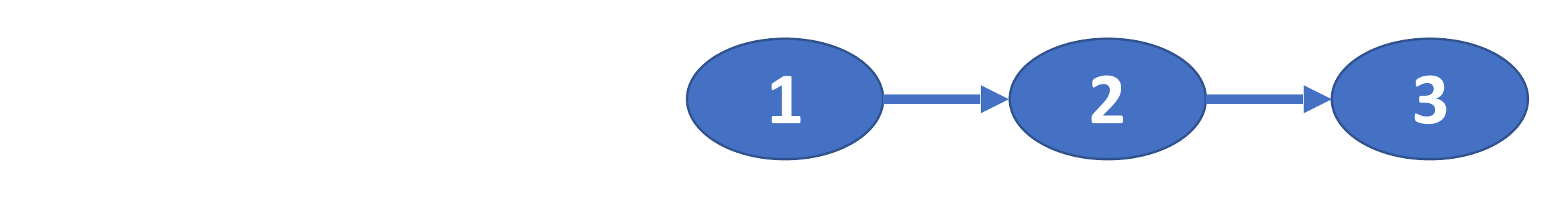}
\caption{} \label{subfig:embthunk1}
\end{subfigure}
\hspace{0.1\textwidth}
\begin{subfigure}[c]{0.4\textwidth}
\includegraphics[width=\textwidth]{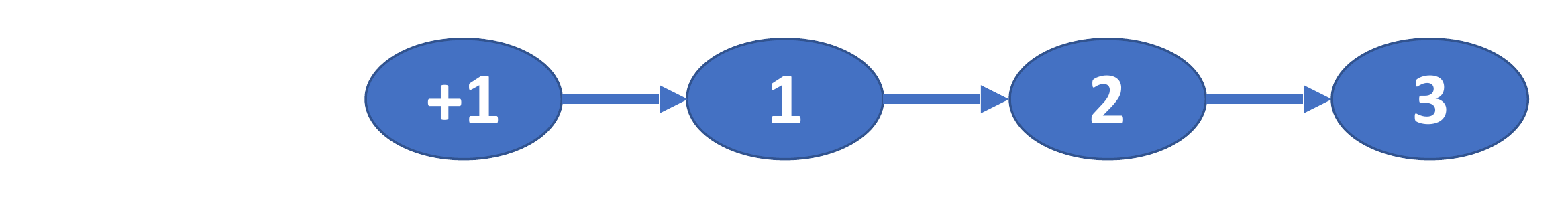}
\caption{} \label{subfig:embthunk2}
\end{subfigure}
\begin{subfigure}[c]{0.4\textwidth}
\includegraphics[width=\textwidth]{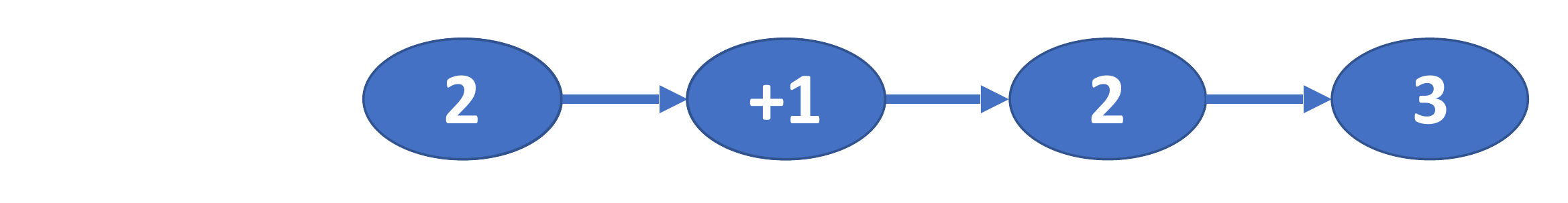}
\caption{} \label{subfig:embthunk3}
\end{subfigure}
\hspace{0.1\textwidth}
\begin{subfigure}[c]{0.4\textwidth}
\includegraphics[width=\textwidth]{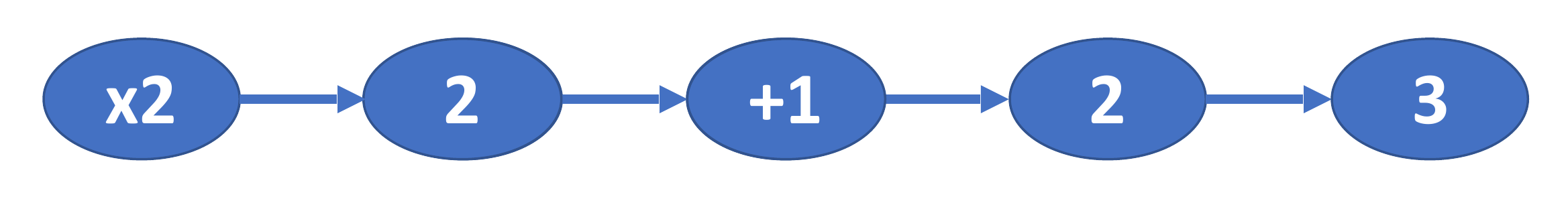}
\caption{} \label{subfig:embthunk4}
\end{subfigure}
\begin{subfigure}[c]{0.4\textwidth}
\includegraphics[width=\textwidth]{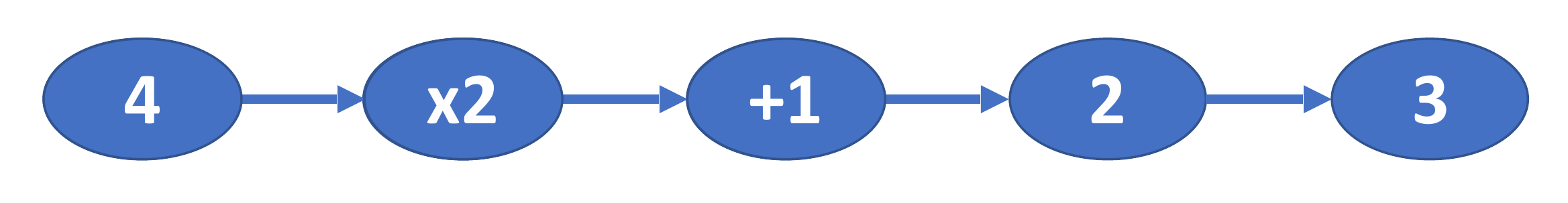}
\caption{} \label{subfig:embthunk5}
\end{subfigure}
\hspace{0.1\textwidth}
\begin{subfigure}[c]{0.4\textwidth}
\includegraphics[width=\textwidth]{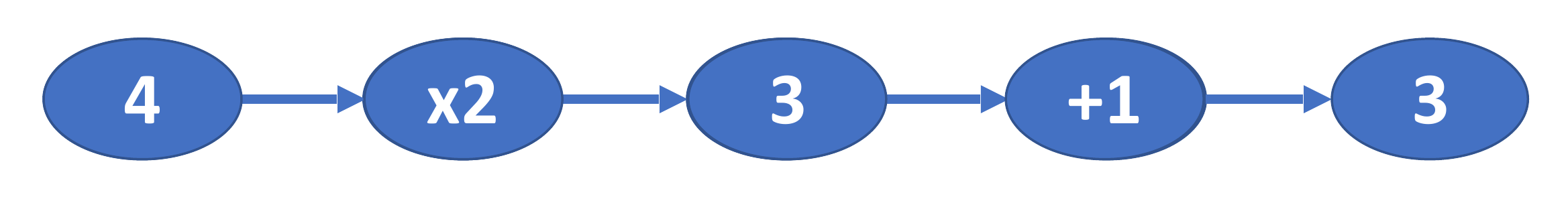}
\caption{} \label{subfig:embthunk6}
\end{subfigure}
\begin{subfigure}[c]{0.4\textwidth}
\includegraphics[width=\textwidth]{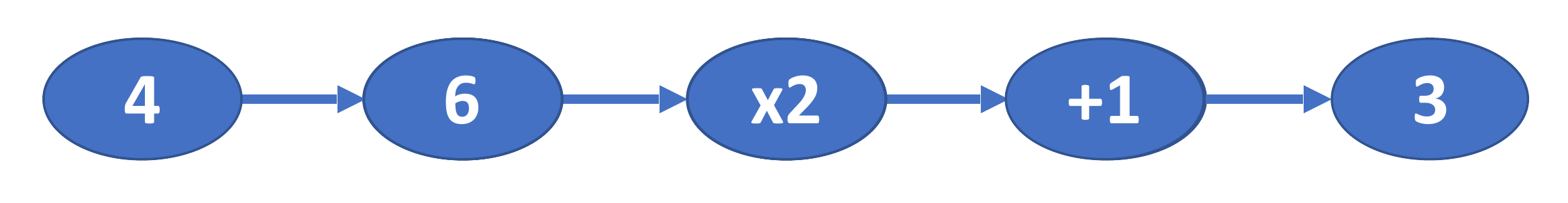}
\caption{} \label{subfig:embthunk7}
\end{subfigure}
\hspace{0.1\textwidth}
\begin{subfigure}[c]{0.4\textwidth}
\includegraphics[width=\textwidth]{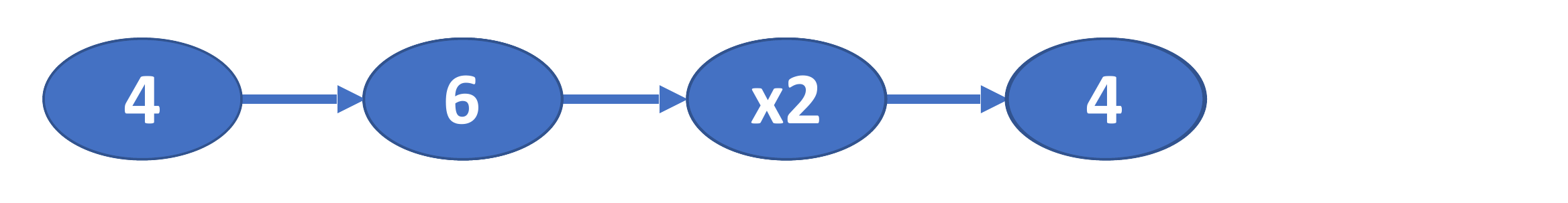}
\caption{} \label{subfig:embthunk8}
\end{subfigure}
\begin{subfigure}[c]{0.4\textwidth}
\includegraphics[width=\textwidth]{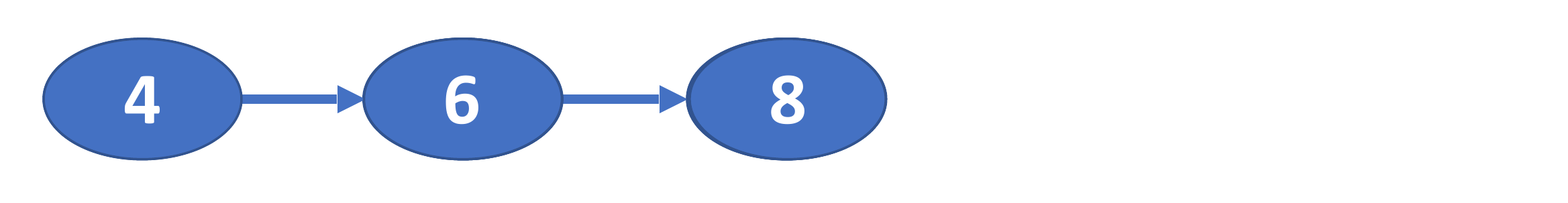}
\caption{} \label{subfig:embthunk9}
\end{subfigure}
\caption{An execution trace of a correct implementation for the compound version of our toy system.}
\label{fig:listToyRight}
\end{figure*}
Returing to our \texttt{addOne}/\texttt{double} example, we now embed the work to be done in the data-structure itself, and it doesn't matter which threads actually carry out the work: there is a consistent ordering on operations as soon as the first operation posts.

We again initialize the system with \texttt{List(1,2,3)} (Figure \ref{subfig:embthunk1}).
Here, \texttt{addOne} arrives first (Figure \ref{subfig:embthunk2}), and manages to process the first node (Figure \ref{subfig:embthunk3}), before \texttt{double} arrives (Figure \ref{subfig:embthunk4}) and catches up (Figure \ref{subfig:embthunk5}).
But now, instead of blindly forging ahead, \texttt{double} ensures the tail of the list is forced after each step, before proceeding with its own work (Figure \ref{subfig:embthunk6} - \ref{subfig:embthunk9}).

\section{Library Design\label{sec:lockfreelib}}
With this conceptual framework in place, we consider the practical question of an implementation based on LLX/SCX.
We will first consider adapting the classic linked list, from the definition provided on lines 2 and 3 of Listing \ref{sublst:adt}.
\begin{listing*}
\begin{lstlisting}[mathescape=true]
def Node[E]::insert(i:Int,e:E,l:Node[E])$\to$Node[E] {
  if (i == 0) { Link(e, l) }
  else {
    l match {
      | Link(h, t) => Link(h,insert(i-1,e,t))
      | Nil => throw IndexOutOfBoundsException
    }
  }
}
\end{lstlisting}
\caption{An \texttt{insert} operation for the \texttt{List} from Listing \ref{sublst:adt}.}
\label{lst:listInsert}
\end{listing*}
We now define an \texttt{insert} operation in Listing \ref{lst:listInsert}.
The \texttt{insert} operation returns a new \texttt{List}, in which an element, \texttt{e}, has been inserted at depth \texttt{i}.
If \texttt{i == 0}, we can simply link in our new element.
Otherwise, we recurse into the tail, (unless the list is \texttt{Nil}).

We observe that the recursive call to \texttt{insert} defines the boundary where our simple linearizable operation must be stitched together into a compound one.

\begin{listing*}
\begin{lstlisting}[mathescape=true]
type Node[E] = Cell(m:Bool,o:Op,data:Imm[E]|Def[E]) // Extra fields save metadata for LLX/SCX
type Imm[E]  = Link(h:E, t:Node[E])
              | Nil
type Def[E]  = Insert(i:Int,e:E,l:Node[E])
              | Remove(i:Int,l:Node[E])
              | Snapshot(l:Node[E])
type Ref[E]  = Final(n:Node[E]) // Same contents
              | Mut(n:Node[E])   // Different behavior
\end{lstlisting}
\caption{Altering the \texttt{List} definition from Listing \ref{sublst:adt} for our new execution model.}
\label{lst:listDefConcurrent}
\end{listing*}
It takes a little more boilerplate to define our lock-free version (Listing \ref{lst:listDefConcurrent}).
The core (Lines 2-3) is the same, but our nodes have been split along two axes.
First, we need to define a special memory cell with the extra fields needed to support LLX/SCX (Line 1).
Second, in addition to our original, immediate nodes, we also define deferred nodes for each operation we want to support (Lines 4-6).
Finally, we define a reference (Lines 7-8), which will provide a more ergonomic interface to LLX/SCX.

\begin{listing*}
\begin{lstlisting}[mathescape=true]
def Insert[E]::apply(provider,ctx)$\to$Imm[E] {
  if (i == 0) { Link(e, l) }
  else {
    val lRef = provider.readStrict(l, ctx)
    lRef() match {
      | Link(h, t) => Link(h,Node(Insert(i-1,e,t)))
      | Nil => throw IndexOutOfBoundsException
    }
  }
}
\end{lstlisting}
\caption{A \texttt{insert} operation for the \texttt{List} from Listing \ref{lst:listDefConcurrent}.}
\label{lst:listInsertConcurrent}
\end{listing*}
Now, instead of simply making recursive calls to \texttt{insert}, our recursion passes through three separate methods: \texttt{Insert[E]::apply} calls \texttt{Provider::readStrict[E]} on a child node, which in turn potentially calls \texttt{Def[E]::force}, which finally loops back to calling the \texttt{apply} method of whatever child node is being read and forced.
Unlike before, this recursive loop is not over the steps to the original call to \texttt{insert}, but rather over all of the deferred nodes which must be forced before the first step of \texttt{insert} can be computed.
Meanwhile, the next step of the \texttt{insert} itself is deferred until some later time.

Taking a closer look, the bulk of \texttt{insert}'s old behavior is now defined in the \texttt{apply} method, called when an \texttt{Insert} node is forced (Listing \ref{lst:listInsertConcurrent}).
Besides that, our new version is almost identical to the old, with two key changes.
First, instead of recursively calling \texttt{insert} directly, we create a new \texttt{Insert} node in our tail, to be invoked later.
Second, we rely on a forced read (Line 4) to fetch the current head of the list.

\begin{listing*}
\begin{lstlisting}[mathescape=true]
def Provider::readStrict[E](n:Node[E],ctx)$\to$Ref[E] {
  this.llx(n,ctx) match {
    | Failed => retry
    | Finalized =>
      n.state match {
        | imm:Imm[E] => Final(now)
        | def:Def[E] => escape // This work is done 
      }
    | Success(cloned) =>
      cloned.state match {
        | imm:Imm[E] => Mut(now)
        | def:Def[E] => this.forgetLast
          this.readStrict(def.force(n,this),ctx)
      }
  }
}
\end{lstlisting}
\caption{Defining the \texttt{readStrict} operation used in Listing \ref{lst:listInsertConcurrent}.}
\label{lst:readStrict}
\end{listing*}
The implementation of that forced read is given in Listing \ref{lst:readStrict}:
\begin{description}
\item[(Line 2)] Begin with an LLX.
\item[(Line 3)] If it fails, retry the current simple linearizable operation.
\item[(Line 6)] If the node is finalized and immediate, return a Final reference.
\item[(Line 7)] If the node is finalized and deferred, it represents a computation that was subsumed as an optimization by a computation further up the tree, and so escape (triggers a retry further up the call stack).
\item[(Line 11)] If the node is active and immediate, return a Mutable reference.
\item[(Lines 12 - 13)] Finally, if the node is active\footnote{In other words: ``not finalized''.} and deferred, wipe the read from the current readset, force its evaluation, and then retry only the current read.
\end{description}

\begin{listing*}
\begin{lstlisting}[mathescape=true]
def Def[E]::force(n:Node[E], provider)$\to$Node[E] {
  val ctx = provider.freshContext
  val thisRef = provider.read(n, ctx)
  thisRef() match {
    | this if !thisRef.final =>
      if ( this instanceof Snapshot[E] ){this.makeFinal}
      thisRef(provider, ctx) = this.apply(provider, ctx)
    | this if thisRef.final => escape // This work is done
    | _ => retry
  }
}
\end{lstlisting}
\caption{The wrapper code that acts as a trampoline for forcing evaluation of a node.}
\label{lst:forceDef}
\end{listing*}
Forcing the evaluation of a deferred node proceeds in five steps (Listing \ref{lst:forceDef}):
\begin{description}
\item[(Line 2)] Begin a new simple linearizable operation, tied to this stack frame (and any recursive calls it makes).
\item[(Line 3)] Perform an LLX, and wrap the result in a \texttt{Ref}.
\item[(Line 5)] Ensure that the memory cell whose contents we're forcing actually contains this node (and that it hasn't already been forced).
\item[(Line 6)] If this operation is a snapshot, we mark the cell to be finalized as part of the next SCX.
\item[(Line 7)] Call the \texttt{apply} method that actually evaluates the new state for the node, and assign the result to our reference.
\end{description}
We note that Line 8 is handling the same type of subsumption as Line 7 of \texttt{readStrict} (Listing \ref{lst:readStrict}).

\begin{listing*}
\begin{lstlisting}[mathescape=true]
def Mut[E]::=(provider,ctx,v:Def[E]|Imm[E])$\to$Node[E] {
  if (provider.scx(n,v,ctx)) { n }
  else { retry }
}

def Final[E]::=(provider,ctx,v:Def[E]|Imm[E])$\to$Node[E] {
  if(provider.vlx(ctx)) { Node(v) }
  else { retry }
}
\end{lstlisting}
\caption{The different assignment operators for the mutable and final references.}
\label{lst:mutFin}
\end{listing*}
Finally, although the assignment through our reference API (Line 7) would here be implemented as an SCX (as we've already checked that the read wasn't finalized), we should discuss how the reference API handles both the mutable and immutable cases (Listing \ref{lst:mutFin}).
The answer is quite simple.
Mutable references try to handle assignment with an SCX, and then return the cell if it was correctly set (Lines 2-3).
Final references handle assignment the functional way: they really just validate the read-set with a VLX (a third primitive described in \cite{BrownLLXSCX}, which validates a sequence of LLXs when no writes are necessary) and allocate a new memory cell (Line 7).

\subsection{Snapshots and Reads}
\begin{listing*}
\begin{lstlisting}[mathescape=true]
def Snapshot[E]::apply(provider,ctx)$\to$Imm[E] {
  val lRef = provider.readStrict(l, ctx)
  if(lRef.isFinal) { lRef }
  else {
    lRef() match {
      | case Nil => Nil
      | case Link(h, t) =>
        val tRef = provider.readLazy(t, ctx)
        val newT = tRef() match { 
          | Snapshot(_) => t // Snapshot is idempotent.
          | _:Imm[E] if tRef.isFinal => t // Snapshot is idempotent.
          | _ => Node(Snapshot(t)) // Real work to do.
        }
        Link(h, newT)
    } 
  }
}
\end{lstlisting}
\caption{A \texttt{Snapshot} operation for the \texttt{List} from Listing \ref{lst:listDefConcurrent}.}
\label{lst:listSnapConcurrent}
\end{listing*}
The heavy lifting to implement the \texttt{Snapshot} operation is performed by special cooperation with \texttt{Def[E]::force} (as seen in Listing \ref{lst:forceDef}); however, it is still necessary to provide a stub implementation for \texttt{Snapshot[E]::apply} (Listing \ref{lst:listSnapConcurrent}), which handles snapshotting each child node when forced.
Special care is taken to avoid duplicating work when a \texttt{Snapshot} has already been performed (Lines 3, 9-13).

Arbitrary read operations may then be implemented by first making a private \texttt{Snapshot}, and accessing every child by means of \texttt{readStrict}.
Read operations which only access the tree up to a constant depth may also be implemented using LLX and VLX; however, the design constraints of Section \ref{sec:shortops} still apply.

\section{Defining the Transformation\label{sec:lfFormal}}
We consider trees whose node definitions are a disjunction of product types, each containing 0 or 1 values, of parametric type \texttt{E}, and 0 or more child nodes:
\begin{lstlisting}[mathescape=true,numbers=none]
type Node[E] = Case$_0$(value:E$^?$, children:Node[E] $\ldots$)
              $\vdots$
              | Case$_n$(value:E$^?$, children:Node[E] $\ldots$)
\end{lstlisting}
We note that our \texttt{List[E]} definition on Lines 1 - 3 of Listing \ref{sublst:adt} satisfies this definition.
We further note, that it does not exclude, for example, a \texttt{Map[K,V]} parameterized in both the key and value types, as conceptually, these can be stored in a single tuple-type, \texttt{(K,V)}.

We also restrict the form of update operations on those nodes, to satisfy four conditions.
\begin{definition}[No Mutation\label{def:nomut}]
The operation must not mutate any state -- in other words, it must be purely functional.
\end{definition}
\begin{definition}[Affine Access\label{def:affacc}]
During a single operation, each node can be passed as an argument to at most 1 function call.
This is equivalent to following affine typing rules for nodes -- as a matter of programmer discipline, if unsupported by the language.~\cite{bernardy2017retrofitting, walker2005substructural} 
\end{definition}
\begin{definition}[Immediate Return\label{def:immret}]
 Every code path through the functions called by the operation must return by constructing a new node, or returning a node which was already read in its body.
In other words, function calls may not appear in tail position.
\end{definition}
\begin{definition}[Fixed Operation Set\label{def:fixedops}]
All supported update operations must be known at compile time.
If there are $k$ operations, then for $0 \le i \le k$, we define \texttt{Node[E]::Operation$_i$} as having a signature consisting of 0 or 1 arguments of arbitrary non-\texttt{Node} type \texttt{A}, followed by 1 or more \texttt{Node}s, and returning a \texttt{Node}.
\begin{lstlisting}[mathescape=true,numbers=none]
def Node[E]::Operation$_i$(arg:A$^?$, trees:Node[E] $\ldots +$)$\to$Node[E]
\end{lstlisting}
\end{definition}
In keeping with Section \ref{sec:shortops}, we also strongly recommend that each function restrict itself to reading from at most a constant number of descendants of its argument nodes, before returning, to avoid starvation.

Then, given a type, \texttt{Cell[T]}, supporting LLX, and SCX operations, we can define a mutable concurrent version of the tree as follows:
\begin{lstlisting}[mathescape=true,numbers=none]
type Node[E] = Cell[Deferred[E] | Immediate[E]]
type Immediate[E] = Case$_0$(value:E$^?$, children:Node[E] $\ldots$)
                   $\vdots$
                   | Case$_n$(value:E$^?$, children:Node[E] $\ldots$)
type Deferred[E] = Operation$_0$(arg:A$_0^?$, children:Node[E] $\ldots +$)
                 $\vdots$
                 | Operation$_k$(arg:A$_k^?$, children:Node[E] $\ldots +$)
\end{lstlisting}
Note that every supported operation is promoted from a function to data type which is a subtype of \texttt{Node}.
This is the reason for the \textbf{Fixed Operation Set} requirement (Definition \ref{def:fixedops}).

Then, for each \texttt{Operation$_i$}, we move the body of the implementation into an \texttt{apply} method, with signature:
\begin{lstlisting}[mathescape=true,numbers=none]
def Operation$_i$::apply(provider:SCXProvider, ctx:LLXContext)$\to$Immediate[E]
\end{lstlisting}
Then, each reference to a \texttt{child}, of type \texttt{Node[E]}, in the original implementation can be replaced with \texttt{provider.readStrict(child, ctx)()}.

Note that the return-type of \texttt{Operation$_i$::apply} is \texttt{Immediate[E]}.
This is the reason for the \textbf{Immediate Return} requirement on the untransformed function definition (Definition \ref{def:immret}).
The significance of this restriction is discussed in Section \ref{sec:lfBottleCommute}

Most tree data structures define some kind of ``root-holder'' (as we saw in Listings \ref{lst:bheaptypes}, \ref{lst:seqbheaphelpers}, and \ref{lst:seqbheap}), providing more structured access to the tree as a whole, and expose only a limited subset of the operations as part of the public API, e.g. in order to maintain certain invariants or other such conveniences.
The root holder's operations can simply construct a \texttt{Deferred[E]} and atomically replace the root (e.g. via CAS).

\subsection{Correctness}
\label{sec:lockfreecorrect}
We now present an informal argument of correctness for trees implemented using this transformation.
In particular, we must show that our concurrent mutable implementations preserve the sequential semantics of the pure functional specification (i.e. are linearizable~\cite{HerlihyW1990}), while remaining lock-free.

\begin{theorem}
The concurrent behavior of trees implemented using the transformation rules in Section \ref{sec:lfFormal} is linearizable according to the sequential definition used as input to the transformation.
\end{theorem}
\begin{proof}
The argument for linearizability is fourfold, and simple.
First, the body of each operation is forbidden from independently mutating any state (Definition \ref{def:nomut}).
Second, the function evaluation trampoline, \texttt{Def[E]::force} (Listing \ref{lst:forceDef}), together with the rewrite of the function body, ensures that each function evaluation step consists of a single sequence of LLXs followed by either an SCX or a VLX, which together constitute a simple linearizable operation.
Third, by explicitly placing the suspension of any lazy recursion into the structure of the tree itself, any other threads must help to force its evaluation before they can proceed past the resulting node.
Finally, the \textbf{Affine Access} restriction (Definition \ref{def:nomut}) enforces a tree topology on non-snapshotted nodes, meaning that a node can only be accessed by a single path from the root, eliminating race conditions wherein two updating threads arrive at the same node while forcing different operations. \qed
\end{proof}

\begin{theorem}
The concurrent behavior of trees implemented using the transformation rules in Section \ref{sec:lfFormal} is lock-free.
\end{theorem}
\begin{proof}
The argument for lock-freedom is twofold, and also simple.
First, LLX/SCX is lock-free, thus each simple linearizable operation is lock-free.~\cite{BrownLLXSCX}
Second, when the result of an LLX is a \texttt{Deferred[E]}, rather than an \texttt{Immediate[E]}, the reading thread is able to call \texttt{Def[E]::force}, and either make progress on its own by successfully forcing the operation, or retry its own operation after discovering that some other thread has (made progress and) successfully forced it in the interim.
Thus each compound linearizable operation is also lock-free. \qed
\end{proof}

We can also make a stronger claim about progress: for any individual \texttt{Operation$_i$}, its execution is \emph{wait-free} if the sequential definition had a finite number of steps.~\cite{WaitFree}
In other words, any thread attempting to force evaluation of a \texttt{Deferred[E]} node is guaranteed that the operation will complete after a finite number of steps \emph{iff} the operation could have been completed in a finite number of steps during a single-threaded execution\footnote{We also assume a finite number of threads are executing simultaneously}.

\begin{theorem}
The concurrent behavior of each \texttt{Operation$_i$} implemented using the transformation rules in Section \ref{sec:lfFormal} is lock-free.
\end{theorem}
\begin{proof}
There are four conditions under which an attempt by a single thread to execute \texttt{Def[E]::force} may fail.
To show that forcing is wait-free, we must show that the number of failures is bounded.
\begin{description}
\item[1. Unsuccessful Root LLX] If the LLX on the node containing the \texttt{Def[E]} to be forced does not return \texttt{Success}, then another thread has already forced or subsumed its execution. Therefore no more retries are required, and the execution time is bounded.
\item[2. Mismatched Root LLX] If the LLX on the node succeeded, but the node no longer contains the \texttt{Def[E]} to be forced, then another thread has already forced its execution. Therefore no more retries are required, and the execution time is bounded.
\item[3. Unsuccessful Child LLX] If an LLX on a child node returns \texttt{Failed}, then another thread has modified it concurrently, and a retry is required; however, when operations are defined according to our rules, this can only happen as the result of a subtree being forced.
None of our operation's subtrees can move from \texttt{Imm[E]} to \texttt{Def[E]} until our operation has been forced.
Therefore, if our operation has a bounded number of subtrees, we can only experience a bounded number of failed child LLXs, and the operation will complete in a bounded number of steps.
If our operation has an unbounded number of subtrees, then our pre-condition on the boundedness of the sequential definition has been violated.
\item[4. Unsuccessful Root SCX] If the SCX on the node containing the \texttt{Def[E]} to be forced does not return \texttt{true}, then either another thread has already forced or subsumed its execution (and the argument is as for cases 1 and 2), or another thread has concurrently forced a child (and the argument is as for case 3).
\end{description}
Thus in all cases, if our preconditions are satisfied, forcing takes a bounded number of steps, and is wait-free.
\qed
\end{proof}

We note further that this means that if each operation on the root holder is wait-free, than the entire data structure is wait-free.
If the operations on the root holder are merely lock-free, and respect the design constraints from Section \ref{sec:nolongops}, and the number of processes is bounded, than the entire data structure is wait-free \emph{in expectation}.

\section{Lazy, Eager, and Randomized Evaluation Strategies}
\label{sec:eagerForce}
As our trees are defined purely functionally, we are free to choose \emph{when} to execute our operations, except insofar as client code requires a return value (so-called \emph{strictness points}).~\cite{Launchbury93anatural,strictnessAnalysis}
Thus far our implementation has defaulted to laziness wherever possible, which despite offering opportunities for optimization, can also make runtime performance quite unpredictable.

We can allow individual data structure operations to opt-in to strict evaluation by modifying the return types of the \texttt{Operation$_i$::apply}s from \texttt{Immediate[E]} to \texttt{(Immediate[E],Maybe[(Node[E],Def[E])]}.
The type \texttt{(Node[E],Def[E])} can be viewed as a restricted form of continuation, consisting of a node to be forced \emph{iff} another thread has not already done so.~\cite{continuations,continuations2}
\begin{listing}
\begin{lstlisting}[mathescape=true]
def Def[E]::force(n:Node[E], provider)$\to$Node[E] {
  val ctx = provider.freshContext
  val thisRef = provider.read(n, ctx)
  thisRef() match {
    | this if !thisRef.final =>
      if ( this instanceof Snapshot[E] ){this.makeFinal}
      val (newThis, maybeCont) = this.apply(provider, ctx)
      val ret = (thisRef(provider, ctx) = newThis)
      
      maybeCont match {
        | Some((nextNode, nextDef)) =>
          nextDef.force(nextNode, provider)
        | None => // Be lazy as before
      }
      
      ret
    | _ => retry
  }
}
\end{lstlisting}
\caption{A modification to Listing \ref{lst:forceDef}, to support opt-in strict evaluation.}
\label{lst:forceDefEager}
\end{listing}
Listing \ref{lst:forceDefEager} shows how \texttt{Def[E]::force} (Listing \ref{lst:forceDef}) can be modified to cooperate with this change.
Lines 7 - 16 here replace the old Line 7.
After applying our operation (Line 7), we first assign the immediate result to \texttt{thisRef} as before (Line 8).
We next check the result included a continuation (Line 10), and, if so, recursively force its evaluation as well (Lines 11-12).
Finally, we return the result of updating \texttt{thisRef}, as before (Line 16).
No other changes are necessary; however, operations on the root-holder may choose to force the root once after posting their work, to ensure their workload is processed as eagerly as possible.

\subsection{Contention Diffusion by Randomized Forcing Orders}
One limitation of the evaluation strategy described so far is that if several threads attempt to force the same \texttt{Def[E]} node simultaneously, they will end up executing each \texttt{readStrict} in the same order, resulting in a performance pathology similar to the \emph{restart convoy} described in \cite{Bobba:2007:PPH:1250662.1250674}.
Although being lock free guarantees that if $p$ threads attempt to execute the same operation 1 of them will make progress, this has the corollary that $(p-1)$ of them will not do useful work, and these remaining $(p-1)$ will retry together on next \texttt{readStrict} in the sequence.

\begin{listing}
\begin{lstlisting}[mathescape=true]
def Provider::readStrictBulk[E](ns:Array[Node[E]],ctx)$\to$Array[Ref[E]] {
  val readSet = Buffer[Ref[E]]()
  val forceSet = Buffer[(Node[E],Def[E])]
  forEach(n $\in$ ns){
    this.llx(n,ctx) match {
      | Failed => retry
      | Finalized =>
        n.state match {
          | imm:Imm[E] => readSet += Final(now)
          | def:Def[E] => escape // This work is done 
        }
      | Success(cloned) =>
        cloned.state match {
          | imm:Imm[E] => readSet += Mut(now)
          | def:Def[E] => forceSet += (n, def)
        }
    }
  }
  if(forceSet.empty){
    readSet
  } else {
    forEach((n,def) $\in$ forceSet){
      def.force(n,this)
    }
    retry
  }
}
\end{lstlisting}
\caption{A bulk version of \texttt{readStrict} (Listing \ref{lst:readStrict}), with randomized forcing order for deferred results.}
\label{lst:readStrictBulk}
\end{listing}
We can alleviate this for operations which perform multiple \texttt{readStrict}s by randomizing the order in which each thread performs the nested calls to \texttt{Def[E]::force}.
We must be slightly careful about how we do this, because although the purely functional setting ensures that each call to \texttt{Def[E]::force} will be free of side effects on the others, the lock-free progress guarantees for LLX/SCX require a consistent ordering among LLXs.~\cite{BrownLLXSCX}
Therefore, we provide a new library function, \texttt{readStrictBulk} (Listing \ref{lst:readStrictBulk}), which given a sequence of \texttt{Node[E]}s, LLXs each in order, and partitions the results into \texttt{Imm[E]}s and \texttt{Def[E]}s.
Lines 4 - 17 of \texttt{readStrictBulk} follow closely the logic of Lines 2 - 15 from \texttt{readStrict} (Listing \ref{lst:readStrict}), except that each LLX result is assigned to a partition, rather than returned immediately.

If the whole read-set is immediate (Line 19), then it is returned (line 20), otherwise, the \texttt{Def[E]}s (the \emph{force-set}) are \texttt{force}d in a random order (Lines 20 - 22).
Rather than forgetting and rereading each LLX'd node from the force-set, we then simply retry the whole operation (Line 25), which is now guaranteed to have only immediate children.

\section{Evaluation: Braun Heaps Redux\label{sec:braunheaplf}}
We implemented a Braun Heap using the library described in the preceding section; and benchmarked it using a setup identical to that of the preceding chapter.
The \texttt{removeMin} operation was rewritten as an $O(log^2 n)$ single-pass operation that lazily leaves behind cleanup tasks embedded in the structure of the heap, sacrificing single-threaded asymptotics in favor of better behavior under contention.

\begin{figure*}
\centering
\includegraphics[width=\textwidth]{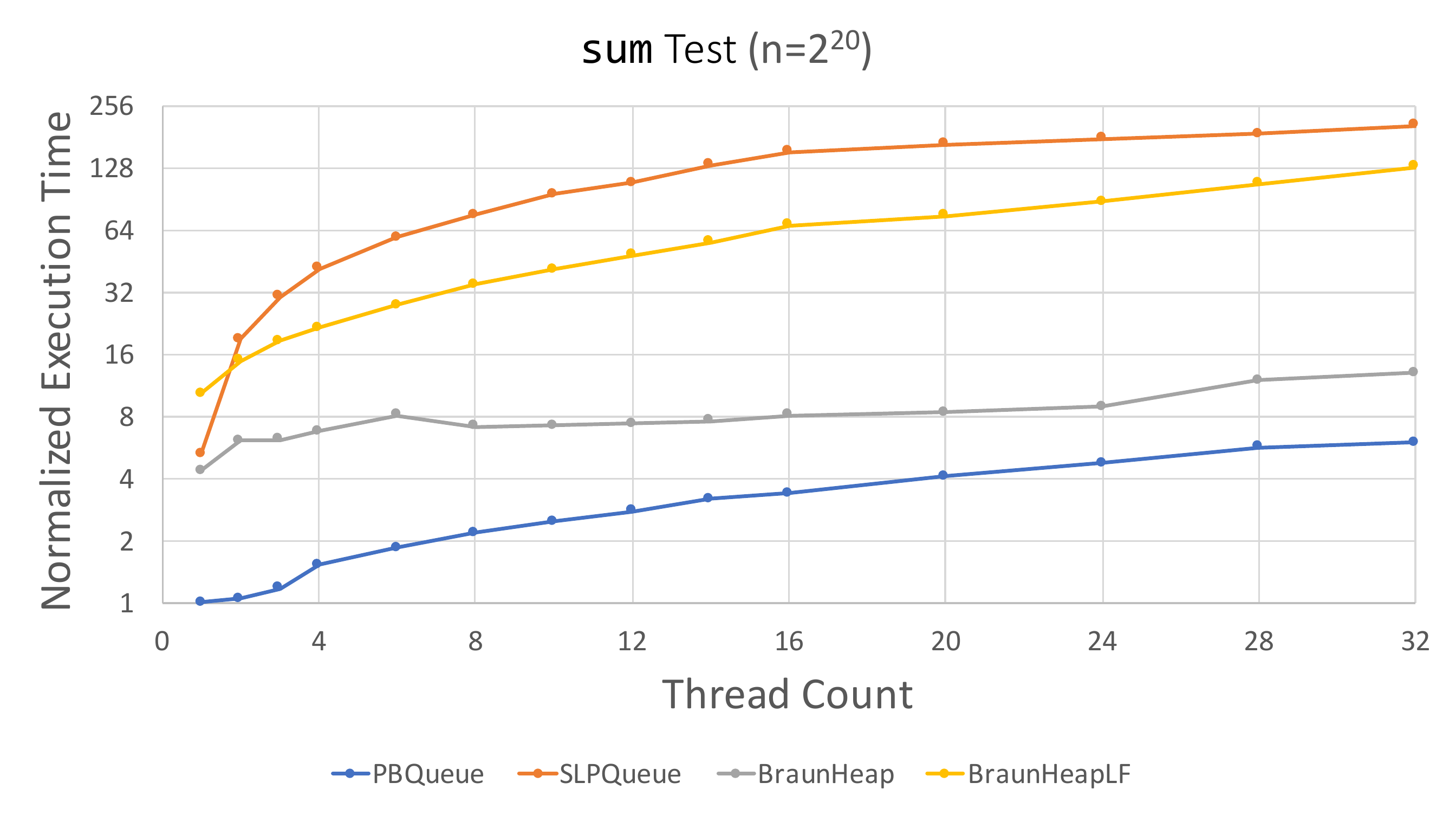}
\caption{The \texttt{sum} benchmark. Vertical axis is log-scale normalized execution times. Smaller is better.}
\label{fig:lfSumTest}
\end{figure*}
\begin{figure*}
\centering
\includegraphics[width=\textwidth]{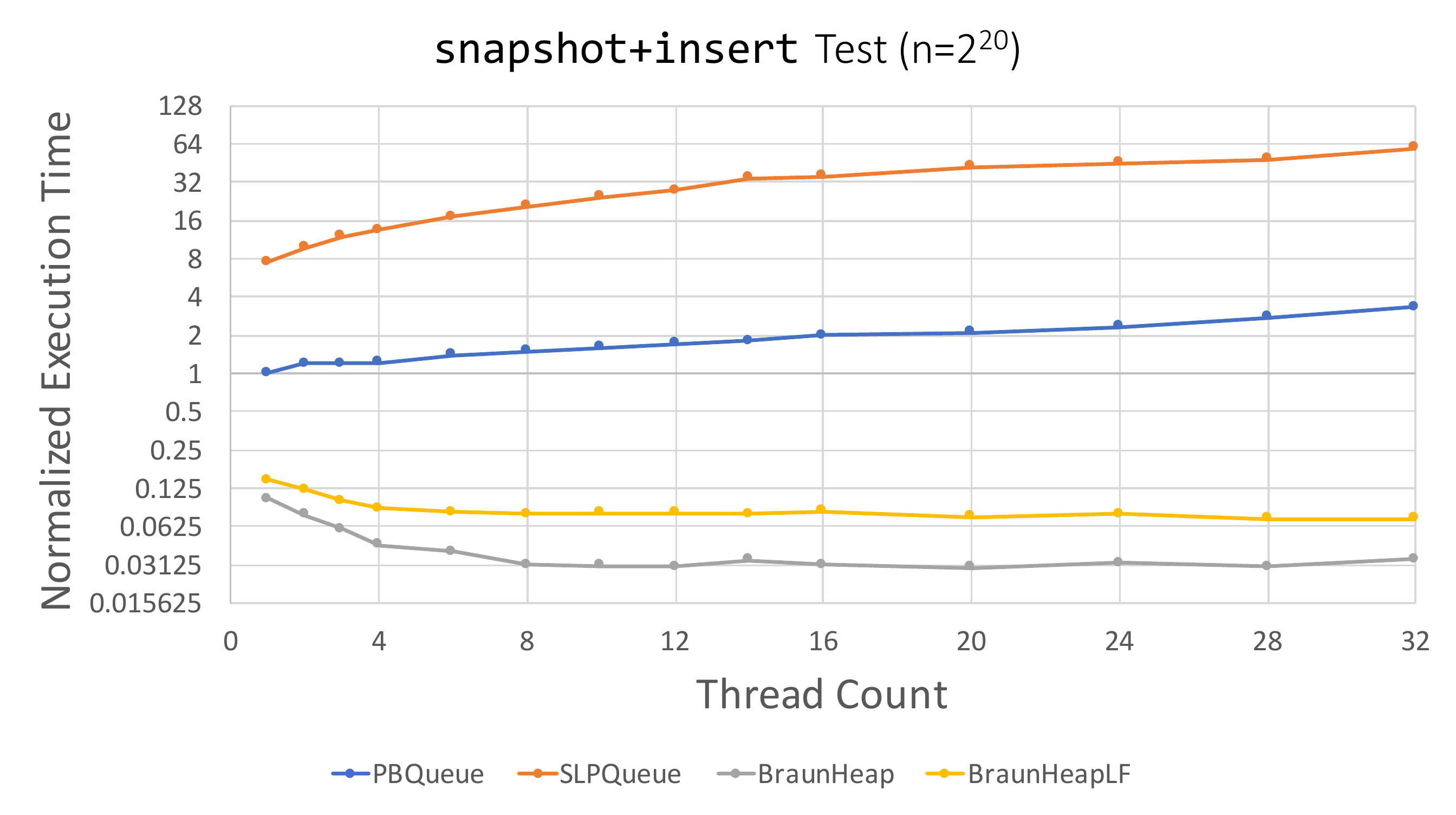}
\caption{The \texttt{snap+insert} benchmark. Vertical axis is log-scale normalized execution times. Smaller is better.}
\label{fig:lfSnapAndInsertTest}
\end{figure*}
\begin{figure*}
\centering
\includegraphics[width=\textwidth]{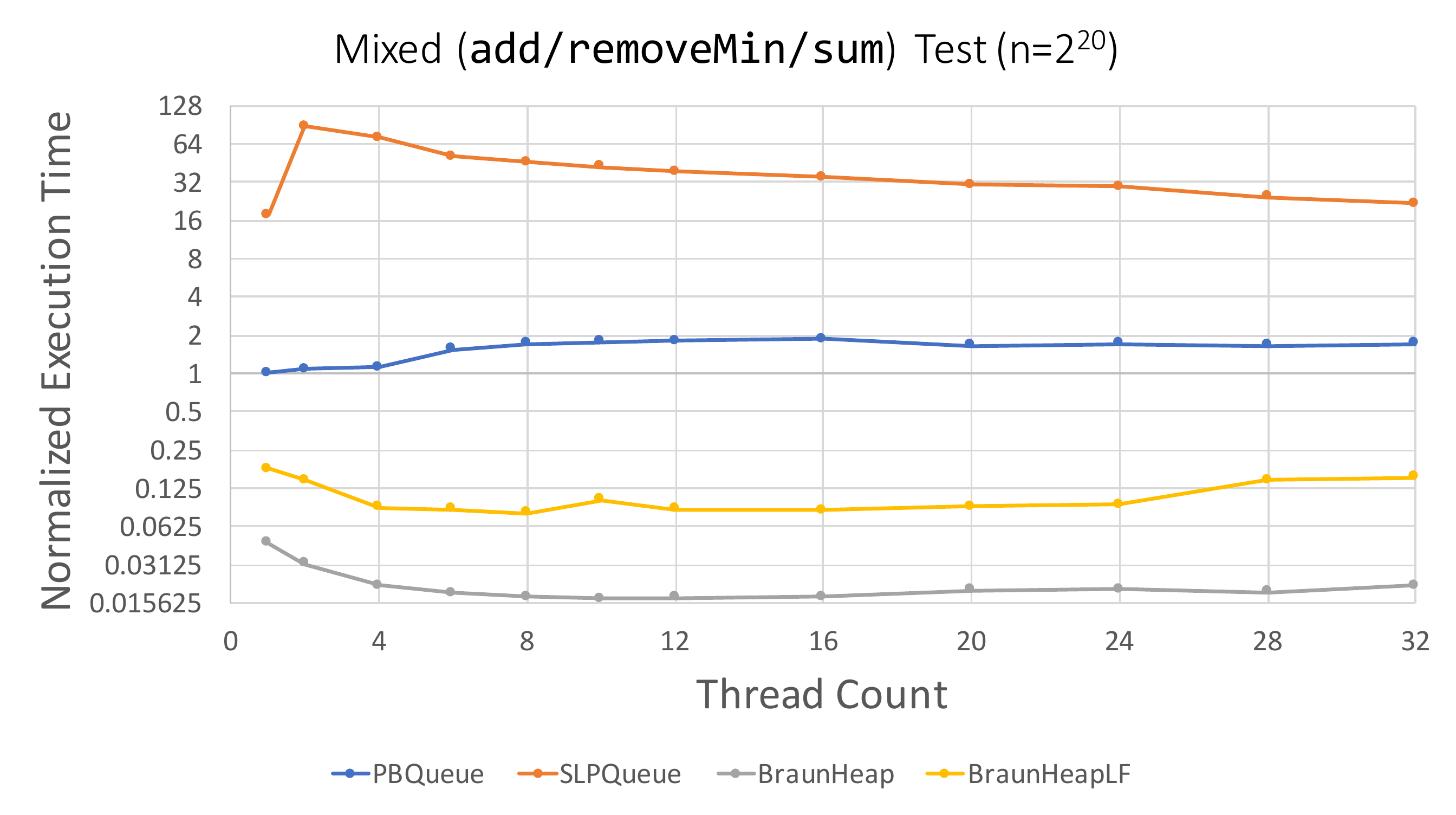}
\caption{The \texttt{add/removeMin/sum} benchmark. Vertical axis is log-scale normalized execution times. Smaller is better.}
\label{fig:lfMixedTest}
\end{figure*}
\begin{figure*}
\centering
\includegraphics[width=\textwidth]{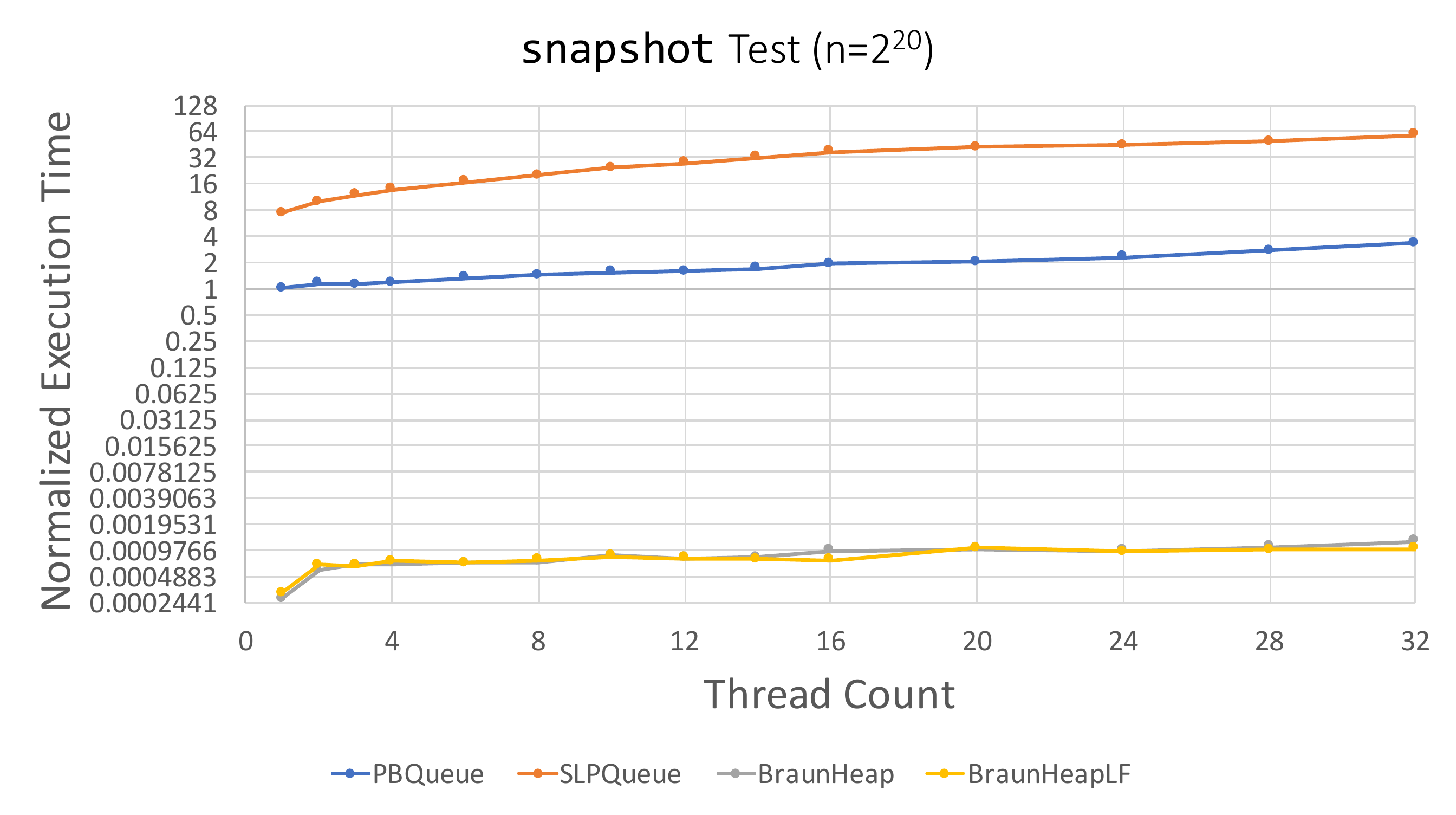}
\caption{The \texttt{snap-only} benchmark. Vertical axis is log-scale normalized execution times. Smaller is better.}
\label{fig:lfSnapOnlyTest}
\end{figure*}
\begin{figure*}
\centering
\includegraphics[width=\textwidth]{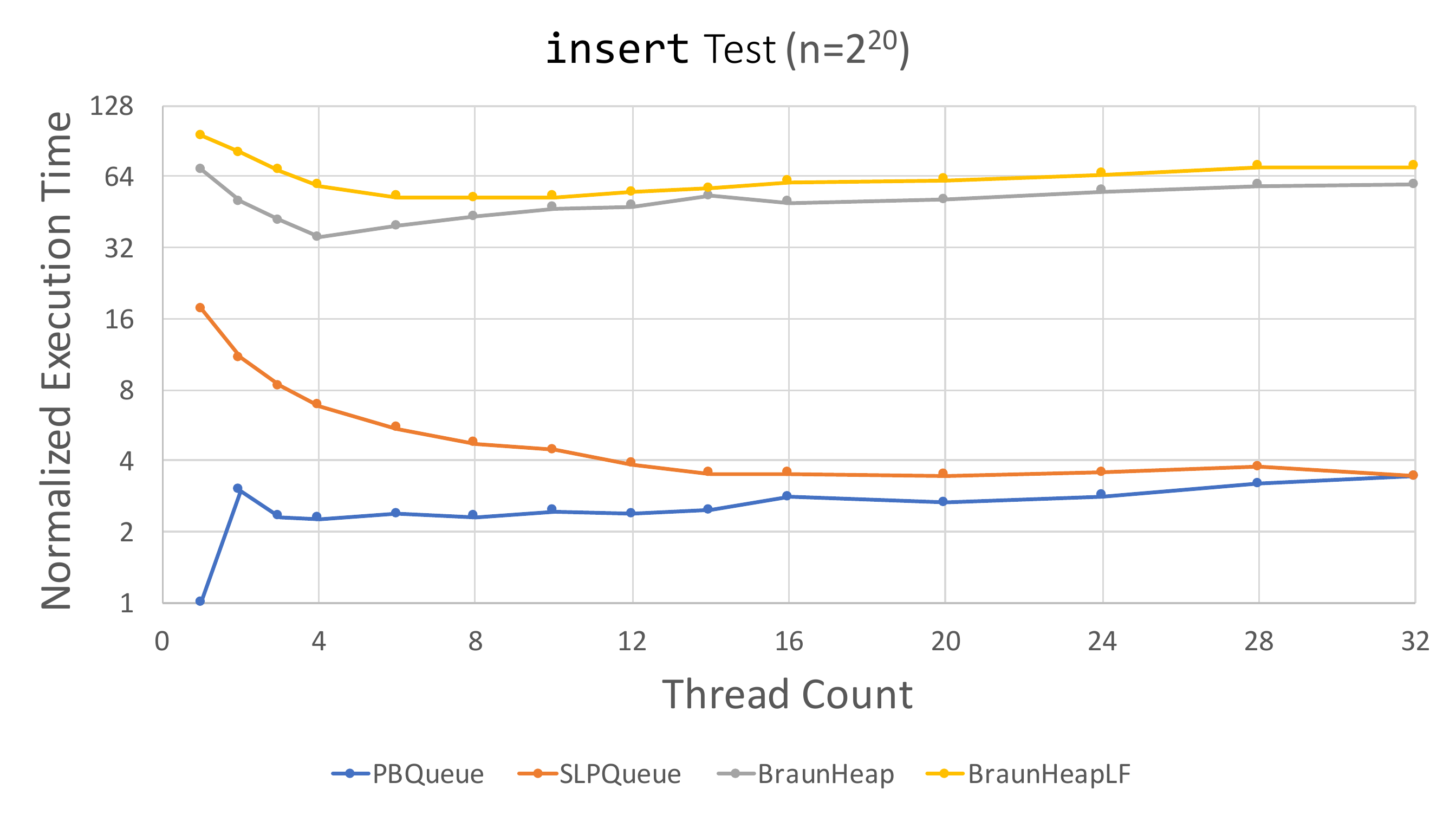}
\caption{The \texttt{insert} benchmark. Vertical axis is log-scale normalized execution times. Smaller is better.}
\label{fig:lfInsertTest}
\end{figure*}
\begin{figure*}
\centering
\includegraphics[width=\textwidth]{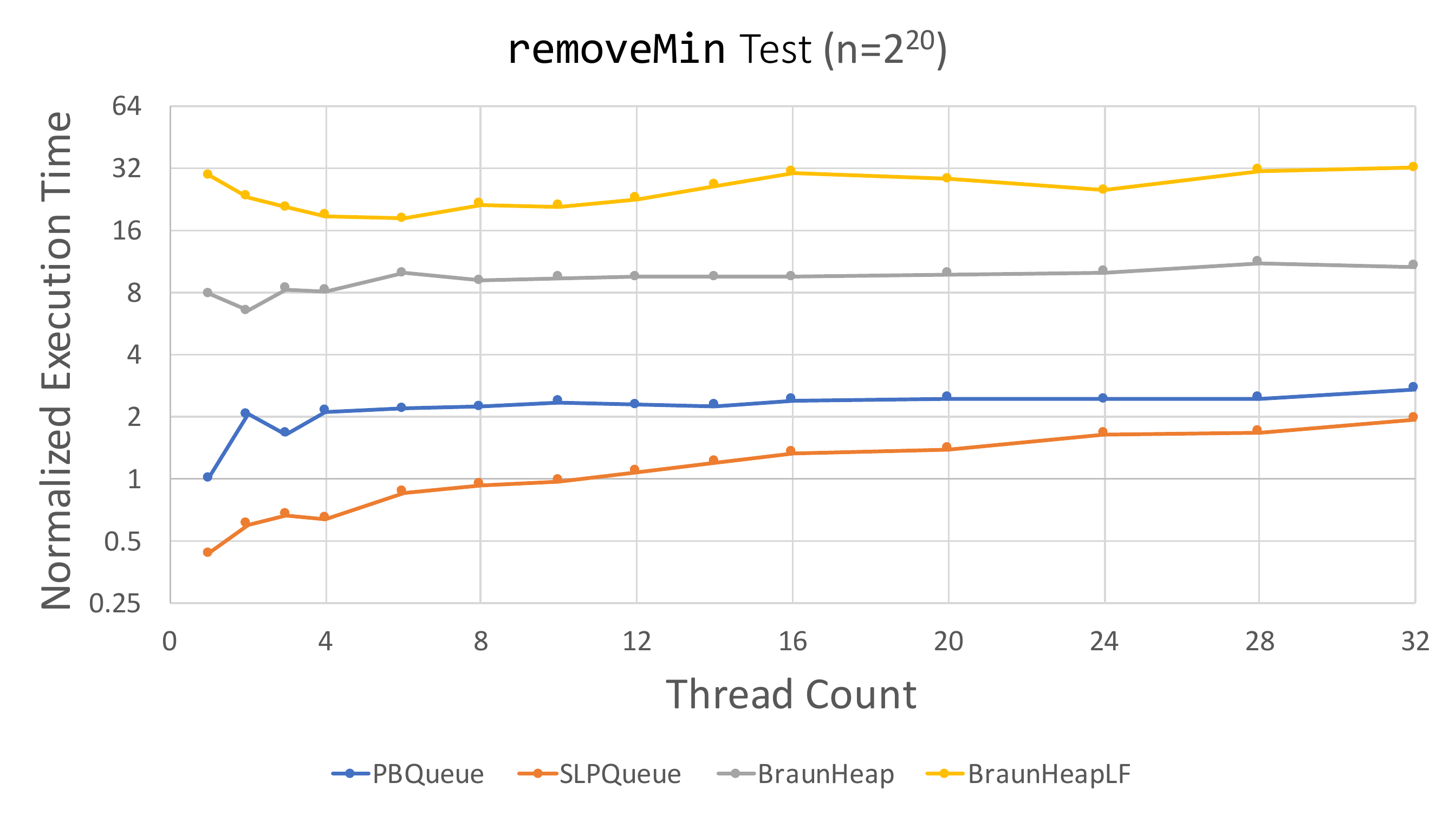}
\caption{The \texttt{removeMin} benchmark. Vertical axis is log-scale normalized execution times. Smaller is better.}
\label{fig:lfremoveMinTest}
\end{figure*}
The results for the \texttt{sum}, \texttt{snap+insert}, \texttt{add/removeMin/sum}, \texttt{snap-only} and \texttt{insert}, \texttt{removeMin} tests are presented in Figures \ref{fig:lfSumTest}, \ref{fig:lfSnapAndInsertTest}, \ref{fig:lfMixedTest}, \ref{fig:lfSnapOnlyTest}, \ref{fig:lfInsertTest}, and \ref{fig:lfremoveMinTest}, respectively.
In most cases, the performance of the lock-free Braun Heap implementation (yellow) closely tracks with the performance of the version based on hand-over-hand locking (grey).
In the case of \texttt{removeMin}, we attribute the performance differences to the asymptotically worse algorithm, although this should be mediated somewhat by the improved potential for concurrency.
In the case of \texttt{sum}, we attribute the performance differences to the increased memory allocation and reclamation overhead from needing to allocate a deferred \texttt{Snapshot} object at each node.
We believe that it should be possible to implement a version of the transformation that relies entirely on the finalized status of LLX/SCX cells to propagate snapshot operations, but leave this to future work.

\subsection{Development Costs}
\begin{figure*}
\centering
\includegraphics[width=\textwidth]{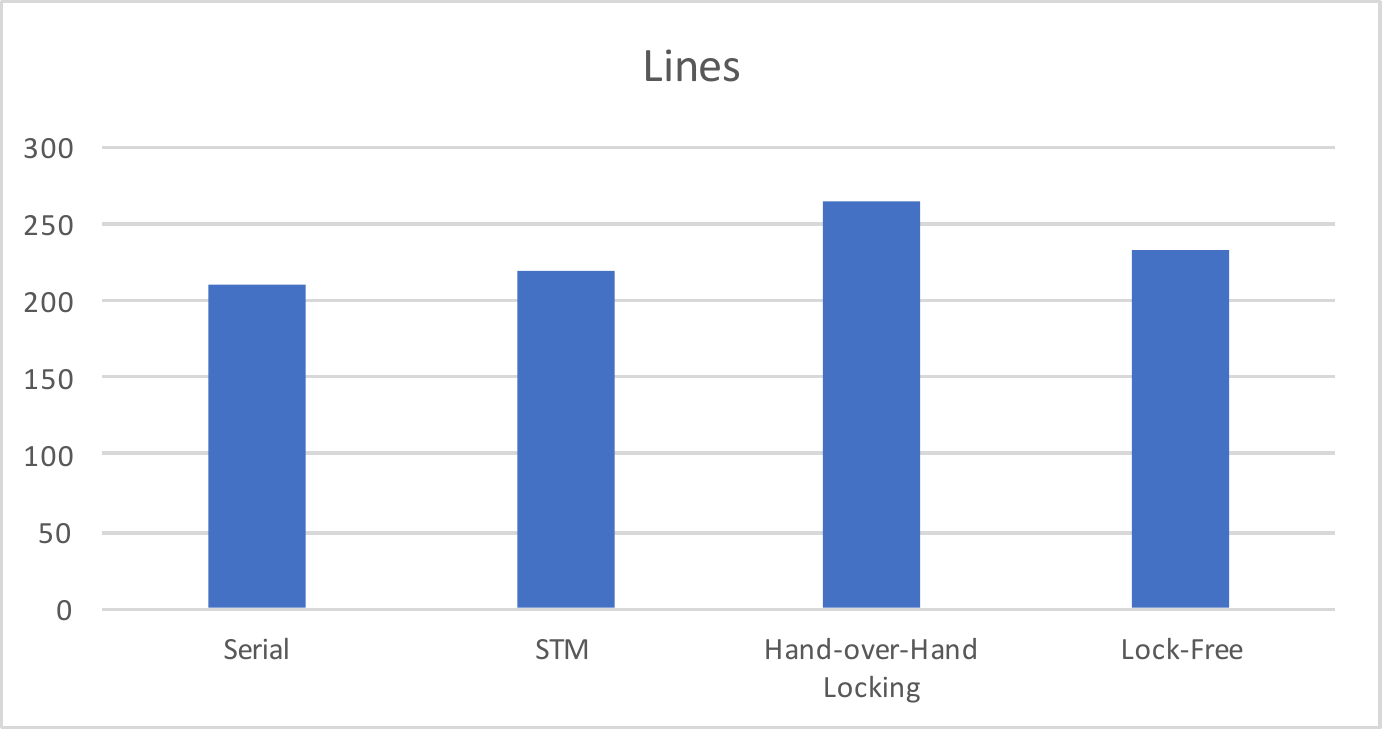}
\caption{Source-code statistics for several Braun Heaps, comparing the line-count of each implementation against a sequential version.}
\label{fig:SrcStatsBHLines}
\end{figure*}
\begin{figure*}
\centering
\includegraphics[width=\textwidth]{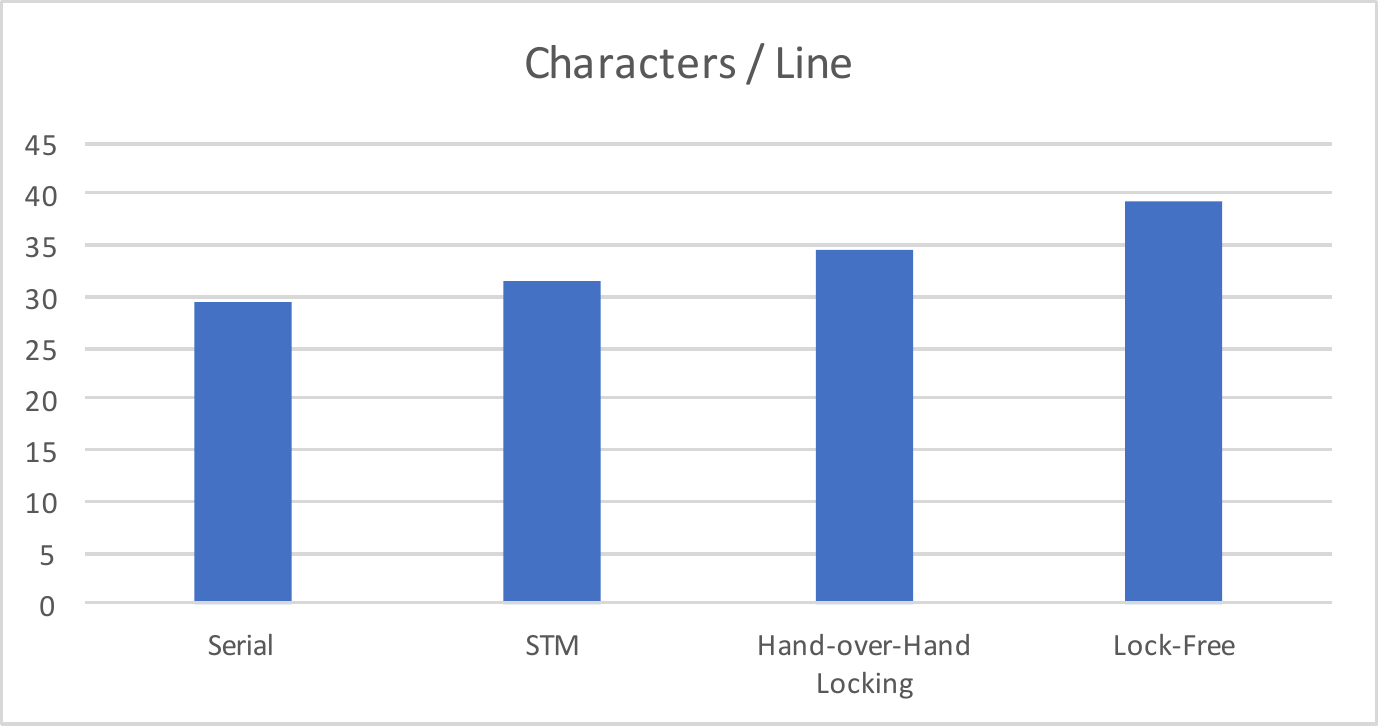}
\caption{Source-code statistics for several Braun Heaps, comparing the average characters / line of each implementation against a sequential version.}
\label{fig:SrcStatsBHCPL}
\end{figure*}
\begin{figure*}
\centering
\includegraphics[width=\textwidth]{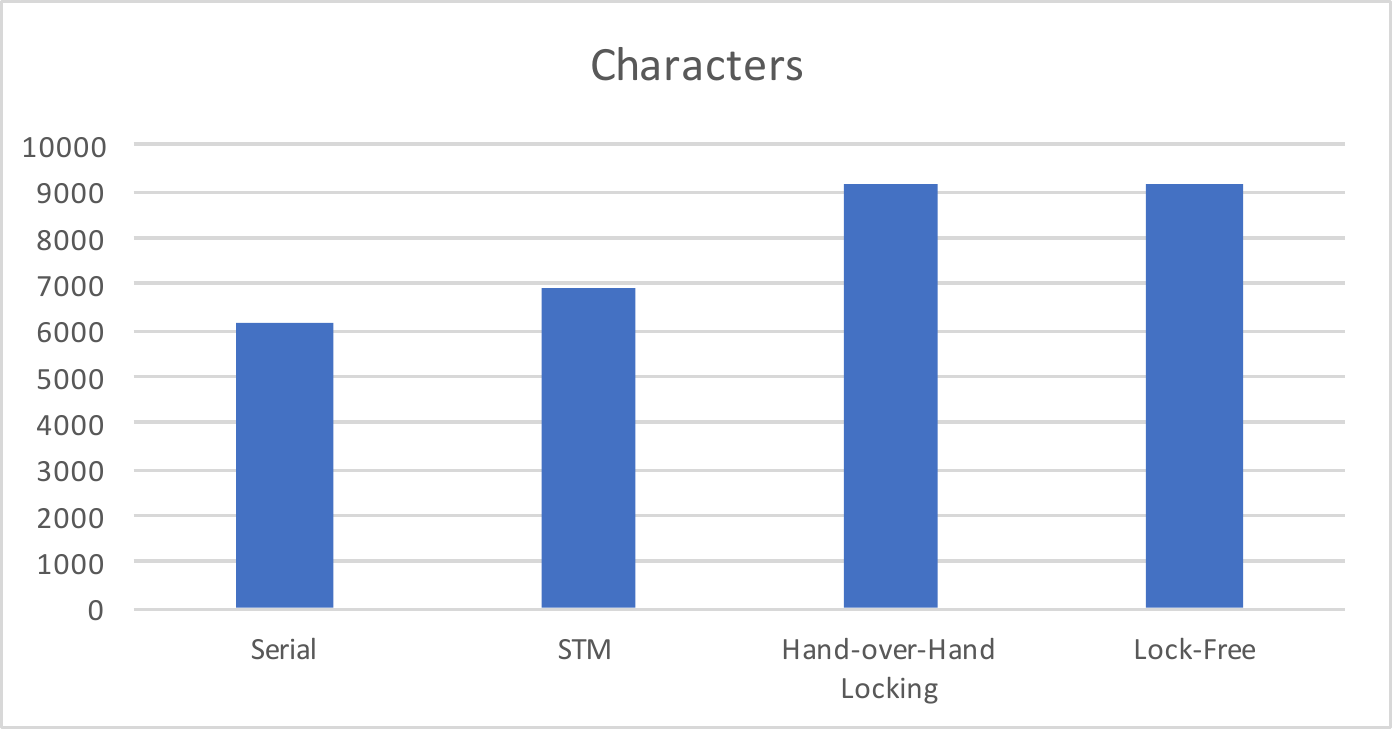}
\caption{Source-code statistics for several Braun Heaps, comparing the character-count of each implementation against a sequential version.}
\label{fig:SrcStatsBHChars}
\end{figure*}
\begin{figure*}
\centering
\includegraphics[width=\textwidth]{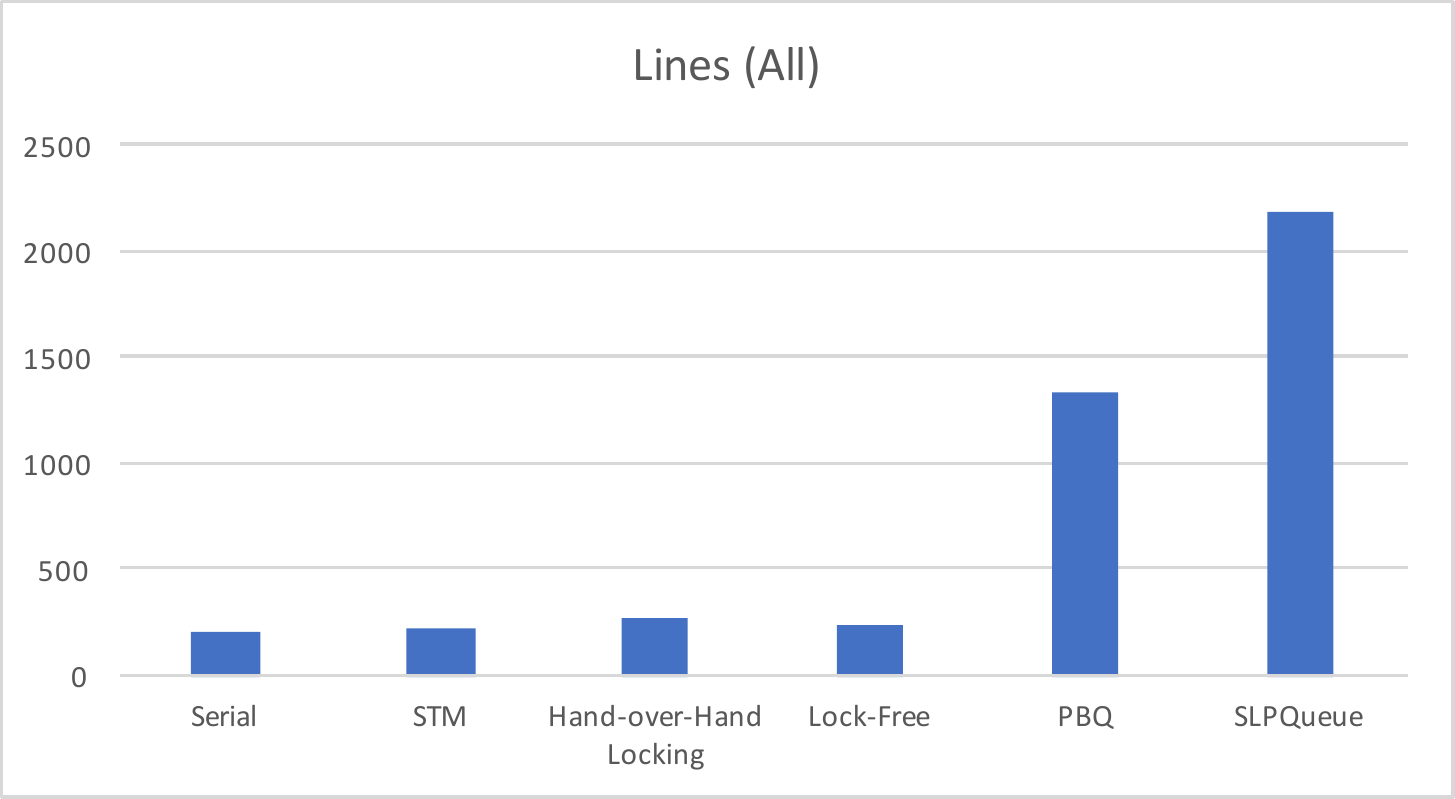}
\caption{Source-code statistics for several Braun Heaps, as well as the \texttt{PriorityBlockingQueue} and \texttt{SkipListPriorityQueue} used for benchmarking, comparing the line-count of each implementation against a sequential version.}
\label{fig:SrcStatsBHLinesAll}
\end{figure*}
Figure \ref{fig:SrcStatsBHLines} demonstrates the minimally intrusive nature of the changes required to implement data structures with this library, as compared to a single-threaded reference implementation, in terms of line count.
They are also compared against a ScalaSTM port of the reference implementation, and the hand-over-hand locking implementation from the preceding chapter.
The direct port to this library (``Lock-Free'') has a smaller line-count than the previous chapter's lock-based implementation (``Hand-over-Hand Locking''), although still larger than serial or STM versions.
While the small line-count overhead reflects minimal changes to declared control-flow (compared to locking), the use of the library does substantially increase the complexity of the type signatures for individual operations, and, as can be seen in Figure \ref{fig:SrcStatsBHCPL}, this is reflected in the increased overhead in average characters per line.
The overall length in character count is nearly identical for both the locking and lock-free versions (see Figure \ref{fig:SrcStatsBHChars}).

In Figure \ref{fig:SrcStatsBHLinesAll} we also compare the Braun Heaps to the priority queues used in benchmarking.
Although a direct comparison is biased, as \texttt{PriorityBlockingQueue} and \texttt{SkipListPriorityQueue} are written in Java rather than Scala, we believe the substantial difference in lengths is nevertheless illustrative of the reduced implementation complexity of our algorithm.

Anecdotally, we are also pleased to report that both the \texttt{List}, used as an example in Listings \ref{lst:listDefConcurrent} and \ref{lst:listInsertConcurrent}, and the \texttt{Braunheap}, benchmarked in this section, were debugged exclusively according to their sequential semantics before passing a concurrent test suite.

\section{Future Work}
The transformation described in this chapter still requires manual programmer intervention, but are well-enough defined that it should be possible to automate at the level of either a macro library or the compiler, as appropriate for the language in question.
Attempting such an implementation, besides requiring a substantial engineering effort, would likely reveal shortcomings in the metaprogramming capabilities of the target language, providing an opportunity for innovative metaprogramming techniques or even proposals for improving the language.

In Section \ref{sec:lockfreelib}, we briefly noted logic on Line 7 of \texttt{readStrict} (Listing \ref{lst:readStrict}) that would handle the case when one or more operations had been subsumed as an optimization; however, we did not make use of that capability in any of our implemented data structures.
Another direction for further research would be to leverage that code path to characterize the performance impact of adapting the Flat-Combining methodology to this setting, or by drawing inspiration from the batched updates of PAM.~\cite{FC, sun2018pam}

A more principled analysis of the benefits of these techniques in terms of the development costs for creating concurrent data structures might introduce our transformation in a pedagogical setting and compare student outcomes to a version of the same curriculum that builds the same concurrent data structures from lower level primitives.

\subsection{Beyond Garbage Collection\label{sec:lockfreenogc}}
The library developed here is built on Dotty, an experimental compiler for the language that will eventually become Scala 3.0.
Like most languages with first class functions and closures, it relies on a garbage collector for memory management, which also handles all issues surrounding concurrency and memory management.
Unfortunately many ``systems'' languages (C, C++) rely on manual memory management or disciplined use of smart pointers, and even those that manage memory automatically (Rust) do not use a garbage collector.
Without the ability of a garbage collector to pause threads and inspect their contents, it is a difficult task to know when it is safe to free nodes from a data structure.
Extant techniques include hazard pointers \cite{Michael:2004:HPS:987524.987595}, Pass-the-Buck \cite{Herlihy:2002:ROP:645959.676129}, (lock-free) reference counting \cite{Detlefs:2001:LRC:383962.384016,SNZI2007},  and epoch-based-reclamation (EBR) \cite{fraser2004practical}.
EBR is preferred by the creators of LLX/SCX; and they developed several new EBR variants in support of their C++ implementation of LLX/SCX.~\cite{brown2017techniques}

Unfortunately, reference counting suffers under heavy contention, and EBR algorithms must be explicitly told when a node is removed from a data-structure.
For data-structures built to exploit structural sharing, node removal is no longer a property which can be inferred locally, and instead is a property of the global state of the program, unless supported by some form of reference counting.

We observe that while a pure reference counting solution must increment and decrement the counter every time a thread acquires or deacquires a reference to a node, the snapshot count, as used in the hand-over-hand locking-based Braun Heap from the preceding chapter, should be modified at a much lower frequency, and provide sufficient information to determine whether a node is globally contained in a data structure or any of its snapshots.

We therefore propose that future work into memory management for lock-free data structures implement and benchmark a hybrid EBR/snapshot-counting algorithm to support lock-free data structures with structural sharing.

	 \chapter{\label{chap:returns}Getting Something In Return}
	   As discussed in the previous chapter, a side effect of \texttt{readStrict} and \texttt{force} (Listings \ref{lst:readStrict} and \ref{lst:forceDef}, respectively), is that the thread which originally posted a deferred operation to the root of the tree will be unable to follow its progress once helpers intervene.
This limits the expressivity of the library to operations which either do not return a value (\texttt{insert}), or which only traverse a constant depth from the root to obtain their return value (\texttt{removeMin}).
Many interesting tree algorithms require that return values be obtainable at any depth in the tree.
Moreover, it is a desirable property for software development, debugging, and testing purposes, that if exceptions are thrown due to users performing data structure operations with illegal inputs, then those exceptions should propagate through the stack of the thread that requested the work, rather than the thread that was only incidentally performing the work.

\section{LLX/SCX with Continuations}
To address both of these shortcomings, we propose modifying the commit protocol for SCX in order to allow it to propagate continuations to the thread whose work is actually being done\footnote{In our library, that is the thread which posted a deferred operation to the root of the tree, and not necessarily even the thread which executed the initial SCX.}.~\cite{continuations,continuations2}
This builds on the work done to support opt-in eager execution modes in Section \ref{sec:eagerForce}.
The continuations will allow that thread to follow the execution of a compound linearizable task through the tree in order to: continue helping with the work it wanted done, detect abnormal conditions and propagate its own exceptions, and retrieve any desired return values.

One reason to make this feature opt-in, rather than required, is that the added expressivity comes at the cost of potentially much worse cache-locality.~\cite{DBLP:conf/spaa/FatourouK11,DBLP:journals/topc/HerlihyL16}

\begin{listing*}
\begin{lstlisting}[mathescape=true]
def helpSCXFirstLoop(op:Op, i: Int, ub: Int)$\to$TriBool {
  if(i == ub){
    Maybe
  } else {
    val node = op.nodes(i)
    val expectedOp = op.ops(i)
    if (!(updateOp(node, expectedOp) = op) && node.op != op) { // if work was not done
      if (op.allFrozen) {
        True
      } else {
        op.state = Aborted
        False
      }
    } else {
      helpSCXFirstLoop(op, i + 1, ub)
    }
  }
}

def helpSCX(op:Op)$\to$Boolean {
  helpSCXFirstLoop(op, 0, op.nodes.length) match {
    | case Maybe =>
      op.allFrozen = true
      forEach(i $\in$ toFinalize) {
        op.nodes(i).marked = true
      }
      op.update()
      op.state = Committed
      true
    | case b:Definite => b.value
  }
}
\end{lstlisting}
\caption{\texttt{helpSCX}, and its subroutine \texttt{helpSCXFirstLoop}.}
\label{lst:helpSCX}
\end{listing*}
Implementing the LLX/SCX primitives in software requires a routine called \texttt{helpSCX}.~\cite{BrownLLXSCX}
In our library, that is implemented as shown in Listing \ref{lst:helpSCX}.
We observe the following about \texttt{helpSCX}.
First, after reaching line 23, the SCX operation in progress has a reached a state where it can no longer fail. 
Second, the results of the SCX are fully visible in memory after the call to \texttt{op.update()} uses a CAS to publish the write to memory (line 27).
Third, once the state of the operation switches from \texttt{InProgress} to \texttt{Committed} (line 28), other SCXs are free to overwrite references to the operation in the metadata of any cells they access, rather than helping it complete.
Finally, the entire \texttt{helpSCX} operation is idempotent: it doesn't matter how many times it is called in a row (or concurrently) with the same arguments, it reaches a steady-state in memory.

Thus, any changes we make to the commit protocol must: first, preserve its idempotency; second, execute after the call to \texttt{op.update()} (line 27) to ensure that all effects are visible in memory; finally, execute before the operation switches from \texttt{InProgress} to \texttt{Committed} (line 28), or risk the requesting thread be stuck waiting on an updated continuation after the previous one has already been invalidated by new threads doing their own work on the same node.

Therefore, we propose to augment the \texttt{Op} data structure (referred to as an SCXRecord in the original paper \cite{BrownLLXSCX}) to also hold a pointer to a simple atomic reference, held by instigating thread, that can hold a continuation, as well as the old continuation stored in that reference at the time of the SCX, and the new continuation which should replace it.
Then, between lines 27 and 28, we add a second CAS, which will update the atomic reference to the continuation iff it hasn't already been updated (thus preserving idempotency).
Finally, we add an outer trampoline to each compound linearizable operation on the data structure root, so that after the work is first posted, the instigating thread will continually poll its continuation, test if it is valid, and attempt to force the deferred node it points to if so.
If the continuation is invalid, it means another thread has already done the work, in which case it can be immediately repolled, with the guarantee that a fresh continuation is already available.
This continues until the continuation contains an exception or an instruction to return instead of further deferred nodes to be forced.

\section{Library Changes}
Allowing the library from Chapter \ref{chap:lockfree} to exploit the modified SCX requires a relatively small set of changes.

First, we change the definition of each \texttt{Operation$_i$} from \texttt{Operation$_0$(arg:A$_0^?$, children:Node[E] $\ldots +$)} to \texttt{Operation$_0$(arg:A$_0^?$, children:Node[E] $\ldots +$, parent:RichOperation$^?$)}.
The optional \texttt{RichOperation}, \texttt{parent}, refers to the root-holder operation that instigated the work, and contains the CAS-cell for it to receive continuations from the modified SCX.
We define a singleton default \texttt{RichOperation}, \texttt{Poor}, for any \texttt{Operation$_i$} that does not wish to receive continuations, allowing data structure authors to choose on a per-operation basis (as they do with eager and lazy evaluation strategies in the preceding chapter).

\begin{listing}
\begin{lstlisting}[mathescape=true]
type Continuation[E,Z] = Continue(node:Node[E], def:Def[E,Z])
                       | Return(value:Z)
                       | Throw(exception:Exception)
                       | Yield

def Def[E,Z]::force(n:Node[E], provider)$\to$Node[E] {
  val ctx = provider.freshContext
  val thisRef = provider.read(n, ctx)
  thisRef() match {
    | this if !thisRef.final =>
      if ( this instanceof Snapshot[E] ){this.makeFinal}
      val (newThis, maybeCont) = this.apply(provider, ctx)
      
      if(this.parent == RichOperation.Poor){
        val ret = (thisRef(provider, ctx) = newThis)
        
        maybeCont match {
          | Continue(nextNode, nextDef) =>
            nextDef.force(nextNode, provider)
          | Yield => // Be lazy as before
          | _ => throw IllegalStateException("Poor operations can't do that")
        }
        
        ret
      } else {
        val c = this.parent.read() 
        c match {
          | Continue(nextNode, nextDef) if ((nextNode == n) && (nextDef == this)) =>
            thisRef(c, k, provider, ctx) = newSelf
          | _ => escape // The work has already been done
        }
      }
    | _ => retry
  }
}
\end{lstlisting}
\caption{Further augmented \texttt{Def[E]::apply}, allowing evaluation to be followed by the instigating thread, and type definitions for the supported continuation states.}
\label{lst:forceDefReturns}
\end{listing}
Second, we further generalize the definition of \texttt{Def[E]::apply}:\\ \texttt{Def[E]::apply(SCXProvider,LLXContext)$\to$(Immediate[E],Maybe[(Node[E],Def[E])]} becomes \texttt{Def[E,Z]::apply(SCXProvider,LLXContext)$\to$(Immediate[E],Continuation[E,Z])}.
The added type-parameter, \texttt{Z}, defines the return type, and may be specialized by each \texttt{Operation$_i$}.
The new type, \texttt{Continuation[E,Z]}, defines a restricted family of supported continuations supported by the a modified \texttt{Def[E]::force}, both shown in Listing \ref{lst:forceDefReturns}.
Lines 15 - 24 correspond to Lines 8 - 16 in the Listing \ref{lst:forceDefEager} (the variant in which we first supported eager evaluation), with an added check to make sure a return value or exception isn't being propagated; however, they have been nested within a conditional that checks if the \texttt{Def[E,Z]} being forced is parented to a \texttt{RichOperation} (Line 14).
If so (Line 25), and the parent's current continuation matches the operation currently being forced (Line 28), then we invoke\footnote{Through a similarly augmented reference API.} our augmented SCX (Line 29).
Otherwise, another thread has already completed the operation currently in progress, and we can escape up the call stack (Line 30).

The informal correctness arguments are the same as for Section \ref{sec:lockfreecorrect}.

\section{Evaluation: HAMTs}
To demonstrate the new expressive power, we implemented a simplified Hash Array Mapped Trie, or HAMT, in 320 lines of Dotty.~\cite{bagwell2001ideal}
The full HAMT algorithm includes an occupancy bitmap with each table to improve cache locality, which we did not support.
Despite these limitations, our HAMT exhibited superior concurrent scaling on a snap-and-insert test (following a similar methodology to the priority queue benchmarks in Chapters \ref{chap:handoverhand} and \ref{chap:lockfree}) than either the Scala standard library's CTrie implementation or a version of the Java standard library's \texttt{ConcurrentSkipListMap} that has been augmented with the lock-free linearizable iterator of Petrank and Timnat.~\cite{ProkopecCTrie,dleaCSLM,petrank2013lock}
\begin{figure*}
\centering
\includegraphics[width=\textwidth]{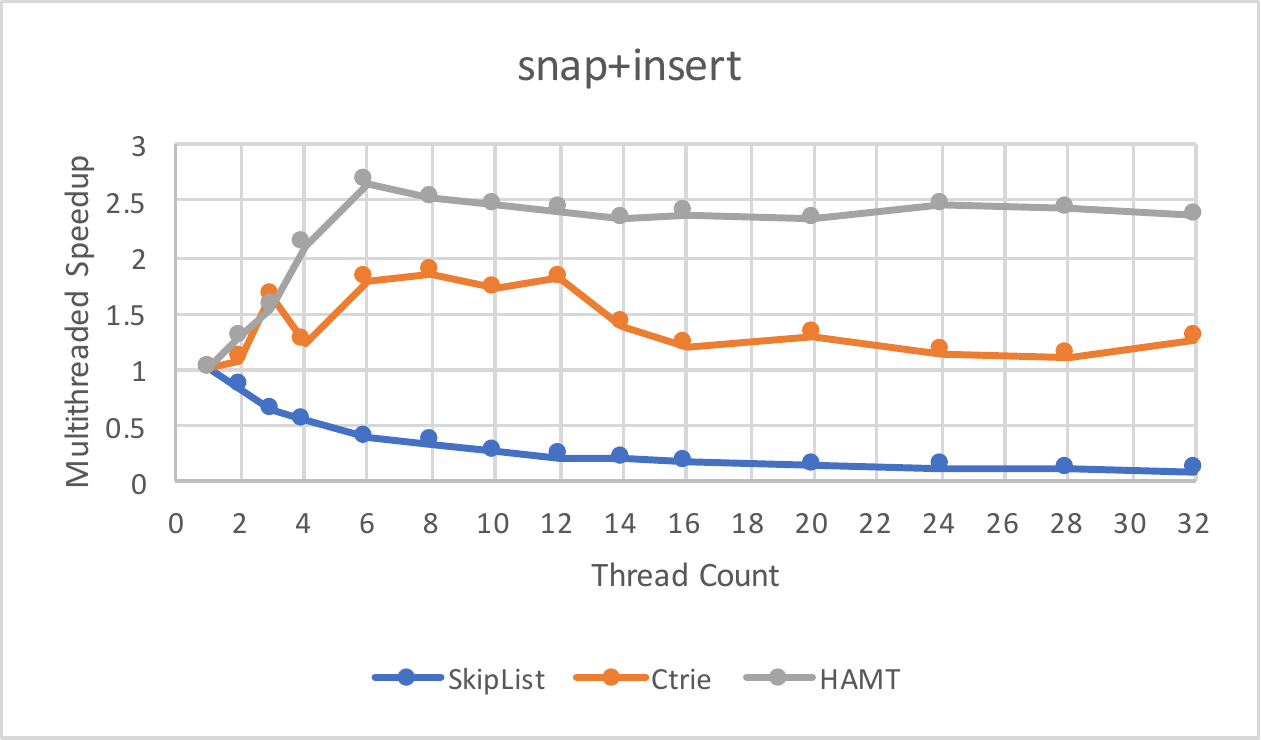}
\caption{The \texttt{snap+insert} benchmark. Vertical axis shows the multithreaded speedup for each data structure. Larger is better.}
\label{fig:lfHAMTscaling}
\end{figure*}

\section{Future Work: Expressivity, Bottle-necks, and Commutativity \label{sec:lfBottleCommute}}
In this chapter, we described a modified commit protocol for the SCX primitive that allows the original thread to follow the execution of a compound operation in the face of ``helping'', and modified our library to support a broader class of operations.
We use this increased expressivity to implement a lock-free concurrent Hash Array Mapped Trie (HAMT~\cite{bagwell2001ideal}); however, one major restriction on expressivity remains.
In Section \ref{sec:lfFormal}, we imposed a restriction that we called \emph{Immediate Return} (Definition \ref{def:immret}) without further explanation.

Consider the problem of augmenting our HAMT so that it supports the operations of a Merkle-tree.~\cite{merkle1982method,becker2008merkle}
After a successful insertion or deletion, we would also need to recalculate the hash of each node we altered along the path from the root: in other words, control-flow would need to propagate rootward from the leaves in addition to leafward from the root.
If we merely alter the type of each \texttt{Operation$_i$::apply} from $(\mathrm{Operation}_i,\mathrm{SCXProvider},\mathrm{LLXContext}) \to \mathrm{Immediate[E]}$\footnote{
We assume \texttt{this} is passed as an implicit first parameter.
} to $(\mathrm{Operation}_i,\mathrm{SCXProvider},\mathrm{LLXContext}) \to \mathrm{Deferred[E]}$, then it would be possible to return a hypothetical \texttt{ComputeHash} that subsequent threads arriving at that node could force.
Unfortunately, this would create a concurrency bottleneck, as each thread, while able to help and maintain lock-freedom, would have to follow the same leafward path, and at each node only one of those threads would be contributing \emph{useful} work (effectively serializing the workload).

In the next chapter, we discuss the ways in which commutativity can be used to characterize conflicting operations and synthesize a Software Transactional Memory (STM) compatible API for existing non-transactional concurrent data structures.
Here, we also propose that future work pursue the question of how commutativity, and other algebraic properties, could be used to further generalize the expressivity of the techniques developed here, without effectively serializing operations that require rootward traversals.
For example, a \texttt{Insert} operation on a Merkle-tree that encounters a deferred \texttt{ComputeHash} might replace it with a modified \texttt{ComputeHash} that has an additional updated subtree, rather than forcing it entirely.

	\chapter{\label{chap:proust}Proust}
	 \newcommand\toolName{Scala{\sc Proust}}

\section{Introduction}
In the preceding chapters, we have shown how to compose low-level synchronization primitives into potentially multi-step operations on concurrent data structures, while preserving the sequential semantics of those data structures.
We have focused on building data structures which support efficient snapshots, and have hinted that these snapshots will enable further composition into more complex programs.
In this chapter we will show how Software Transactional Memory (STM) runtimes can exploit efficient snapshots, in conjunction with commutativity, to address some of their traditional shortcomings.
We observe that software transactions provide a general, but heavy-weight, mechanism for composing concurrent operations within a specialized runtime environment.

Modern STM systems typically perform
synchronization on the basis of \emph{read-write} conflicts:
two transactions conflict if they access the same memory location,
and at least one access is a write.
It is well understood that this technique works poorly for contended
data objects because  operations that could have correctly executed
concurrently are deemed to conflict,
causing unnecessary rollbacks and serialization.

Some prior works have aimed at this problem and found solutions to
some cases.
\emph{Transactional boosting}~\cite{HerlihyK2008}
centers around constructing a transactional  ``wrapper'' for legacy
thread-safe concurrent data structures.
Designing a boosting wrapper requires identifying which operations commute,
as well as providing operation inverses.
Boosting  can take advantage of existing
thread-safe libraries, so there is no need to ``re-invent the wheel'', but
requires duplicating some STM
functionality in the form of abstract locks, inverses,
and customized commit and abort handlers.
\emph{Transactional Predication}~\cite{BronsonCCOn2010} is another
technique for defining highly-concurrent transactional objects.
This technique ``maps'' semantic conflicts onto read-write conflicts
handled by an underlying STM.
Predication can exploit highly-optimized mechanisms provided by off-the-shelf STM systems, but it applies only to sets and maps.
\emph{Software transactional objects}~\cite{HermanIHTKLS2016} (STO)
is an STM design that provides built-in primitives to track conflicts among
arbitrary operations, not just read-write conflicts.
Similarly,
\emph{Transactional Data Structure   Libraries}~\cite{Spiegelman:2016:TDS:2908080.2908112}
describe techniques for building libraries of transaction-aware data structures.
The latter two works do not support migration of legacy libraries.
Existing concurrent data structure libraries are often large and highly optimized,
and it is appealing to build on that code rather than reinvent it.

Despite these advances, we still lack a unified approach to building transactional systems that exploit both the conflict resolution of state-of-the-art STM systems, as well as the high performance of off-the-shelf concurrent abstract data type (ADT) implementations.
What if we had, say, an optimistic STM that performed lazy (commit-time) write/write conflict detection, and we wanted to use an off-the-shelf concurrent priority queue that supported efficient snapshots, but had no efficient inverse for \texttt{insert}\footnote{Recall that Boosting relied on a lazy deletion mechanism to support priority queues, since \texttt{removeMin()/y} is inverted by \texttt{insert(y)}, but the reverse is not necessarily true.}.
These kinds of combinations are out of reach of existing strategies and we new abstractions must be developed to support them.
There are several challenges.
How can we provide \emph{flexibility} to programmers whose base objects may be better suited to either eager or lazy updates, and whose workloads may be better suited to either pessimistic or optimistic synchronization?
How can we support \emph{optimistic} operations on off-the-shelf concurrent ADT implementations?
How can we cope with the fact that different STMs offer
different conflict detection/resolution strategies?
How can we use our objects along-side traditional STM memory operations?
How can we be sure that opacity~\cite{opacity}\footnote{Roughly, opacity implies serializability, but with the added condition that even aborted transactions observe a consistent state.} is maintained across these many possible configurations?

This chapter introduces two new key concepts---\emph{conflict abstractions}
and \emph{alternate histories}---which, together, enable programmers to
build transactional systems addressing all of the challenges above.
Conflict between ADT operations is typically understood in terms of
commutativity specifications~\cite{commu,steele,Kulkarni2009,HerlihyK2008} which are implementation-independent, but aren't easily translated into code.
Instead, we introduce \emph{conflict abstractions}, which map elements of an object's abstract state to synchronization primitives (either locks or STM-managed memory locations).
Threads that wish to perform an ADT operation $o.m(\bar{x})$ must synchronize through read or write access to the appropriate primitives (corresponding to the abstract state touched by $o.m(\bar{x})$).
We observe that the high-level ADT operations will commute when their low-level operations on the conflict abstraction primitives commute.
In this way, we can leverage STMs to perform abstract conflict detection via concrete memory locations.
Meanwhile, as compared to reasoning directly about commutativity, conflict abstractions are intuitively easier for programmers to specify because they pertain more directly to each method's effect on the abstract state.

Depending on the operations supported by the ADT, programmers may wish to perform updates either eagerly or lazy.
For Boosting-like eager updates, operations must be provided with inverses, which will be applied (in reverse order) in case of an abort due to conflicts. 
Some ADTs have operations which are not easily inverted, and the programmer will prefer to simply wait and apply changes at commit time.
Where non-void return values are in play, this requires the ability to predict the effects of operations which have not yet been applied.
To this end, we introduce the idea of \emph{alternate histories}, to encompass approaches, based either on inverses/undo-logs or on \emph{shadow copies}/replay-logs, allowing threads to speculatively modify an object while ensuring the updates can be viewed by only one transaction, and discarded if necessary.

Finally, we discuss how these ingredients can be combined to construct a
transactional system. We describe choices that a programmer can make,
along with an algorithm which wraps 
data structure invocations, using conflict abstractions and alternate histories
appropriately in a manner that depends on the guarantees of the
off-the-shelf STM system.
We have incorporated these ideas into a new transactional object
system called \emph{Proust}\footnote{This name is a
portmanteau of \emph{predication} and \emph{boosting}, both
influential prior works. The name is also an \emph{hommage} to Marcel
Proust, an author famous for his exploration of the complexities of
memory.}, built on top of ScalaSTM.
\emph{Proust}, unlike predication, can support objects of arbitrary abstract
type, including, for example, priority queues and non-zero indicators.
\emph{Proust}, unlike boosting, allows objects to be integrated with the underlying STM,
to take advantage of well-engineered STM conflict-detection mechanisms. 

\subsection{Contributions}
\begin{description}
\item[Conflict abstractions] provide a novel way to concretely realize an
  abstract data type's semantic notions of conflict to efficiently cooperate with a generic software transactional memory
  run-time (Section~\ref{section:conflict}).
  This style of conflict analysis reveals shortcomings in the expressivity of existing
    STM implementations to distinguish between pure and impure writes (Sections~\ref{section:eval} and~\ref{sec:concl}).
\item[Alternate histories] allow individual transactions
  to make private speculative updates to highly-concurrent black-box objects (Section~\ref{section:shadow}).
\item[A methodology and algorithm] based on conflict abstractions and alternate histories, that
    allows one to combine off-the-shelf ADTs with differing synchronization (optimistic or pessimistic) and update
(eager vs lazy) strategies
  (Section~\ref{section:algorithm}).
\item[\toolName{}] a prototype built on top of ScalaSTM,
  showing scalability competitive with existing specialized approaches,
  but with a much wider range of applicability
  (Sections~\ref{section:impl} and~\ref{section:eval}).
\end{description}

\begin{figure}
  \includegraphics[width=3.2in]{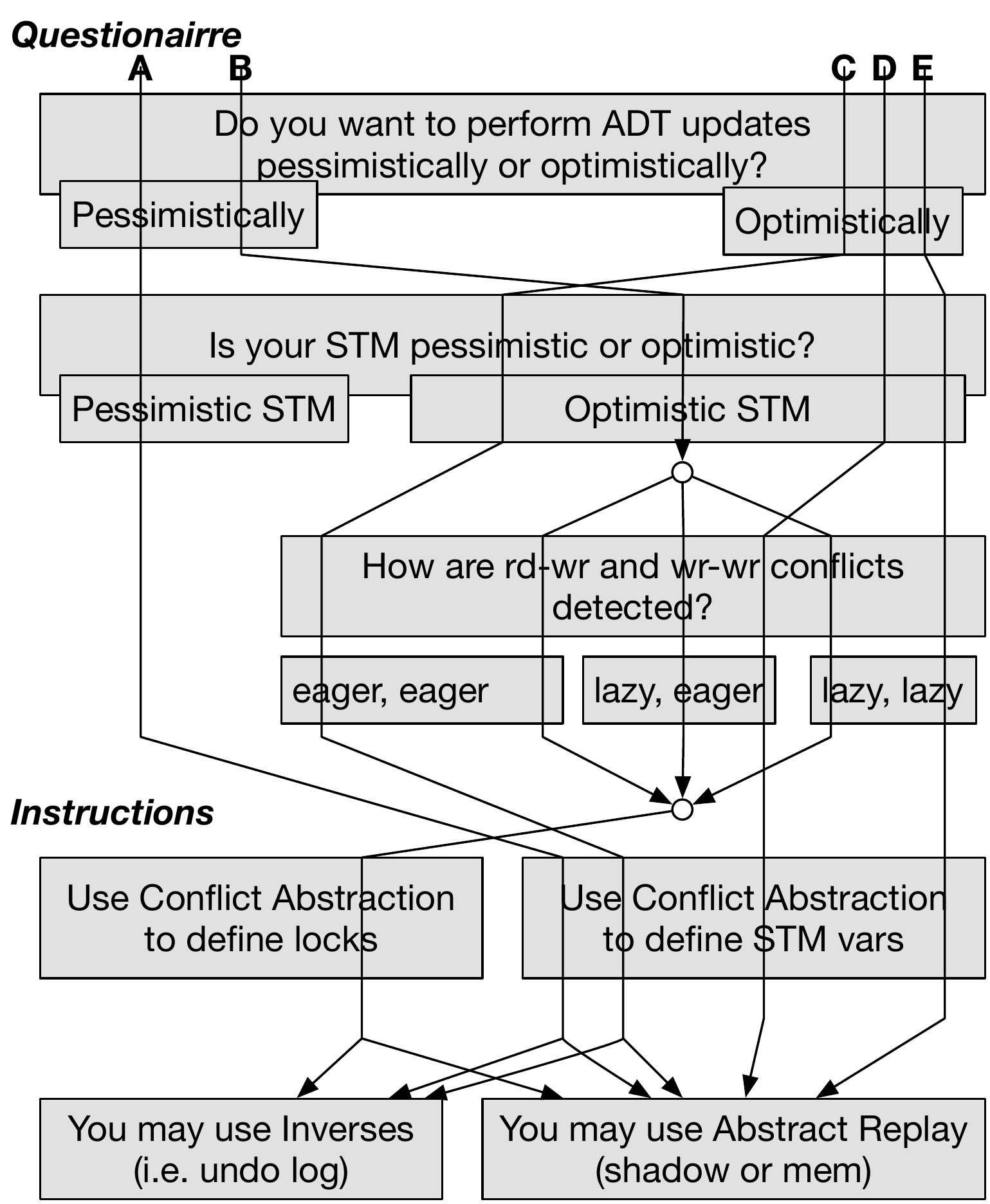}
  \caption{\label{fig:flowchart} Mapping the Proust design space.}
\end{figure}

\section{Overview}

We now highlight the key ideas of this chapter with two example
concurrent ADTs, a priority queue and a non-zero indicator (NZI), and describe how to simultaneously use them in an STM system while enjoying their native performance,

Our goal is to provide a unifying abstraction that is nonetheless able to
leverages these black-box data-structures and 
black-box STMs.
One challenge is that different STMs offer different kinds of sychronization strategies (pessimism versus optimism) and perform conflict detection differently (eager versus lazy). 
For example, as outlined by Dragojevic, \emph{et al.}~\cite{dragojevic2009stretching},
conflict detection is different in several popular STMs:
(i) SwissTM and CCSTM are eager write/write but lazy read/write
(ii) TL2 is lazy write/write and lazy read/write, and
(iii) TinySTM is eager write/write and eager read/write.
Therefore, at the outset, we ask the programmer:
\begin{enumerate}
\item What ADTs are you using?
\item For each ADT, would you like to update it eagerly or lazily?
\item Is your STM pessimistic or optimistic?
\item If optimistic, does it detect read/write conflicts eagerly or lazily?
\item If optimistic, does it detect write/write conflicts eagerly or lazily?
\end{enumerate}
Our research goal was to develop concepts that would provide programmers with
a path toward a workable system, regardless of the answers to these questions.
A flow-chart diagram is given in Figure~\ref{fig:flowchart}, summarizing
how the programmer proceeds.
The upper portion consists of  a series of questions for the programmer, and
the lower portion consists of  instructions for how to proceed. Each vertical
line (labeled {\bf A}, {\bf B}, {\bf C}, etc.) represents a different scenario. For example, in scenario {\bf D}, the user wishes to perform ADT updates optimistically,
and he/she is using an optimistic STM with lazy read/write conflict detection
and eager write/write conflict detection.
In the rest of this section we will discuss
the unifying concepts (\emph{conflict abstractions} and \emph{alternate histories})
which are the instructions illustrated in Figure~\ref{fig:flowchart}.

\subsection{\label{section:comm2ca}From Commutativity to Conflict Abstractions}

Let us first consider an example ADT: a priority queue supporting the
three operations
\texttt{min()/x}, \texttt{removeMin()/x}, and \texttt{insert(x)}.
We assume that the programmer already has a concurrent implementation.
Moreover, like in
transactional boosting~\cite{HerlihyK2008}, we will first require the programmer to be aware which operations commute under which circumstances.
We say that two ADT operations  \emph{commute} provided that they lead to the same final state and return the same values, regardless of the order in which they are applied. 
As a reminder, Table \ref{tbl:pqcommutes} provides the commutativity of priority queue operations.
\begin{table}
\centering
  \begin{tabular}{|l||c|c|c|}
    \hline
     & \texttt{min()/$x$} & \texttt{removeMin()/$x$} & \texttt{insert($x$)} \\
    \hline
    \hline
    \texttt{min()/$y$} & true & false & $y \le x$  \\
    \hline
    \texttt{removeMin()/$y$} & false & false & $y \le x$  \\
    \hline
    \texttt{insert($y$)} & $y \ge x$ & $y \ge x$ & true\\
    \hline
  \end{tabular}
  \caption{The commutativity specification for a priority queue.}
  \label{tbl:pqcommutes}
\end{table}
As an example, \texttt{insert(42)} always commutes with \texttt{removeMin()/1} because the value inserted (42) was greater than the minimum value (1) in the priority queue.

While commutativity specifications benefit from being largely independent of the implementation, they are difficult to translate into program source code.
For example, the \emph{abstract locks} of transactional boosting are intended to reflect commutativity, but their priority queue implementation relies on a single read/write lock (where inserts are considered reads unless altering the min), and does not support a \texttt{min} operation.
We observe that directly translating commutativity specifications into code would see a quadratic blowup in the number of constraints

A principal idea of this chapter is to instead map the ADT's abstract state to a concrete summary, either with pessimistic locks as in boosting, or with STM memory references as in predication.
Threads summarize the ADT operations they plan to perform---a sort of \emph{digest}---through a few read/write operations on the synchronization primitives of choice.
The primitives in this digest are chosen to reflect various aspects of the object's abstract state (e.g. a priority queue's minimum value, size, and multiset).
This digest, if written correctly, is such that whenever the ADT operations being
performed by two threads do not commute, operations on the digest primitives will be found to conflict.
This mapping of abstract state to synchronization primitives, and the rules for which to read and which to write---as a function of the ADT operation being performed---is what we call a \emph{conflict abstraction}.
Table \ref{tbl:pqueueCA} shows a possible conflict abstraction for the priority queue ADT.

\newcommand\ite[3]{\texttt{if }(#1)\ #2\texttt{ else } #3}
\newcommand\ifthenr[2]{\texttt{if }(#1)\ #2}
\newcommand\elser[1]{\texttt{else } #1}

\begin{table}
\centering

  \[\begin{array}{lcl}
      \texttt{min()}/x &:& rd(v_{min})\\
      \texttt{removeMin}()/x  &:& wr(v_{decr}); wr(v_{min})\\
      \texttt{insert}(x) &:& wr(v_{incr}); \\
                       && \ifthenr{x<\texttt{min()})}{wr(v_{min})}\\
                       && \elser{rd(v_{min})}\\
      \texttt{size()}/n &:& rd(v_{decr}); rd(v_{incr})\\
    \end{array}\]
\caption{A Conflict Abstraction for Priority Queue, using CA STM vars: $v_{min}, v_{incr}, v_{decr}$.}
\label{tbl:pqueueCA}
\end{table}

In this conflict abstraction (CA), we will use STM-managed variables,
which we will call $v_{min}$, $v_{incr}$, and $v_{decr}$.
Intuitively, $v_{min}$ summarizes
whether operations are somehow dependent upon the minimum element.
Writing to variable $v_{incr}$ summarizes whether the operation increases the size
of the queue, while reading to $v_{incr}$ indicates that the operation is sensitive
to whether the size will increase. $v_{decr}$ is similar.
Notice that if we take \emph{any} initial state, and consider
\emph{any} pair of ADT operations, if the CA operation rules are followed,
then the STM will detect some kind of conflict on at least one of the \emph{concrete}
memory locations $v_{min}, v_{incr}, v_{decr}$.

Let's say that we have operations $T_1:\texttt{removeMin()}/42$
and $T_2:\texttt{insert}(1)$. In general these operations do not commute because
the element being inserted is less than the current minimum value so, depending
on the order of the operations, $T_1$ will observe different values (and the
final state of the ADT will be different).
Following the CA operation rules, $T_1$ will write $v_{decr}$ and write
$v_{min}$. Meanwhile, $T_2$ will write $v_{incr}$ and either read or
write $v_{min}$, depending on the ADT's current \texttt{min}.
(Here \texttt{min} is a another ADT method, which will itself perform
a read on $v_{min}$.)
Off-the-shelf STMs will detect some kind of conflict, e.g., a write/read or write/write
conflict on $v_{min}$. (Later we will address whether
write/read and write/write conflict detection happens eagerly or lazily.)
On the other hand, let's assume that initially $33$ is the minimal
element of the priority queue and consider two commutative operations:
$T_1:\texttt{insert}(42)$ and $T_2:\texttt{min}/33$.
In thise case $T_1$ will write $v_{incr}$ and read $v_{min}$, while $T_2$ will
read $v_{min}$. An off-the-self STM won't detect any conflicts (two reads
on $v_{min}$ don't conflict), corresponding to the fact that these abstract
ADT operations commute.

Let's consider a second example ADT: a counter that is capable of
non-zero indication (NZI), as inspired by~\cite{SNZI2007}. Like the priority
queue, this is a straight-forward ADT, yet cannot be handled by transactional
predication~\cite{BronsonCCOn2010}. NZI provides three operations: \texttt{inc()},
\texttt{dec()/$p$}, \texttt{zero()$/p$}, where \texttt{dec()} returns a flag
indicating if the operation failed because the NZI was already zero. The
commutativity is given in \ref{tbl:commuteNZI}.
\begin{table}
  \begin{tabular}{|l|c|c|c|}
    \hline
    &  \texttt{inc()} & \texttt{dec()$/p$} & \texttt{zero()$/p$} \\
    \hline
    \hline
    \texttt{inc()} & true &  $\lnot p$ & $\lnot p$  \\
    \hline
    \texttt{dec()$/q$} &  $\lnot q$ & $q = p$ & $q=p$  \\
    \hline
    \texttt{zero()$/q$} & $\lnot q$ & $q=p$ & true \\
    \hline
    \end{tabular}
    \caption{Commutativity specification for the NZI data type.}
    \label{tbl:commuteNZI}
\end{table}
This commutivity specification says that two \texttt{inc()} operations are
independent, as are two \texttt{zero()} operations. Naturally, an \texttt{inc()}
may alter the return value of \texttt{zero()} and \texttt{dec()} further
complicates matters. In these cases, commutativity depends on the return values
of \texttt{dec()} and \texttt{zero()}. Once again, we cannot yet leverage
off-the-shelf STMs for transactional conflict detection because the commutativity
specification is not a format readily understood by STMs; however, Table \ref{tbl:caNZI} provides  a conflict abstraction that corresponds to this commutativity:
\begin{table}
  \[\begin{array}{lcl}
      \texttt{inc()} &:& \ite{\texttt{zero()}}{wr(v_{zero})}{rd(v_{zero})}\\
      \texttt{dec()}/q &:& \ifthenr{\texttt{willBeZero()}}{wr(v_{zero})}\\
      && \elser{rd(v_{zero})}\\
      \texttt{zero()}/q &:& read(v_{zero})\\
    \end{array}\]
    \caption{{Conflict Abstraction rules for a Non-Zero Indicator (NZI)}, using CA STM vars: $v_{zero}$.}
    \label{tbl:caNZI}
\end{table}

\noindent
For NZI, we were able to use a \emph{single} STM memory location $v_{zero}$ to
summarize the abstract conflict. As we discuss in Section~\ref{section:conflict}, one can
construct a CA differently, depending on the ADT and how finely grained one
would like to characterize conflict. Notice that we have used \texttt{zero()}
and \texttt{willBeZero()}, functions that depend not only the NZI ADT's current state, but also on potential future states, to characterize its commutativity.
The Proust algorithm, outlined in Section~\ref{section:algorithm}, is able to support such use cases by absorbing the CA, as a transitive dependency, of one operation into the CA of any operations that reference it in the definition of their own CA.
Taking an example of $T_1:\texttt{inc()}$ and $T_2:\texttt{zero()}$, it is easy
to see that an STM will detect conflict on $v_{zero}$, depending on whather the
NZI is zero.
Finally, note that conflict abstractions can also be used, in the case of
pessimistic synchronization, as a basis for defining abstract locks. Each
variable can instead be a read/write lock and the conflict abstraction specification
indicates whether the lock should be acquired in read or write mode.

\subsection{\label{section:althist}Support for Alternate Histories of Black-Box Objects}
As we discuss below and in the next sections of this chapter, conflict abstractions provide a unified approach to determining when concurrent operations are compatible and when they are in conflict.
They do not, in isolation, provide a unified approach to speculatively executing such operations (and committing or aborting as necessary).

To that end, we also describe two approaches for supporting \emph{``alternate histories''} on top of black-box ADTs.

When a particular ADT (or ADT/STM combo) is well-suited to \emph{eager} updates, then a transactional wrapper must ensure every speculative operation is undone when a transaction aborts.
On the other hand, when a transaction involving such a wrapper commits, no further work is necessary.
Such a wrapper can easily be implemented when every operation has an efficient inverse (ala Boosting - cite).

On the other hand, when a particular ADT (or ADT/STM combo) is well-suited to \emph{lazy} updates, then a transactional wrapper must predict the result of each speculative operation, effectively reaching forward in time to see how an alternate history would play out.
When a transaction aborts, a wrapper of this variety has no further work, because the underlying data structure has not been altered.
On the other hand, when a transaction commits, the wrapper must ensure that every speculative operation is finally applied to the underlying object.
Such a wrapper can easily be implemented when the ADT supports an efficient snapshot operation (the use of which we refer to as a \emph{shadow copy}), or when it has simple semantics which can be efficiently emulated by \emph{memoizing} certain calls.

The memoization approach works well for ADTs such as maps and sets: the result of \texttt{m.set(a,x)} followed by \texttt{m.get(a)} (a read-only operation) is \texttt{x}.
We note that the result of \texttt{m.set(a,x)} followed by \texttt{m.set(a,y)} (a read-write operation) is also \texttt{x}.
Thus, for a single transaction, the results of mutating operations can be predicted by memoizing read-only operations (and the mutation can be retargeted to the memoization table rather than the wrapped data structure).

On the other hand, for more complex data structures like NZIs or Priority Queues, there is no obvious memoization scheme based on ADT operations, but a \emph{shadow copy} can be used to achieve the same effect.
Shadow copies are also helpful in providing the ``peek'' methods such
as \texttt{min} (priority queue) and \texttt{zero}/\texttt{willBeZero} (NZI) discussed above.
Shadow copies let us determine ahead of time if the operation in question will change the result of \texttt{zero()} before and after the invocation.

\subsection{How do I use \emph{Proust}?}

With conflict abstractions and alternate histories, we now have a path to
construct STM systems from black-box ADTs and STMs. The process is
as follows:
\begin{enumerate}
\item For each ADT, determine commutativity of operations.
  Commutativity conditions are known for many existing
  ADTs~\cite{HerlihyK2008,pldi11} and can also be synthesized~\cite{tacas18}.
\item For each ADT, define Conflict Abstraction (see Section~\ref{section:conflict}).
\item Use flowchart to determine, based on your STM and preferences, whether your CA will be realized using locks or STM variables\footnote{For pessimistic STMs, these turn out to be the same thing.}.
\item Use flowchart to determine, based on your STM and preferences, and the data structure to be wrapped, whether you can use inverses or shadow copies.
\item Use our wrapper/algorithm (discussed below and in Section~\ref{section:algorithm}) configured based on these decisions.
\end{enumerate}

We describe an algorithm (Section~\ref{section:algorithm}) that can be used once the above choices are made.  The algorithm operates as a method denoted \textsf{proust\_apply}$(o.m,\bar{\alpha})$, which is a wrapper around an invocation of method $o.m$ on a vector of arguments $\bar{\alpha}$. The algorithm treats the conflict abstraction as a basis for locking (in the case where pessimistic synchronization is desired) or for performing read/write operations on STM locations (in the case where optimistic synchronization is desired). It also address recovery by incorporating inverses or alternate histories, as appropriate. Finally, our approach ensures that opacity is preserved, even in the case of lazy conflict detection. 

We conclude with a discussion of our implementation, \toolName{} (Section~\ref{section:impl}) and an evaluation (Section~\ref{section:eval}). We have built \toolName{} on top of ScalaSTM~\cite{bronson2010ccstm}. We ran a series of benchmarks on Maps and Priority Queues. Our results demonstrate that black-box ADT implementations can be used on top of high-performance STMs. Moreover, we can obtain performance that is on the order of transactional predication, without that approach's restricted expressivity.

\paragraph{Pure vs Impure Writes}
We also note that there is a bit of an impedance mismatch between the rules for applying a conflict abstraction and the semantics provided by extant STMs.
Every STM of which we are aware provides only for impure writes, which assume an implicit read paired with every write (and thus writes always conflict with each other).
We observe here that in Fig \ref{tbl:pqcommutes}, the \texttt{W(MSet)} would cause \texttt{insert} and \texttt{removeMin} to always conflict if interpreted as impure writes to an STM location.
This has no natural mapping to transactional predication, as it doesn't represent any kind of collection.

\subsection{Related Work}

In Section 1, we noted prior works
transactional boosting~\cite{HerlihyK2008}, transactional
predication~\cite{BronsonCCOn2010}, and 
optimistic boosting~\cite{Hassan2014OTB}.
While these three prior works were a source of inspiration,
each of them tackled only some aspects of the general problem.
The concepts of conflict abstractions and alternate histories described
here are novel.
Additionally, these works did not provide a road map for constructing
a system, accounting for differences in the guarantees provided
by the underlying STMs.

Transactional boosting~\cite{HerlihyK2008} was limited to the
pessimistic case, and used inverse operations to cope with aborts.
Abstract locks can be seen as a limited form of conflict abstractions,
but were described in the prior work in a merely declarative way
(i.e. user must define locks such that two non-commutative operations
will have at least one lock in common overlap). The
concept of conflict abstraction described herein is a form of intermediary
between ADT commutativity and abstract locks. Moreover, boosting
did not discuss optimism at all.
Optimistic boosting~\cite{Hassan2014OTB} exploits the fact that
many concurrent objects have a read-only traversal phase and describes methods
of composing multiple operations into a single operation. This approach is optimistic
in the sense that conflict detection/locking is deferred, and requires
white-box access to the ADT implementation.

Transactional predication~\cite{BronsonCCOn2010} was able to exploit STMs for
transactional conflict detection, but doesn't support black-box ADTs.
For example, predication cannot support even the basic priority queue and NZI examples discussed above. Moreover, predication does not cover pessimism nor does it cover all optimistic conflict detection strategies (eager/eager, eager/lazy, etc.).

\emph{Transactional object implementations.}
Several recent works have aimed at developing 
data-structure \emph{implementations} from the ground-up so that they 
are amenable to a transactional setting.
Herman et al.~\cite{HermanIHTKLS2016} recently
described a way of implementing transactional data-structures.
They build on top of a core infrastructure that provides
operations on version numbers and abstract tracking sets that
can be used to make object-specific decisions at commit time.
Similar work by Spiegelman et al.~\cite{Spiegelman:2016:TDS:2908080.2908112} describes
how to build data-structure libraries using traditional STM
read/write tracking primitives. In this way, the implementation
can exploit these STM internals.
Elizarov et al.~\cite{Elizarov:2019:LLT:3293883.3301491} show preliminary results describing a protocol allowing lock-free data structures to support lock-free transactions; however, again this requires specialized implementations of each data structure.

Unlike these prior works, our aim is to
reuse existing linearizable objects and exploit the decades of hard-work
and ingenuity that went into their implementations.

\emph{Other.}
Early work on exploiting commutativity for concurrency control
includes Korth~\cite{korth},
Weihl~\cite{weihlcommu},
CRDTs~\cite{Shapiro2011},
and Galois~\cite{Kulkarni2009}.
Some false conflicts in STMs can be alleviated by other escape mechanisms such as
open nesting~\cite{opennested},
elastic transactions~\cite{felber2009}, and
transactional collection classes~\cite{Carlstrom2007}.
Other mechanisms that exploit commutativity for STM systems include
automatic semantic locking~\cite{Golan-Gueta2014}.
Dimitrov et al.~\cite{DimitrovRVK2014} described a method for translating
commutativity formulae into a format that
could be used by a dynamic race detection tool.

\newcommand{\powset}{\raisebox{.15\baselineskip}{\Large\ensuremath{\wp}}}

\section{Conflict Abstractions}
\label{section:conflict}
The principal challenge for any type-specific transactional object
implementation is how to map type-specific notions of
conflict into a low-level synchronization framework.
Like others~\cite{BronsonCCOn2010,HerlihyK2008,Koskinen:2010,KP:PLDI2015},
we identify type-specific synchronization conflicts with a \emph{failure to commute}:
two operations commute if applying them in either order yields the
same return values and the same final object state.
In this section, we describe \emph{conflict abstractions} which permit 
transactional conflict detection, without exposing the internals of a black-box
object. Our approach representis aspects of the object's abstract state as STM-managed
memory locations, while the ADT implementation itself remains in memory not managed by the STM.

\emph{Background definitions.} We will use the following definitions throughout
this chapter.
$\mathcal{M}$ are the set of object \emph{methods} $o.m,o.n$, etc.
$\mathcal{A}$ are method argument \emph{values} denoted, for example, as
 $\bar{\alpha}$ where each value $\alpha$ corresponds to the argument in $\bar{x}$ (as denoted earlier in this chapter). An \emph{invocation} is an application of a method to a vector of arguments,
$o.m(\bar{\alpha}), o.n(\bar{\beta})$, etc. 
$\Sigma_o$ is the abstract object state space for object $o$.
We also assume that the object provides (or can be extended to provide) 
various read-only methods that permit a transaction to query aspects of the
object's abstract state, such as $o.\texttt{size}()$, etc.

\emph{Conflict abstractions.} 
As discussed in Section \ref{section:comm2ca}, a conflict abstraction (CA) is a way
of summarizing the effects of black-box object methods using a series of synchronization primitives. More precisely,

\begin{definition}[Conflict abstraction]
A conflict abstraction is a pair $(X,f)$ where $X$ is a finite set of
variables and $f : \mathcal{M} \rightarrow \mathcal{A} \rightarrow \Sigma \rightarrow Z$
where $Z \subseteq X \times \{ \textsf{rd,wr} \}$.
\end{definition}	

Intuitively, a conflict abstraction first has a set of abstract locations $X$
(which may represent STM-managed memory or locks).
We say the \emph{cardinality} of a conflict abstraction $N$ 
is the size of $|X|$. For a given object method $o.m(\bar{\alpha})$ with
arguments $\vec{\alpha}$ and object state $\sigma_o$, the conflict abstraction
function $f$ indicates which abstract locations $X$ are to be accessed and
the appropriate mode (read or write).
That is, $f(o.m, \bar{\alpha}, \sigma_o)$ returns a list of synchronization-primitive/mode pairs.

Conflict abstractions are then used as depicted in the following diagram.
\begin{center}
\includegraphics[width=3.2in]{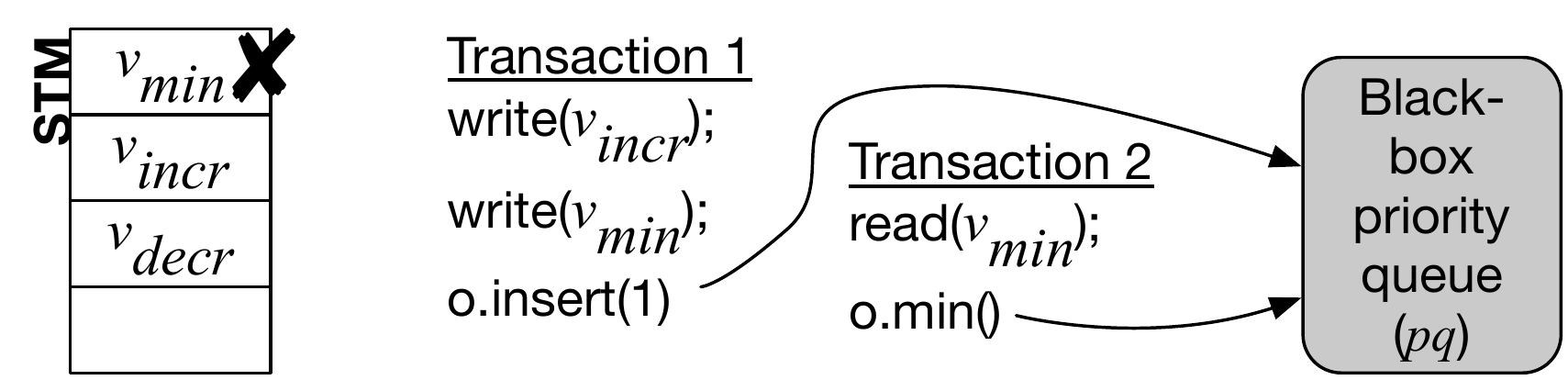}
\end{center}
In this example, $f(o.insert, [1], \sigma_{pq}) = \{ (v_{incr}, \textsf{wr}), (v_{min}, \textsf{wr}) \}$
provided that $1 < o.\textsf{min}()$
and $f(o.min, [], \sigma_{pq}) = \{ (v_{min}, \textsf{rd}) \}$.
Note $o.\textsf{min}()$ appears in the CA of $o.\textsf{insert}()$, so $o.\textsf{insert}()$'s CA depends on $o.\textsf{min}()$'s.

In this way, we can leverage an STM to perform our transactional conflict detection, even though the ADT is
treated as black-box.
In the above example, the STM will detect a read/write conflict
on $v_{min}$\footnote{Section~\ref{section:algorithm} discusses how these conflict abstractions
can be used in STMs, even when those STMs have varying characteristics such as
lazy conflict detection.}.
Thus, conflict abstractions enable efficient STMs to do the work of conflict detection.

Conflict abstractions also have several benefits over conflict strategies based on 
abstract locks~\cite{HerlihyK2008} or commutativity alone. 
The format of a conflict abstraction is more algorithmic and less declarative than prior strategies.
A programmer will already have at least an intuitive understanding of the black-box object's abstract state, and it is easier to translate this into a series of STM locations and read/write operations.
This avoids the need to think about pair-wise reasoning (as in commutativity or abstract locks) upfront: one instead simply considers the effects of each
operation independently.
Later, one can verify the correctness of their conflict abstraction through pair-wise reasoning (see discussion below).

Notice that a conflict abstraction can be more fine-grained or more coarse-grained
with respect to how it represents the object's abstract state. The most 
trivial coarse-grained conflict abstraction would have cardinality $1$ and use a single STM location $x$, and map all
read-oriented object methods to read $x$ and map all object mutator methods to write $x$.
While simple and correct, the downside is of course that concurrency may be lost.
The choice of granularity (cardinality) is often specific to the data structure and the workload. 
Regardless, it is important that the conflict abstraction be correct:

\begin{definition}[Correctness]\label{def:ca}
A conflict abstraction $(X,f)$
is \emph{correct} provided that for every
$m(\bar{\alpha})$ and $n(\bar{\beta})$ that do not commute, and
every $\sigma_o$, there exists some
$(v,m_1)\in f(o.m,\bar{\alpha},\sigma_o)$ and $(v,m_2)\in f(o.n,\bar{\beta},\sigma_o)$
where either $m_1 = \textsf{wr}$ or $m_2 = \textsf{wr}$.
\end{definition}

\noindent
Intuitively, a conflict abstraction is correct if, for any pair of
non-commutative method invocations, there will be some location
that has either a read/write or write/write conflict.
Notice that this definition is independent of the guarantees of the
STM. In Section~\ref{section:algorithm} we discuss how 
CAs can be used regardless of the characteristics
of the STM.

\emph{Verifying Conflict Abstractions.}
Existing
software verification tools can verify the correctness of a conflict
abstraction.
Specifically, the question of correctness can be reduced to satisfiability,
fit for reasoning with SAT/SMT tools. 
First, to reason about correctness, we \emph{do not need} the
actual implementation of the thread-safe concurrent objects.
Instead, it is sufficient to work with a \emph{model} (or sequential
implementation) of the abstract data type.
As done previously~\cite{BKT:EC22015}, it is easy to  
model a variety of ADTs in SMT.

Once we have modeled object methods $m$ and $n$, we further
model conflict abstractions. SMT reasoning then proceeds 
by asserting the following series of constraints:
\begin{enumerate*}
  \item Method $m$ performs its conflict abstraction reads/writes.
  \item Method $m$ performs its data-structure operation.
  \item Method $n$ performs its conflict abstraction reads/writes.
  \item No read/write or write/write conflict occurs.
  \item Method $n$ performs its data-structure operation.
\end{enumerate*}
We now need to ensure that the resulting state is the same as it would
have been if the operations executed in the opposite order. Using different
variable names for the intermediate states, we then assert the 
other order ($n$ before $m$).
Finally, we assert that the results (return values and final state) were
different and check whether this is satisfiable.
If it is not satisfiable, then the conflict abstraction is correct.

\section{Shadow Copies}
\label{section:shadow}
Previous approaches such as boosting~\cite{HerlihyK2008} relied on inverse operations to cleanup an aborted transaction, allowing updates to be made \emph{eagerly}.
Unfortunately, this approach was coupled with pessimistic conflict resolution, where execution would wait when a conflict was detected.
In an optimistic setting, transactions execute as if they will not encounter conflicts, and abort or retry when they are eventually detected.
If these are detected early enough, the eager update strategy still works, but this approach is not compatible with all STMs, and so lazy updates may be preferable in many situations.

We now present techniques for lifting linearizable objects into a transactional setting using \emph{lazy} updates.
A key challenge is that a transaction must be able to observe the results of speculative updates to shared objects, without those updates becoming visible to other transactions until the commit succeeds.

To implement lazy updates, we use \emph{replay logs} to queue updates, and only apply them when it is known the transaction will commit, behind the STM's native locking mechanisms.
If, instead, the transaction aborts, its log is dropped.

When an operation is logged, the transaction which executed it must also be able to obtain the value returned by that operation.
To obtain these values, transactions must utilize a \emph{shadow copy} of that data structure.
We describe two different approaches for implementing such functionality.

\paragraph{Memoization.}
For some data-structures (e.g. sets or maps), the results of an operation (even an update) can be computed purely from the initial state of the wrapped data-structure, or from the arguments to other pending operations.
In these cases, we may implement shadow copies by memoization.
Repeated operations to the same key can be cached in a transaction-local table, and queried, to determine the results of the next operation on that key.
If the key is not present, it's state can be determined by reading the unmodified backing data structure.

We implemented this approach in our \textsf{LazyHashMap}, using Java's
\textsf{ConcurrentHashMap} as the underlying data-structure.

\paragraph{Snapshots.}
For many data structures, memoization will be insufficient.
A more general approach uses the fast-snapshot semantics provided by many concurrent data structures~\cite{petrank2013lock,ProkopecSnapQueue,BronsonPCBST,ProkopecCTrie}.
The first time a transaction attempts to perform an update, a snapshot is made, and all further updates are performed on that snapshot.
Whenever a transaction commits, any changes to the snapshot are replayed onto the shared copy.

We implemented two data-structures this way: \textsf{LazyTrieMap} (based on Scala'a \textsf{TrieMap}) and \textsf{LazyPriorityQueue} (based the concurrent Braun heap from Chapter \ref{chap:handoverhand}).

\newcommand\klet{{\bf let}}
\newcommand\kmatch{{\bf match}}
\newcommand\kwith{{\bf with}}
\newcommand\kif{{\bf if}}
\newcommand\kthen{{\bf then}}
\newcommand\kreturn{{\bf return}}
\begin{listing}
\begin{lstlisting}[mathescape=true]
def proust_apply[Obj,Args,R](o:Obj,o.m:Args$\to$R,$\bar{\alpha}$:Args,T:Txn) $\to$ R {
  val cas_locs = f(o.m,$\bar{\alpha}$,$\sigma_o$) 
  val acquired = cas_locs.map($\lambda$ (v, mode) {
    (cfg.tsync($o$), mode) match {
      | (Pess,_) => lock(v,mode); v
      | (Opt,Rd) => stm_read(v); v
      | (Opt,Wr) => stm_write(v); v
    }
  })
     
  val rv = cfg.updstrat(o) match {
    | Eager => $o.m(\bar{\alpha})$ 
    | Lazy => Predict($o.m$,$\bar{\alpha}$)
  }
  cfg.updstrat(o) match {
    | Eager => T.onAbort($o.m^{-1}(\bar{\alpha})$)
    | Lazy =>
      T.onCommit($o.m(\bar{\alpha})$)
      if(cfg.tsync(o) == Opt) {
        forEach(v $\in$ acquired) {
          stm_read(v)
        }
      }
  }
  rv
}
\end{lstlisting}
\caption{Pseudo-code for the Proust algorithm, turning a CA and object configuration into an STM-friendly wrapper.}
\label{lst:apply}
\end{listing}

\section{The Proust Algorithm}
\label{section:algorithm}
In this section we describe the Proust algorithm for implementing a transactional wrapper for highly-concurrent objects based on conflict abstractions (Sections~\ref{section:comm2ca}) and alternate histories (Section~\ref{section:althist}).

Before this algorithm can be applied, several things must be determined.
First, a conflict abstraction must be defined for the ADT, as described in Section~\ref{section:conflict}.
After following the flowchart in Figure~\ref{fig:flowchart}, and comparing the allowed configurations to the set of operations supported by the ADT, two further decisions remain: whether the wrapper will manage alternate histories through shadow copies (Section~\ref{section:shadow}) and replay logs, or through inverses and undo logs (as in Boosting); and whether the wrapper will rely on optimistic or pessimistic synchronization.

Once these decisions have been made, the wrapper can be constructed, by using the Proust algorithm, defined in Listing~\ref{lst:apply}, to invoke each supported operation.
The algorithm, \texttt{proust\_apply}, must execute in the context of a transaction $T$. 
A system running Proust also will also have a \emph{configuration} \textsf{cfg}, indicating
the user's choices as to the per-object update strategy (\textsf{cfg.updstrat}$(o)$) as well as the
transactional synchronization to be used (\textsf{cfg.tsync}$(o)$).

On line 2, we invoke the conflict abstraction $f$ to obtain a list of synchronization primitives corresponding to abstract state elements (and for each element, whether $o.m$ is a read or a write).
Then on lines 3-8, we either acquire the resulting lock\footnote{
It is assumed that the read/write locks here are transaction-aware, and will be automatically released on either commit or abort.
} (for pessimistic synchronization under an optimistic STM), or perform a read or write operation on the CA memory locations (for all other use cases).
Lines 10-12 compute the return value $rv$ either by eagerly applying $o.m$ directly, or by predicting the outcome on a shadow copy of $o$, depending on the update strategy.
Before returning, the log entries are made to either undo (line 15) or replay (line 16) the operation.
Finally, lines 17-18 handle the edge case\footnote{
With eager conflict detection, this would have happened as soon as the conflict occurred, and with pessimistic synchronization the locks provide stronger protection.
} where optimistic updates are being applied lazily: it is possible that a conflict has occurred since the STM references for the abstract state were accessed on lines 6 and 7, and performing a fresh dummy read for each state element gives the STM a chance to intervene before opacity can be violated.

We now argue that transactional objects using this algorithm/wrapper,
conflict
  abstractions (Section~\ref{section:conflict}) and shadow copies (Section~\ref{section:shadow})
  satisfy opacity, regardless of the guarantees of the underlying STM.
\emph{Opacity}~\cite{opacity} is a correctness condition (simplifying somewhat),
ensuring that committed transactions appear to have executed in a serial order,
and that aborted transactions observed consistent memory states at all times.
Modern STMs typically guarantee opacity.
  While the primary contribution of this chapter involve the concepts, implementation, and
  evaluation, we nonetheless sketch arguments for opacity.
In consideration of opacity, we refer to the \texttt{proust\_apply} algorithm in Listing~\ref{lst:apply},
and consider the effect of \texttt{proust\_apply} in each possible scenario ({\bf A}, {\bf B}, etc.) in the flow chart in Figure~\ref{fig:flowchart}.

\emph{1. Pessimistic synchronization}
In Listing~\ref{lst:apply}, the conflict abstraction specifies
locks, which are acquired on Line 5.
The argument for opacity is largely that  of boosting~\cite{HerlihyK2008}, which was shown~\cite{KP:PLDI2015} to satisfy opacity.
Regardless of update strategy, only commutative operations may 
proceed concurrently, so transactions cannot observe inconsistent state from other uncommitted transactions.
The distinction here is that conflict abstractions provide a concrete intermediary concept, rather than the declarative abstract locking condition of~\cite{HerlihyK2008}.

\emph{2. Optimistic STMs.}
In a setting where the STM operates optimistically, there is an opportunity to synchronize the ADT optimistically, using the wrapper discussed above; however, challenges arise when conflicts are detected lazily, and care must be taken to maintain opacity. 
If STM conflict detection is done \emph{lazily}, a lazy update strategy will be used for the wrapper.
This is necessary, intuitively, because until commit, there may be a conflict that STM has not detected yet, so uncommitted changes to shared state are potentially observable. Thus we must be able to discard speculative data structure mutations without other transactions observing them.
If STM conflict detection is \emph{eagerly}, a lazy or eager update strategy can be used because the STM will abort as soon as it notices a conflict.
\section{Implementation \& Evaluation}

\label{section:impl}
\begin{figure*}[!htb]\centering
\small
\setlength\tabcolsep{2.5pt}
\begin{tabular}{cl|cccc}
& & \multicolumn{4}{c}{Fraction of operations that are writes ($u$)}\\
& & 0.25 & 0.5 & 0.75 & 1 \\
\hline
\multirow{2}{*}{\rotatebox{90}{Ops per transaction ($o$)\hspace{2em}}} & & & & \\
&
\rotatebox{90}{\hspace{.5in}1} &
\includegraphics[width=1.45in,trim={0.75cm 0.25cm 0.75cm 0},clip]{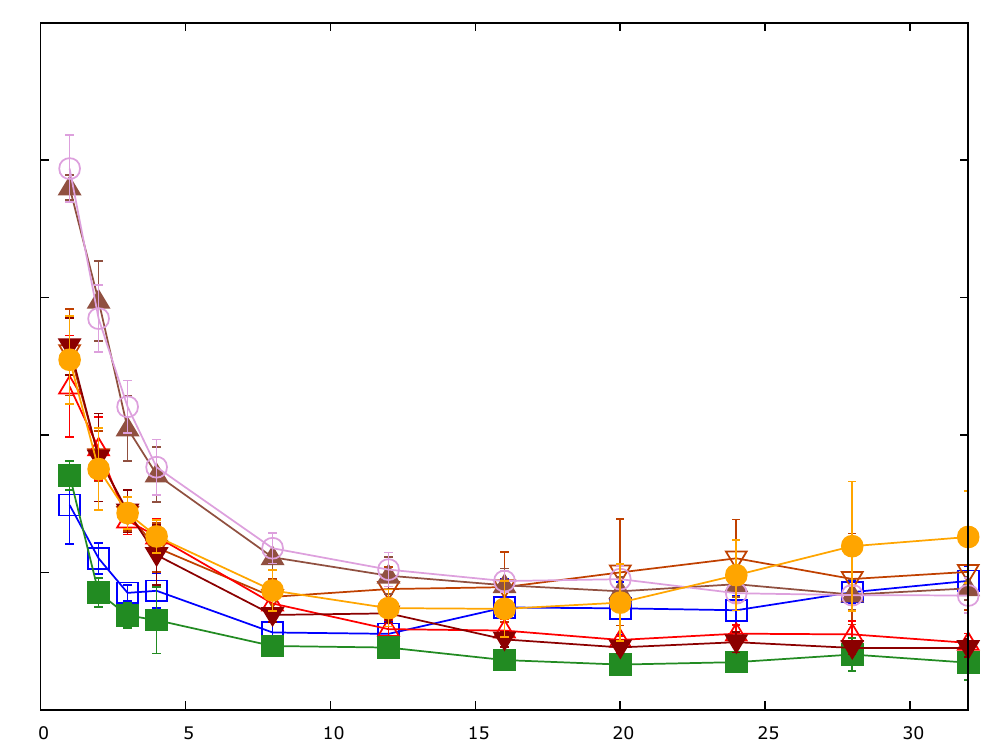} &
\includegraphics[width=1.45in,trim={0.75cm 0.25cm 0.75cm 0},clip]{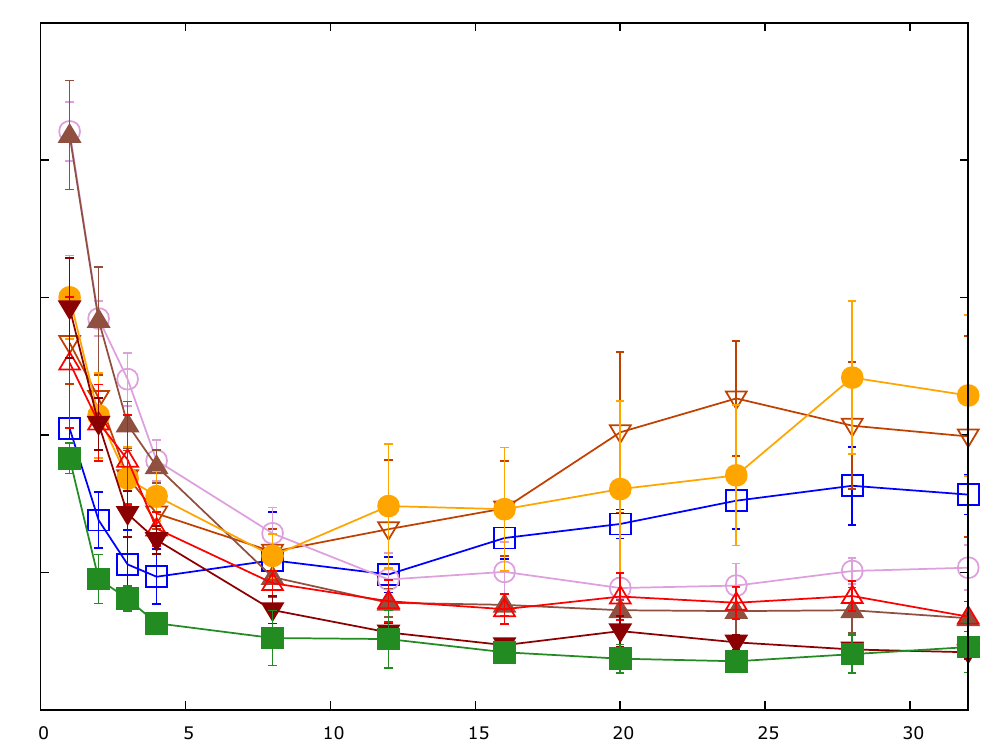} &
\includegraphics[width=1.45in,trim={0.75cm 0.25cm 0.75cm 0},clip]{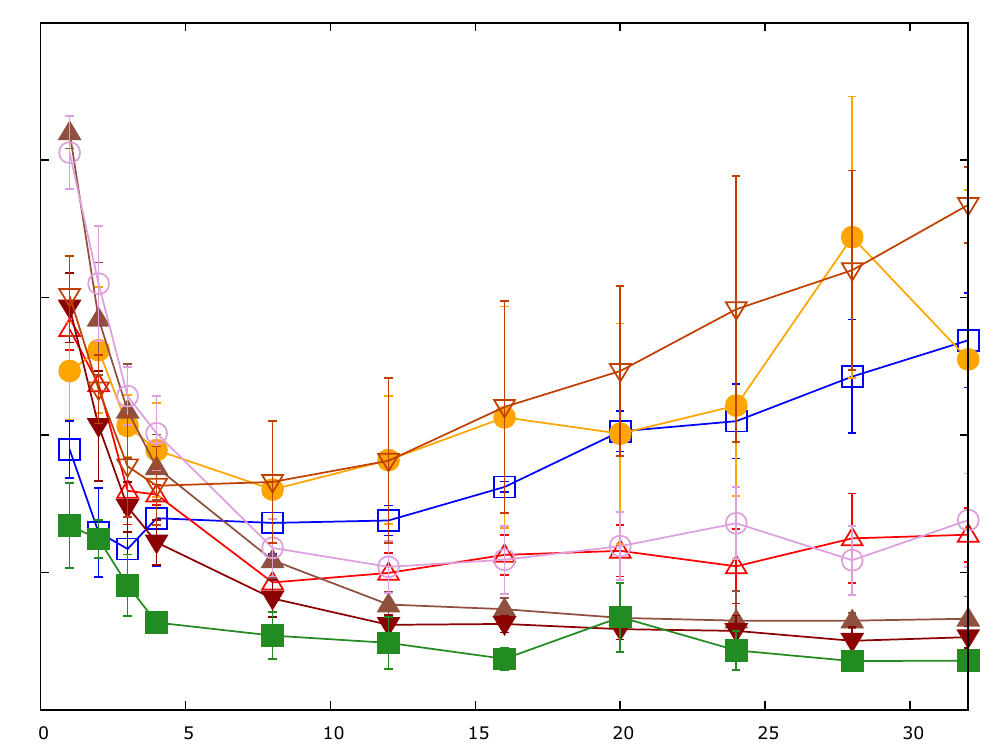} &
\includegraphics[width=1.45in,trim={0.75cm 0.25cm 0.75cm 0},clip]{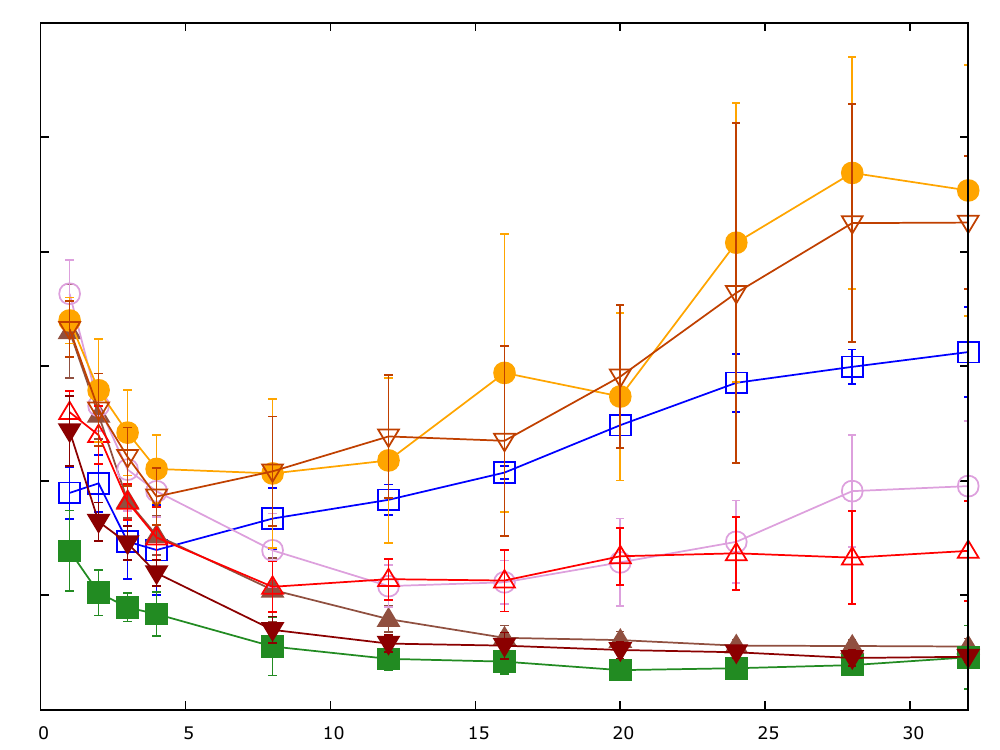} \\&
\rotatebox{90}{\hspace{.5in}16} &
\includegraphics[width=1.45in,trim={0.75cm 0.25cm 0.75cm 0},clip]{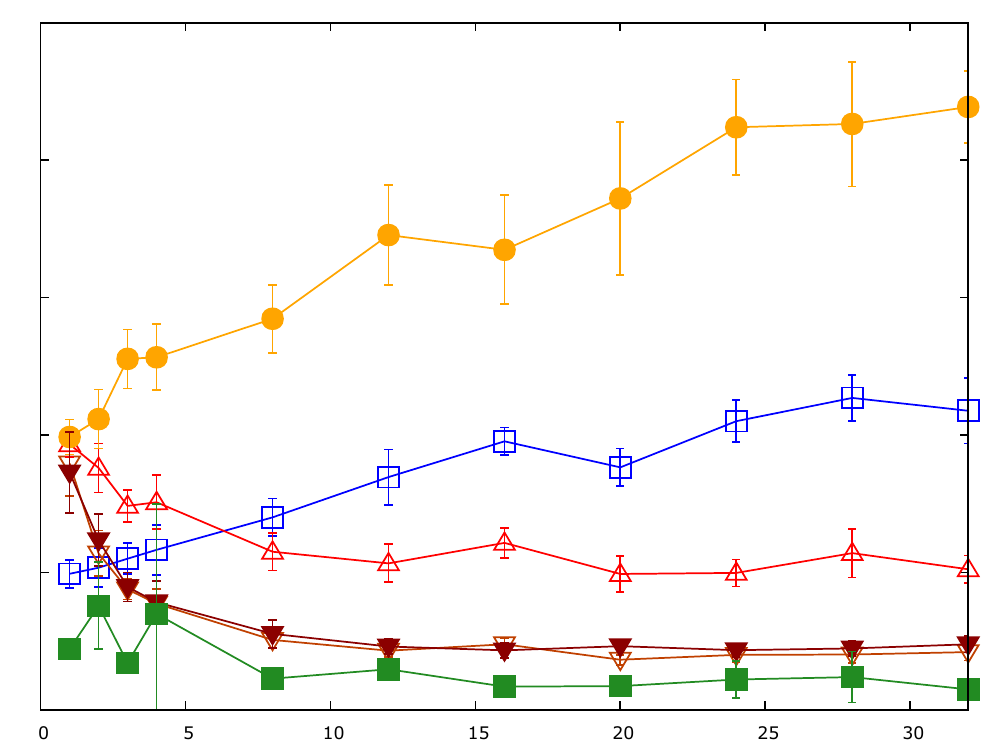} &
\includegraphics[width=1.45in,trim={0.75cm 0.25cm 0.75cm 0},clip]{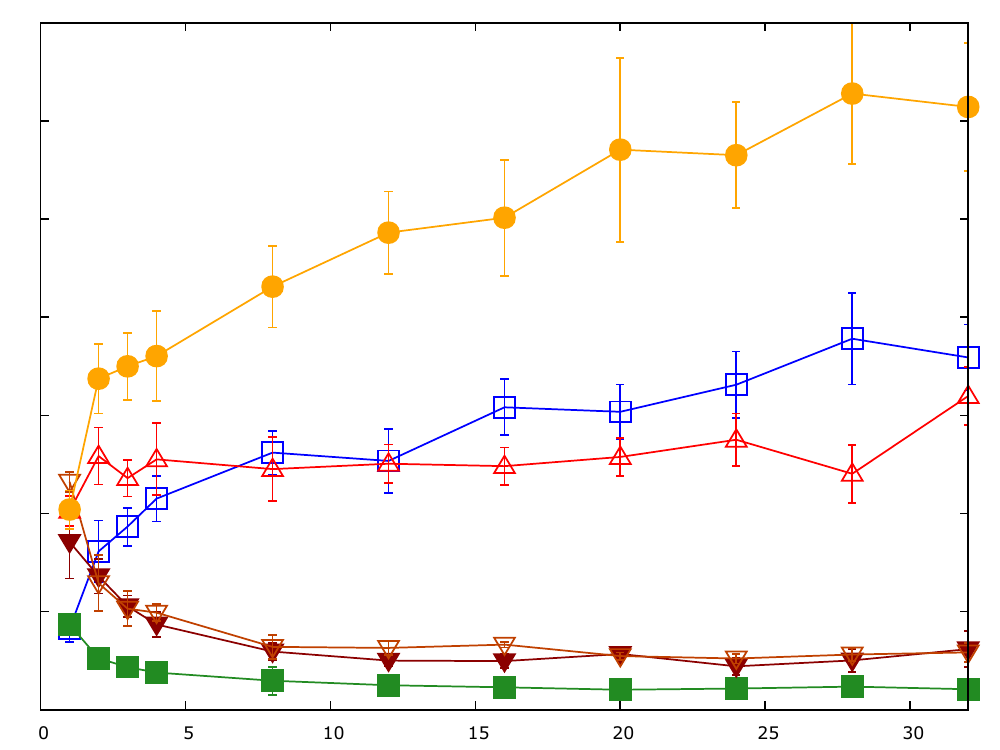} &
\includegraphics[width=1.45in,trim={0.75cm 0.25cm 0.75cm 0},clip]{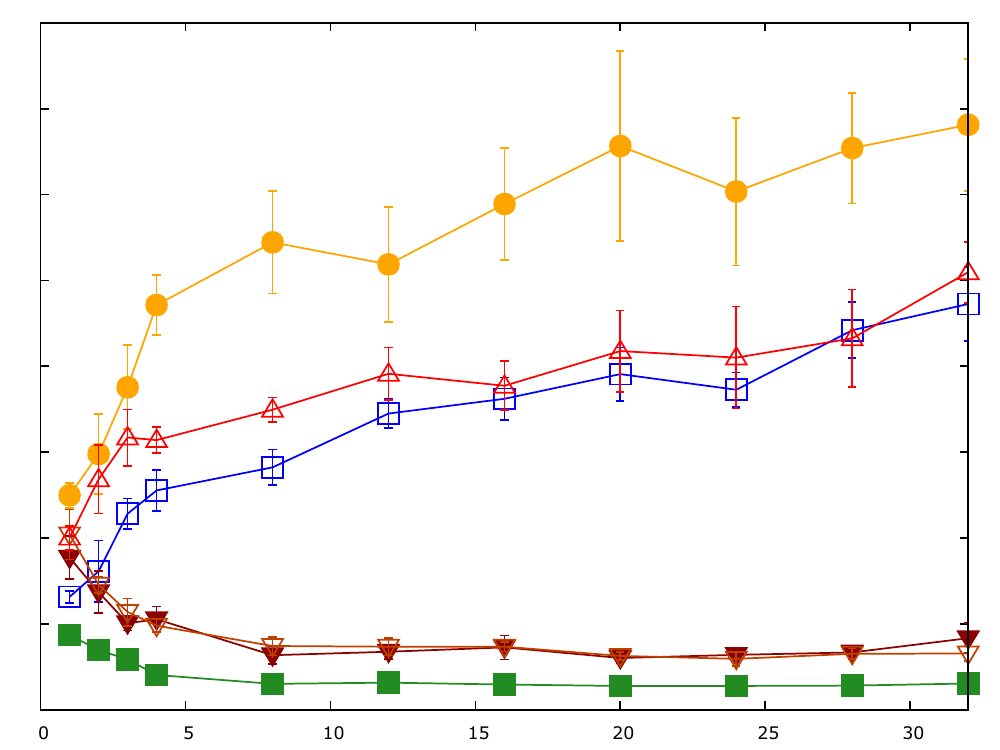} &
\includegraphics[width=1.45in,trim={0.75cm 0.25cm 0.75cm 0},clip]{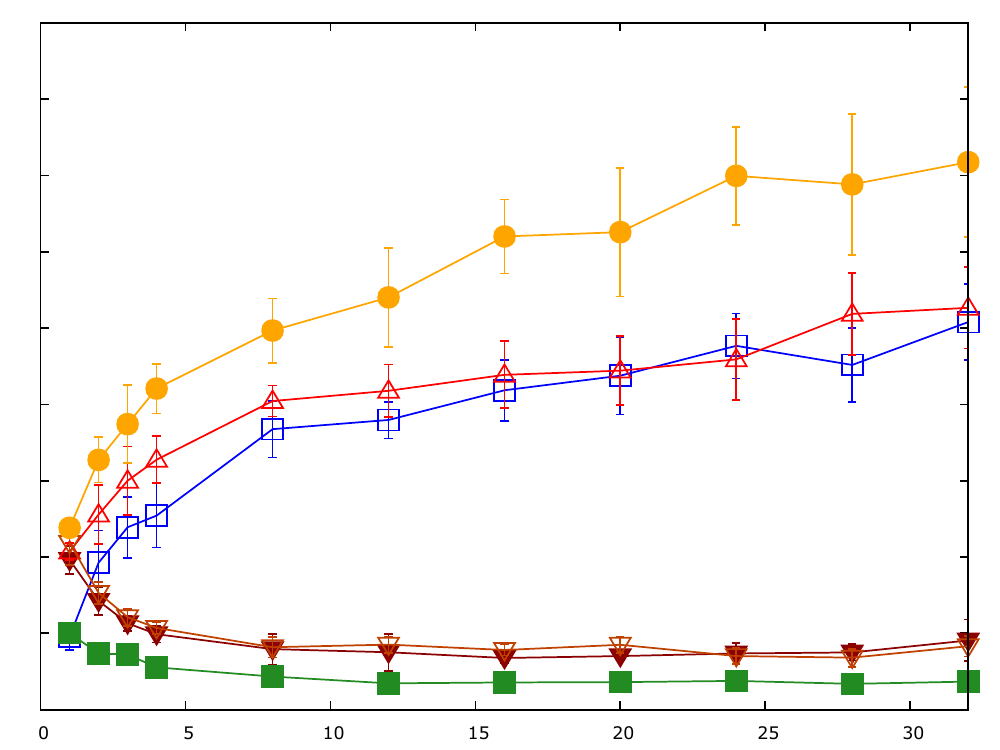} \\
& & \multicolumn{4}{c}{\includegraphics[width=4.0in]{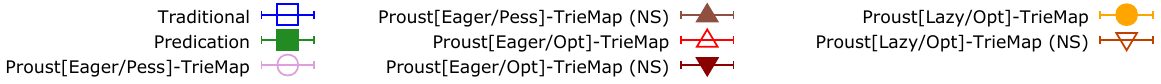}}\\
\end{tabular}
\caption{\label{fig:mapThroughput} Time to process $10^{6}$ operations on concurrent maps (\emph{smaller is better}), varying \%-updates and \#ops/txn. For each chart, the x-axis is the number of threads from 0 to 32 and the y-axis is the average time in milliseconds from 0 to 250. The (NS) variants disabled \texttt{size()}.}
\end{figure*}
\begin{figure*}[!htb]\centering
\small
\setlength\tabcolsep{2.5pt}
\begin{tabular}{cl|cccc}
& & \multicolumn{4}{c}{Fraction of operations that are writes ($u$)}\\
& & 0.25 & 0.5 & 0.75 & 1 \\
\hline
\multirow{1}{*}{\rotatebox{90}{Ops per transaction ($o$)\hspace{1em}}} & & & & \\
&
\rotatebox{90}{\hspace{.5in}256} &
\includegraphics[width=1.45in,trim={0.75cm 0.25cm 0.75cm 0.5cm},clip]{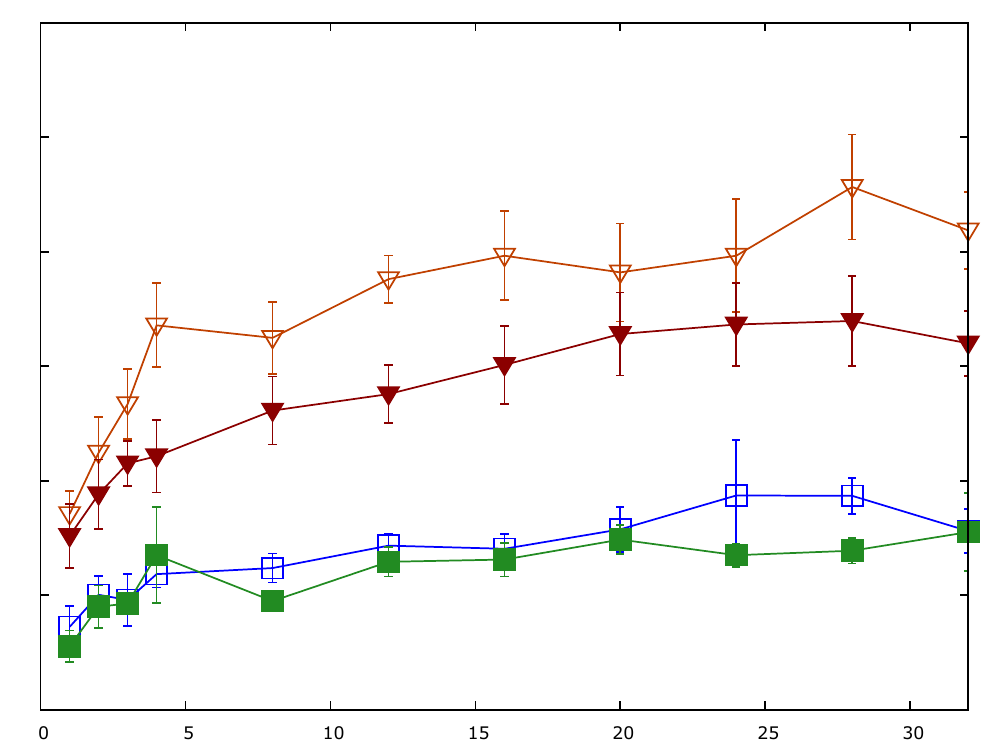} &
\includegraphics[width=1.45in,trim={0.75cm 0.25cm 0.75cm 0.5cm}]{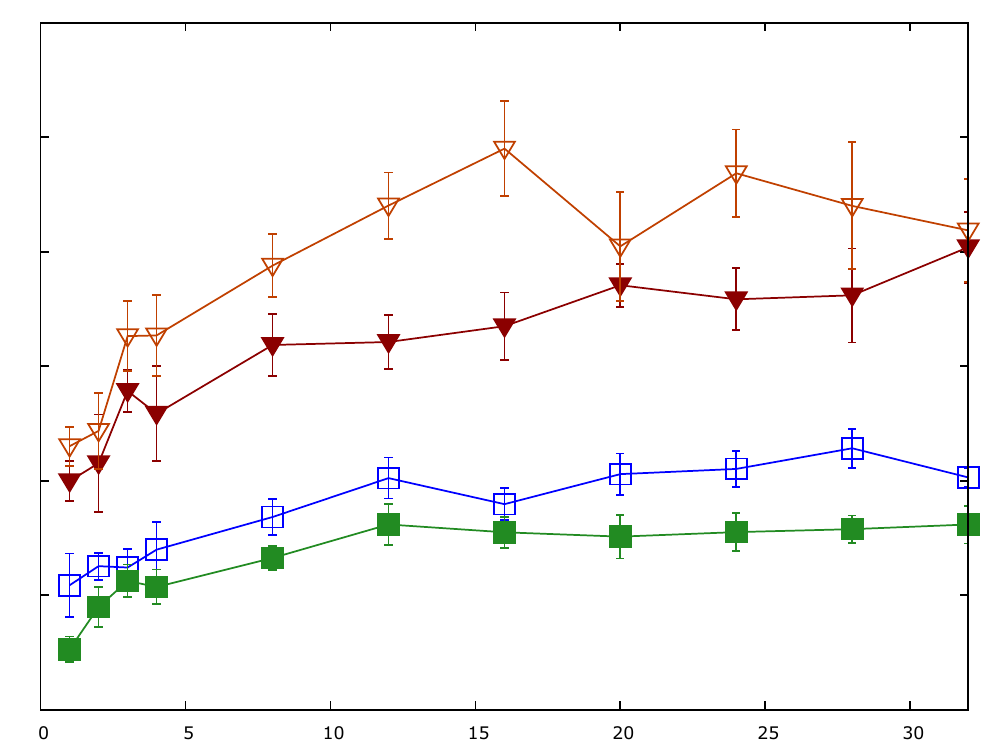} &
\includegraphics[width=1.45in,trim={0.75cm 0.25cm 0.75cm 0.5cm}]{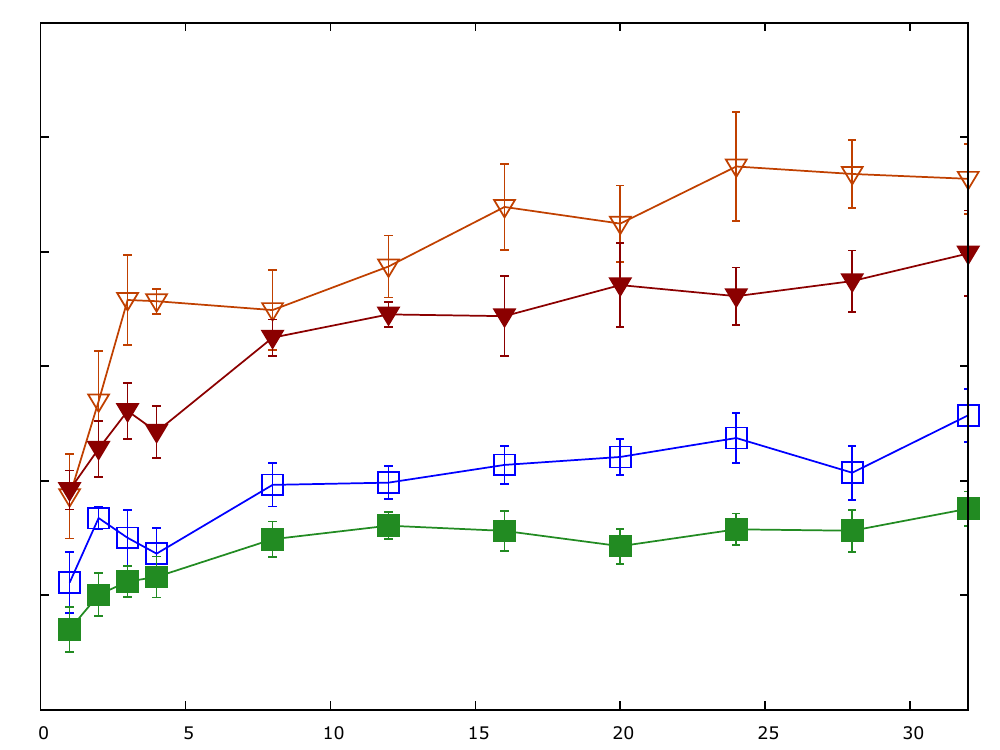} &
\includegraphics[width=1.45in,trim={0.75cm 0.25cm 0.75cm 0.5cm}]{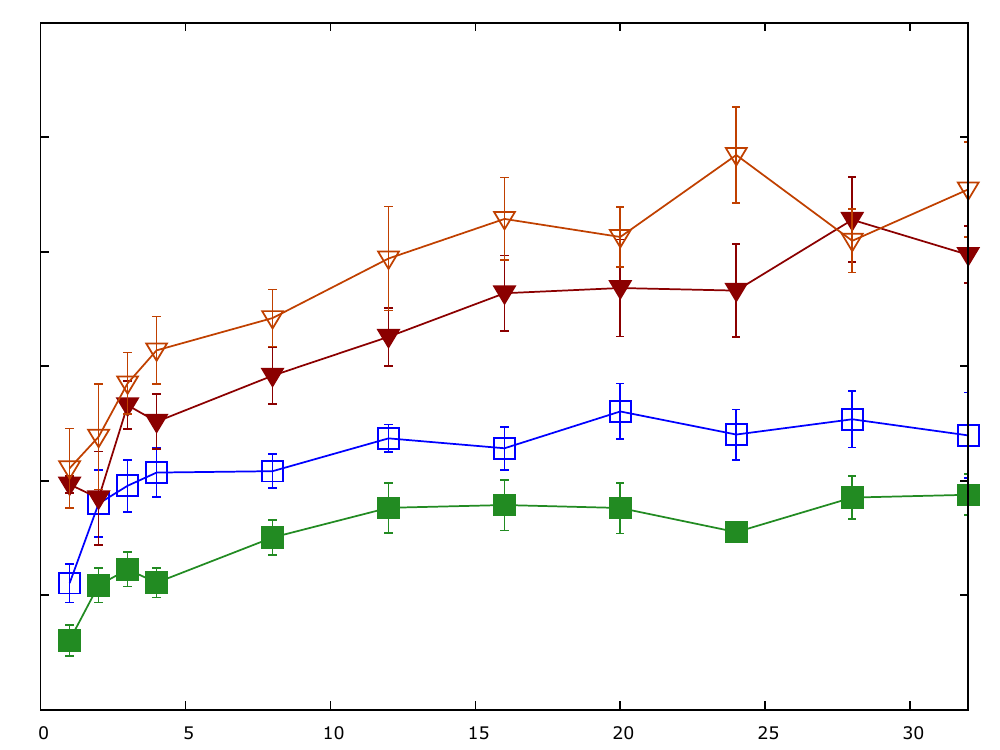} \\
& & \multicolumn{4}{c}{\includegraphics[width=2.7in]{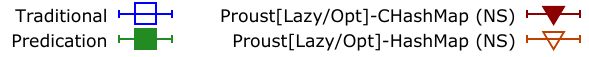}}\\
\end{tabular}
\caption{\label{fig:replayRewindCosts} Memoizing shadow copies allow updates of the same entry to be combined, providing a substantial decrease in execution time. \emph{Smaller is better}.}
\end{figure*}
\begin{figure*}[!htb]\centering
\small
\setlength\tabcolsep{2.5pt}
\begin{tabular}{cl|cccc}
& & \multicolumn{4}{c}{Fraction of operations that are writes ($u$)}\\
& & 0.25 & 0.5 & 0.75 & 1 \\
\hline
\multirow{1}{*}{\rotatebox{90}{Ops per transaction ($o$)\hspace{2em}}} & & & & \\
&
\rotatebox{90}{\hspace{.5in}16} &
\includegraphics[width=1.45in,trim={0.75cm 0.25cm 0.75cm 0},clip]{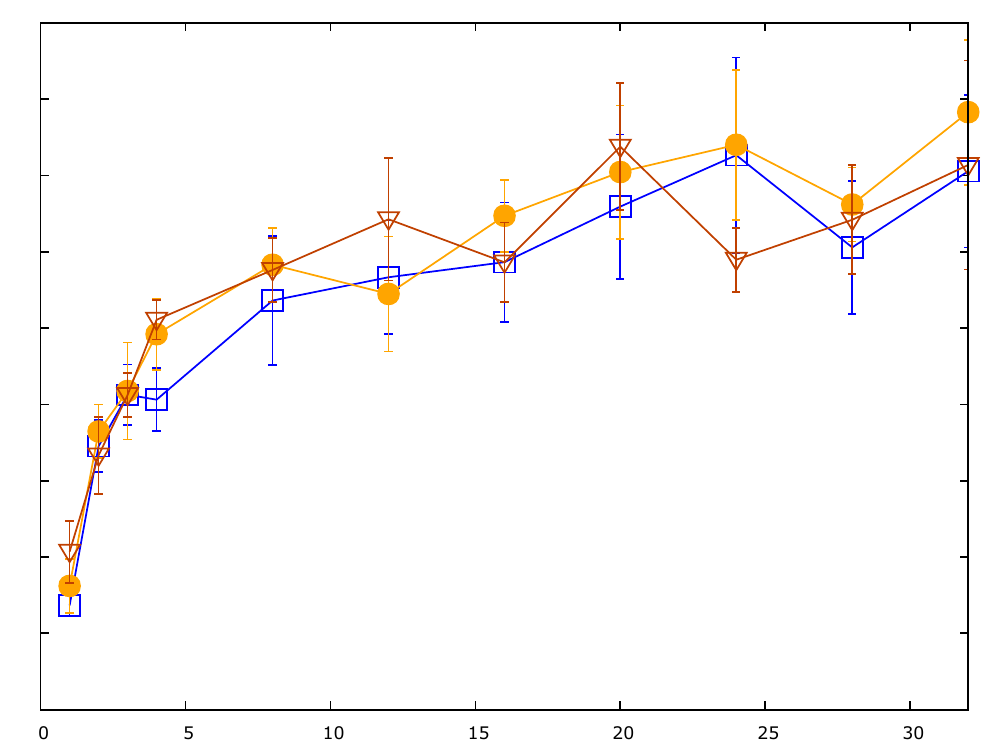} &
\includegraphics[width=1.45in,trim={0.75cm 0.25cm 0.75cm 0},clip]{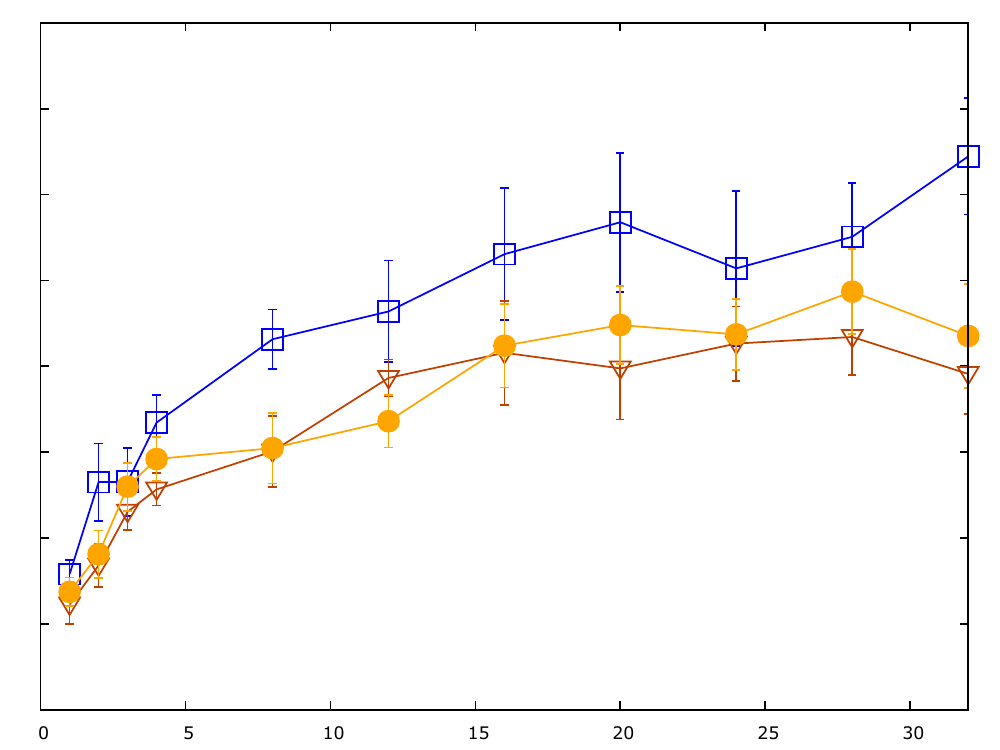} &
\includegraphics[width=1.45in,trim={0.75cm 0.25cm 0.75cm 0},clip]{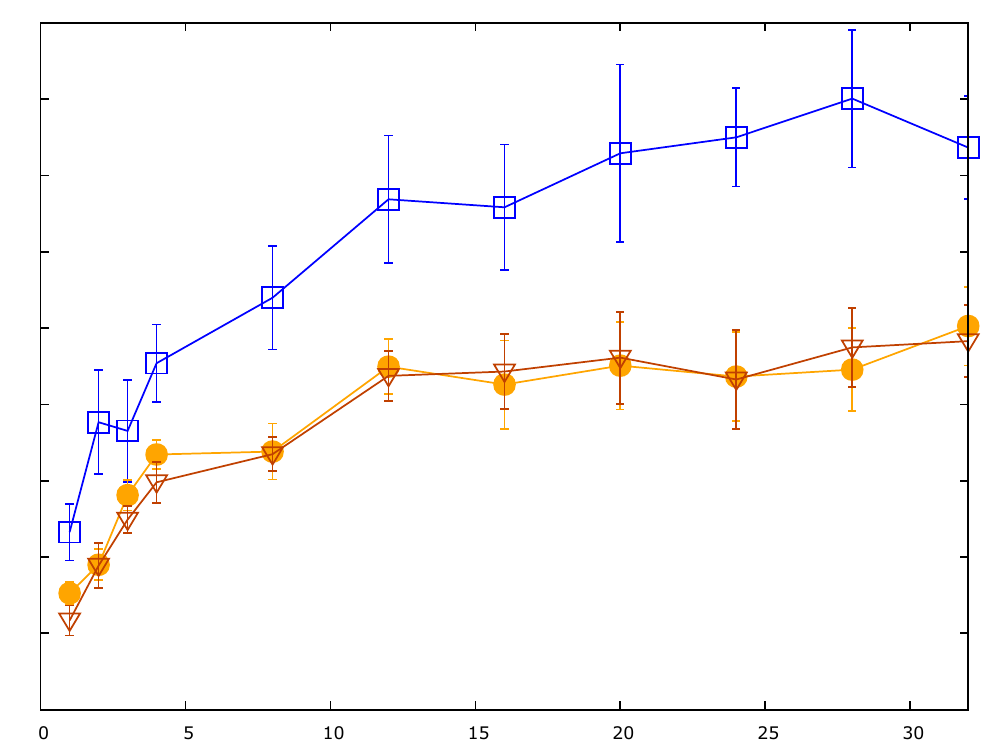} &
\includegraphics[width=1.45in,trim={0.75cm 0.25cm 0.75cm 0},clip]{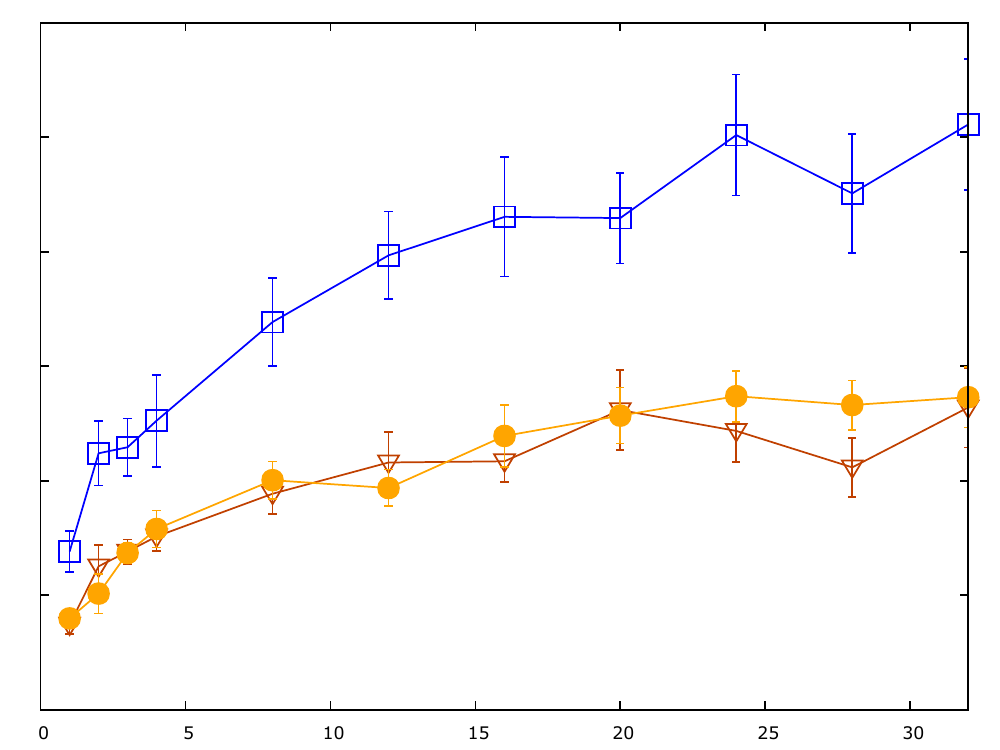} \\
& & \multicolumn{4}{c}{\includegraphics[width=4.0in]{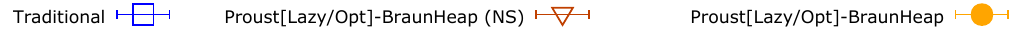}}\\
\end{tabular}
\caption{\label{fig:pqThroughput} Time to process $10^{6}$ operations on concurrent priority queues (\emph{smaller is better}), varying \%-updates and \#ops/txn.  For each chart, the x-axis is the number of threads from 0 to 32 and the y-axis is the average time in milliseconds from 0 to 2400. The (NS) variants disabled \texttt{size()}.}
\end{figure*}
\begin{figure}[!htb]
\centering
\includegraphics[width=1.5in]{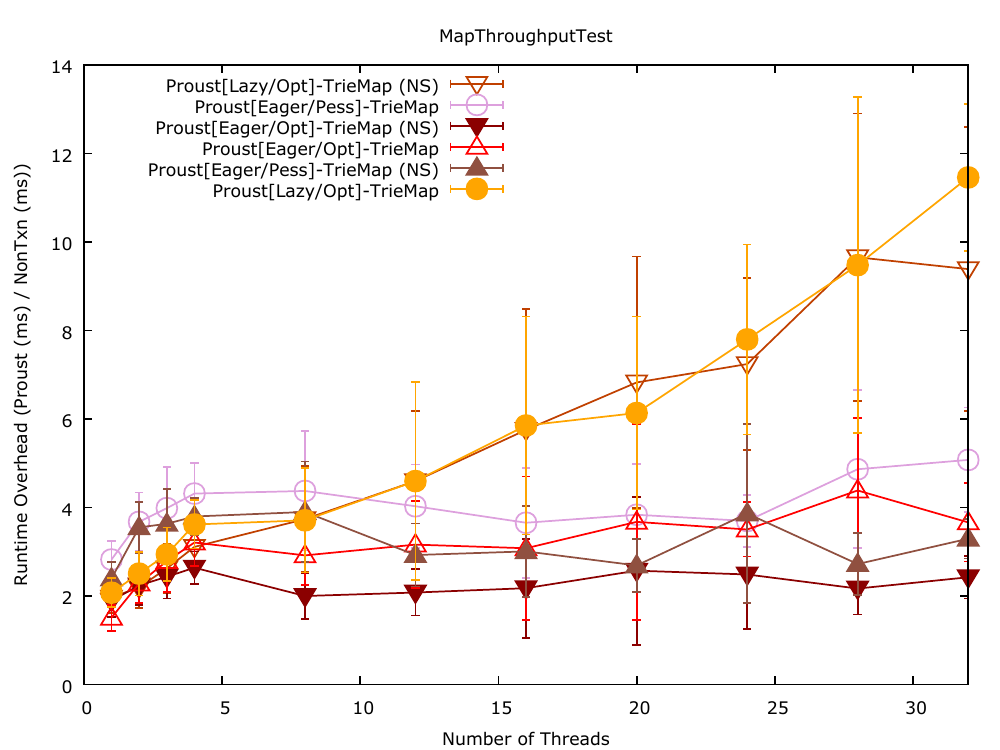}
\includegraphics[width=1.5in]{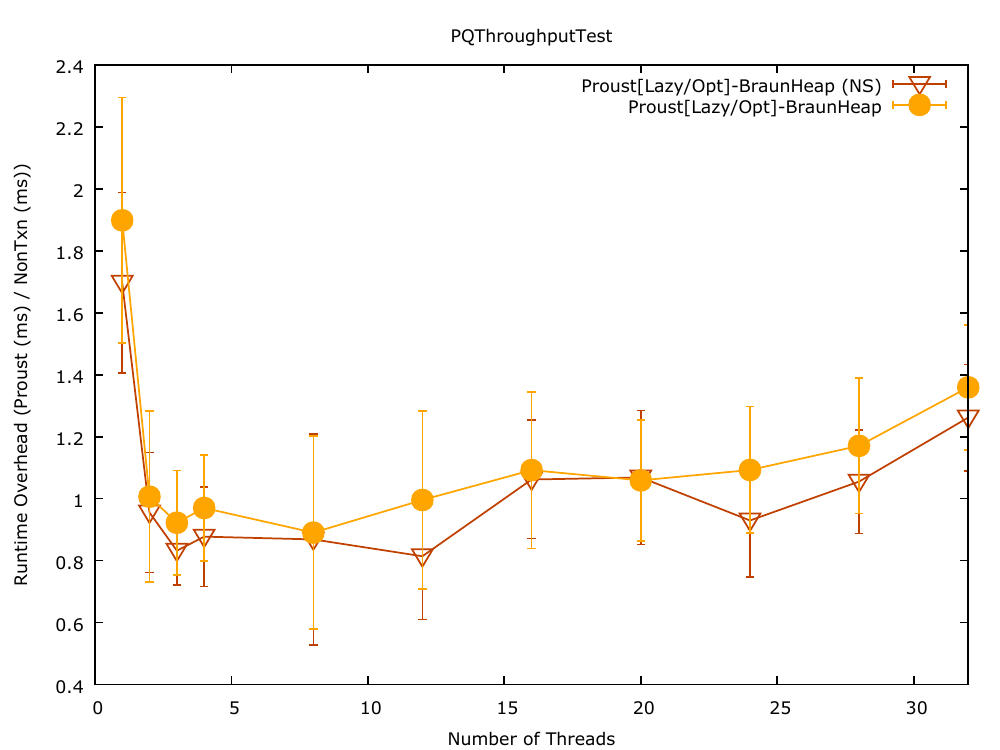}
\caption{\label{fig:overheads} Overhead of the transactional wrapper (relative to the base data structure) for different configurations of Proust ($o=1$).
On the left are transactional maps and on the right are transactional priority queues. \emph{Smaller is better.}}
\end{figure}
We have developed a library implementation of Proust on top of ScalaSTM, which we call \toolName.
\toolName{} provides an API implementing abstract locks, replay logs for both shadow-copy techniques, and lock allocator policies of both the optimistic\footnote{We note that the default backend for ScalaSTM is CCSTM which uses a mixed conflict detection strategy. ~\cite{bronson2010ccstm} One consequence of this is that eager/optimistic objects will not satisfy opacity out of the box.
We nonetheless felt it was useful to provide them as ScalaSTM supports pluggable backends, and some practical applications may endure opacity violations without ill-effect.} and pessimistic varieties.
It also provides a number of wrapped Proustian data structures out of the box, including both transactional maps and transactional priority queues, which can be used as-is, or serve as example code for developers to create their own wrappers.

\subsection{\label{section:eval}Evaluation}
\paragraph{Maps} 
We benchmarked several \toolName{} map wrappers (including variants with and without a \texttt{size()} operation) against both predication and a traditional pure-STM hash map, with a setup similar to that used by Bronson, et al. for predication~\cite{BronsonCCOn2010}\footnote{
We ran our experiments on an Amazon EC2 \texttt{m4.10xlarge} instance (\url{https://aws.amazon.com/blogs/aws/the-new-m4-instance-type-bonus-price-}\\\url{reduction-on-m3-c4/}), which has 40 vCPUs and 160 GB of RAM.
}.
For each experiment, we performed $10^{6}$ random operations on a shared map, split across $t$ threads, with $o$ operations per transaction.
A $u$ fraction of the operations were writes (evenly split between \texttt{put} and \texttt{remove}), and the remaining $(1- u)$ were \texttt{get} operations.
We varied $t$, $o$, and $u$ to achieve different levels of contention\footnote{We did not vary key range as in the predication paper, as garbage collection was not a focus of this implementation}.
For each configuration, we warmed up the JVM for $10$ executions, then timed each of the following $10$ executions, garbage collecting in between to reduce jitter, and reported the mean and standard deviation.

Three notes about the experimental setup. 
First, we were only able to achieve a weak coupling with the CCSTM contention manager for the pessimistic experiments, and found that under the artificially high contention seen in these experiments, longer transaction times could lead to live-lock, as the STM lacked required information about the instigating (non-STM) memory accesses.
For this reason, we only show the pessimistic results in the initial $o=1$ experiments.
Second, though the Eager/Optimistic configuration does not satisfy opacity under the CCSTM backend for ScalaSTM, we benchmarked it anyway, and did not observe any instances where this violated correctness (notably our benchmark makes no explicit control flow decisions based on the results of map accesses, and ScalaSTM performs an abort and retry if it ever observes an unchecked exception).
It seems likely that a performance penalty was paid for late detection of inconsistent memory accesses, and we believe this speaks well to the potential performance of Eager/Optimistic wrappers on STMs where they satisfy opacity.
Third, substantial performance differences between the standard and \texttt{(NS)} wrapper variants illustrate the previously discussed impedance mismatch between ``pure'' writes in a conflict abstraction and ``impure'' writes provided by extant STMs.

The experimental results depicted in Figure \ref{fig:mapThroughput}
display the effects of several competing trends. 
Intuitively, Proust's performance scales much better than the traditional STM implementation as contention increases, due to varying $t$ and $u$ (though we are consistently outperformed by the highly engineered predication implementation\footnote{Predication as a technique is specialized to maps and sets, in essence embedding their conflict abstraction as the member elements of the backing collection and allowing frequent updates to the same element to avoid updating the concrete state of the backing data structure.}); however, increasing values of $o$ have a negative influence on the relative performance of the Proust wrappers.
Intuitively, this is to be expected, as our log sizes (either to undo or replay) are proportional to the number of updates performed, whereas predication and traditional implementations replay with time proportional to the number of unique memory locations updated, and as $o$ increases, so does the probability that multiple writes will alter the same location.

An optimization for memoization-, rather than snapshot-, based shadow copies is to apply only the final state of each abstract state element; resulting in Figure~\ref{fig:replayRewindCosts}.

The overhead of the wrapper, relative to the base data structure can be seen in Figure~\ref{fig:overheads}.

\paragraph{Priority Queues}
We used a nearly identical experimental setup to compare the runtimes of two priority queues based on Braun heaps (one traditional STM implementation and one wrapper around the snapshottable concurrent implementation from Chapter \ref{chap:handoverhand}).
The writes were split evenly between \texttt{insert} and \texttt{removeMin} operations.

The experimental results depicted in Figure \ref{fig:pqThroughput} show that across a variety of conditions, the Proustian queue was competitive with, or outperformed, the traditional implementation.
In general, run times were substantially longer than for the map throughput test, as the min-element is subject to heavy contention; however, unlike for the map tests, the effects of additional operations per transaction were less pronounced, as most contention is discovered early in the transaction.

The overhead of the wrapper, relative to the base data structure can be seen in Figure~\ref{fig:overheads}.

\section{Conclusions}
\label{sec:concl}
We introduced two new concepts---conflict abstractions and shadow copies---for transactional object systems.
These concepts unify prior works (e.g. predication and boosting) and lead to a broader design space for incorporating existing fine-tuned highly-concurrent objects into a transactional setting, permitting use of different synchronization and update strategies to selectively optimize wrapped data structures for different STMs and different expected work-loads.

Benchmarks show we outperform, or are competitive with, pure STM solutions, and use of extant data structures libraries achieves that performance without writing implementations from scratch.
While we are outperformed by Predication on the map throughput tests, we believe that our utility as a tool for wrapping arbitrary data structures will encourage use beyond sets and maps.

\paragraph{Future Work} On the experimental front, Proust could be implemented for other STMs.
More theoretically: \textbf{(a)} an extension of our log-combining optimization from memoized replays to snapshot replays and undo logs would further improve performance, or, alternatively, shadow copies based on confluently persistent data structures could even be merged without an explicit log;~\cite{FIAT200316}
\textbf{(b)} the use of conflict abstractions to describe commutativity and synchronization reveals a use-case for STMs to support ``pure writes'', allowing them to match the expressivity of handcrafted locks, but we know of no STMs providing such an operation;
\textbf{(c)} automatic verification techniques (such as those discussed in Section~\ref{section:conflict}) might be
used as a building-block for an automatic synthesis technique, and we believe it would be interesting to apply the approach of counter-example guided inductive synthesis
(CEGIS)~\cite{GJTV:PLDI2011,IGIS:OOPSLA2010,BCKGM:PPoPP2013,solar2008},
using SAT/SMT counter-examples as the basis for constructing $f$, allowing conservative estimates of commutativity to be parameterized by the size of a condensed state-space.

	\chapter{\label{chap:smartcontracts}An Algorithm for Concurrent Execution of Smart-Contracts}
	
\newcommand\BallotcontentionMiner{1.57}
\newcommand\BallotcontentionValidator{1.57}
\newcommand\BallotblocksizeMiner{1.35}
\newcommand\BallotblocksizeValidator{1.40}
\newcommand\SimpleAuctioncontentionMiner{1.20}
\newcommand\SimpleAuctioncontentionValidator{1.21}
\newcommand\SimpleAuctionblocksizeMiner{1.49}
\newcommand\SimpleAuctionblocksizeValidator{1.53}
\newcommand\EtherDoccontentionMiner{0.76}
\newcommand\EtherDoccontentionValidator{1.77}
\newcommand\EtherDocblocksizeMiner{1.00}
\newcommand\EtherDocblocksizeValidator{1.57}
\newcommand\TokencontentionMiner{1.73}
\newcommand\TokencontentionValidator{1.72}
\newcommand\TokenblocksizeMiner{1.63}
\newcommand\TokenblocksizeValidator{1.63}
\newcommand\MixedcontentionMiner{1.70}
\newcommand\MixedcontentionValidator{1.89}
\newcommand\MixedblocksizeMiner{1.50}
\newcommand\MixedblocksizeValidator{1.62}
\newcommand\overallMinerSpeedup{1.39}
\newcommand\overallValidatorSpeedup{1.59}

\section{Introduction}
The preceding chapters used techniques inspired by functional programming to ease the implementation of efficient concurrent data structures.
In this chapter we leverage the techniques from Chapter \ref{chap:proust} to improve the performance of mostly-sequential operations on a distributed persistent data structure.
In particular, we focus on the timely topics of \emph{blockchains} and \emph{smart contracts}.
We will define these terms in more detail below, but we encourage the reader to adopt the following perspective:
from an API perspective, a blockchain is just a persistent map or dictionary, augmented with a linked list providing programmatic access to the prior history.
Furthermore, executing a smart contract is just running an STM transaction to apply one or more updates to that map.

Cryptocurrencies such as Bitcoin~\cite{bitcoin} or
Ethereum~\cite{ethereum} are very much in the news.
Each is an instance of a \emph{distributed ledger}:
a publicly-readable tamper-proof record of a sequence of events.
Simplifying somewhat, early distributed ledgers, such as Bitcoin's, work like this:
\emph{clients} send
\emph{transactions}\footnote{Following blockchain terminology,
  a transaction is a payment or set of payments,
  not an atomic unit of synchronization as in databases or transactional memory.}
to \emph{miners},
who package the transactions into \emph{blocks}.
Miners repeatedly \emph{propose} new blocks to be applied to the ledger,
and follow a global consensus protocol to agree on which blocks are chosen.
Each block contains a cryptographic hash of the previous block,
making it difficult to tamper with the ledger.
The resulting distributed data structure, called a \emph{blockchain},
defines the sequence of transactions that constitutes the distributed
ledger\footnote{This description omits many important issues,
such as incentives, forking, and fork resolution.}.

Modern blockchain systems often interpose an additional software layer between
clients and the blockchain.
Client requests are directed to scripts, called \emph{smart contracts},
that perform the logic needed to provide a complex service,
such as managing state, enforcing governance, or checking credentials.
Smart contracts can take many forms,
but here we will use (a simplified form of) the Ethereum model~\cite{ethereum}.

A smart contract resembles an object in a programming language.
It manages long-lived \emph{state}, which is encoded in the blockchain.
The state is manipulated by a set of \emph{functions},
analogous to \emph{methods} in many programming languages.
Functions can be called either directly by clients or indirectly by other
smart contracts.
Smart contract languages are typically Turing-complete.
To ensure that function calls terminate,
the client is charged for each computational step in a function call.
If the charge exceeds what the client is willing to pay,
the computation is terminated and rolled back.

When and where is smart contract code executed?
There are two distinct circumstances.
Each smart contract is first executed by one or more \emph{miners},
nodes that repeatedly propose new blocks to append to the blockchain.
When a miner creates a block,
it selects a sequence of user requests
and executes the associated smart contract code for each Ethereum transaction
in sequence,
transforming the old contract state into a new state.
It then records both the sequence of transactions and the new state in
the block, and proposes it for inclusion in the blockchain.

Later, when the block has been appended to the blockchain,
each smart contract is repeatedly re-executed by \emph{validators}:
nodes that reconstruct (and check) the current blockchain state.
As a validator acquires each successive block,
it replays each of the transactions' contract codes to check that the
block's initial and final states match.
Each miner validates blocks proposed by other miners,
and older block are validated by newly-joined miners,
or by clients querying the contract state.
Code executions for validation vastly exceed code executions for mining.

Existing smart contract designs limit throughput because they admit no concurrency.
When a miner creates a block,
it assembles a sequence of transactions,
and computes a tentative new state by executing  those transactions' smart contracts
serially, in the order they occur in the block.
A miner cannot simply execute these contracts in parallel,
because they may perform conflicting accesses to shared data,
and an arbitrary interleaving could produce an inconsistent final state.
For Bitcoin transactions, it is easy to tell in advance when two transaction conflict,
because input and output data are statically declared.
For smart contracts, by contrast,
it is impossible to tell in advance whether two contract executions will conflict,
because the contract language is Turing-complete.

Miners are rewarded for each block they successfully append to the blockchain,
so they have a strong incentive to increase throughput by parallelizing
smart contract executions.
We propose to allow miners to execute contract codes in parallel by adapting
techniques from Software Transactional Memory (STM)~\cite{Herlihy:2003:STM:872035.872048}:
treating each invocation as a speculative atomic action.
Data conflicts, detected at run-time,
are resolved by delaying or rolling back some conflicting invocations.
Treating smart contract invocations as speculative atomic actions
dynamically ``discovers'' a \emph{serializable} concurrent schedule,
producing the same final state as a serial schedule
where the contract functions were executed in some one-at-a-time order.

But what about later validators?
Existing STM systems are \emph{non-deterministic}:
if a later validator simply mimics the miner by
re-running the same mix of speculative transactions,
it may produce a different serialization order and a
different final state,
causing validation to fail incorrectly.
Treating contract invocations as speculative transactions
improves miners' throughput,
but fails to support deterministic re-execution as required by validators.

Notice, however,
that the miner has already ``discovered'' a serializable concurrent schedule for those transactions.
We propose a novel scheme where the miner records that successful schedule,
along with the final state,
allowing later validators to replay that same schedule in a concurrent but deterministic way.
Deterministic replay avoids many of the the miner's original synchronization costs,
such as conflict detection and roll-back.
Over time,
parallel validation would be a significant benefit because validators perform
the vast majority of contract executions.
Naturally,
the validator must be able to check that the proposed schedule
really is serializable.

This chapter makes the following contributions.
\begin{itemize}
\item A way for miners to speculatively execute smart contracts in parallel.
  We adapt techniques from \emph{transactional boosting}~\cite{HerlihyK2008}
  to permit non-conflicting smart contracts to execute concurrently.

\item A way for miners to capture the resulting parallel execution in the form of a
  \emph{fork-join}~\cite{BlumofeJKLRZ1995} schedule to be executed by validators,
  deterministically, verifiably, and in parallel.

\item A formal model %
    and proof that the validator's execution is equivalent to that of the miner.
  
\item A prototype implementation, built on the Java virtual machine and ScalaSTM~\cite{scalastm}.
  An evaluation using smart contract examples drawn from the Solidity documentation
  yields an overall speedup of \overallMinerSpeedup{}x for miners,
  and \overallValidatorSpeedup{}x for validators with three concurrent threads of execution.
\end{itemize}

\section{Blockchains and Smart Contracts}
\label{sec:model}

\begin{listing*}
\begin{lstlisting}[escapeinside={`}{`}]
contract Ballot {`\label{ln:defBallot}`
  mapping(address => Voter) public voters; `\label{ln:defVoters}`
  // more state definitions
  function vote(uint proposal) {`\label{ln:vote}`
    Voter sender = voters[msg.sender];
    if (sender.voted)
      throw;
    sender.voted = true;
    sender.vote = proposal;
    proposals[proposal].voteCount += sender.weight;
  }
  // more operation definitions
}
\end{lstlisting}
\caption{Part of the Ballot contract.}
\label{lst:ballot}
\end{listing*}

In Bitcoin and similar systems,
transactions typically have a simple structure,
distributing the balances from a set of input accounts to a set of
newly-created output accounts.
In Blockchains such as Ethereum, however,
each block also includes an explicit \emph{state} capturing the cumulative
effect of transactions in prior blocks.
A Transaction is expressed as executable code,
often called a \emph{smart contract},
that modifies that state.
Ethereum blocks thus contain both transactions' smart contracts
and the final state produced by executing those contacts.

The contracts themselves are stored in the blockchain as byte-code instructions for 
the Ethereum virtual machine (EVM).
Several higher-level languages exist for writing smart contracts.
Here, we describe
smart contracts as expressed in the Solidity language~\cite{solidity}.

\newcommand\vvv[1]{\lstinline`#1`}

Listing~\ref{lst:ballot} is part of the source code for an example smart
contract that implements a ballot box~\cite{solidityexamples}.
The owner
initializes the contract with a list of proposals and gives
the right to vote to a set of Ethereum addresses.
Voters cast their votes for a particular proposal,
which they may do only once.
Alternatively, voters may delegate their vote.
The \vvv{contract} keyword declares the smart contract (Line~\ref{ln:defBallot}).

The contract's persistent state is recorded in \emph{state variables}. For \vvv{Ballot}, the persistent state includes fields of scalar type such as the owner (omitted for lack of space).
State variables such as \vvv{voters} (declared on
Line~\ref{ln:defVoters}) can also use the built-in Solidity type
\vvv{mapping} which, in this case, associates each voter's
\vvv{address} with a \vvv{Voter} data structure (declaration omitted
for brevity). 
The keys in this mapping are of built-in type \vvv{address},
which uniquely identifies Ethereum accounts (clients or other contracts).
These state variables are the persistent state of the contract.

Line~\ref{ln:vote} declares contract \emph{function}, \vvv{vote}, to
cast a vote for the given proposal. 
Within a function there are transient \emph{memory} and \emph{stack} areas
such as \vvv{sender}.
The function \vvv{vote} first recovers the \vvv{Voter} data from the
contract's state by indexing into the \vvv{voters} mapping using the
sender's address \vvv{msg.sender}.
The \vvv{msg} variable is a global variable containing data about the
contract's current invocation.
Next, the \vvv{sender.vote} flag is checked to prevent multiple votes.
Note that sequential execution is critical:
if this code were na{\"\i}vely run in parallel,
it would be vulnerable to a race condition permitting double voting.
Ethereum contract functions can be aborted at any time via
\vvv{throw}, as seen here when a voter is detected attempting
to vote twice.
The \vvv{throw} statement causes the contract's transient state and
tentative storage changes to be discarded.
Finally, this \vvv{Ballot} contract also provides functions to register voters, delegate
one's vote, and compute the winning proposal.  The complete Ballot
example is available elsewhere%
\footnote{\url{http://solidity.readthedocs.io/en/develop/solidity-by-example.html}}.

\paragraph{Execution Model: Miners and Validators.}
When a miner prepares a block for inclusion in the blockchain,
it starts with the ledger state as of the chain's most recent block.
The miner selects a sequence of new transactions,
records them in the new block,
and executes them, one at a time,
to compute the new block's state.
The miner then participates in a consensus protocol to decide
whether this new block will be appended to the blockchain.

To ensure that each transaction terminates in a reasonable number
of steps, each call to contract bytecode comes with an explicit limit on
the number of virtual machine steps that a call can take.
(In Ethereum,
these steps are measured in ``gas'' and clients pay a fee to the miner that
successfully appends that transaction's block to the blockchain.)

After a block has been successfully appended to the blockchain,
that block's transactions are sequentially re-executed
\emph{by every node in the network}
to check that the block's state transition was computed honestly and correctly.
(Smart contract transactions are deterministic,
so each re-execution yields the same results as the original.)
These \emph{validator} nodes do not receive fees for re-execution.

To summarize,
a transaction is executed in two contexts:
once by miners before attempting to append a block to the blockchain,
and many times afterward by validators checking that each block in the blockchain
is honest.
In both contexts, each block's transactions are executed sequentially in block order.

\section{Speculative Smart Contracts}
\label{sec:boosting}
This section discusses how miners can execute contract codes
concurrently.
Concurrency for validators is addressed in the next section.

Smart contract semantics is \emph{sequential}:
each miner has a single thread of control that executes one EVM instruction at a time.
The miner executes each of the block's contracts in sequence.
One contract can call another contract's functions,
causing control to pass from the first contract code to the second, and back again.
(Indeed, misuse of this control structure has been the source of
well-known security breaches~\cite{theDao}.)
Clearly,
even sequential smart contracts must be written with care,
and introducing explicit concurrency to contract programming languages
would only make the situation worse.
We conclude that concurrent smart contract executions must be
\emph{serializable}: indistinguishable, except for execution time,
from a sequential execution.

There are several obstacles to running contracts in parallel.
First, smart contract codes read and modify shared storage,
so it is essential to ensure that concurrent contract code executions do not result
in inconsistent storage states.
Second,
smart contract languages are Turing-complete,
and therefore it is impossible in general to determine statically whether
contracts have data conflicts. 

We propose that miners execute contract codes as \emph{speculative actions}.
A miner schedules multiple concurrent contracts to run in parallel.
Contracts' data structures are instrumented
to detect synchronization conflicts at run-time,
in much the same way as mechanisms like transactional boosting~\cite{HerlihyK2008}.
If one speculative contract execution conflicts with another,
the conflict is resolved either by delaying one contract until the other completes,
or by rolling back and restarting one of the conflicting executions.
When a speculative action completes successfully,
it is said to \emph{commit}, and otherwise it \emph{aborts}.

\paragraph{Storage Operations.}
We assume that, as in Solidity,
state variables are restricted to predefined types such as scalars,
structures, enumerations, arrays, and mappings.
A \emph{storage operation} is a primitive operation on a state
variable.
For example, binding a key to a value in a mapping,
or reading from a variable or an array are storage operations.
Two storage operations \emph{commute} if executing them in either
order yields the same result values and the same storage state.
For example, in the \vvv{address}-to-\vvv{Voter} \vvv{Ballot} mapping in
Listing~\ref{lst:ballot}, binding Alice's address to a vote of
42 commutes with binding Bob's address to a vote of 17, but does not
commute when deleting Alice's vote.
An \emph{inverse} for a storage operation is another operation
that undoes its effects.
For example,
the inverse of assigning to a variable is restoring its prior value,
and the inverse of adding a new key-value pair to a mapping is to
remove that binding, and so on.
The virtual machine system can provide all storage operations with inverses.

The virtual machine is in charge of managing concurrency for state
variables such as mappings and arrays.
Speculation is controlled by two run-time mechanisms,
invisible to the programmer,
and managed by the virtual machine:
\emph{abstract locks}, and \emph{inverse logs}.

Each storage operation has an associated abstract lock.
The rule for assigning abstract locks to operations is simple:
if two storage operations map to distinct abstract locks,
then they must commute.
Before a thread can execute a storage operation,
it must acquire the associated abstract lock.
The thread is delayed while that lock is held by another thread%
\footnote{For ease of exposition,
abstract locks are mutually exclusive,
although it is not hard to accommodate shared and exclusive modes.}.
Once the lock is acquired, 
the thread records an \emph{inverse operation} in a log,
and proceeds with the operation.

If the action commits,
its abstract locks are released and its log is discarded.
If the action aborts,
the inverse log is replayed,
most recent operation first,
to undo the effects of that speculative action.
When the replay is complete, the action's abstract locks are released.

The advantage of combining abstract locks with inverse logs is that
the virtual machine can support very fine-grained concurrency.
A more traditional implementation of speculative actions might
associate locks with memory regions such as cache lines or pages,
and keep track of old and versions of those regions for recovery.
Such a coarse-grained approach could lead to many false conflicts,
where operations that commute in a semantic sense are treated as
conflicting because they access overlapping memory regions.
In the next section,
we will see how to use abstract locks to speed up verifiers.
  
When one smart contract calls another,
the run-time system creates a \emph{nested} speculative action,
which can commit or abort independently of its parent.
A nested speculative action inherits the abstract locks held by its parent,
and it creates its own inverse log.
If the nested action commits,
any abstract locks it acquired are passed to its parent,
and its inverse log is appended to its parent's log.
If the nested action aborts,
its inverse log is replayed to undo its effects,
and any abstract locks it acquired are released.
Aborting a child action does not abort the parent,
but a child action's effects become permanent only when the parent commits.
The abstract locking mechanism also detects and resolves deadlocks,
which are expected to be rare.

The scheme described here is \emph{eager},
acquiring locks, applying operations, and recording inverses.
An alternative \emph{lazy} implementation could buffer changes to
a contract's storage, applying them only on commit.

A miner's incentive to perform speculative concurrent execution
is the possibility of increased throughput,
and hence a competitive advantage against other miners.
Of course,
the miner undertakes a risk that synchronization conflicts among
contracts will cause some contracts to be rolled back and re-executed,
possibly delaying block construction,
and forcing the miner to re-execute code not compensated by client fees.
Nevertheless,
the experimental results reported below suggest that even a small
degree of concurrent speculative execution pays off,
even in the face of moderate data conflicts.

\section{Concurrent Validation}
\label{sec:replay}

The speculative techniques proposed above for miners are no help for validators.
Here is the problem:
miners use speculation to discover a concurrent schedule for a block's transactions,
a schedule equivalent to some sequential schedule, except faster.
That schedule is constructed non-deterministically,
depending on the order in which threads acquired abstract locks.
To check that the block's miner was honest,
validators need to reconstruct the same (or an equivalent) schedule
chosen by the miner.

Validators need a way to deterministically reproduce the miner's concurrent schedule.
To this end, we extend abstract locks to track dependencies,
that is, who passed which abstract locks to whom.
Each speculative lock includes a \emph{use counter} that keeps track
of the number of times it has been released by a committing action during the
construction of the current block.
When a miner starts a block, it sets these counters to zero.

When a speculative action commits,
it increments the counters for each of the locks it holds,
and then it registers a \emph{lock profile} with the VM recording the
abstract locks and their counter values. 

When all the actions have committed,
it is possible to reconstruct their common schedule by comparing their
lock profiles.
For example, consider three committed speculative actions, $A$, $B$,
and $C$.
If $A$ and $B$ have no abstract locks in common, they can run concurrently.
If an abstract lock has counter value $1$ in $A$'s profile
and $2$ in $C$'s profile,
then $C$ must be scheduled after $A$.

\begin{algorithm}
\caption{$\textsc{MineInParallel}(T)$ - Mine in parallel}
\label{alg:mine}
\begin{algorithmic}[1]
  \Require A set of contract transactions $T$
  \Ensure A serial order $S$ of transactions and a happens-before graph $H$ of the locking schedule
  \Function{MineInParallel}{$B$}
  \State {Initialize log $L$ for recording locking operations}
  \parState {Execute all transactions $t \in T$ in parallel, recording locking activity in $L$}
  \State {Generate happens-before graph $H$ from $L$}
  \State {Create the serial ordering $S$ via a topological sort of $H$}
  \State {\Return $(S, H)$}
  \EndFunction
\end{algorithmic}
\end{algorithm}

\begin{algorithm}
\caption{$\textsc{ConstructValidator}(S, H)$ - Construct a parallel validator}
\label{alg:validate}
\begin{algorithmic}[1]
  \Require A set of contract transactions $T$ and the happens-before graph $H$ from the miner
  \Ensure A set of fork-join tasks ensuring parallel execution according to $H$
  \Function{ConstructValidator}{$B$}
  \parState {Initialize a mapping $F$ from each transaction $t \in T$ to its fork-join task $f$}
  \parState {Create the happens-after graph $H'$ by reversing the edges of $H$}
  \ForAll {$t \in S$}
  \parState {$B \gets$ all transactions $u \in H'$ that happen immediately before $t$, i.e., its outedges}
  \parState {Create a fork-join task $f$ for $t$ that first joins with all tasks in $B$, i.e.,}
\begin{lstlisting}[numbers=none,mathescape=true]
       $f \gets $ for ($b$ in $B$) { $F$.get($b$).join() } execute($t$)
\end{lstlisting}
  \State {Save the new fork-join task in $F$, i.e., $F.\texttt{put}(t, f)$}
  \EndFor
  \State \Return the value set of $F$, the fork-join tasks
  \EndFunction
\end{algorithmic}
\end{algorithm}

A miner includes these profiles in the blockchain along with usual information.
From this profile information,
validators can construct a \emph{fork-join} program that
deterministically reproduces the miner's original, speculative schedule.
Algorithm~\ref{alg:mine} provides a high-level sketch of the operation
of the miner.  By logging the locking schedule during parallel
execution, the miner generates a happens-before graph of transactions
according to the order in which they acquire locks and commit.  A
valid serial history is produced from a topological sort of this
graph.
Algorithm~\ref{alg:validate} constructs
the validator by scanning through the list of actions as they
appear in the serial history.  A fork-join task is created for each
action and stored for lookup by its identifier.  Each
task will first lookup and join any tasks that must precede it
according to the locking schedule before executing the action
itself.

The resulting fork-join program is not speculative,
nor does it require inter-thread synchronization other than forks and joins.
The validator is not required to match the miner's level of parallelism:
using a work-stealing scheduler~\cite{BlumofeJKLRZ1995},
the validator can exploit whatever degree of parallelism it has available.
The validator does not need abstract locks,
dynamic conflict detection,
or the ability to roll back speculative actions,
because the fork-join structure ensures that conflicting actions never
execute concurrently.

To check that the miner's proposed schedule is correct,
the validator's virtual machine records a trace of
the abstract locks each thread would have acquired,
had it been executing speculatively.
This trace is thread-local,
requiring no expensive inter-thread synchronization.
At the end of the execution,
the validator's VM compares the traces it generated with the lock
profiles provided by the miner.
If they differ, the block is rejected.

Miners have an incentive to publish a block's fork-join schedule along
with the block to induce other miners to build on that block.
If a miner publishes an incorrect schedule,
the error will be detected and that block rejected.
A miner could publish a correct schedule equivalent to,
but less parallel than the schedule it discovered,
it would have no motive to do so because a less parallel schedule
makes that block less attractive than competing blocks with more
parallel schedules,
and the miner will be rewarded only if the other miners choose to build on that block.
Because fork-join schedules are published in the blockchain,
their degree of parallelism is easily evaluated.

\newcommand\Ts{\mathbf{T}}
\newcommand\As{\mathbf{A}}
\newcommand\tx[1]{\texttt{tx}\ #1}
\newcommand\Lstate[1]{\{ #1 \}}
\newcommand\step[1]{\;\underrightharpdown{#1}\;}
\newcommand\fwd[2]{#1 \step{\textsf{fwd}} #2}
\newcommand\bwd[2]{#1 \step{\textsf{bwd}} #2}
\newcommand\unpushed[1]{\textsf{Unpushed}\ #1}
\newcommand\pushed[1]{\textsf{Pushed}\ #1}
\newcommand\tstep[3]{#1 \lightning (#2,#3)}
\newcommand\allows{\;\textsf{allows}\;}
\newcommand\conditionsXXX[3]{%
  \begin{array}{ll}%
    \text{({\it i})}- & #1\\%
    \text{({\it ii})}- & #2\\%
    \text{({\it iii})}- & #3\\%
    \end{array}}
\newcommand\skipt{\texttt{skip}}
\newcommand\nothingtext{fin}
\newcommand\nothing[1]{\textsf{\nothingtext}(#1)}

\newcommand\Ops{\textsf{Op}}
\newcommand\lock{\ell}
\newcommand\localR{R}

\section{Correctness}

Concurrent calls to smart contract functions might leave persistent storage
in an inconsistent state not possible after a serial execution.
In this section we show that, instead, every concurrent execution permitted by our
proposal is equivalent to some sequential execution.
Because miners are free to choose the order in which contracts appear in a block,
any sequential execution will do.
We further show that the executions of validators are equivalent to their corresponding
miners.

Our correctness arguments in this section build on the Push/Pull
model~\cite{KP:PLDI2015}, which is expressive enough to model the
transactional boosting~\cite{HerlihyK2008} algorithm that our miners
execution. Next, we build on top of Push/Pull with a new model that
captures validators' behavior and show that it simulates
Push/Pull. This is not enough, however. We finally show that the
validator's concurrent behavior is equivalent to the miner's
concurrent behavior, even though each may simulate different serial
executions. %

\subsection{Preliminaries}

We now establish some preliminaries and, for the benefit of the reader, we provide a short background on the Push/Pull model. The 
Push/Pull model is capable of characterizing a wide range of transactions and witnessing their
serializability/opacity. It decomposes state into thread-local components and a shared component, each represented as \emph{logs of operations}. 
Concurrent transactions use the {\sc Push} rule to share operations from their \emph{local log} into
the \emph{single shared log} (or {\sc Unpush} to rollback). Meanwhile, other threads may
{\sc Pull} those effects from the shared into their local view.
This model is serializable and, in some cases, satisfies opacity.
As we will now explore, one benefit of this semantic model is that most of the elaborate
reasoning (coinduction, simulation relations, invariants, etc.) necessary for
proving the correctness of a transactional algorithm is contained
within the semantic model, and we don't need to redo these steps in this chapter.

For the most part, to understand
this section it is not necessary to understand the inner workings of Push/Pull.
We begin with some definitions:
$$\begin{array}{lll}
\tx{c} &\in C\qquad&\text{Finite set of transaction code}\\
\tau &\in T& \text{Set of unique transaction identifiers.}\\
o.m(\vec{x})&\in \Ops & \text{Object methods/operations}\\
id &\in Ids\;\;\;\;\;\;& \text{Operation identifiers.}\\
\end{array}$$
Each thread executes code $c \in C$ from some programming language that
includes
lock acquisition \texttt{lock}($\ell$),
method calls such as $o.m(\vec{x})$
(consisting of object name $o$, method name $m$, and a vector of arguments $\vec{x}$), and a \skipt\ statement.
We use the notation $\tx{c}$ to mean that the code $c$ is a single (unnested) transaction. Reducing $c$ to, say, $c'$, is denoted as reducing $\tx{c}$ to $\tx{c'}$, indicating the transactional context for the reduction\footnote{The Push/Pull model is more general, allowing code outside of a transaction.}. For our purposes, there will be $N$ transactions and each will have a unique id $\tau\in T$.
We treat all operations as occurring on a single object. Disjoint objects are easily modeled as disjoint sections of a single objects.
In addition to the thread-local and shared logs of operations (formalized below), we also permit additional thread-local state, used to model primitive variables as well as arguments and return values for method calls. This thread-local state space is denoted $\Sigma$. Thread code $c$ can also perform updates to this thread-local state, but we omit details for simplicity.

As in~\cite{KP:PLDI2015}, we abstract away the programming
  language with a couple functions:
\[\begin{array}{ll}
\tstep{c}{o.m(\vec{x})}{c'} : &
\begin{minipage}[t]{3.3in}
  Code $c$ can be reduced to the
  pair $(o.m(\vec{x}),c')$ where $o.m(\vec{x})$ is a next reachable method
  call in the reduction of $c$, with remaining code $c'$.
\end{minipage}\\
\nothing{c} : &
\begin{minipage}[t]{3.3in}
  This predicate is true provided that there is a
  reduction of $c$ to $\skipt$ that does not encounter more work, e.g. a method call or a local operation.

\end{minipage}
\end{array}\]
The definition of $\tstep{c}{o.m(\vec{x})}{c'}$ can easily be instantiated for
simple programming languages.

\newcommand\locks{\vec{\ell}}
\subsection{Miners} %
Recall that, in the absence of our approach to concurrent mining,
a miner would assemble a collection of $N$ transactions to execute
in a serial order:
$$\begin{array}{ccc}
    \underbrace{\tx{c_1}} &\;\;\;;\;\;\; ... \;\;\;;\;\;\;& \underbrace{\tx{c_N}}\\
    \tau_1 & & \tau_N
  \end{array}
$$
In Section~\ref{sec:boosting}, we described how these $N$ transactions
can instead be executed speculatively. For simplicity, assume that
each of the $N$ transactions is executed in parallel.
The execution of a transaction $\tau_i$ reduces $c_i$, and whenever it
reaches an operation $o.m(\vec{x})$, it will have first acquired a set of abstract locks $\locks
\subseteq \mathbb{L}$ that satisfies the \emph{boosting lock
rule}~\cite{HerlihyK2008}:
\begin{definition}[Transactional boosting~\cite{HerlihyK2008}]\label{def:boosting}
   Before executing an operation $o.m(\vec{x})$, the transaction must
   acquire a set of locks $\locks$ such that for every concurrent operation
   $o.n(\vec{y})$ of an uncommitted transaction that has acquired
   locks $\locks'$, if $o.m(\vec{x})$ does not commute with\footnote{For brevity, we rely on the reader's intuition of commutativity rather than reiterating the formalization here.}
   $o.n(\vec{y})$, then $\locks\cap \locks' \neq \emptyset$.
\end{definition}
That is, if the operation conflicts with another concurrent operation,
then there will be at least one lock $\ell$ that is acquired by both
transactions.  Thread \emph{commits} once it has reduced $\tx{c_i}$
completely to \skipt. At commit, the thread releases all held abstract locks.

We model miner execution using Push/Pull. A configuration in
the Push/Pull model has a component $\{c,\sigma,L\}$ to represent each
thread, executing code $c$ with a primitive variable state $\sigma$. The final component
$L$ is a log (list) of the operations (e.g. $[o.m_1(\vec{x_1}),o.m_2(\vec{x_2}),...]$) performed thus
far by the transaction. This log is only viewable locally. We denote the
set of all thread components as $\Ts$, treating it as a list and using
notations such as $\Ts_1 \cdot \{ c_i, \sigma_i, L_i \} \cdot \Ts_2$ to
focus on the $i$-th transaction. The Push/Pull configuration also has
a single shared log $G$ which is a sequence of operations that have
been shared to the global view. The rules {\sc Push}, {\sc Unpush}, {\sc Pull}, {\sc Unpull}
in~\cite{KP:PLDI2015} ferry operations between local logs and the shared log,
and correspond to different logical stages in a transactional system.

We denote the miner transition relation as $\Ts,G \leadsto_M \Ts',G'$ and it is
an instance of the Push/Pull model ({\it i.e.} $\leadsto_M \subseteq \leadsto_{PP}$ where $\leadsto_{PP}$ is the main transition relation in~\cite{KP:PLDI2015}). The initial configuration is:
$$
   \{ \tx{c_1},\sigma_0,L_0\} \cdots \{ \tx{c_N},\sigma_N,L_N\}, G_0
$$
Here $\sigma_i$ is a transaction-local initial state, each log $L_i=\emptyset$, and the initial shared state is represented as an empty log of operations $G_0=\emptyset$.
Let $\Pi_M$ denote the sets of traces (sequences of configurations) of $\leadsto_M$.
As noted by Satoshi Nakanishi~\cite{minersDeterministic}, code executed by
miners must be deterministic. Therefore, we assume that 
$\tstep{c}{o.m(\vec{x})}{c'}$ is deterministic and that every operation $o.m(\vec{x})$ is deterministic.

The \emph{atomic machine}~\cite{KP:PLDI2015} executes transactions
in any serial order. A configuration of the atomic machine is denoted $\As,l$ where $\As$ is a list of code/local-state pairs $(c_i,\sigma_i)$.
The atomic transition system (Fig. 3 of~\cite{KP:PLDI2015}) is 
denoted $\As,l \xrightarrow{A} \As',l'$. The semantics of the atomic machine are far simpler: the machine nondeterministically selects one transaction at a time, running each to completion. $l$ accumulates the operations performed by each transaction.

\begin{lemma}[Miner serializability~\cite{KP:PLDI2015}]\label{lemma:miner}
  Miners $\leadsto_M$ simulate atomic $\xrightarrow{A}$.
  \begin{proof}By showing that miner transitions satisfy the
    conditions of all Push/Pull rules. 
    Condition ({\it ii}) of the {\sc Push} rule
    is satisfied because miners follow Def.~\ref{def:boosting}.
  \end{proof}
\end{lemma}

\subsection{Miners emit happens-before}

Our miners emit an \emph{event log} $E$, which is a sequence, each element
being an event:
\[\begin{array}{ll}
  (\tau, \textsf{begin})&\text{Beginning of txn $\tau$}\\
  (\tau, \textsf{lock}\ \ell)&\text{Acquire lock $\ell$}\\
  (\tau, \textsf{op}\ o.m(\vec{x}))\;\;\;&\text{Begin $o.m(\vec{x})$}\\
  (\tau, \textsf{commit})&\text{Commit of txn $\tau$, release all locks}\\
\end{array}\]
From this event log collected across all threads, we construct the happens-before
relation. 
Let $e <_E e'$ denote that event $e$ occurs before event $e'$ in log $E$.
A  happens before relation, denoted $\;\sqsubseteq\; : T \times T$ is a partial order on the execution of transactions. (A more fine-grained relation is possible, that orders individual operations rather than transactions.)
From the event log, our system constructs the relation:
$$
\tau' \sqsubseteq \tau
\;\;\Longleftrightarrow\;\;
  (\tau',\textsf{commit}) <_E (\tau,\textsf{commit})
 \wedge\; \exists \ell. (\tau',\ell)\in E\wedge (\tau,\ell)\in E
$$
Intuitively, this means that transaction $\tau'$ is ordered before $\tau$
if it commits first and there is at least one lock $\ell$ that both
transactions acquire.
Notice that transactions that acquire the same locks will be
  ordered in $\sqsubseteq$ even if they were not concurrent in $E$.
Indeed, in the following lemma, we prove that the miner machine orders all
conflicting transactions.

\begin{lemma}\label{lemma:allOrdered}
  For all $\tau_1$ that executes operation $o.m(\vec{x})$ and 
  $\tau_2$ that executes operation $o.n(\vec{y})$ such that
  $o.m(\vec{x})$ does not commute with $o.n(\vec{y})$, if a miner executes
  $\tau_1$ and $\tau_2$ (among others), then either $\tau_1\sqsubseteq\tau_2$ or $\tau_2\sqsubseteq\tau_1$.
  \begin{proof}By induction on the event log and miner transition system.
    Due to Def.~\ref{def:boosting}, since $o.m(\vec{x}) $ does not commute with $ o.n(\vec{y})$,
    there will be some lock $\ell$ acquired first by both $\tau_1$ and $\tau_2$.
    If the transactions are non-concurrent, then they will be ordered accordingly.
    Otherwise, the first one to acquire $\ell$ will be ordered before the other.
    \end{proof}
    
\end{lemma}

\subsection{Validator}

We now describe a transition system that models the validator. The transition system is a conservative extension of Push/Pull~\cite{KP:PLDI2015}, constrained to abide by the happens-before relation $\sqsubseteq$, but free to perform operations on the shared state without acquiring abstract locks or checking for conflicts. 
The validator transition relation is denoted $\Ts,G
\leadsto^\sqsubseteq_V \Ts',G'$. It is the same as $\leadsto_{PP}$, but modified in two ways:

\begin{enumerate}
\item We remove condition ({\it ii}) from the {\sc Push} rule. Roughly speaking,
  this condition requires that a thread can only perform an operation if it commutes with
  all active operations of uncommitted transactions. By removing this condition,
  operations can be ferried from the local log to the shared log
  without checking commutativity with concurrent operations.
\item The {\sc Step} rule selects a single transaction and lets it take a single
  step forward. We modify it, adding the condition ({\it i}) below:
\end{enumerate}
  $$
  \infer{
    \Ts_1\cdot\{ \tx{c}, \sigma, L \} \cdot \Ts_2, G
    \leadsto
    \Ts_1\cdot\{ \tx{c''}, \sigma'', L'' \} \cdot \Ts_2, G'
  }{
    \begin{array}{ll}
      \text{({\it i})}- & \forall \tx{c'} \text{ s.t. } \tx{c'} \sqsubseteq \tx{c}.\; c'=\skipt\\
      \text{({\it ii})}-\;\;\; & \{ \tx{c}, \sigma, L \} \step{dir} \{ \tx{c''}, \sigma'', L'' \}
    \end{array}
  }
  $$
\noindent
We abuse notation here, conflating/equating a transaction id $\tau$ with the code $\tx{c}$ that it executes. Hence the quantification above gives us the current code $\tx{c'}$ being executed by each $\tau'$ such that $\tau' \sqsubseteq \tau$.
This change to the {\sc Step} rule delays each transaction $\tau$ until each 
predecessor has completed, effectively implementing
  a thread \texttt{join} operation. 
Finally, the initial configuration is the same as the initial configuration of the miner:
$$\begin{array}{cccl}
\underbrace{\{  \tx{c_1}, \sigma_0, L_0 \}} & \cdots & \underbrace{\{ \tx{c_N}, \sigma_N, L_N \}}, & G_0\\
\tau_1 & \cdots & \tau_n & \\
\end{array}$$
With each $L_i=\emptyset$ and $G_0=\emptyset$. Let $\Pi^\sqsubseteq_V$ denote the sets of traces of $\leadsto^\sqsubseteq_V$.
We now show that the validators are serializable.

\newcommand\lPushed{\small\textsf{pushed}}
\newcommand\gUncommitted{\textsf{gUCmt}}
\renewcommand\pushed[1]{\lfloor #1 \rfloor_{\lPushed}}
\newcommand\mine[1]{\lfloor #1 \rfloor^{\lUnpushed}_{\lPushed}}
\newcommand\pulled[1]{\lfloor #1 \rfloor_{\lPulled}}
\newcommand\gcommitted[1]{\lfloor #1 \rfloor_{\gCommitted}}
\newcommand\guncommitted[1]{\lfloor #1 \rfloor_{\gUncommitted}}

\begin{lemma}[Validator serializability]\label{lemma:validator}
  For all $\sqsubseteq$, validators $\leadsto^\sqsubseteq_V$ simulate
  atomic machine $\xrightarrow{A}$.
  \begin{proof}
    By showing that the validator machine simulates the Push/Pull model~\cite{KP:PLDI2015} which, in turn, simulates $\xrightarrow{A}$.
    The Validator-to-Push/Pull simulation relation is $Id$. 
    Two rules that have been modified, so there are two cases:
    \begin{enumerate}
    \item {\sc Step}. The relation is maintained because transitions have only been removed: the $\sqsubseteq$ constraint has been added.
    \item {\sc Push}. Condition ({\it ii}) has been removed in $\leadsto_V$, so we must show that it is satisfied in $\leadsto_{PP}$.  Condition ({\it ii}) requires that $\guncommitted{G_1} \setminus \pushed{L_1 \cdot L_2} \;\blacktriangleleft\; o\!p$. This means that when the validator performs an operation $o\!p$, that it must commute with all other operations of active transactions. This is ensured by Lemma~\ref{lemma:allOrdered} in concert with (1) above.
    \end{enumerate}
    \end{proof}
\end{lemma}

Notice that for $N$ transactions, there are $N!$ executions of atomic machine $\xrightarrow{A}$.
However, since some transactions may be independent of each other ({\it i.e.} none of their
operations conflict), not all of these traces are unique. We now define equality
between traces of $\xrightarrow{A}$, which will allow us to show that some
permutations of traces are equivalent to each other.

\begin{definition}[Trace Equality] Two (complete) traces  $\pi = \As,l \xrightarrow{A}^{*} \As',l'$ and
  $\pi' = \As,l \xrightarrow{A}^{*} \As'',l''$
  are equivalent, denoted $\pi \simeq \pi'$ provided that:
  \begin{enumerate}
  \item The final shared states are equal: $l' \preceq l''$ and $l'' \preceq l'$.
  \item Corresponding local states in $\As'$ and $\As''$ are equal.
  \end{enumerate}
\end{definition}

\noindent
Above we have used the asymmetric log relation $\preceq$ that lets one
compare logs in terms of observational equivalence (more detail in~\cite{KP:PLDI2015}).
$\sqsubseteq$ defines an equivalence class over these executions:

\begin{lemma}[Permutation of atomic traces]\label{lemma:permute}
  For every $\pi,\pi'\in \Pi_A$, if $\pi$ and $\pi'$ are ordered by
  $\sqsubseteq$, then $\pi \simeq \pi'$.
  \begin{proof}
    Assume not. Then $\pi$ is not equal to $\pi'$, so they reach final
    configurations $\As,l$ and $\As',l'$ respectively and one of the two cases must hold:
    \begin{enumerate}
    \item $l\neq l'$. There must be two transactions $\tau,\tau'$ that conflict and
      $\tau$ is before $\tau'$ in $\pi$, but 
      $\tau'$ is before $\tau$ in $\pi'$. By Lemma~\ref{lemma:allOrdered}, $\sqsubseteq$ places some order on all pairs of conflicting transactions. Contradiction.
    \item $\sigma_i \neq \sigma'_i$ for some $(\skipt,\sigma_i) \in \As$ and $(\skipt,\sigma'_i)\in\As'$.
      Recall that, in both traces, transaction $\tx{c_i}$ begins from the same state $\sigma_0$.
      If $\sigma_i \neq \sigma'_i$, then there is some operation $o.m(\vec{x})$ in $\tx{c_i}$ that
      does not commute with some other operation $o.n(\vec{y})$ in another transaction.
      However, by Lemma~\ref{lemma:allOrdered}, $\sqsubseteq$ places some order on all pairs of conflicting transactions. Contradiction.
      
          \end{enumerate}
  \end{proof}
\end{lemma}

\begin{figure}
  \centering
  \includegraphics[width=\columnwidth]{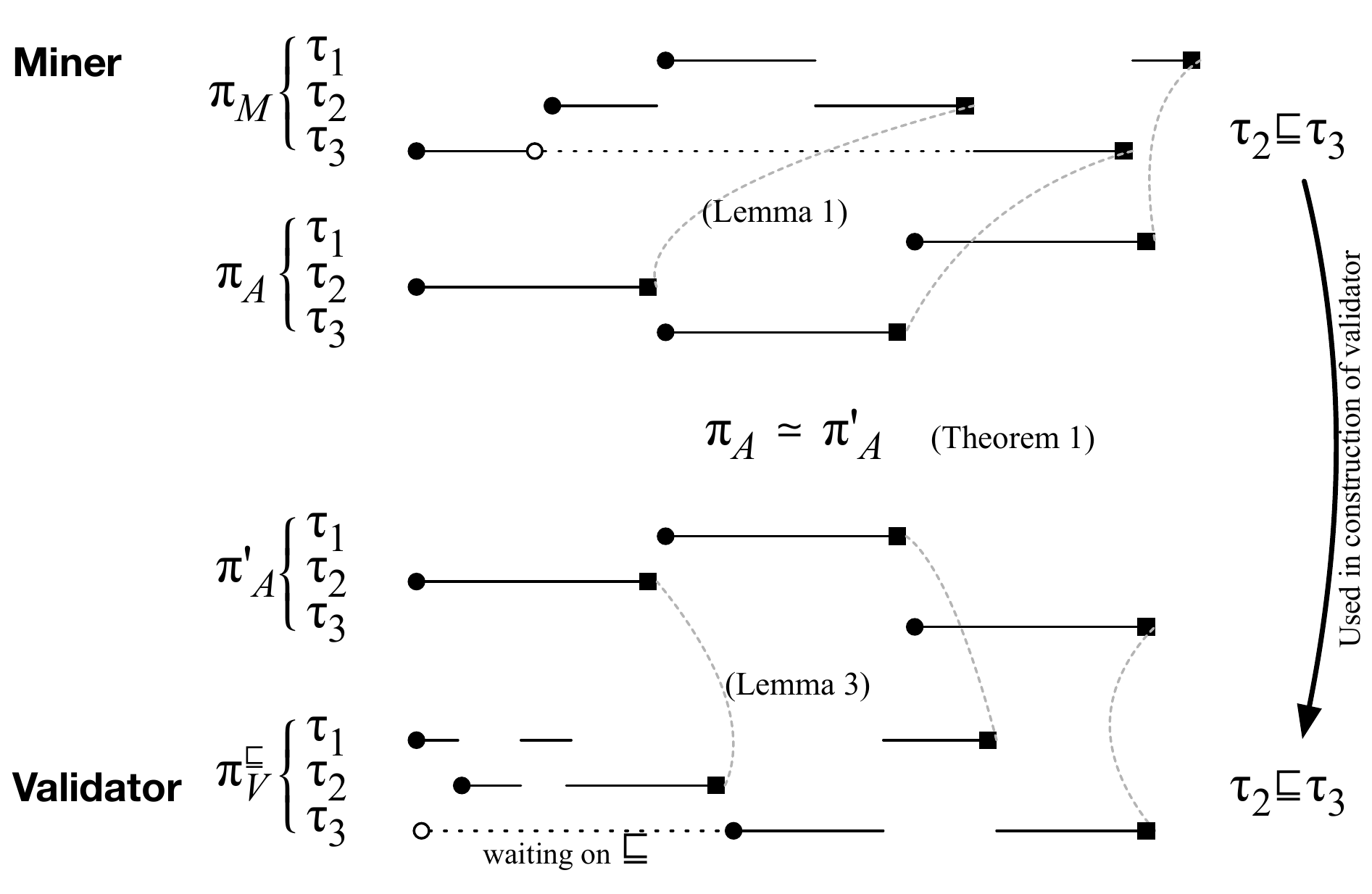}
  \caption{\label{fig:traces} An example with three transactions: $\tau_1,\tau_2,\tau_3$,
    demonstrating the equivalence between
    a miner trace $\pi_M$ and a validator trace $\pi_V$, via their corresponding
    atomic traces $\pi_A$ and $\pi'_A$, respectively. Horizontal solid lines indicate
    execution; horizontal dotted lines indicate blocking on a lock or waiting for
    $\sqsubseteq$; and curved dashed lines show commit-order consistency.
    $\pi_A$ and $\pi'_A$ are equivalent
    because the transactions that conflict $\tau_2$ and $\tau_3$ are ordered in both
    traces: $\tau_2\sqsubseteq\tau_3$. }
\end{figure}

We now show that the validator trace is equivalent to the miner trace. When applying the
Push/Pull results to obtain an equivalent serial trace for a given interleaved trace, we
again use the asymmetric log relation $\preceq$ (lifted from logs to traces)
because, technically, the relation is about
observational inclusion rather than equality.

\begin{theorem}
  For every miner trace $\pi_M \in \Pi_M$ emitting happens-before for $\sqsubseteq$ and validator trace $\pi_v \in \Pi^\sqsubseteq_V$, let $\pi_A$ be such that $\pi_M \preceq \pi_A$ (Lemma~\ref{lemma:miner}) and
  let $\pi'_A$ be such that $\pi_V \preceq \pi'_A$ (Lemma~\ref{lemma:validator}), then
  $\pi_A \simeq \pi'_A$.
  \begin{proof}
    By Lemma~\ref{lemma:permute}, it suffices to show that $\sqsubseteq$ orders all traces in 
    $\Pi_M$ and $\Pi^\sqsubseteq_V$.   
    \end{proof}
\end{theorem}

In summary, although the validator is constrained by the miner's happens
before relation, the validator is not constrained to the miner's commit order.
We have shown that they are nonetheless equivalent.

\section{Implementation}
Because the EVM is not multithreaded,
our prototype uses the Java Virtual Machine (JVM).
Speculative actions are executed by the Scala Software Transactional
Memory Library (ScalaSTM~\cite{scalastm}). 

Examples of smart contracts were translated from Solidity into Scala,
then modified to use the concurrency libraries.
Each function from the Solidity contract is turned into a speculative
transaction by wrapping its contents with a ScalaSTM \vvv{atomic}
section.
Solidity \vvv{mapping} objects are implemented as boosted hashtables,
where key values are used to index abstract locks.
Additionally, solidity \vvv{struct} types were translated into immutable
case classes.
Methods take a \var{msg} field to emulate 
Solidity contracts' global state,
which includes details of the transaction,
addresses of participants, and so on.
Scalar fields are implemented as a single boosted mapping.
The Solidity \var{throw} operation,
which explicitly rolls back a contract execution,
is emulated by throwing a Java runtime exception caught by the miner.

In our prototype,
abstract locks are implemented via interfaces exported by ScalaSTM,
relying on the native deadlock detection and resolution mechanisms in ScalaSTM.
Miners manage concurrency using Java's \vvv{ExecutorService}.
This class provides a pool of threads and runs a collection of
\vvv{callable} objects in parallel.
A block of transactions in Ethereum is implemented as a set of
\vvv{callable} objects passed to the thread pool.
To generate locking profiles from the parallel execution,
we instrument smart contracts to log when atomic sections start and end,
as well as calls to boosted operations.
From the log, we can encode the locking schedule as a
happens-before graph for the validator.
The validator transforms this happens-before graph into a fork-join program.
Each transaction from the block is a fork-join task that first joins with
all tasks according to its in-edges on the happens-before graph.

\section{Virtual Machine Comparisons}

While our approach is implemented on the Java Virtual Machine (JVM)
the two most popular blockchains supporting smart contracts have their
own custom VMs: the Ethereum Virtual Machine (EVM) and Hyperledger
Fabric.  In this section we briefly discuss the differences between
our implementation on the JVM and what modification to support
concurrency would be necessary in the EVM.

\subsection{EVM vs JVM}
Briefly leaving aside the issue of concurrency, there are a number of architectural differences between the EVM and the JVM, potentially raising the concern that our experimental results would not generalize to a hypothetical multithreaded EVM.

For example, the EVM's native word size is 256 bits to simplify interactions with cryptographic hashes; however, both the JVM and the modern CPUs likely to be executing either EVM or JVM bytecode have native word sizes of either 32 or 64 bits.
Thus there is likely to be a large constant factor slowdown to execute simple arithmetic operations on the EVM as compared to the JVM.
In absolute terms, this means that gains of even moderate concurrency are likely to be amplified on the EVM.
For example, a 4x speedup of a 500ms operation would save 375ms; whereas on a faster architecture, the same operation might only take 125ms to begin with, and a 4x speedup would save only 93.75ms.
In relative terms, a constant factor in the cost of basic operations makes absolutely no difference in the degree of concurrency.

\subsection{A Multithreaded EVM}

To implement the our approach in the EVM, several modifications for
concurrency necessary.  Notably, however, the EVM instruction set need
not change, and developers can use existing smart contracts.  Since we
base our concurrency on boosting, the contract developers need not
implement any explicit concurrency.

The first modification necessary is to make the EVM multithreaded.
The EVM needs to be able to take all transaction simultaneously and
fork multiple threads.  In the JVM this is already supported with the
Executor service.  Similar functionality is required for the EVM.  To
enable safe speculative concurrency, the EVM needs abstract locks for
memory access.  The EVM provides a flat, dictionary-style memory model
for persistent storage, which requires per-key abstract locks for
fine-grained concurrency.

To support validators requires not only changes the EVM but also
changes to the block format.  After recording the locking schedule
from executing smart contract concurrently, the EVM needs to export
the happens-before graph of contract executions for use by validators.
To record this schedule on the blockchain requires an expanded block
format that can efficiently store the edges of the happens-before
graph.  The size of this entry in the block format should be
adjustable to the size of the graph.
The EVM also needs to be extended to support fork-join execution.  The
JVM provides the machinery to create and run fork-join programs, and
the EVM requires similar functionality.

\subsection{Ethereum vs Hyperledger}
In Hyperledger Fabric, a validator executes a block's transactions
concurrently against the most recent "world state", and collects the
transactions' read and write sets. It then orders the transactions and
sequentially checks whether each successive transaction's read set is
current.  If so, the validator installs that transaction's write set,
and if not, it discards the transaction and tries it
later~\footnote{https://hyperledger-fabric.readthedocs.io/en/release-1.2/readwrite.html}.
This technique, however, does not address the problem of helping
Ethereum validators to replay transactions in parallel.  Validators in
Fabric do not need to replay transactions because it uses BFT
consensus, not proof-of-work.  Another minor difference is that
conflicting transactions are discarded rather than scheduled.

\newcommand\benchstamp{dist} %
\newcommand\benchstampt[1]{exp/\benchstamp-#1}

\section{Experimental Evaluation}
Our goal is to improve throughput for miners and validators by
allowing unrelated contracts to execute in parallel.
To evaluate this approach,
we created a series of benchmarks for sample
contracts that vary the number of transactions and their degree of conflict.
These benchmarks are conservative,
operating on only one or a few contracts at a time and permitting higher
degrees of data conflict than one would expect in practice.

Our experiments are designed to answer two questions.
(1) For a given amount of data conflict, how does speedup
change over increasing transactions?  We expect to see more speedup as
the number of transactions increases, limited by the number of cores
available on the underlying hardware.
(2) How does the speedup change as data conflict increases?
For low data conflict,
we expect our parallel miner to perform better than serial.
But as data conflict increases,
we expect a drop-off in speedup,
limited by core availability.

\subsection{Contract Selection}

The evaluate our approach, we selected four contracts adapted from
real-world examples of how smart contract are used in practice.  Two
contract, Ballot for voting and SimpleAuction for an auction, The code
was adapted from contract examples in the Solidity
documentation~\cite{solidityexamples}.  Two are taken from actual
contract implementations used in practice: EtherDoc, a document
management system, and Token, an implementation of the ERC-20 token
contract
standard~\footnote{\url{https://theethereum.wiki/w/index.php/ERC20_Token_Standard}}.

These contract were chosen to represent the wide variety of use-cases
for smart contracts, ranging from financial applications with Token,
collaborative activities with Ballot and SimpleAuction, as well as
utilities with EtherDoc.  Because of the variety of applications, the
smart contracts use and share data in very different ways, and
therefore we expect different speedup profiles during testing.  For
instance, we don't expect double-voting in Ballot to cause a great of
a slowdown as transfering tokens to the same account with Token.  In
the former, contention is only with the voter's own data while the
latter causes contention for a single user's data.

The speedups are not meant for comparison between contracts. Rather
they are to demonstrate the range of performance improvements that our
approach will achieve across a variety of contract use-cases.

\subsection{Benchmarks}

There are five benchmarks, one for each of the example contracts we
implemented, Ballot, SimpleAuction, EtherDoc, and Token, as well as one Mixed
benchmark containing transactions from all other contracts.  For each benchmark, our
implementation is evaluated on blocks containing between 10 and 400
transactions with 15\% data conflict, as well as blocks containing 200
transactions with data conflict percentages ranging from 0\% to 100\%
data conflict.
The data conflict percentage is defined to be the percentage
of transactions that contend with at least one other transaction for
shared data.  As we will see, the impact of data conflict on speedup
depends on the contract implementation.

These benchmarks are conservative.  For all
benchmarks besides Mixed, the entire block operates on the same
contract, calling only one or two methods.  In reality, mined blocks
contained transactions on unrelated contracts and accounts.  While the
theoretical maximum number of transactions per block is currently
around 200 transactions\footnote{A transaction costs 21,000 gas plus the gas for
the computation~\cite{wood}.  The gas limit on block 3,110,235
(latest as of writing) was 4,005,875, a maximum close to 200.}, we
test a wide range from 10 to 400.  This maximum increases and decreases
over time, as determined by miner
preference~\cite{ethereumdesign}.
In practice, the number of transactions can be far fewer per block,
e.g., when there are costly transactions.  For testing speedup over
number of transactions, we fix the data conflict rate at 15\%, though we
expect that blocks in practice rarely have very much internal data conflict.
While we did not measure data conflict in the existing blockchain,
our approach implemented in EVM could be used to collect such data on
an existing blockchain.  For testing speedup as data conflict increases,
we fix the number of transactions per block to 200, the current
theoretical maximum.

\paragraph{Ballot.}
This contract is an example voting application from the Solidity
documentation~\cite{solidityexamples}
and is described in Section~\ref{sec:model}.  For all benchmarks, the
contract is put into an initial state where voters are already
registered.  All block transactions for this benchmark are requests to
vote on the same proposal.  To add data conflict, some voters attempt to
double-vote, creating two transactions that contend for the same voter
data.  100\% data conflict occurs when all voters attempt to vote twice.

\paragraph{SimpleAuction.}
This contract, also from the Solidity
documentation~\cite{solidityexamples}
implements an auction. There is a single owner who initiates the
auction, while any participant can place bids with the \texttt{bid()}
method. A \vvv{mapping} tracks how much money needs to be returned to which
bidder once the auction is over.  Bidders can \texttt{withdraw()}
their money.  For the benchmarks, the contract state is initialized by
several bidders entering a bid.  The block consists of transactions
that withdraw these bids.  Data conflict is added by including new
bidders who call \texttt{bidPlusOne()} to read and increase the
highest bid.  The rate of data conflict depends on how many bidders are bidding at
the same time, thus accessing the same highest bidder. 100\% data conflict
happens when all transactions are \texttt{bidPlusOne()} bids.

\paragraph{EtherDoc.}
EtherDoc\footnote{\url{https://github.com/maran/notareth}}
is a ``Proof of Existence'' decentralized application (DAPP) that
tracks per-document metadata including hashcode and owner.
It permits new document creation,
metadata retrieval, and ownership transfer.  For the benchmarks, the
contract is initialized with a number of documents and owners.
Transactions consist of owners checking the existence of the document
by hashcode.  Data conflict is added by including transactions that
transfer ownership to the contract creator.  As with SimpleAuction,
all contending transactions touch the same shared data, so we expect a
faster drop-off in speedup with increased data conflict than Ballot.
100\% data conflict happens when all transactions are transfers.

\paragraph{Token.}
Token is based on an implementation of a standard token
contract\footnote{\url{https://github.com/OpenZeppelin/zeppelin-solidity/blob/master/contracts/token/StandardToken.sol}}.
It allows a contract owner to create an initial set of tokens,
maintains account balances for users participating in the contract,
and provides methods to transfer tokens to between users.  For the
benchmarks, the contract is initialized with a set of tokens and a
number of users.  Transactions consist of users approve token
transfers back to the owner account.  Data conflict is added by
including transactions that perform the actual transfer of tokens to
the contract owner, which compete for shared state, i.e., the owner's
account balance.

\paragraph{Mixed.}
This benchmark combines transactions on the above smart contracts in
equal proportions, and data conflict is added the same way in equal
proportions from their corresponding benchmarks.

\subsection{Adding Contention}

In order to introduce and measure speedup under contention, we add
varying percentages of conflicting transactions.  The contract runners
take a conflict percentage parameter, which controls the percentage of
conflicting transactions vs. the non-conflicting transactions.

The source of contention is different for each contract, because each
uses shared data in a different way.  In all contracts, conflicting
transactions are those that attempt to write to the same shared data.
Depending on the contract, however, the contention that a single may
conflict with only one or a few other transaction or all other
transactions.  We expect this result in large differences in speedups
between the different contracts.
This means that speedup graphs are not comparable between different
contracts.  This is expected, because the contracts represent very
different ways in which data sharing occurs in a smart contract.

The contracts have been chosen to represent different, common
use-cases for smart contracts.
For example, an auction contract keeping track of the highest bid
causing any bid transactions to contend for the same contract.  A
voting contract that protects against double-votes would only see
contention for a particularly voter's data if that voter attempts to
double vote rather than.  In this case, the contract may still have a
speedup at 100\% contention, because the conflicting transactions are
not all contending for the same data.

Conflicting transactions are added for each contract in the following
ways:
\paragraph{Ballot.}
At 0\% contention, all transactions are separate voters, each voting
once.  Contention is added by having voters attempt to double-vote.
The percentage of contention is a measure of how many transactions
conflict with at least one other tranaction.  At 100\% contention, all
the voters are attempting to double-vote.  If there are 100
transactions, there are 50 voters, each issues two voting
transactions.  Not all transactions are contending for the same data,
rather there are 50 separate voting entries, each with two
transactions contending for them.

\paragraph{SimpleAuction.}
Before the experiment, the contract state is initialized by entering a
bid for each user.  When running the experiment at 0\% contention, all
transactions are to remove the bid for each user, which incurs no
conflict for any transaction.  Contention is increased by replacing a
percentage of the transactions with a transaction that increments a
user's bid.  Incrementing a bid conflicts always conflicts with all
other increment transactions, because incrementing requires modifying
the field that tracks the current highest bid for the entire contract.
At 100\% contention, all transactions are to increment bids and
contend for the same shared state.

\paragraph{EtherDoc.}
EtherDoc is initialized with a set of user's, each of which owns a
single document.  At 0\% contention, all transactions are a check for
the existence of each user's document by that user.  Since these are
reads of separate data, no contention exist.  Contention is added by
including transactions that transfer ownership of the document to the
same user (the contract owner).  The transfer transactions all
conflict with each other, since the owner's account data is touched by
all the conflicting transactions.  At 100\% contention, all
transactions are document transfers to the owner's account.

\paragraph{Token.}
This contract is initialized with a set of users, each having the same
account balance.  At 0\% contention, the transactions for the
experiment consist only of each user approving a token transfer to the
same account, the owner account.  The approve/transer convention is a
common use-case for transferring tokens between accounts: rather than
user A sending money to user B, user A approves a certain amount to
transfer and user B initiates the send.  Approving a transaction to
the same user incurs no conflict, because it only affects the send
account, not the receive account. Contention is added by including
transactions that transfer tokens to the same owner account.  At 100\%
contention, half of the transactions are approve the transfer and half
are to perform the actual transfer the owner account.  Because only
half the transactions conflict with each other we would still expect a
speedup even at 100\% contention.

\subsection{Block Size and Contention}

We ran our experiments on a 4-core 3.07GHz Intel Xeon W3550 with 12 GB
of memory running Ubuntu 16.  All of our experiments run on the Java
Virtual Machine (JVM) with JIT compilation disabled.  Parallel mining
and validation are run with a fixed pool of three threads, leaving one
core available for garbage collection and other system
processes/threads.

For each benchmark, blocks were generated for each combination of the
number of transactions and data conflict percentage.  Each block is run
on the parallel miner, the validator, and a serial miner that runs the
block without parallelization.  The serial results serve as the
baseline for computing speedup.  The running time is collected five
times and the mean and standard deviation are measured.  All runs are
given three warm-up runs per collection.

\begin{figure*}\centering
\begin{tabular}{cc}
\includegraphics[width=3.2in]{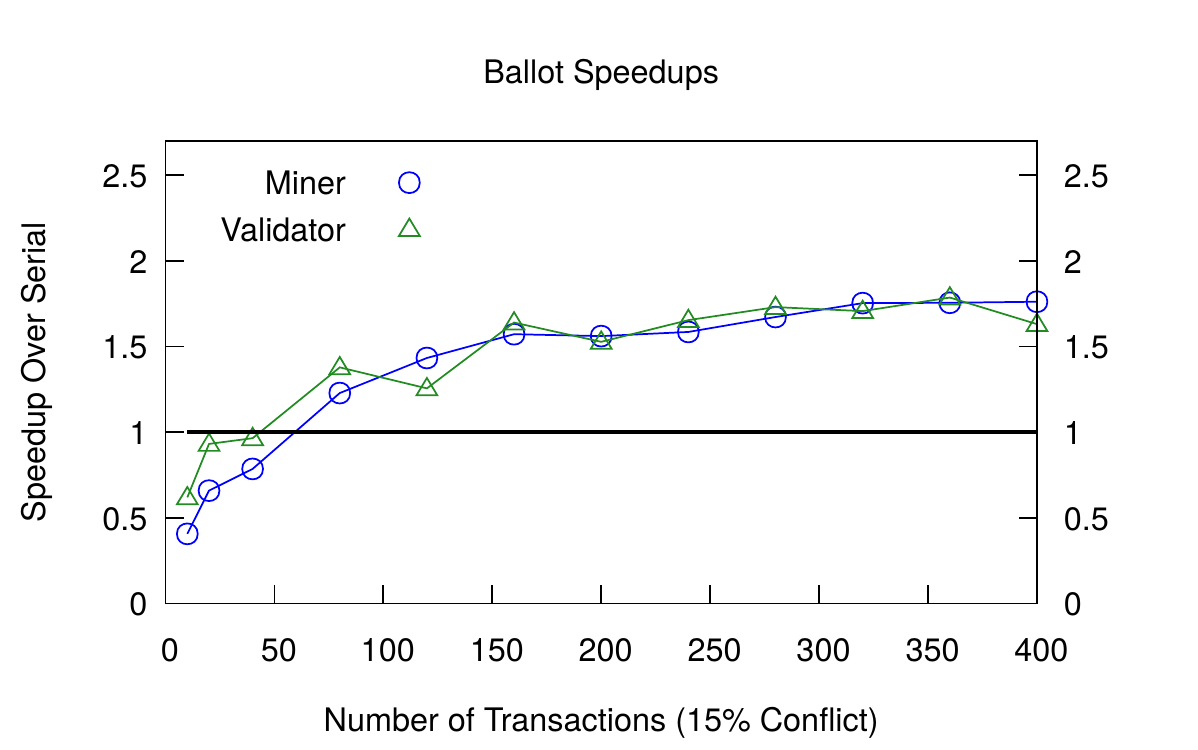} &
\includegraphics[width=3.2in]{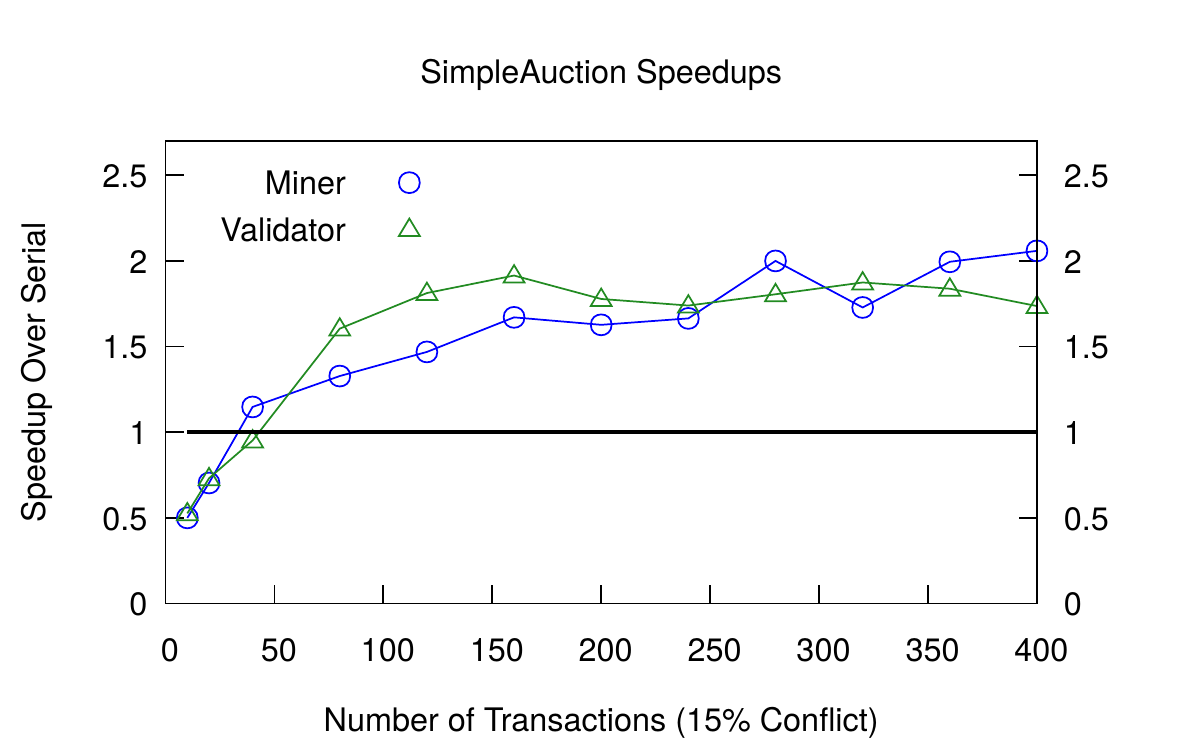} \\
\includegraphics[width=3.2in]{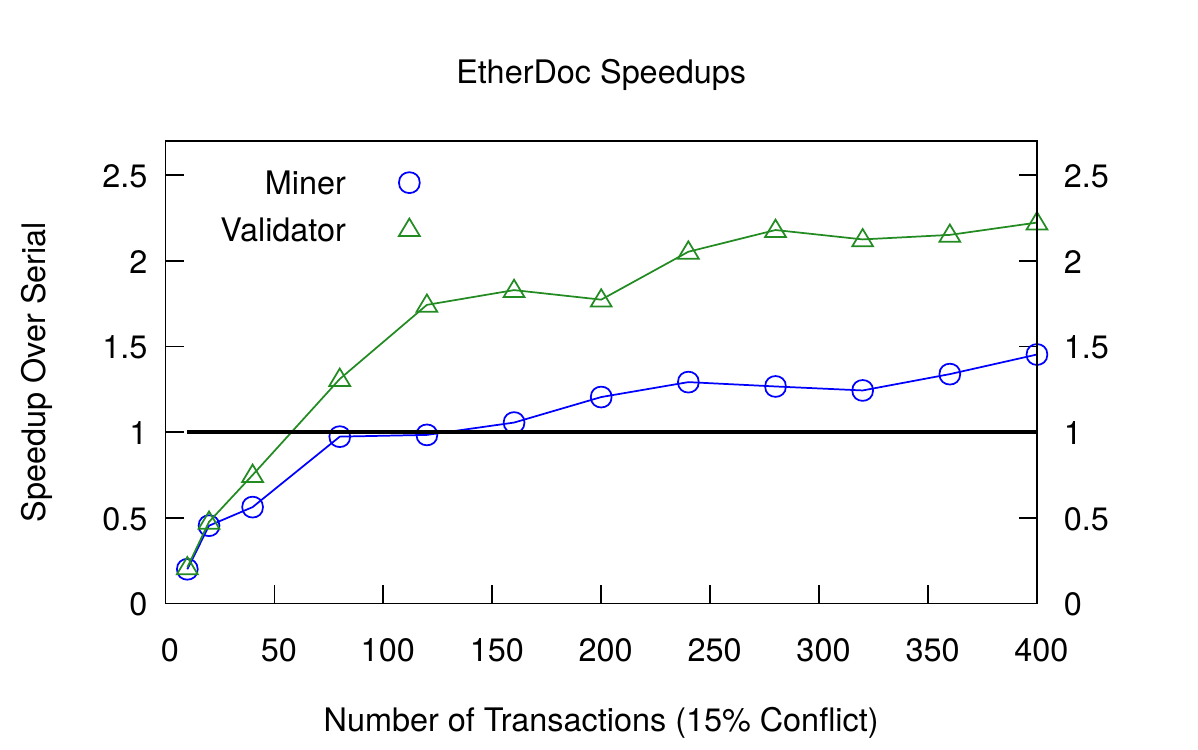} &
\includegraphics[width=3.2in]{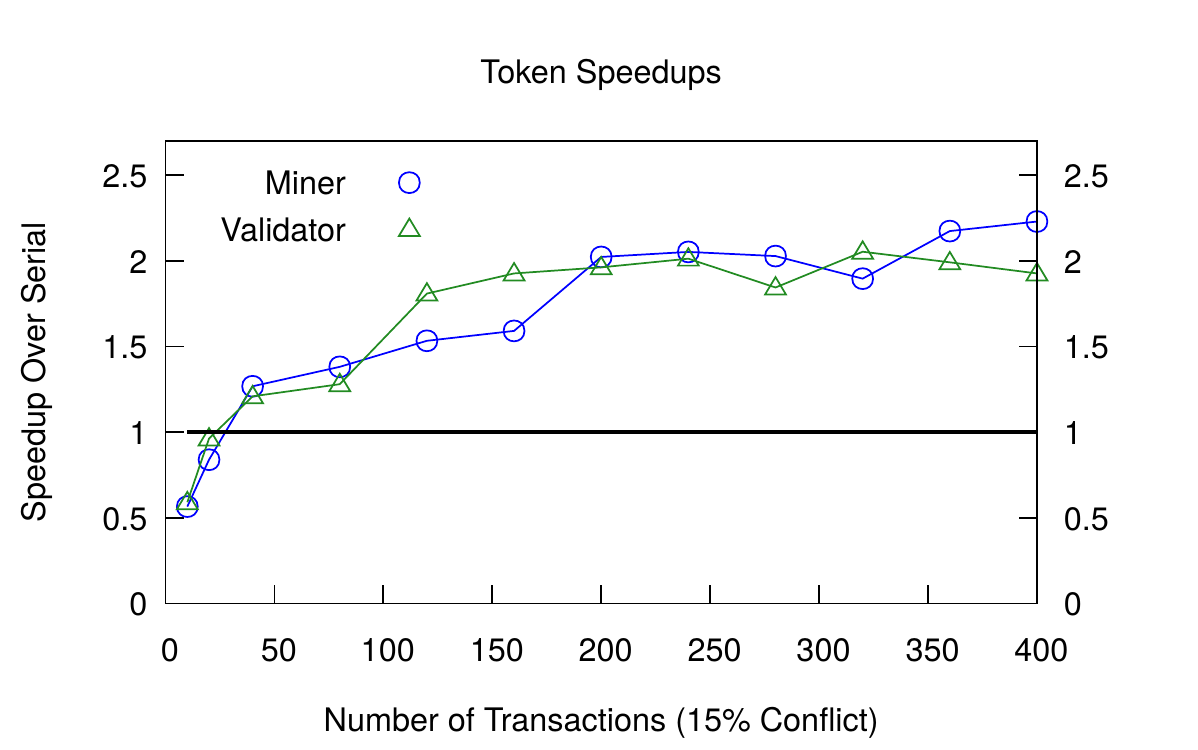} 
\end{tabular}
\caption{\label{fig:speedupsize} The speedup as block size increases of the miner and validator
  versus serial mining for Ballot, SimpleAuction, EtherDoc, and Token
  benchmarks.}
\end{figure*}

\begin{figure*}\centering
\begin{tabular}{cc}
\includegraphics[width=3.2in]{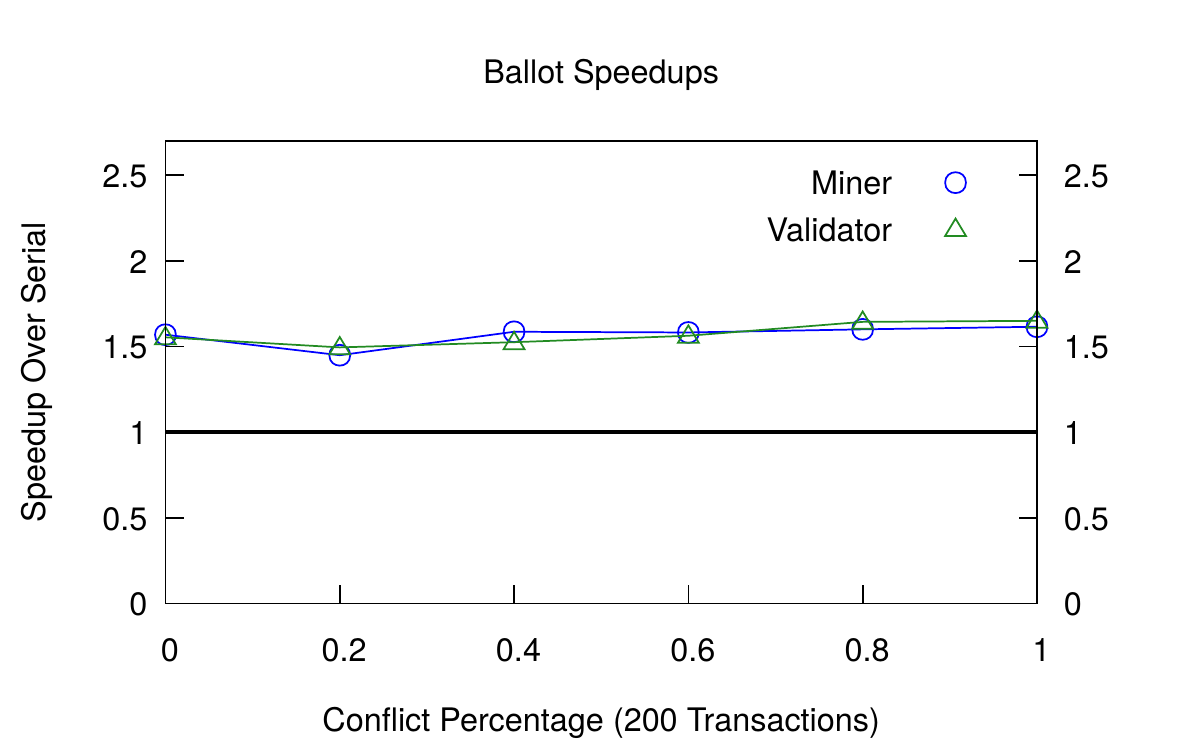} &
\includegraphics[width=3.2in]{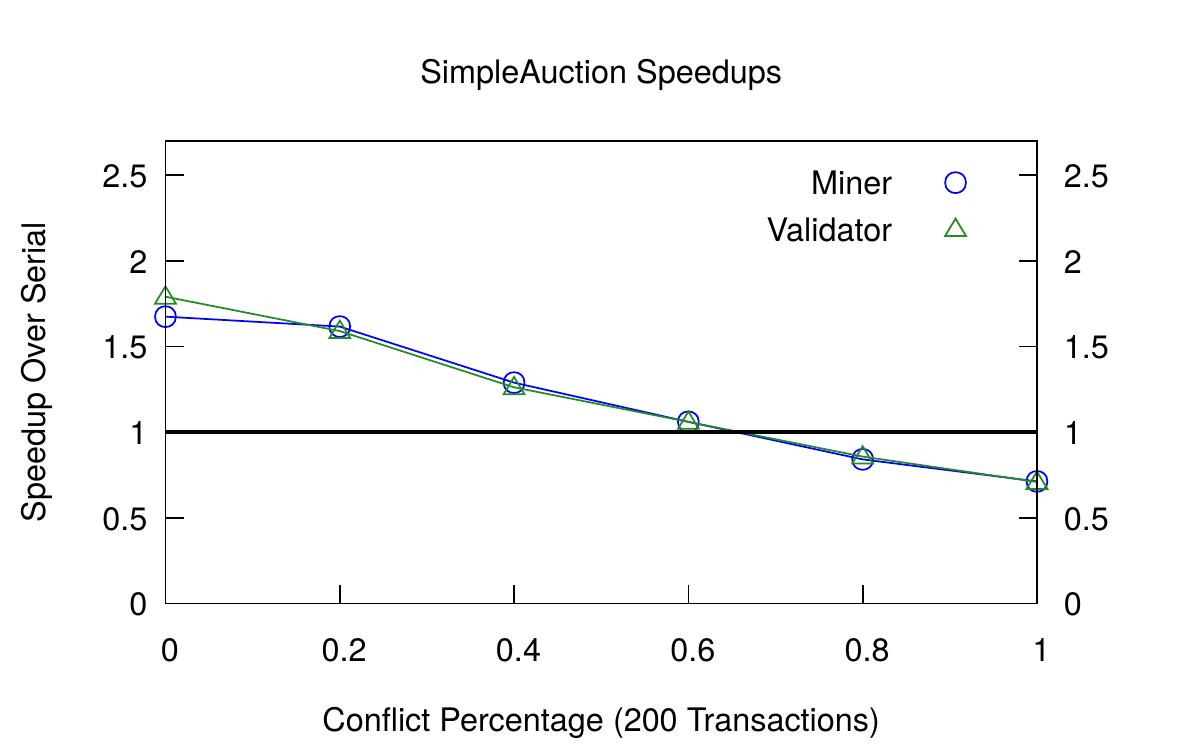}\\
\includegraphics[width=3.2in]{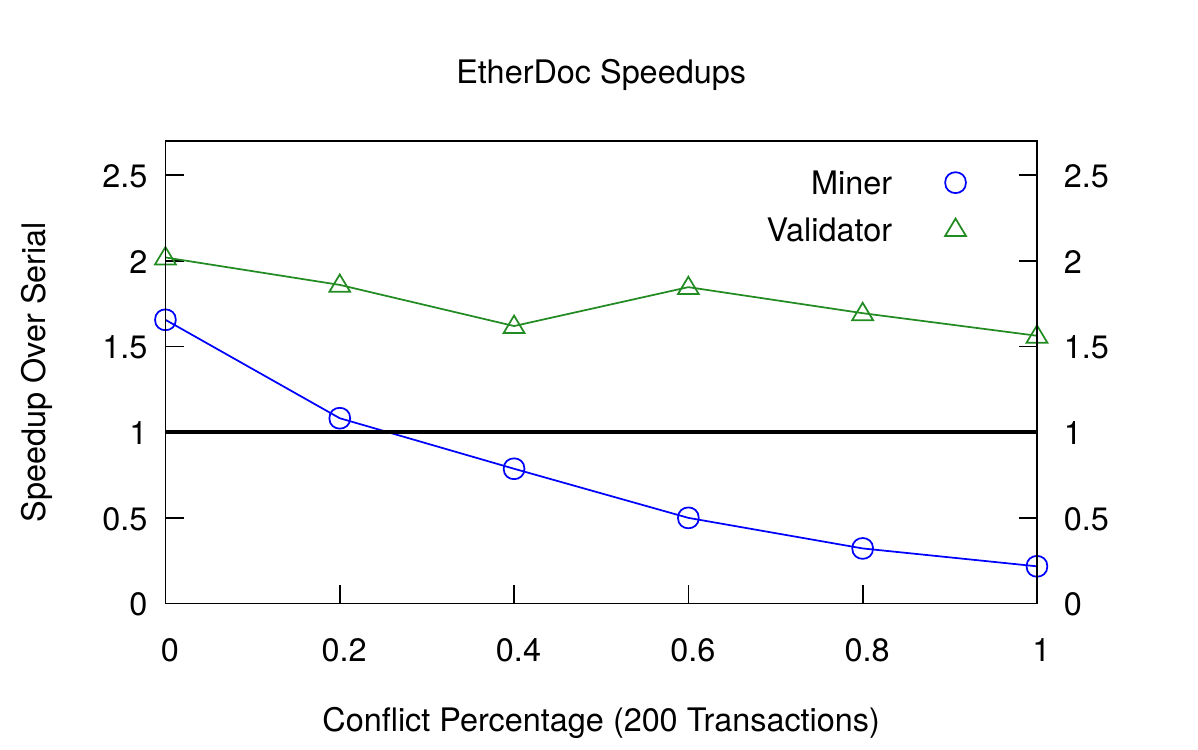}&
\includegraphics[width=3.2in]{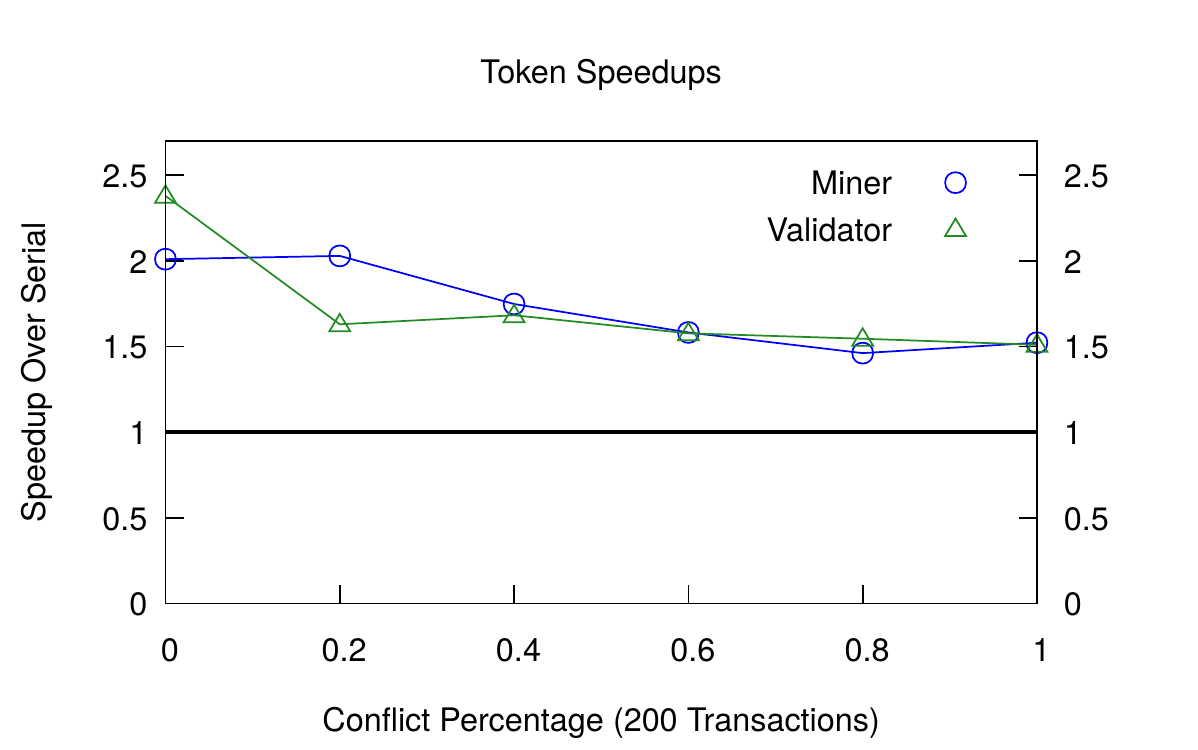}
\end{tabular}
\caption{\label{fig:speedupconflict} The speedup as data conflict increases of the miner and validator
  versus serial mining for Ballot, SimpleAuction, EtherDoc, and Token
  benchmarks.}
\end{figure*}

\begin{figure*}\centering
\begin{tabular}{cc}
\includegraphics[width=3.2in]{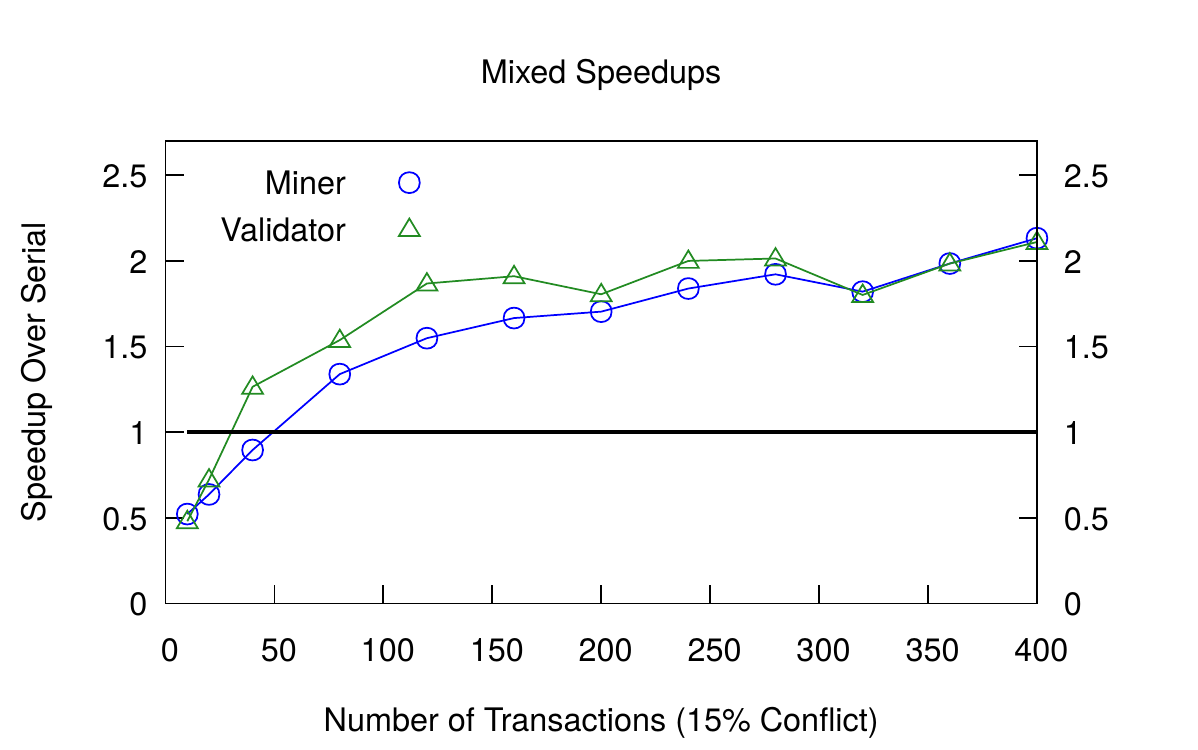} &
\includegraphics[width=3.2in]{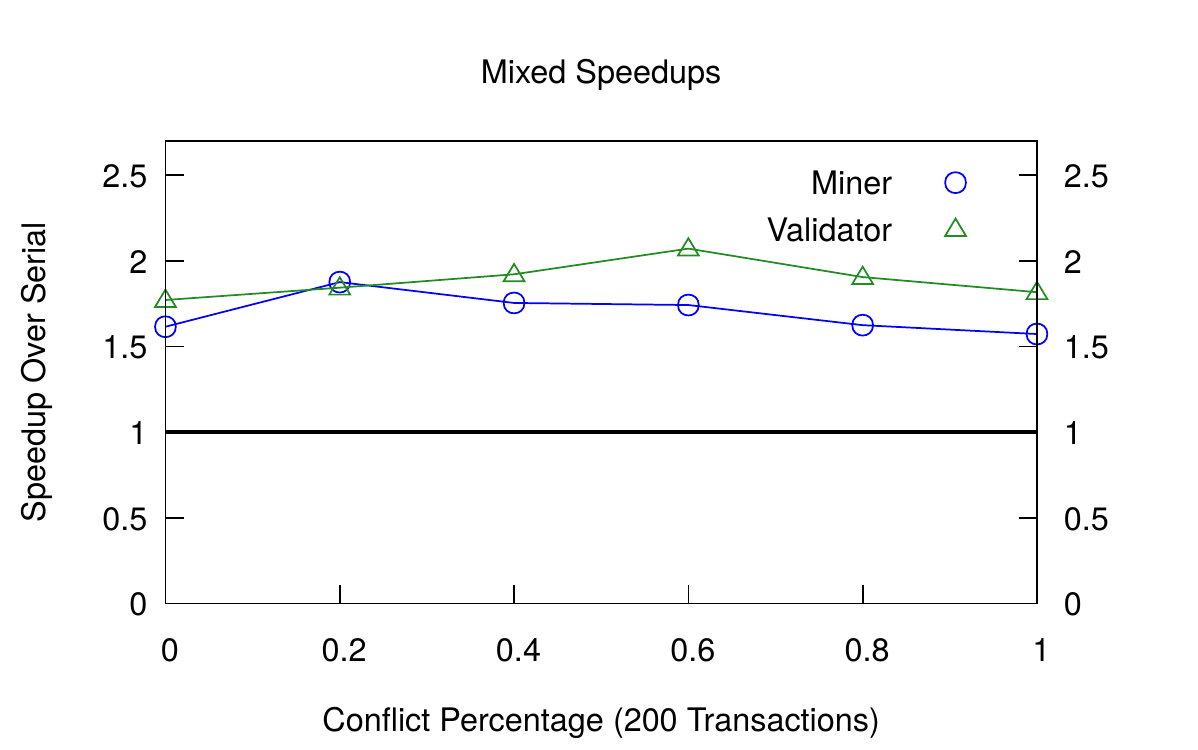}
\end{tabular}
\caption{\label{fig:speedupmixed} The speedup for the Mixed benchmark.
  The first chart is the speedup as block size increases, while the
  second is the speedup as data conflict increases}
\end{figure*}

Figures~\ref{fig:speedupsize} and~\ref{fig:speedupconflict} show the speedup of the parallel miner and
validator relative to the serial miner for all benchmarks.  (The running times
with mean and standard deviation can be found in Figures~\ref{fig:devsize} and~\ref{fig:devconflict}).
Figures~\ref{fig:speedupsize} plots the speedup over the
number of transactions in the block at a fixed data conflict percentage
of 15\%.  The speedup for all benchmarks follows roughly the same
pattern.  For low numbers of transactions, there is no speedup and
even some slowdown.  This is likely due to data conflict as well as the
overhead of multithreading.  For over around 50 transactions, there is
a speedup that increases to about 2x, in line with expectations from a
thread pool of size three.  EtherDoc is an exception, seeing less than
1.5x speedup.  The validator generally has a higher speedup than the
parallel miner.  This is because the parallel miner has done the hard
work of finding data conflict and produced a locking schedule for the
validator to follow.

The charts in Figure~\ref{fig:speedupconflict} plot the speedup as the
data conflict percentage increases for fixed blocks of 200 transactions.
As data conflict increases, the miner's speedup reduces from 2x to close
to serial as many transactions touch shared data.  The validator also
starts at around 2x with no data conflict, but goes down to about 1.5x,
again benefiting from the work of the parallel miner.

Ballot's parallel
mining hovers around 1.5x speedup, suffering little from the extra
data conflict.  Ballot's speedup does not drop to 1x, because the conflicting transactions are not contending for the same data.  Data conflict in SimpleAuction and EtherDoc, however, has an
expectedly higher impact, because each contending transaction touches
the same data.  Figure~\ref{fig:speedupmixed} shows the the Mixed benchmark.  This provides a more realistic view of
a block by combining transactions from unrelated contracts.  Even though EtherDoc
reduces parallelism under high data conflict, when mixed with other transactions, the parallel
miner can still gain a substantial speedup.  The mixed speedup is effectively results in an average of the speedups of all contracts.

\begin{table}

\begin{center}
\begin{tabular}{l|rr|rr}
 & \multicolumn{2}{c|}{\vvv{Miner}} & \multicolumn{2}{c}{\vvv{Validator}} \\
 & Conflict & Size & Conflict & Size \\
\hline
Ballot & \BallotcontentionMiner{}x & \BallotblocksizeMiner{}x & \BallotcontentionValidator{}x & \BallotblocksizeValidator{}x \\
SimpleAuction & \SimpleAuctioncontentionMiner{}x & \SimpleAuctionblocksizeMiner{}x & \SimpleAuctioncontentionValidator{}x & \SimpleAuctionblocksizeValidator{}x \\
EtherDoc & \EtherDoccontentionMiner{}x & \EtherDocblocksizeMiner{}x & \EtherDoccontentionValidator{}x & \EtherDocblocksizeValidator{}x \\
Token & \TokencontentionMiner{}x & \TokenblocksizeMiner{}x & \TokencontentionValidator{}x & \TokenblocksizeValidator{}x \\
Mixed & \MixedcontentionMiner{}x & \MixedblocksizeMiner{}x & \MixedcontentionValidator{}x & \MixedblocksizeValidator{}x \\
\end{tabular}
\end{center}

\caption{The average speedups for each benchmark.}
\label{tab:speedups}
\end{table}

The average of speedups of all benchmarks is \overallMinerSpeedup{}x for the
parallel miner and \overallValidatorSpeedup{}x for the validator.
Table~\ref{tab:speedups} shows the average speedups for each benchmark.

\subsection{Threads}

\begin{figure*}\centering
\begin{tabular}{cc}
\includegraphics[width=3.45in]{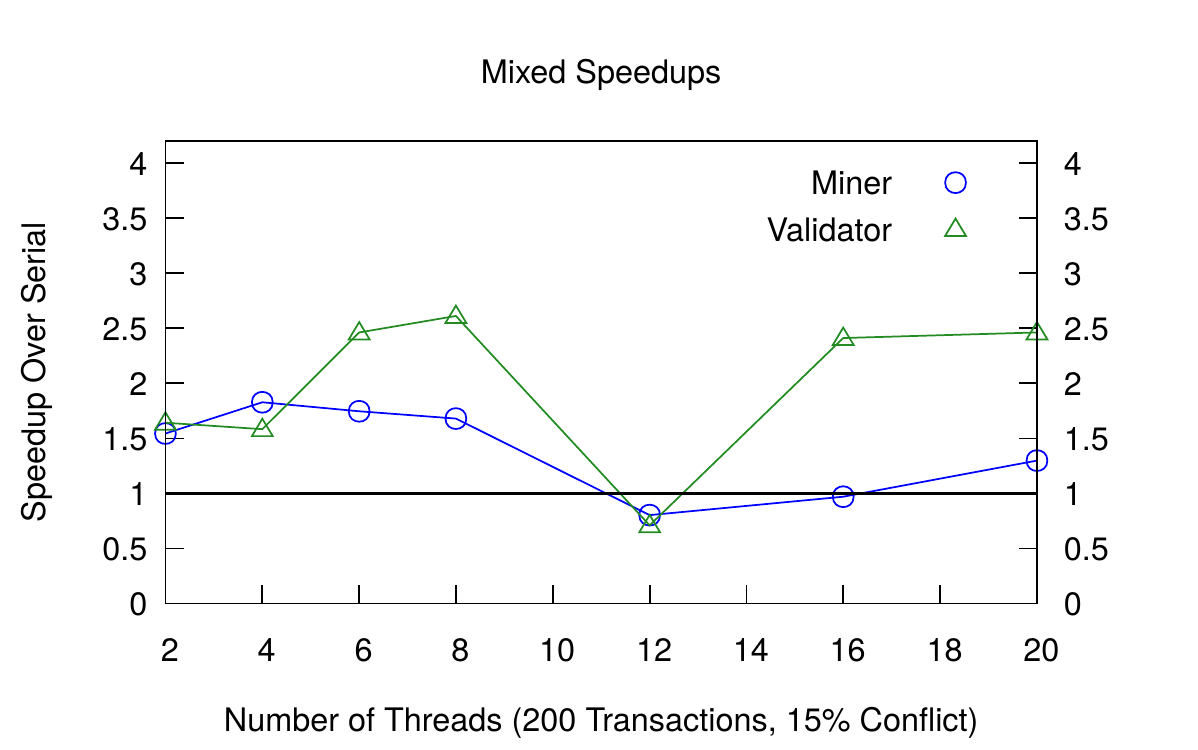}\\
\end{tabular}
\caption{\label{fig:speedupthreads} The speedup as the number of threads increase of the miner and validator
  versus serial mining for the Mixed
  benchmarks.}
\end{figure*}

Figure~\ref{fig:speedupthreads} shows how the speedup changes as the
number of threads increases from two to twenty.  This experiment was
run on a more powerful machine that than the previous section, which
contains 24 processors, each a 6-core Intel(R) Xeon(R) E5645 2.40GHz
processor.  Since this was using a different machine, the results
cannot be compared exactly to those in Figure~\ref{fig:speedupmixed}.
At four threads with 200 transactions and 0.15 conflict, however, the
speedup is comparable, about 1.5x.

As the number of threads grows, the speedup for mining declines, the
opposite of our expectation.  For validation, there is a greater
speedup, reaching a maximum of about 2.5x, compared to
Figure~\ref{fig:speedupmixed}.  This speedup plateaus, except for the
outlier at 12 threads.

The results are erratic.  The experiment with 12 threads had a steep
dropoff in speedup.  These unexpected results may be due to the use of
a multiprocessor machine rather than a multicore machine.  We observed
unpredictable assignment of threads to processors: some maxed out a
single CPU, while others were spread across CPU cores.  While
averaging smoothed this effect somewhat, the nondeterminism was great
enough to cause the outlier at 12 threads.  Since each processor had 6
cores, this appears to explain the dropoff near 6 threads.
Concurrency between processors, rather than cores, apparently incurred
a high enough overhead to not only prevent an increase in speedup, but
reduce the speedup, particularly at 12 threads.

\subsection{Discussion}
These results show that speculative concurrent execution speeds up
mining when threads are occupied and the data conflict rate is not too high.
Data conflicts among transactions in the same block is likely to be
infrequent over the long term.
(Miners could also choose transactions so as to reduce the likelihood of conflict,
say by including only those contracts that operate on disjoint data sets.)
Due to limited hardware,
our experiments used only three concurrent threads,
but even this modest level of concurrency showed a benefit.
Concurrent hardware has proved effective for speeding up solutions to proof-of-work puzzles,
and now similar investments could speed up smart contract execution and validation.

Concurrent smart contract execution speeds up miners by enabling them
to construct blocks faster before appending them to the chain.  But in
permissionless blockchains, the bulk of miner time is spent computing
proof-of-work for the block after construction.  Concurrent smart
contracts, however, still provide a big win for validators.
Validators spend much time executing transactions, while validating
the proof-of-work by miners is fast.  Ethereum is transitioning to
proof-of-stake to reduce the computational burden imposed by
proof-of-work.  Additionally, permissioned blockchains eschew
proof-of-work, so concurrent smart contract execution provides even
more of a boost to throughput to these blockchains.

\begin{figure*}
\centering
\begin{tabular}{cc}
\includegraphics[width=2.8in]{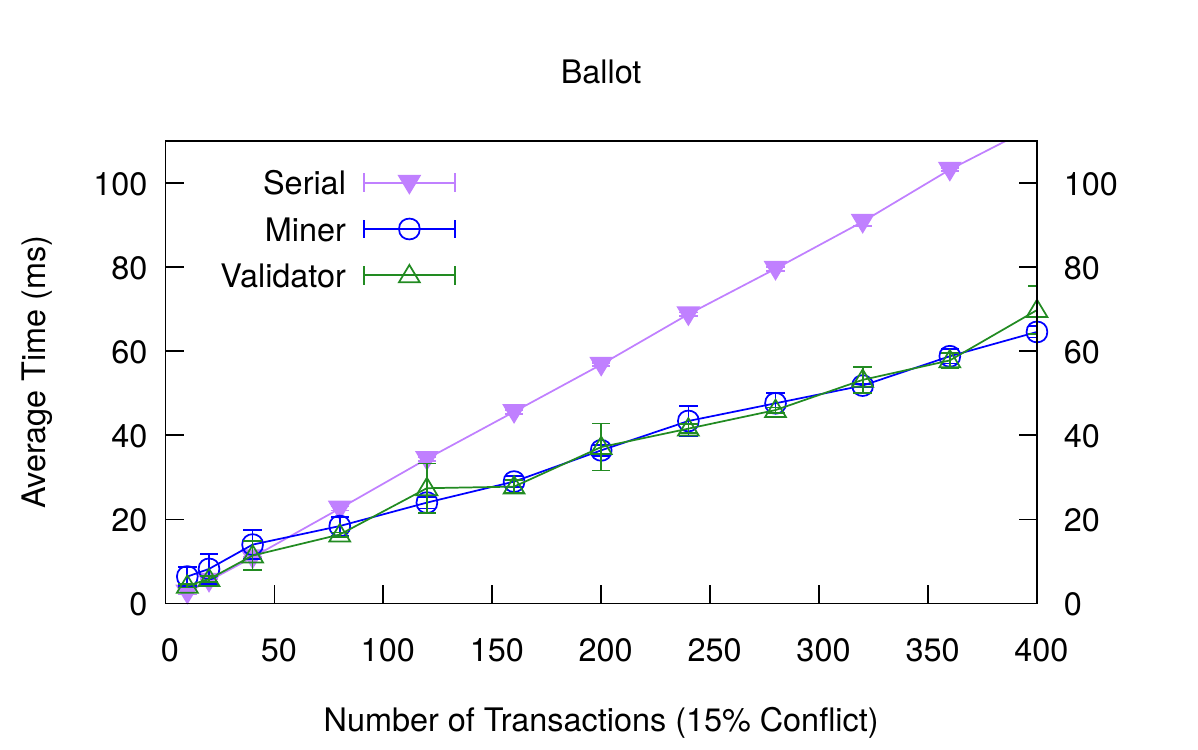}
\includegraphics[width=2.8in]{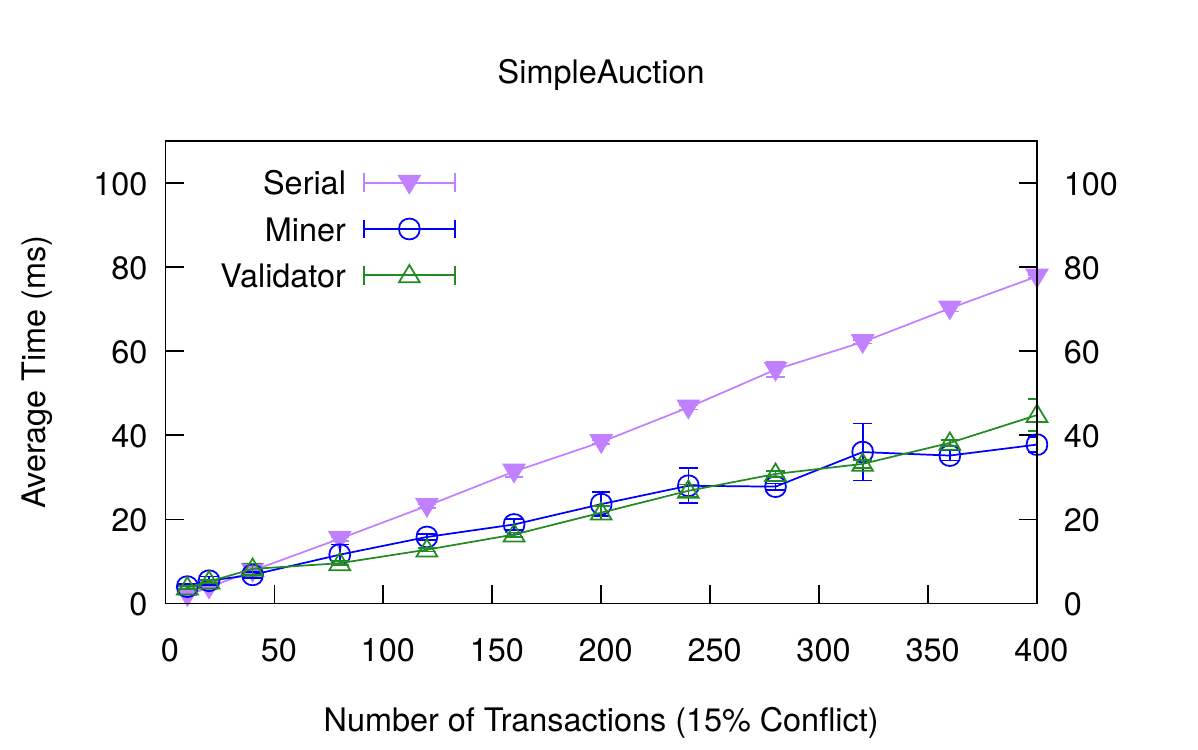} \\
\includegraphics[width=2.8in]{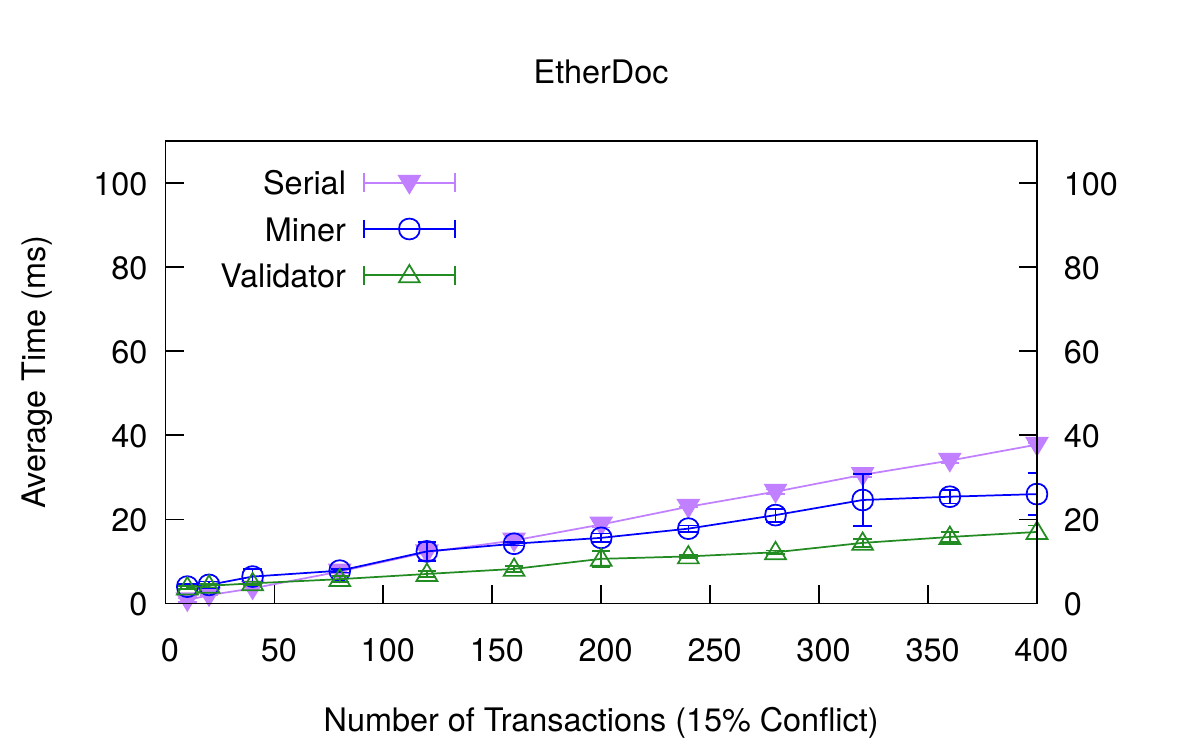} \\
\includegraphics[width=2.8in]{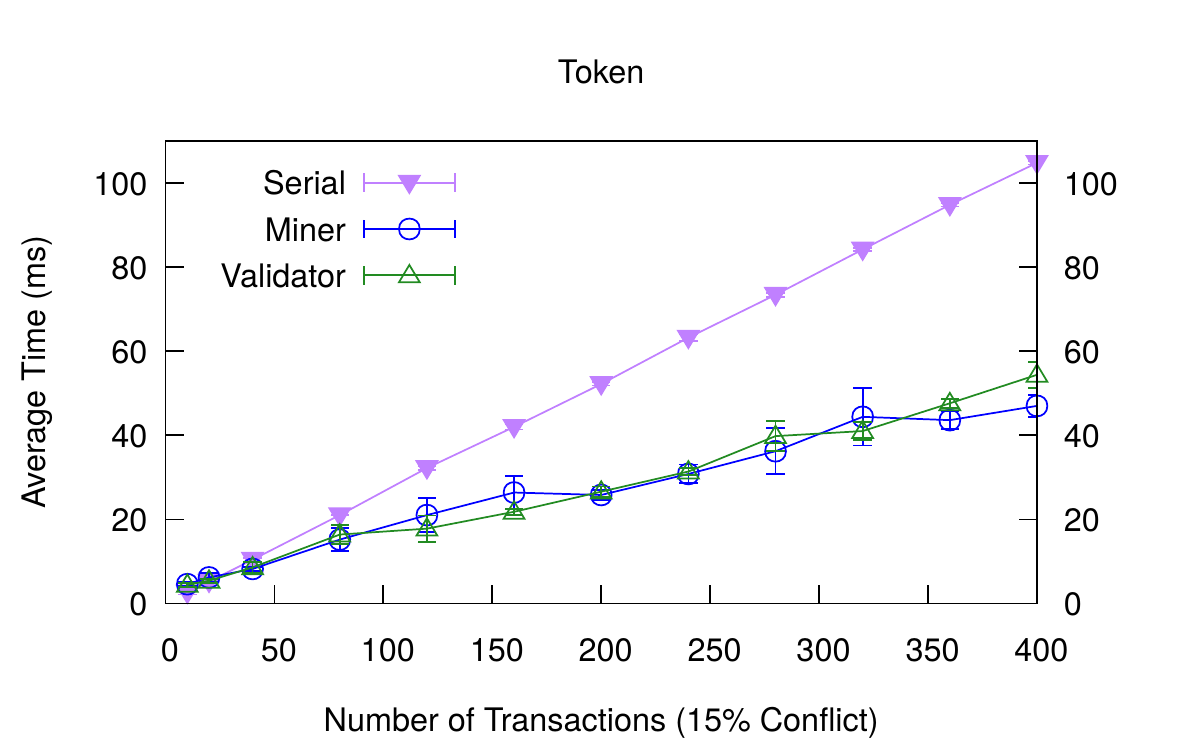}
\includegraphics[width=2.8in]{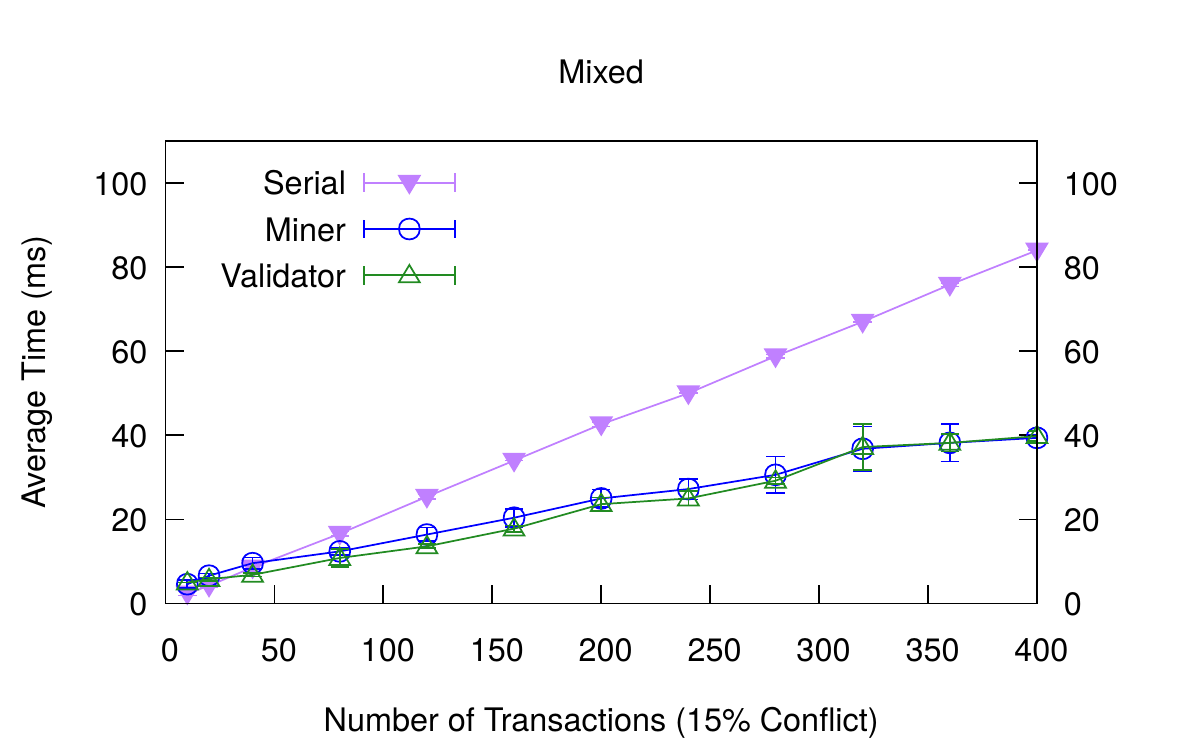}
\end{tabular}
\caption{\label{fig:devsize} Mean and standard deviation of benchmark running times as block size increases.}
\end{figure*}

\begin{figure*}
\centering
\begin{tabular}{cc}
\includegraphics[width=2.8in]{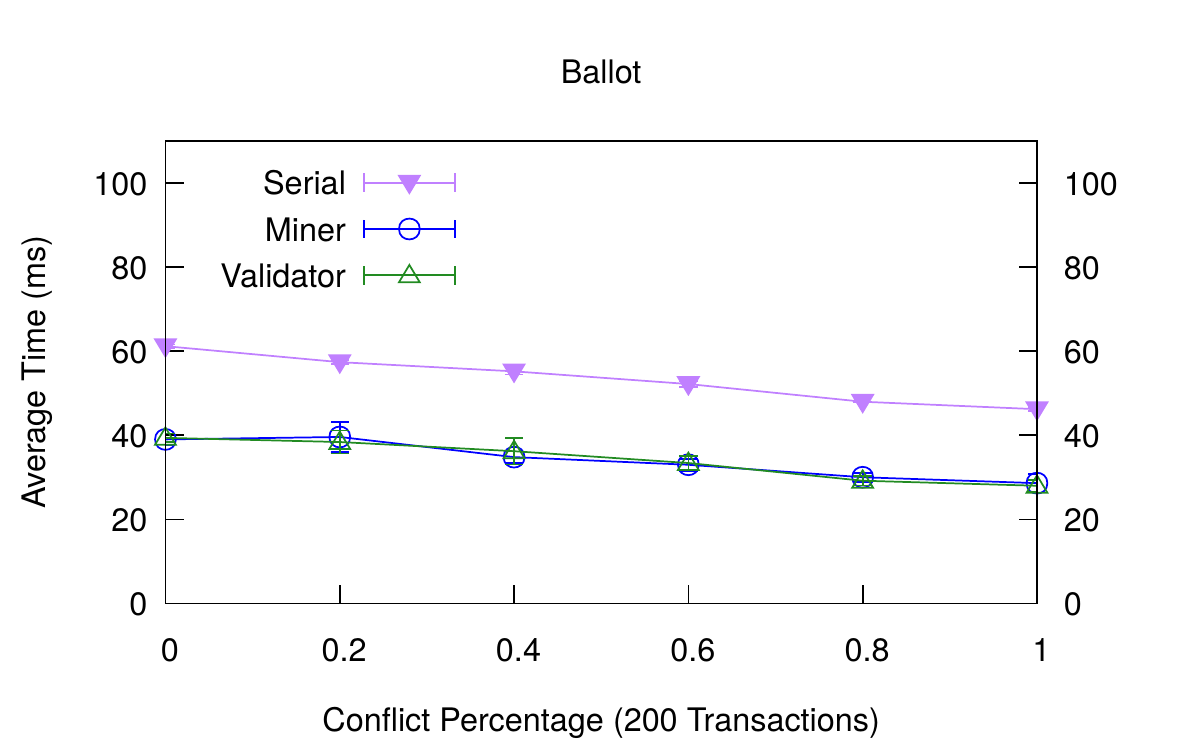} 
\includegraphics[width=2.8in]{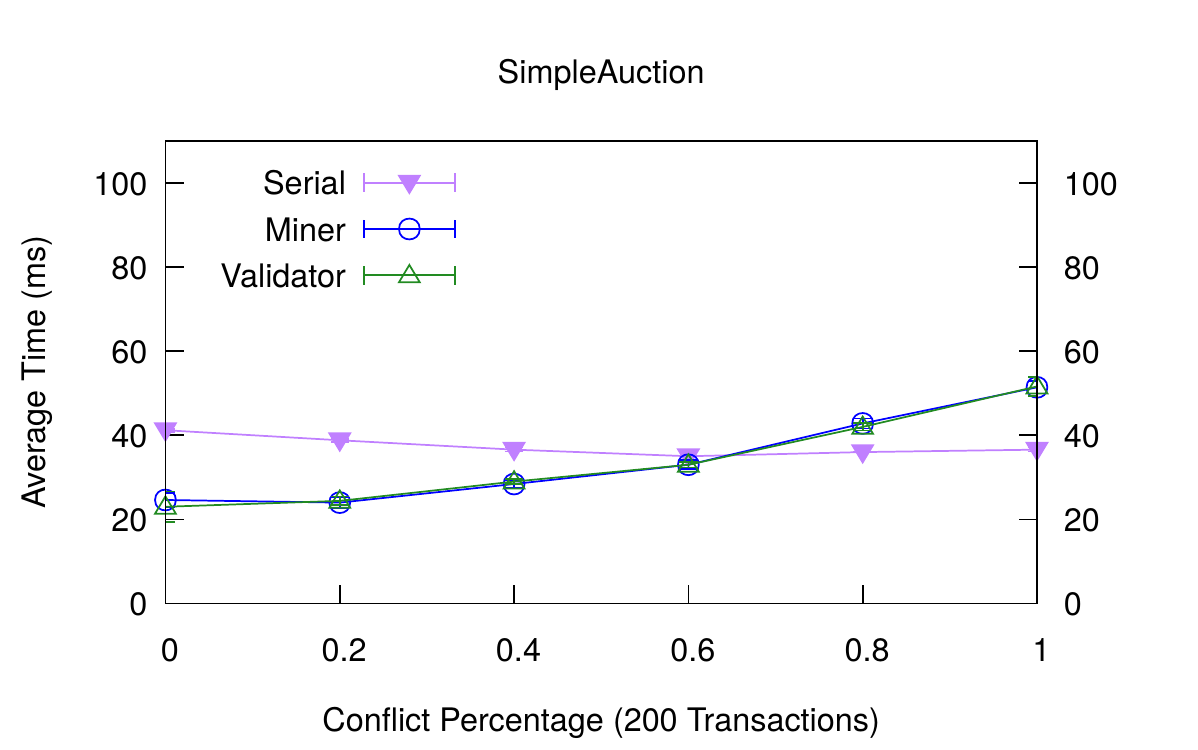} \\
\includegraphics[width=2.8in]{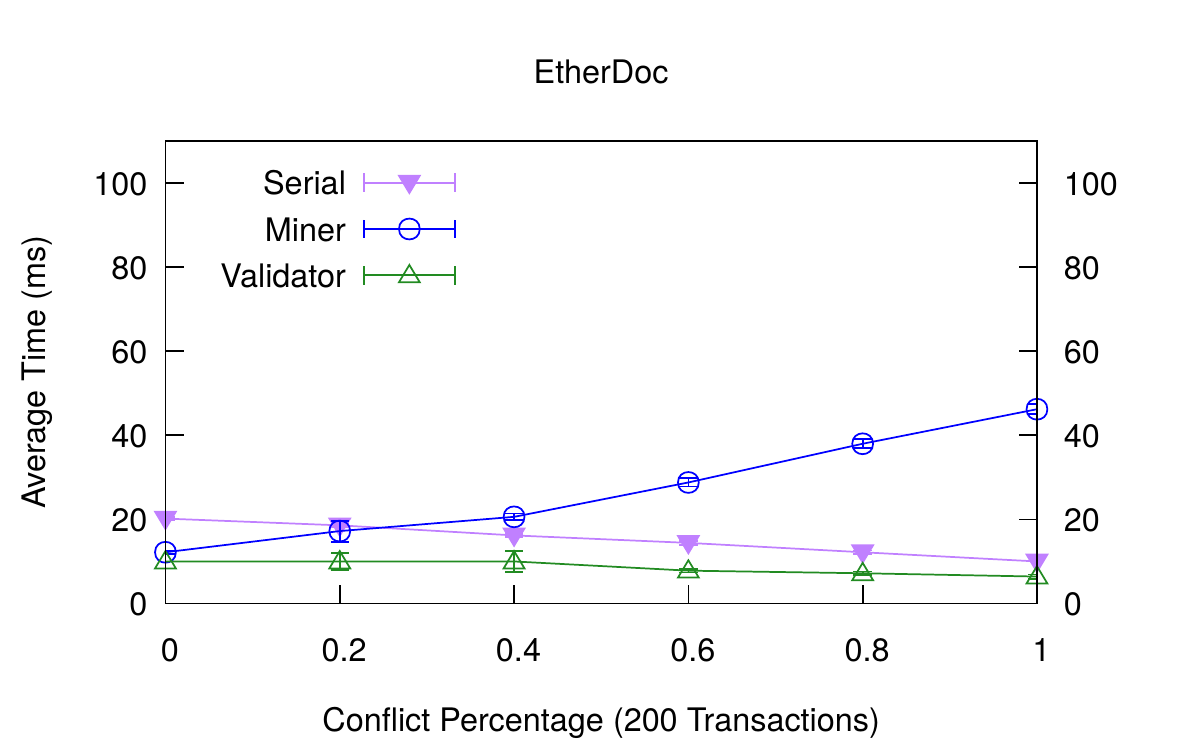} \\
\includegraphics[width=2.8in]{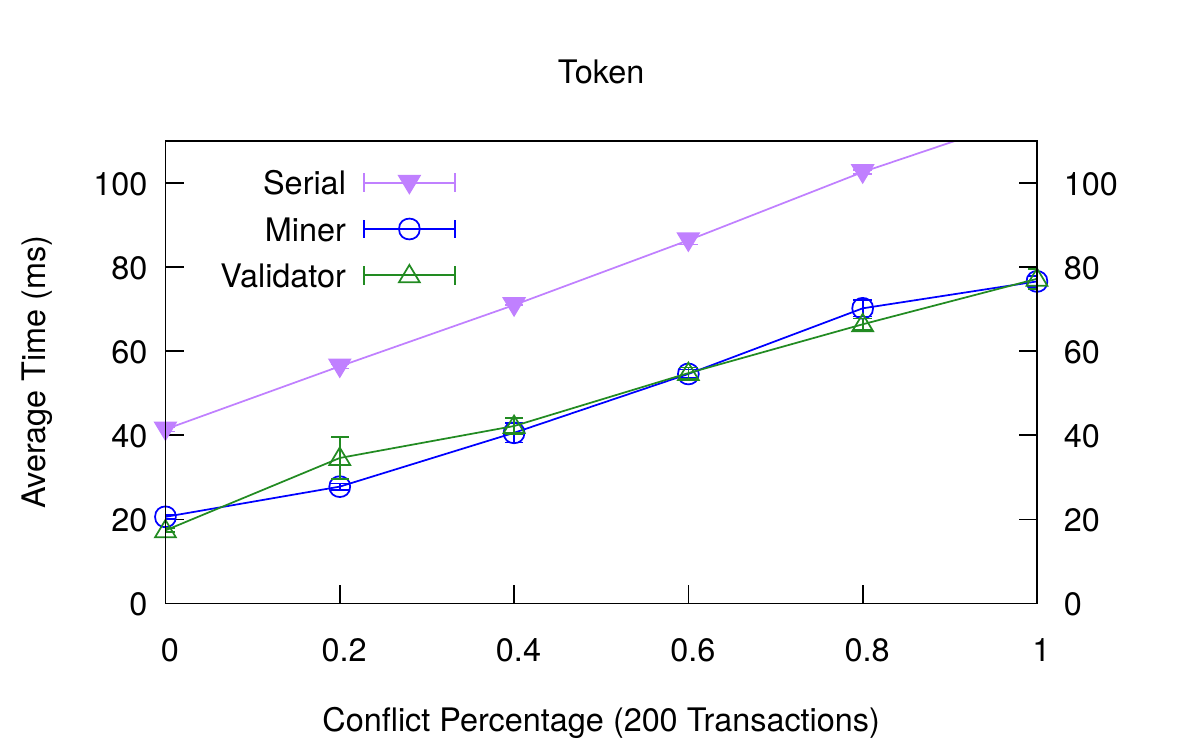}
\includegraphics[width=2.8in]{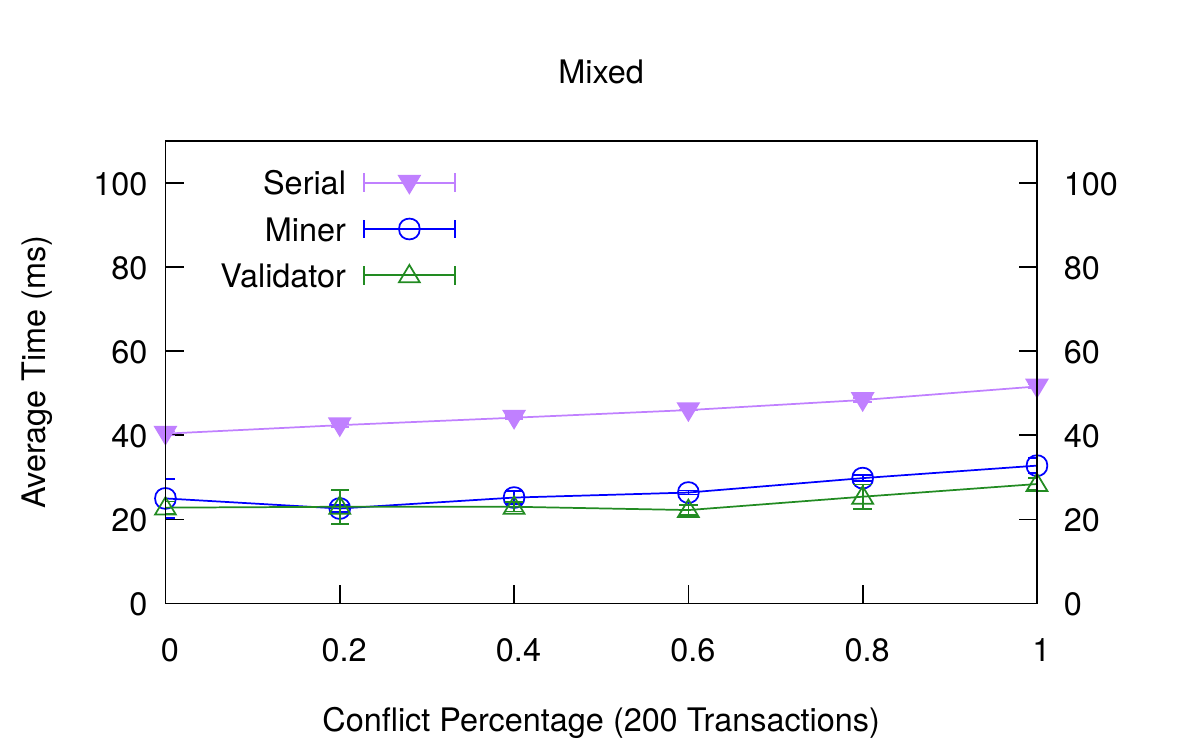}
\end{tabular}
\caption{\label{fig:devconflict} Mean and standard deviation of benchmark running times as data conflict size increases.}
\end{figure*}

\section{Related Work}
The notion of smart contracts can be traced back to an article by Nick
Szabo in 1997~\cite{Szabo1997}.
Bitcoin~\cite{bitcoin} includes a scripting language whose expressive
power was limited to protect against non-terminating scripts.
The Ethereum blockchain~\cite{ethereum} is perhaps the most widely used smart
contract platform,
employing a combination of a Turing-complete virtual machine protected from
non-termination by charging clients for contract running times.
Solidity~\cite{solidity} is the most popular programming language for
programming the Ethereum virtual machine.

Luu \emph{et al.}~\cite{DBLP:conf/ccs/LuuCOSH16} identify a number of
security vulnerabilities and pitfalls in the Ethereum smart contract
model.
Luu \emph{et al.}~\cite{Luu:2015:DIC:2810103.2813659} also identify
perverse incentives that cause rational miners sometimes to accept
unvalidated blocks.
Delmolino \emph{et al.}~\cite{Delmolino2016}
document common programming errors observed in smart contracts.
The Hawk~\cite{Kosba2015HawkTB} smart contract system is designed to protect the privacy of participants.

As noted, many of the speculative mechanisms introduced here were
adapted from \emph{transactional boosting}~\cite{HerlihyK2008},
a technique for transforming thread-safe linearizable objects into
highly-concurrent transactional objects.
Boosting was originally developed to enhance the concurrency provided by
software transactional memory
systems~\cite{Herlihy:2003:STM:872035.872048} by exploiting
type-specific information.
Other techniques that exploit type-specific properties to enhance
concurrency in STMs include
\emph{transactional predication}~\cite{BronsonCCOn2010}
and \emph{software transactional objects}~\cite{HermanIHTKLS2016}.

There are other techniques for deterministically reproducing a prior concurrent execution.
See Bocchino \emph{et al.}~\cite{Bocchino:2009:PPM:1855591.1855595} for a survey.

Cachin et al. discuss non-deterministic execution of smart contracts
in the context of BFT-based permissioned blockchains~\cite{cachin17}.

\section{Conclusion}

We have shown that one can exploit multi-core architectures to
increase smart contract processing throughput for both miners and validators.
First, miners execute a block's contracts speculatively and in parallel,
resulting in lower latency whenever the block's contracts lack data conflicts.
Miners are incentivized to include in each block an encoding of the
serializable parallel schedule that produced that block.
Validators convert that schedule into a deterministic, parallel
fork-join program that allows them to validate the block in parallel.
Even with only three threads,
a prototype implementation yields overall
speedups of \overallMinerSpeedup{}x for miners and
\overallValidatorSpeedup{}x for validators on representative
smart contracts.

Although our discussion has focused on ``permisionless'' systems where anyone can participate,
the mechanisms proposed here would also be useful for ``permissioned'' systems,
such as Hyperledger~\cite{Hyperledger},
where participants are controlled by an authority such as an organization or consortium.
For example,
in a permissioned blockchain based on Practical Byzantine Fault-Tolerance (PBF)~\cite{PBFT},
the leader might use speculative execution to discover a concurrent schedule for a block,
while particpants in the PBFT protocol would use the concurrent
schedule to validate the block before voting.

Future work includes adding support for multithreading to
the Ethereum virtual machine, in much the same way as today's Java virtual machines.
Our proposal for miners only is compatible with current smart contract systems such as Ethereum,
but our overall proposal is not,
because it requires including scheduling metadata in blocks
and incentivizing miners to publish their parallel schedules.
It may well be compatible with a future ``soft fork'' (backward
compatible change), a subject for future research.

In addition to a multithreaded VM, we see room for
advancement in programming language support for smart contracts.
Designing a language that lends itself to
finer-grained concurrency will increase the success of speculative
execution thereby increasing throughput.  It would also be useful for
the language to provide better control of concurrency, helping the
smart contract developer maximize throughput while avoiding
concurrency pitfalls.

	\chapter{\label{chap:conclusion}Concluding Remarks}
	This dissertation advances our knowledge of concurrent data structures along several frontiers.

We began, in Chapter \ref{chap:intro}, with the thesis statement that unifying insights from the machine and $\lambda$ communities would enable compositional reasoning about (concurrent) software which is nonetheless efficiently realizable in hardware, thereby also making more efficient use of the human developers of that software.

Justifying this assertion, in Chapter \ref{chap:handoverhand}, we show that persistent data structures naturally allow transient implementations which nevertheless allow efficient snapshots, thereby allowing developers of client code to choose for themselves when to preserve previous versions of a data structure.
We also show that there is a natural correspondence between points where an immutable data structure must perform fresh allocations and points where a mutable concurrent data structure must perform synchronization.
In Chapters \ref{chap:lockfree} and \ref{chap:returns}, we show that by treating the input/output relationship of a function as a separate concern from when it executes, we can automate the composition of complex, recursive tree-traversals and data structure updates across thread boundaries, with strong progress guarantees, and even diffuse contention by randomizing execution orders.
We further show that careful choice of the type signatures for our transformed tree operations can restrict programmers from even expressing certain classes of concurrency bottlenecks. 
We use these insights to build concurrent data structures which outperform extant implementations on a variety of workloads, despite requiring substantially less source-code to implement.

In Chapter \ref{chap:proust} we show how the efficient snapshots from the earlier chapters can be used in conjunction with a Software Transactional Memory runtime to allow the composition of arbitrary operations on concurrent data structures without having to modify their underlying implementation, while avoiding the structural conflicts inherent when building a data structure directly with STM primitives.
In Chapter \ref{chap:smartcontracts} we used these same techniques to improve the mining and validation times of blocks in an Ethereum-syle blockchain network with smart contracts, when miners and validators have access to multicore hardware.

In addition to these contributions, each of Chapters \ref{chap:lockfree} - \ref{chap:smartcontracts} concludes with discussion of numerous avenues for future work.
	\bibliography{references}

\end{document}